\documentclass[5p]{elsarticle}       
\usepackage[utf8]{inputenc}
\usepackage{amsfonts}
\usepackage{amsmath}
\usepackage{multicol}
\usepackage{amssymb}
\usepackage{graphicx}
\usepackage{mathtools}
\usepackage{xcolor}
\usepackage{amsthm}

\usepackage{algorithm}
\usepackage{algorithmicx}
\usepackage{algpseudocode}

\usepackage{colortbl}
\usepackage{arydshln}
\usepackage{booktabs}
\usepackage{hhline}

 \usepackage{breqn}
\graphicspath{{./}{./Figures/}}
\usepackage{enumitem}
\newcommand{\subscript}[2]{$#1 _ #2$}

\usepackage{units}

\algnewcommand\algorithmicoutput{\textbf{Output:}} 
\algnewcommand\Output{\item[\algorithmicoutput]}
\algnewcommand\algorithmicinput{\textbf{Input:}} 
\algnewcommand\Input{\item[\algorithmicinput]}

\DeclareMathOperator*{\maxflow}{max-flow}

\newcounter{theExample}
\newenvironment{Example}{
\par\smallskip\small\refstepcounter{theExample}%
\noindent\hspace*{-0em}\textbf{Example~\arabic{theExample}}:~%
\leftskip0em\ignorespaces%
}{
\par\smallskip
}
\newcounter{theRemark}
\newenvironment{Remark}{
\par\smallskip\refstepcounter{theRemark}%
\noindent%
\textbf{Remark~\arabic{theRemark}}:~%
\ignorespaces%
}{
\par\smallskip
}

\begin{document}

\title{Graph Signal Processing -- Part I: Graphs, Graph Spectra, and Spectral Clustering}

\author[etf]{\!Ljubi\v{s}a \!Stankovi\'{c}}
\ead{ljubisa@ucg.ac.me}
\author[ic]{\!Danilo \!Mandic}
\ead{d.mandic@imperial.ac.uk}
\author[etf]{\!Milo\v{s} \!Dakovi\'{c}} 
\ead{milosb@ucg.ac.me} 
\author[etf]{\!Milo\v{s} \!Brajovi\'{c}}
\ead{milosb@ucg.ac.me}
\author[ic]{\!Bruno \!Scalzo}
\ead{bruno.scalzo-dees12@imperial.ac.uk}
\author[ic]{\!Anthony G. \!Constantinides} 
\ead{a.constantinides@imperial.ac.uk}

\address[etf]{University of Montenegro, Podgorica, Montenegro }
\address[ic]{Imperial College London, London, United Kingdom }

\date{Received: date / Accepted: date}

\begin{abstract}
The area of Data Analytics on graphs promises a paradigm shift as we  approach information processing of classes of data, which are typically acquired on irregular but structured domains (social networks, various ad-hoc sensor networks). Yet, despite its long history, current approaches mostly focus on the optimization of graphs themselves, rather than on directly inferring learning strategies, such as detection, estimation, statistical and probabilistic  inference, clustering and separation from signals and data acquired on graphs. To fill this void, we first revisit graph topologies from a Data Analytics point of view, and establish a taxonomy of graph networks through a linear algebraic formalism of graph topology (vertices, connections, directivity). This serves as a basis for spectral analysis of graphs, whereby the eigenvalues  and eigenvectors of graph Laplacian and adjacency matrices are shown to convey physical meaning related to both graph topology and higher-order graph properties, such as cuts, walks, paths, and neighborhoods. Through a number of carefully chosen examples, we demonstrate that the isomorphic nature of graphs enables the basic properties and descriptors to be preserved throughout  the data analytics process, even in the case of reordering of graph vertices, where classical approaches fail. Next, to illustrate estimation strategies performed on graph signals, spectral analysis of graphs is introduced through eigenanalysis of mathematical descriptors of graphs and in a generic way. Finally, a framework for vertex clustering and graph segmentation is established based on graph spectral representation (eigenanalysis) which illustrates the power of graphs in various data association tasks. The supporting examples demonstrate the promise of Graph Data Analytics in  modeling structural and functional/semantic inferences. At the same time, Part I serves as a basis for Part II and Part III which deal with theory, methods and applications of processing Data on Graphs and Graph Topology Learning from data.    

\end{abstract}

\maketitle
\tableofcontents
\setcounter{tocdepth}{3}


\section{Introduction}

Graph signal processing is a multidisciplinary research area, the roots of which can be traced back to the 1970s \cite{NC,TC,AC}, but which has witnessed a rapid resurgence. The recent developments, in response to the requirements posed by radically new classes of data sources,  typically embark upon the classic results on graphs as irregular data domains, to address completely new paradigms of \textquotedblleft information on graphs\textquotedblright and \textquotedblleft signals on graphs\textquotedblright. This has resulted in advanced and physically meaningful  solutions in manifold applications \cite{grady2010discrete,ray2012graph,Tutorialicassp2017,krim2015geometric,jordan1998learning}. While the emerging areas of Graph Machine Learning (GML) and Graph Signal Processing (GSP)  do comprise the classic methods of optimization of graphs themselves \cite{bunse1988singular,grebenkov2013geometrical,bapat1996laplacian,o2016eigenvectors,fujiwara1995eigenvalues,maheswari2016some,jordan2004graphical}, significant progress has been made towards redefining basic data analysis paradigms (spectral estimation, probabilistic inference, filtering, dimensionality reduction, clustering, statistical learning), to make them amenable for direct estimation of signals on graphs \cite{mouraaa2018graph,VetterliBook,sandryhaila2013discrete,ekambaram2014graph,sandryhaila2014discrete,sandryhaila2014big,shuman2013emerging,hamon2016extraction,chen2014signal,gavili2017shift, wainwright2008graphical}.  Indeed, this is a necessity in numerous practical scenarios where the signal domain is not designated by equidistant instants in time or a regular grid in a space or a transform domain. Examples include modern Data Analytics for e.g. social network modeling or in smart grid -- data domains which are typically irregular and, in some cases, not even related to the notions of time or space, where ideally, the data sensing domain should also reflect domain-specific properties of the considered system/network; for example, in social or web related networks, the sensing points and their connectivity may be related to specific individuals or topics, and their links, where, processing on irregular domains therefore requires the consideration of data properties other than time or space relationships. In addition, even for the data sensed in well defined time and space domains,  the new contextual and semantic-related relations between the sensing points, introduced through graphs, promise to equip  problem definition with physical relevance, and consequently provide new insights into  analysis and enhanced data processing results. 

In applications where the data domain is conveniently defined by a graph (social networks, power grids, vehicular networks, brain connectivity), the role of classic temporal/spatial sampling points is assumed by graph vertices -- the nodes -- where the data values are observed, while the edges between vertices designate the existence and nature of  vertex
connections (directionality, strength).  In this way, graphs are perfectly well equipped to exploit the fundamental relations among both the measured data and the underlying graph topology; this inherent ability to incorporate  physically relevant data properties has made GSP and GML key technologies in the emerging field of Big Data Analytics (BDA). Indeed, in applications defined on irregular data domains, Graph Data Analytics (GDA) has been proven to offer a quantum step forward from the classical time (or space) series analyses \cite{cvetkovic1980spectra,cvetkovic1985developments,cvetkovic2011selected,brouwer2011spectra,Chung1997,jones2013spectra,mejia2017spectral,stankovic2017LLLvertex,stankovic2019vertex}, including the following aspects

\begin{itemize}
\item
Graph-based data processing approaches
can be applied not only to technological, biological, and social networks, but  also they can lead to both improvements of the existing and even to the creation of radically new methods in classical signal processing and machine learning \cite{lu2014non,dong2012clustering,horaud2009short,hamon2016relabelling,masoumi2017spectral,masoumi2016spectral,stankovic2017vertex,stankovic2018reduced}. 

\item The involvement of graphs makes it possible for the classical sensing domains of time and space (that may be represented as a  linear or circular graph) to be structured in a more advanced way, e.g., by considering the connectivity of sensing points from a signal similarity or sensor association point of view.  
\end{itemize}

The first step in graph data analytics is to decide on the properties of the graph as a new signal/information domain, however, while the data sensing points (graph vertices) may be well defined by the application itself, that is not the case with their connectivity (graph edges), where 
\begin{itemize}
\item In the case of the various computer, social, road, transportation and electrical networks,  the vertex connectivity is often naturally defined, resulting in an exact underlying graph topology. 
\item In many other cases, the data domain definition in a graph form becomes part of the problem definition itself,  as is the case with, e.g., graphs for sensor networks, in finance or smart cities.  In such cases, a vertex connectivity scheme needs to be determined based on the properties of the sensing positions or from the acquired data, as e.g. in the estimation of the temperature field in meteorology \cite{LNDM}.
\end{itemize}

 This additional aspect of the definition of an appropriate graph structure is of crucial importance for a meaningful and efficient application of the GML and GSP approaches.


With that in mind, this tutorial was written in response to the urgent need of multidisciplinary data analytics communities for a seamless and rigorous transition from classical data analytics to the corresponding paradigms which operate directly on irregular graph domains. To this end, we start our approach from a review of basic definitions of graphs and their properties, followed by a physical intuition and step-by-step introduction of graph spectral analysis (eigen-analysis). Particular emphasis is on eigendecomposition of graph matrices,  which serves as a basis for mathematical formalisms in graph signal and information processing. As an example of the ability of GML and GSP to generalize standard methodologies for graphs, we elaborate upon a step-by-step introduction of Graph Discrete Fourier Transform (GDFT), and show that it simplifies into standard Discrete Fourier Transform (DFT) for directed circular graphs; this also exemplifies the generic nature of graph approaches. Finally, spectral vertex analysis and spectral graph segmentation are elucidated as the basis for the understanding of relations among distinct but physically meaningful regions in graphs; this is demonstrated on examples of regional infrastructure modeling, brain connectivity, clustering, and dimensionality reduction.     

\section{Graph Definitions and Properties}

Graph theory has been established  for almost three centuries as a branch in mathematics, and has become a staple methodology in science and engineering areas including chemistry,  operational research, electrical and civil engineering, social networks, and computer sciences. The beginning of graph theory applications in electrical engineering can be traced back to the mid-XIX century with the introduction of Kirchoff's laws.  Fast forward two centuries or so, the analytics of data acquired on graphs has become a rapidly developing research paradigm in Signal Processing and Machine Learning \cite{grady2010discrete,ray2012graph,Tutorialicassp2017,krim2015geometric}.

\subsection{Basic Definitions}

\medskip\noindent\textit{Definition:} A graph $\mathcal{G}=\{\mathcal{V},\mathcal{B}\}$ is defined as a set of vertices, $\mathcal{V}$, which are connected by a set of edges, $\mathcal{B}\subset
\mathcal{V} \times \mathcal{V}$, where the symbol $\times$ denotes a direct product operator. 

\begin{figure}[h]
\centering
\includegraphics[]{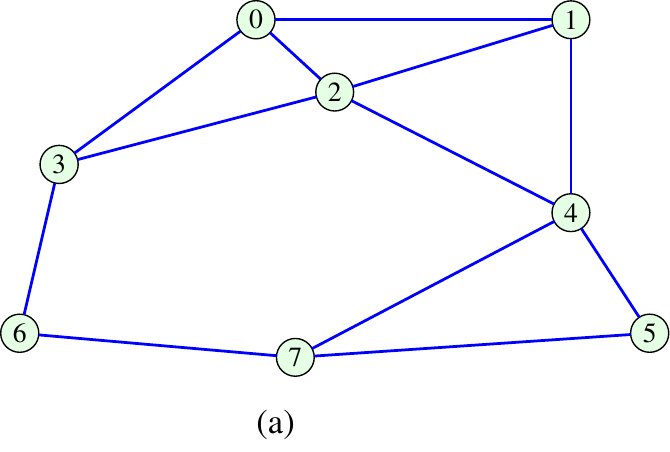}
\hfill
\includegraphics[]{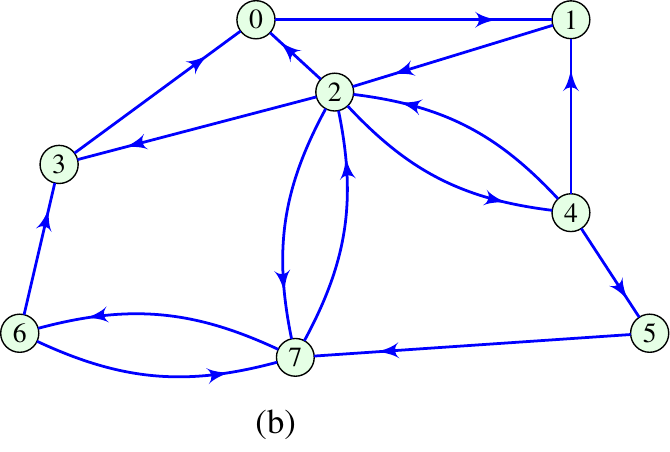}
\caption{Basic graph structures. (a) Undirected graph and (b) Directed graph.}
\label{GSPb_ex1a}
\end{figure}

Examples of graph topologies with $N=8$ vertices, with $$\mathcal{V}=\{0,1,2,3,4,5,6,7\}$$  are  presented in Fig. \ref{GSPb_ex1a}, along with the corresponding edges. The vertices are usually depicted as points (circles) and the edges as lines that connect the vertices. More formally, a line between the vertices $m$ and $n$ indicates the existence of an edge between vertices $m$ and $n$, that is, $(m,n)\in \mathcal{B}$, so that, for example, the graph from Fig. \ref{GSPb_ex1a}(b) can be described as
\begin{gather*}
  \mathcal{V}    =  \{{0,1,2,3,4,5,6,7}\} \\
  \mathcal{B}    \subset  \{{0,1,2,3,4,5,6,7}\} \times  \{{0,1,2,3,4,5,6,7}\}  \\
  \mathcal{B}  =  \{ (0,\!1)\!, \!(1,\!2)\!,\!(2,\!0)\!, \!(2,\!3)\!,\!(2,\!4)\!, \!(2,\!7)\!,\!(3,\!0)\!, \\ \!(4,\!1)\!,\!(4,\!2)\!, \!(4,\!5)\!,\!(5,\!7)\!,\!(6,\!3),\!(6,\!7),\!(7,\!2),\!(7,\!6)\}.
\end{gather*}

Regarding the directionality of vertex connections, a graph can be undirected and directed, as illustrated respectively in Fig. \ref{GSPb_ex1a}(a) and Fig. \ref{GSPb_ex1a}(b). 

\medskip\noindent\textit{Definition:} A graph is undirected if the edge connecting a vertex $m$ to a vertex $n$ also connects the vertex $n$ to the vertex $m$, for all $m$ and $n$. 

In other words, for an undirected graph, if $(n,m)\in \mathcal{B}$ then also $(m,n) \in \mathcal{B}$, as in the case, for example, with edges $(1,2)$ and $(2,1)$ in Fig. \ref{GSPb_ex1a}(a). For directed graphs, in general, this property does not hold, as shown in Fig. \ref{GSPb_ex1a}(b). Observe, for example, that the edge $(2,1)$ does not exist, although the edge  $(1,2)$ connects vertices $1$ and $2$. 
Therefore, undirected graphs can be considered as a special case of directed graphs.  

For a given set of vertices and edges, a graph can be formally represented by its \textit{adjacency matrix}, $\mathbf{A}$, which 
describes the vertex connectivity; for $N$ vertices $\mathbf{A}$ is an
$N\times N$ matrix.  

\medskip\noindent\textit{Definition:} The elements $A_{mn}$ of the adjacency matrix $\mathbf{A}$ assume values $A_{mn} \in \{0,1\}$. The value $A_{mn}=0$ is assigned  if the vertices $m$ and $n$ are not connected with an edge, and $A_{mn}=1$ if these vertices are connected, that is
$$
A_{mn}\, {\overset{def}{=}} \,
\begin{cases}
1, & \text{ if }  (m,n) \in \mathcal{B} \\
0, & \text{ if }  (m,n) \notin \mathcal{B}.
\end{cases}
$$

Therefore, the respective adjacency matrices, $\mathbf{A_{\operatorname{un}}}$ and $\mathbf{A_{\operatorname{dir}}}$, for the undirected and directed graphs from Fig. \ref{GSPb_ex1a}(a) and (b) are given by
\begin{gather}
\!\!\mathbf{A_{\operatorname{un}}}=
\begin{array}{cr}
& \\
{
\color{blue}
\begin{matrix}
\text{\footnotesize 0}\\
\text{\footnotesize 1}\\
\text{\footnotesize 2}\\
\text{\footnotesize 3}\\
\text{\footnotesize 4}\\
\text{\footnotesize 5}\\
\text{\footnotesize 6}\\
\text{\footnotesize 7}\\
\end{matrix}
} &
\begin{bmatrix}
\ 0\  & \ 1\  & \ 1\  & \ 1\  & \ 0\  & \ 0\  & \ 0\  & \ 0\  \\
\ 1\  & \ 0\  & \ 1\  & \ 0\  & \ 1\  & \ 0\  & \ 0\  & \ 0\  \\
\ 1\  & \ 1\  & \ 0\  & \ 1\  & \ 1\  & \ 0\  & \ 0\  & \ 0\  \\
\ 1\  & \ 0\  & \ 1\  & \ 0\  & \ 0\  & \ 0\  & \ 1\  & \ 0\  \\
\ 0\  & \ 1\  & \ 1\  & \ 0\  & \ 0\  & \ 1\  & \ 0\  & \ 1\  \\
\ 0\  & \ 0\  & \ 0\  & \ 0\  & \ 1\  & \ 0\  & \ 0\  & \ 1\  \\
\ 0\  & \ 0\  & \ 0\  & \ 1\  & \ 0\  & \ 0\  & \ 0\  & \ 1\  \\
\ 0\  & \ 0\  & \ 0\  & \ 0\  & \ 1\  & \ 1\  & \ 1\  & \ 0\ 
\end{bmatrix} \\
& 
{
\color{blue}
 \begin{matrix}
 \text{\footnotesize 0}\ &
\ \text{\footnotesize 1\ }\  &
\ \text{\footnotesize 2}\  &
\ \text{\footnotesize 3}\  &
\ \text{\footnotesize 4}\  &
\ \text{\footnotesize 5\ }\  &
\ \text{\footnotesize 6}\  &
\ \text{\footnotesize 7}  &
\end{matrix}
}
\end{array}\!\!, \label{matA1a}
\end{gather}
\begin{gather}
\hspace{-1.8mm}   \mathbf{A}_{\operatorname{dir}}  = \!
 \begin{array}{cr}
 & \\
 {
 	\color{blue}
 	\begin{matrix}
 	\text{\footnotesize 0}\\
 	\text{\footnotesize 1}\\
 	\text{\footnotesize 2}\\
 	\text{\footnotesize 3}\\
 	\text{\footnotesize 4}\\
 	\text{\footnotesize 5}\\
 	\text{\footnotesize 6}\\
 	\text{\footnotesize 7}\\
 	\end{matrix}
 } &  \!
\begin{bmatrix}
\ 0\  & \ 1\  & \ 0\  & \ 0\  & \ 0\  & \ 0\  & \ 0\  & \ 0\  \\
\ 0\  & \ 0\  & \ 1\  & \ 0\  & \ 0\  & \ 0\  & \ 0\  & \ 0\  \\
\ 1\  & \ 0\  & \ 0\  & \ 1\  & \ 1\  & \ 0\  & \ 0\  & \ 1\  \\
\ 1\  & \ 0\  & \ 0\  & \ 0\  & \ 0\  & \ 0\  & \ 0\  & \ 0\  \\
\ 0\  & \ 1\  & \ 1\  & \ 0\  & \ 0\  & \ 1\  & \ 0\  & \ 0\  \\
\ 0\  & \ 0\  & \ 0\  & \ 0\  & \ 0\  & \ 0\  & \ 0\  & \ 1\  \\
\ 0\  & \ 0\  & \ 0\  & \ 1\  & \ 0\  & \ 0\  & \ 0\  & \ 1\  \\
\ 0\  & \ 0\  & \ 1\  & \ 0\  & \ 0\  & \ 0\  & \ 1\  & \ 0\ 
\end{bmatrix}
\end{array}\!\!\!. \label{AdjMtxFirs}
\end{gather} 

 Adjacency matrices not only fully reflect the structure arising from the topology of data acquisition,  but also they admit the usual feature analysis through linear algebra, and can be sparse, or exhibit some other interesting and useful matrix properties. 
 
 \begin{Remark} The adjacency matrix of an undirected graph is symmetric, that is,  $$\mathbf{A}=\mathbf{A}^T.$$
\end{Remark}

Since a graph is fully determined by its  adjacency matrix, defined over a given set  of vertices, any change in vertex  ordering will cause the corresponding changes in the adjacency matrix. 

 \begin{Remark} Observe that a vertex indexing scheme does not change the graph itself (graphs are isomorphic domains), so that the relation between adjacency matrices of the original and renumerated graphs, $\mathbf{A}_1$ and $\mathbf{A}_2$ respectively, is straightforwardly defined using an appropriate permutation matrix, $\mathbf{P}$, in the form
\begin{equation}
\mathbf{A}_2=\mathbf{P}\,\mathbf{A}_1\mathbf{P}^T.
\label{PermMat}
\end{equation}
Recall that a permutation matrix has exactly one nonzero element equal to unity, in each row and in each column.
 \end{Remark}
In general, the edges can also convey information about a relative importance of their connection, through a weighted graph. 

 \begin{Remark} The set of weights, $\mathcal{W}$, corresponds morphologically to the set of edges, $\mathcal{B}$, so that a weighted graph is a generic extension of an unweighted graph. It is commonly assumed that edge weights are nonnegative real numbers; therefore, if weight $0$ is associated with a nonexisting edge, then the graph can be described by a weight matrix,
$\mathbf{W}$, similar to the description by the adjacency matrix $\mathbf{A}$. 
 \end{Remark}
\medskip\noindent\textit{Definition:} A nonzero element in the weight matrix
$\mathbf{W}$,
$W_{mn}\in\mathcal{W}$, designates both an edge between the vertices $m$ and $n$ and the corresponding weight.  The value $W_{mn}=0$ indicates no edge connecting the vertices $m$ and $n$. \textit{The elements of a weight matrix are nonnegative real numbers.} 

\begin{figure}[ptb]
	\centering
	\includegraphics[]{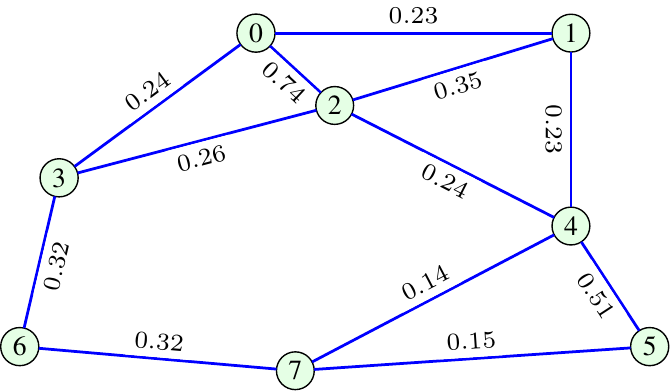}
	\caption{Example of a weighted graph.}
	\label{GSPb_ex2}
\end{figure}

Fig. \ref{GSPb_ex2} shows an example of a weighted undirected graph, with the corresponding weight matrix given by
\begin{equation}
\small
\setlength{\arraycolsep}{5pt}
\mathbf{W}= \!\!
\begin{array}{cr}
& \\
{
	\color{blue}
	\begin{matrix}
	\text{\footnotesize 0}\\
	\text{\footnotesize 1}\\
	\text{\footnotesize 2}\\
	\text{\footnotesize 3}\\
	\text{\footnotesize 4}\\
	\text{\footnotesize 5}\\
	\text{\footnotesize 6}\\
	\text{\footnotesize 7}\\
	\end{matrix}
} &  \!\! \!\!
\begin{bmatrix*}[r]
  0  &  0.23 & 0.74  &  0.24&   0   &   0   &   0   & 0 \\
0.23 &   0   & 0.35  & 0 	& 0.23  &   0   &   0   & 0  \\
0.74 & 0.35  &   0   & 0.26 &   0.24&   0   &   0   &   0   \\
0.24 & 0 	 & 0.26  &   0  & 0 	&   0   &  0.32 & 0  \\
  0  &  0.23 & 0.24  & 0 	&  0    & 0.51  & 0     & 0.14 \\
  0  &    0  &   0   & 0 	& 0.51  &   0   &   0   &   0.15   \\
  0  &    0  &   0   & 0.32 & 0     & 0     &   0   &   0.32   \\
0    & 0 	 &  0    & 0    & 0.14	& 0.15 	&  0.32 &   0   \\
\end{bmatrix*}.
 \\
& 
{
	\color{blue}
	\begin{matrix*}
	\text{\footnotesize 0 \hspace{3.3mm}} &
	\text{\footnotesize 1 \hspace{3.3mm}}  &
	\text{\footnotesize 2 \hspace{3.3mm}} &
	\text{\footnotesize 3 \hspace{3.3mm}} &
	\text{\footnotesize 4 \hspace{3.3mm}} &
	\text{\footnotesize 5 \hspace{3.3mm}} &
	\text{\footnotesize 6 \hspace{3.3mm}} &
	\text{\footnotesize 7 \hspace{2.5mm}}
	\end{matrix*}
}
\end{array}\label{WeightMatr}
\end{equation}  
 
In this sense, the adjacency matrix $\mathbf{A}$ can be considered as a special
case of the weight matrix $\mathbf{W}$, whereby all nonzero weights are equal to unity.  
It then follows that the weighting matrix of undirected graphs is also  symmetric,
\begin{equation}
 \mathbf{W}=\mathbf{W}^T, \label{simW}
\end{equation}
while, in general, for directed graphs this property does not hold. 

\medskip\noindent\textit{Definition:} A degree matrix, $\mathbf{D}$, for an undirected graph is a diagonal matrix with elements,
$D_{mm}$, which are equal to the sum of weights of all edges connected to the vertex $m$, that is, the sum of elements in its $m$-th row
$$
D_{mm}\, {\overset{def}{=}} \,\sum_{n=0}^{N-1} W_{mn}.
$$

 \begin{Remark} For an unweighted and undirected graph, the value of the element $D_{mm}$ is equal to the number of edges connected to the $m$-th vertex.
 \end{Remark}
\noindent\textbf{Vertex degree centrality.}  The degree centrality of a vertex is defined as the number of vertices connected to the considered vertex with a single edge, and in this way it models the importance of a given vertex. For undirected and unweighted graphs, the vertex degree centrality of a vertex is equal to the element, $D_{mm}$, of the degree matrix.   

\begin{Example} For the undirected weighted graph from Fig. \ref{GSPb_ex2}, the degree matrix is given by
\begin{equation}
\small
\setlength{\arraycolsep}{2.5pt}
\mathbf{D}=
\begin{array}{cr}
& \\
{
	\color{blue}
	\begin{matrix}
	\text{\footnotesize 0}\\
	\text{\footnotesize 1}\\
	\text{\footnotesize 2}\\
	\text{\footnotesize 3}\\
	\text{\footnotesize 4}\\
	\text{\footnotesize 5}\\
	\text{\footnotesize 6}\\
	\text{\footnotesize 7}\\
	\end{matrix}
} &
\begin{bmatrix}
1.21 & 0 & 0 & 0 & 0 & 0 & 0 & 0 \\
0 & 0.81 & 0 & 0 & 0 & 0 & 0 & 0 \\
0 & 0 & 1.59 & 0 & 0 & 0 & 0 & 0 \\
0 & 0 & 0 & 0.82 & 0 & 0 & 0 & 0 \\
0 & 0 & 0 & 0 & 1.12 & 0 & 0 & 0 \\
0 & 0 & 0 & 0 & 0 & 0.66 & 0 & 0 \\
0 & 0 & 0 & 0 & 0 & 0 & 0.64 & 0 \\
0 & 0 & 0 & 0 & 0 & 0 & 0 & 0.61 \\
\end{bmatrix}. \\
& 
{
	\color{blue}
	\begin{matrix}
     \text{\footnotesize 0 \hspace{3.5mm}} &
	 \text{\footnotesize 1 \hspace{3.5mm}}  &
 	 \text{\footnotesize 2 \hspace{3.5mm}} &
 	 \text{\footnotesize 3 \hspace{3.5mm}} &
	 \text{\footnotesize 4 \hspace{3.5mm}} &
	 \text{\footnotesize 5 \hspace{3.5mm}} &
	 \text{\footnotesize 6 \hspace{3.5mm}} &
	 \text{\footnotesize 7 \hspace{3.5mm}}
	\end{matrix}
}
\end{array}\!\! \label{DFTegMatrix}
\end{equation} 
\end{Example}

Another important descriptor of graph connectivity is the graph Laplacian matrix, $\mathbf{L}$, which combines the weight matrix and the degree matrix. 

\medskip\noindent\textit{Definition:} The Laplacian matrix is defined as
\begin{equation} 
 \mathbf{L}\, {\overset{def}{=}} \,\mathbf{D}-\mathbf{W}, \label{LapDef}
 \end{equation} 
where $\mathbf{W}$ is the weighting matrix and $\mathbf{D}$ the diagonal degree matrix with elements $D_{mm}=\sum_n W_{mn}$. The elements of a Laplacian matrix are nonnegative real numbers at the diagonal positions, and nonpositive real numbers at the off-diagonal positions.  

For an undirected graph, the Laplacian matrix is symmetric, $\mathbf{L}=\mathbf{L}^T$, for example, the graph Laplacian for the weighted graph from Fig. \ref{GSPb_ex2} is given by
\begin{equation}
\small
\setlength{\arraycolsep}{2.5pt}
\mathbf{L}=
\begin{bmatrix*}[r]
 1.21 & -0.23 & -0.74 & -0.24 & 0 & 0 & 0 & 0 \\
-0.23 & 0.81 & -0.35 & 0 & -0.23 & 0 & 0 & 0\\
-0.74 & -0.35 & 1.59 & -0.26 & -0.24 & 0 & 0 & 0\\
-0.24 & 0 & -0.26 & 0.82 & 0 & 0 & -0.32 & 0\\
0 & -0.23 & -0.24 & 0 & 1.12 & -0.51 & 0 & -0.14\\
0 & 0 & 0 & 0 & -0.51 & 0.66 & 0 & -0.15\\
0 & 0 & 0 & -0.32 & 0 & 0 & 0.64 & -0.32\\
0 & 0 & 0 & 0 & -0.14 & -0.15 & -0.32 & 0.61

\end{bmatrix*}. \label{LaplacianSCe}
\end{equation}  

 For practical reasons, it is often advantageous to use the normalized Laplacian, defined as 
\begin{equation} \mathbf{L}_N\, {\overset{def}{=}} \,\mathbf{D}^{-1/2}(\mathbf{D}-\mathbf{W})\mathbf{D}^{-1/2}=\mathbf{I}-\mathbf{D}^{-1/2}\mathbf{W}\mathbf{D}^{-1/2}. \label{LapNor}
\end{equation} 

\begin{Remark}The normalized Laplacian matrix is symmetric for undirected graphs, and has all diagonal values equal to $1$, with its trace equal to the number of vertices $N$.
\end{Remark} 
Other interesting properties, obtained through Laplacian normalization, shall be described later in the application context.
 
 One more form of the graph Laplacian is the so called \textbf{random-walk Laplacian}, defined as
\begin{equation} \mathbf{L}_{RW}\, {\overset{def}{=}} \,\mathbf{D}^{-1}\mathbf{L}=\mathbf{I}-\mathbf{D}^{-1}\mathbf{W}. \label{LapRW}
\end{equation}
The random-walk graph Laplacian is rarely used, since it has lost the symmetry property of the original graph Laplacian for undirected graphs, $\mathbf{L}_{RW} \ne \mathbf{L}^T_{RW}$.

\noindent\textbf{Vertex-weighted graphs.} Most of the applications of graph theory are based on edge-weighted graphs, where edge-weighting is designated by the weighting matrix, $\mathbf{W}$. Note that the weighting can be also  introduced into graphs based on vertex-weighted approaches (although rather rarely), whereby a weight is assigned to each vertex of a graph. To this end, we can use a diagonal matrix, $\mathbf{V}$, to define the vertex weights $v_i$, $i=0,1,\dots,N-1$, with one possible (the Chung/Langlands, \cite{chung1996combinatorial}) version of the vertex-weighted graph Laplacian, given by 
\begin{equation} \mathbf{L}_V\, {\overset{def}{=}} \,\mathbf{V}^{1/2}\mathbf{L}\mathbf{V}^{1/2}. \label{LapNorV}
\end{equation} 
Observe that for $\mathbf{V}=\mathbf{D}^{-1}$, the vertex-weighted graph Laplacian in (\ref{LapNorV}) reduces to the standard edge-weighted normalized graph Laplacian in (\ref{LapNor}).

\subsection{Some Frequently Used Graph Topologies}

When dealing with graphs, it is useful to introduce the following taxonomy of graph topologies.

\begin{enumerate}
 \item {\bf Complete graph.} A graph is complete if there exists an edge between every pair of its vertices. Therefore, the adjacency matrix of a complete graph has elements $A_{mn}=1$ for all $m\ne n$, and  $A_{mm}=0$, that is, no self-connections are present. Fig.~\ref{fig:spec-graph}(a) gives an example of a complete graph.

\begin{figure}
\centering
	\includegraphics[]{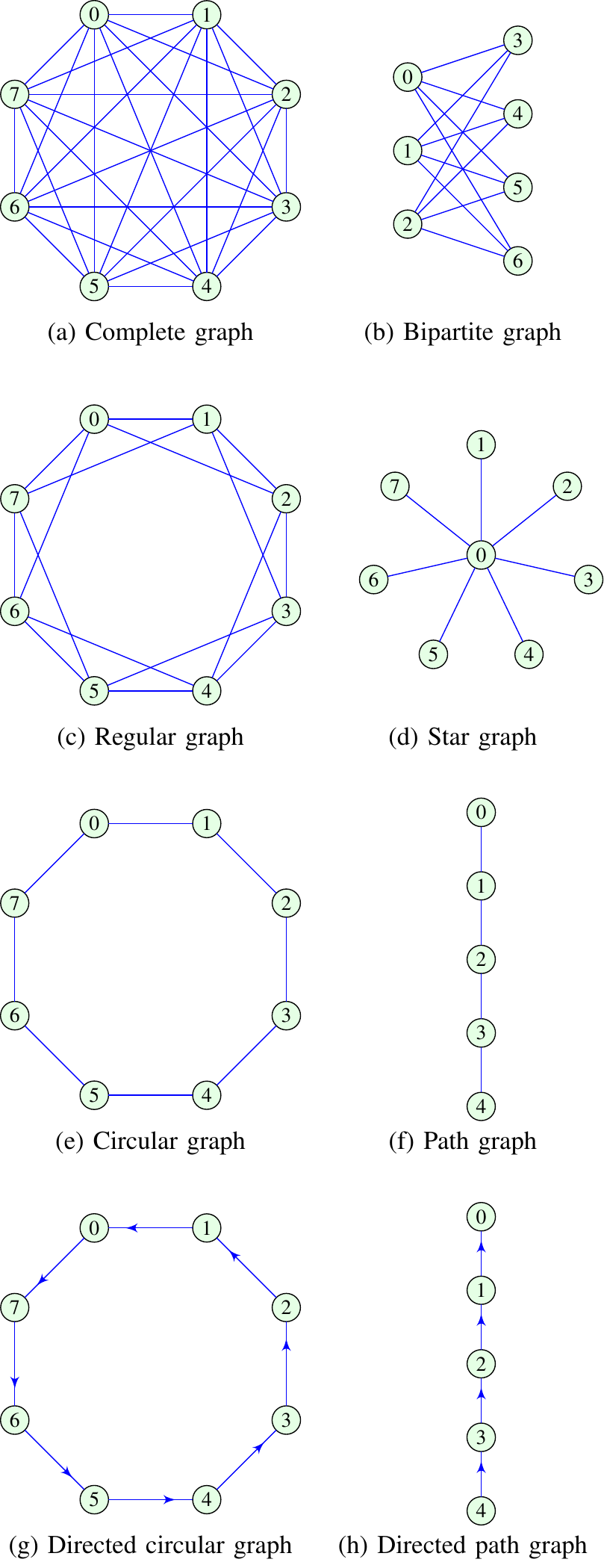}
	\caption{Typical graph topologies. (a) Complete graph with 8 vertices. (b) Complete bipartite graph. (c) Regular graph whereby each vertex is connected to 4 vertices. (d) Star graph. (e) Circular graph. (f) Path graph. (g) Directed circular graph. (h) Directed path graph.}
	\label{fig:spec-graph}
\end{figure}

\bigskip \item {\bf Bipartite graph.} A graph for which the graph vertices, $\mathcal{V}$, can be partitioned into two disjoint subsets, $\mathcal{E}$  and $\mathcal{H}$, whereby $\mathcal{V}=\mathcal{E} \cup \mathcal{H}$ and $\mathcal{E} \cap \mathcal{H} = \emptyset$, such that there are no edges between the vertices within the same subset $\mathcal{E}$  or $\mathcal{H}$, is referred to as a bipartite graph.  Fig.~\ref{fig:spec-graph}(b) gives an example of a bipartite undirected graph with $\mathcal{E}=\{0,1,2\}$ and $\mathcal{H}=\{3,4,5,6\}$, whereby all edges designate only connections between the sets $\mathcal{E}$  and $\mathcal{H}$. Observe also that the graph in Fig.~\ref{fig:spec-graph}(b) is a complete bipartite graph, since all possible edges between the sets $\mathcal{E}$  and $\mathcal{H}$  are present.

For convenience of mathematical formalism, if vertex ordering is performed in a such way that all vertices belonging to $\mathcal{E}$ are indexed before the vertices belonging to $\mathcal{H}$, then the resulting adjacency matrix can be written in a block form
\begin{equation}
\mathbf{A}
=\begin{bmatrix}
\mathbf{0} &  \mathbf{A}_{\mathcal{E}\mathcal{H}}\\
\mathbf{A}_{\mathcal{H}\mathcal{E}} &  \mathbf{0}
\end{bmatrix}, \label{bipartA}
\end{equation} 
where the submatrices $\mathbf{A}_{\mathcal{E}\mathcal{H}}$  and $\mathbf{A}_{\mathcal{H}\mathcal{E}}$ define the respective connections between the vertices belonging to the sets $\mathcal{E}$  and  $\mathcal{H}$. Observe that for an undirected bipartite graph, $\mathbf{A}_{\mathcal{E}\mathcal{H}}=\mathbf{A}_{\mathcal{H}\mathcal{E}}^T$.
Bipartite graphs are also referred to as Kuratowski graphs, denoted by $K_{N_{\mathcal{E}},N_{\mathcal{H}}}$, where $N_{\mathcal{E}}$ and $N_{\mathcal{H}}$ are the respective numbers of vertices in the sets $\mathcal{E}$ and $\mathcal{H}$.
	 It is important to mention that a complete bipartite graph with three vertices in each set, $\mathcal{H}$ and $\mathcal{E}$, is referred to as \textit{the first Kuratowski graph}, denoted by  $K_{3,3}$, which may be used to define conditions for a graph to be planar (more detail is given in the sequel).  
	
 {\bf Multipartite graph.}  A generalization of the concept of bipartite graph is  a multipartite ($M$-partite) graph for which the vertices are partitioned into $M$ subsets, whereby each edge connects vertices that belong to one of $M$ different subsets.

\bigskip \item  {\bf Regular graph.} An unweighted graph is said to be regular (or $\mathcal{J}$-regular) if all its vertices exhibit the same degree  of connectivity, $\mathcal{J}$.  In other words, the number of edges connected to each vertex is $\mathcal{J}$. An example of a regular graph with $\mathcal{J}=4$ is given in Fig.~\ref{fig:spec-graph}(c). From (\ref{LapDef}) and (\ref{LapNor}), the Laplacian and the normalized Laplacian of a $\mathcal{J}$-regular graph  are
\begin{equation} \mathbf{L}=\mathcal{J}\,\mathbf{I}-\mathbf{A} \,\,\,\,  \textrm{ and } \,\,\,\, \mathbf{L}_N=\mathbf{I}-\frac{1}{\mathcal{J}}\mathbf{A}. \label{regulGGGG}
\end{equation}

	\bigskip \item {\bf Planar graph.} A graph that can be drawn on a two-dimensional plane without the crossing of its edges is called planar.
	
	For example, if the edges $(0,2)$, $(2,4)$, $(4,6)$, and $(6,0)$  in the regular graph from Fig.~\ref{fig:spec-graph}(c) are plotted as arches outside the circle defined by the vertices, all instances of edge crossing will be avoided and such graph presentation will be planar.

\bigskip \item {\bf Star graph.} This type of graph has one central vertex that is connected to all other vertices, with no other edges present. An example of  star graph is given in Fig.~\ref{fig:spec-graph}(d). Observe that a star graph can be considered as a special case of a complete bipartite graph, with only one vertex in the first set, $\mathcal{E}$. The vertex degree centrality for the central vertex of a star graph with $N$ vertices is therefore $N-1$.

\bigskip \item {\bf Circular graph.} A graph is said to be circular if its every vertex is of the degree $\mathcal{J}=2$. This graph is also a regular graph with $\mathcal{J}=2$. An example of a circular graph with $8$ vertices is given in Fig.~\ref{fig:spec-graph}(e). 

\bigskip \item {\bf Path graph.} A series of connected vertices defines a path graph, whereby the first and the last vertex are of connectivity degree $\mathcal{J}=1$, while all other vertices are of the connectivity degree $\mathcal{J}=2$. An example of a path graph with $5$ vertices is presented in Fig.~\ref{fig:spec-graph}(f).

\bigskip \item {\bf Directed circular  graph.} A directed graph is said to be circular if each vertex is related to only one predecessor vertex and only one successor vertex.  An example of a circular directed graph with $8$ vertices is given in Fig.~\ref{fig:spec-graph}(g), with the adjacency matrix
\begin{gather}
\!\!\mathbf{A}=
\begin{array}{cr}
& \\
{
	\color{blue}
	\begin{matrix}
	\text{\footnotesize 0}\\
	\text{\footnotesize 1}\\
	\text{\footnotesize 2}\\
	\text{\footnotesize 3}\\
	\text{\footnotesize 4}\\
	\text{\footnotesize 5}\\
	\text{\footnotesize 6}\\
	\text{\footnotesize 7}\\
	\end{matrix}
} & \!\! \!\!
\begin{bmatrix}
\ 0\  & \ 0\  & \ 0\  & \ 0\  & \ 0\  & \ 0\  & \ 0\  & \ 1\  \\
\ 1\  & \ 0\  & \ 0\  & \ 0\  & \ 0\  & \ 0\  & \ 0\  & \ 0\  \\
\ 0\  & \ 1\  & \ 0\  & \ 0\  & \ 0\  & \ 0\  & \ 0\  & \ 0\  \\
\ 0\  & \ 0\  & \ 1\  & \ 0\  & \ 0\  & \ 0\  & \ 0\  & \ 0\  \\
\ 0\  & \ 0\  & \ 0\  & \ 1\  & \ 0\  & \ 0\  & \ 0\  & \ 0\  \\
\ 0\  & \ 0\  & \ 0\  & \ 0\  & \ 1\  & \ 0\  & \ 0\  & \ 0\  \\
\ 0\  & \ 0\  & \ 0\  & \ 0\  & \ 0\  & \ 1\  & \ 0\  & \ 0\  \\
\ 0\  & \ 0\  & \ 0\  & \ 0\  & \ 0\  & \ 0\  & \ 1\  & \ 0\ 
\end{bmatrix} \\
 & 
{
	\color{blue}
	\begin{matrix}
	\text{\footnotesize 0} \ &
	\ \text{\footnotesize 1}\  &
	\ \text{\footnotesize 2}\  &
	\ \text{\footnotesize 3}\ \ &
	\ \text{\footnotesize 4}\  &
	\ \text{\footnotesize 5}\  &
	\ \text{\footnotesize 6}\  &
	\ \text{\footnotesize 7}  &
	\end{matrix}
}
\end{array}\!\!. \label{AdjMtxFirsDir}
\end{gather} 

\begin{Remark}\label{ReCir}  The adjacency matrix of any directed or undirected circular graph is a circulant matrix. 
\end{Remark} 

\bigskip \item {\bf Directed path graph.} A directed path graph is defined by a series of vertices  connected in one direction, whereby the first and the last vertex do not have a respective predecessor or successor. An example of a directed path graph with $5$ vertices is presented in Fig.~\ref{fig:spec-graph}(h).

\end{enumerate}

\begin{Remark} Path and circular graphs (directed and undirected) are of particular interest in Data Analytics, since their domain properties correspond to classical time or space domains.
	Therefore, any graph signal processing or machine learning paradigm which is developed for path and circular graphs is equivalent to its corresponding standard time and/or spatial domain  paradigm. 
\end{Remark}

\subsection{Properties of Graphs and Associated Matrices}

The notions from  graph analysis that are most relevant to the processing of data on graphs are:
\begin{enumerate}[label=\subscript{M}{{\arabic*}}:] 
	
\item \textit{Symmetry:} For an undirected graph, the matrices $\mathbf{A}$, $\mathbf{W}$, and $\mathbf{L}$ are all symmetric.

\item \textit{A walk} between a vertex $m$ and a vertex $n$ is a connected sequence of edges and vertices that begins at the vertex $m$ and ends at the vertex $n$. 
Edges and vertices can be included in a walk more than once. 

\textit{The length of a walk}  is equal to the number of included edges in unweighted graphs. 
The number of walks of the length $K$, between a vertex $m$ and a vertex $n$,  is equal to the value of the $mn$-th element of the matrix $\mathbf{A}^K$, which can be proved through mathematical induction, as follows \cite{duncan2004powers}: 

(i) The elements,  $A_{mn}$, of the adjacency matrix $\mathbf{A}$,  by definition, indicate if there is a walk of length $K=1$ (an edge, in this case) between the vertices $m$ and $n$ in a graph; 

(ii) Assume that the elements of matrix $\mathbf{A}^{K-1}$ are equal to the number of walks of length $K-1$, between two arbitrary vertices $m$  and $n$; 

(iii) The number of walks of length $K$ between two vertices, $m$ and $n$, is equal to the number of all walks of length $K-1$, between the vertex $m$ and an intermediate vertex $s$, $s\in \mathcal{V}$, which is indicated by the element at the position $ms$ of the matrix $\mathbf{A}^{K-1}$, according to the assumption in (ii), for all $s$ for which there is an edge from vertex $s$ to the destination vertex $n$. If an edge between the intermediate vertex $s$ and the final vertex $n$ exists, then $A_{sn}=1$. This means that the number of walks of  length $K$ between two vertices $m$ and $n$ is obtained as the inner product  of the $m$-th row of $\mathbf{A}^{K-1}$ with the $n$-th column in $\mathbf{A}$, to yield the element $mn$ of matrix $\mathbf{A}^{K-1}\mathbf{A}=\mathbf{A}^{K}$.

\begin{Example} Consider the vertex 0 and the vertex 4 in the graph from Fig. \ref{fig:walks}, and only the walks of length $K=2$. The adjacency matrix for this graph is given in (\ref{matA1a}). There are two such walks ($0 \to 1 \to 4$ and $0 \to 2 \to 4$), so that the element $A^2_{04}$ in the first row and the fifth column of matrix $\mathbf{A}^2$,  is equal to $2$, as designated in bold font in the matrix $\mathbf{A}^2$ below, 
\begin{equation}
\mathbf{A}^2=
\begin{array}{cr}
& \\
{
	\color{blue}
	\begin{matrix}
	\text{\footnotesize 0}\\
	\text{\footnotesize 1}\\
	\text{\footnotesize 2}\\
	\text{\footnotesize 3}\\
	\text{\footnotesize 4}\\
	\text{\footnotesize 5}\\
	\text{\footnotesize 6}\\
	\text{\footnotesize 7}\\
	\end{matrix}
} 
\begin{bmatrix}
 3 & 1 & 2 & 1 & \mathbf{2} & 0 & 1 & 0\\
1 & 3 & 2 & 2 & 1 & 1 & 0 & 1\\
2 & 2 & 4 & 1 & 1 & 1 & 1 & 1\\
1 & 2 & 1 & 3 & 1 & 0 & 0 & 1\\
2 & 1 & 1 & 1 & 4 & 1 & 1 & 1\\
0 & 1 & 1 & 0 & 1 & 2 & 1 & 1\\
1 & 0 & 1 & 0 & 1 & 1 & 2 & 0\\
0 & 1 & 1 & 1 & 1 & 1 & 0 & 3
\end{bmatrix}, \\
{
	\color{blue}
	\begin{matrix}
&	\text{\footnotesize \ \  0} &
	 \text{\footnotesize 1 }  &
	 \text{\footnotesize 2}  &
	 \text{\footnotesize 3}  &
	 \text{\footnotesize 4}  &
	 \text{\footnotesize 5 }  &
	 \text{\footnotesize 6}  &
	 \text{\footnotesize 7 } &
	\end{matrix}
}
\end{array}\!\!
\end{equation}
thus indicating $K=2$ walks between these vertices.

\begin{figure}[tb]
\centering
\includegraphics[]{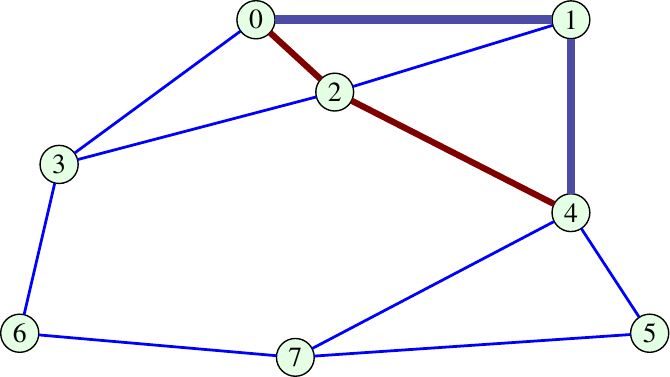}
\caption{Walks of length $K=2$ from vertex 0 to vertex 4 (thick blue and brown lines).}
\label{fig:walks}
\end{figure}

\end{Example}

\bigskip \item
\textit{The number of walks} between the vertices $m$ and $n$, that are of  length not higher than $K$, is given by the $mn$-th element  of the matrix 
\begin{equation}
\mathbf{B}_K=\mathbf{A}+\mathbf{A}^2+\cdots+\mathbf{A}^K, \label{sumwalks}
\end{equation}
that is, by a value in its $m$-th row and $n$-th column.  In other words, the total number of walks is equal to the sum of all walks, which are individually modeled by $\mathbf{A}^k$, $k=1,2,\dots,K$, as stated in property $M_2$.

\bigskip \item
\textit{The $K$-neighborhood}  of a vertex is defined as a set of vertices that are reachable from this vertex in walks whose length is up to $K$. For a vertex $m$, based on the property $M_2$, the $K$-neighborhood is designated by the positions and the numbers of non-zero elements in the $m$-th row of matrix $\mathbf{B}_K$ in (\ref{sumwalks}). The $K$-neighborhoods of vertex $0$ for $K=1$ and $K=2$ are illustrated in Fig. \ref{fig:neighborhood}.

\begin{figure}[tb]
\centering
\includegraphics[]{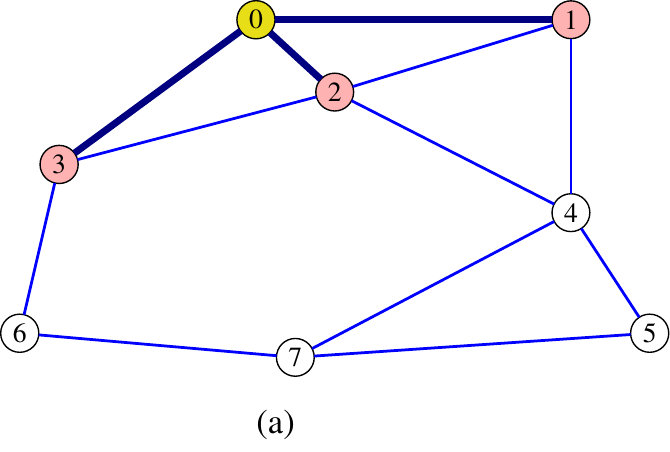}
\hfill
\includegraphics[]{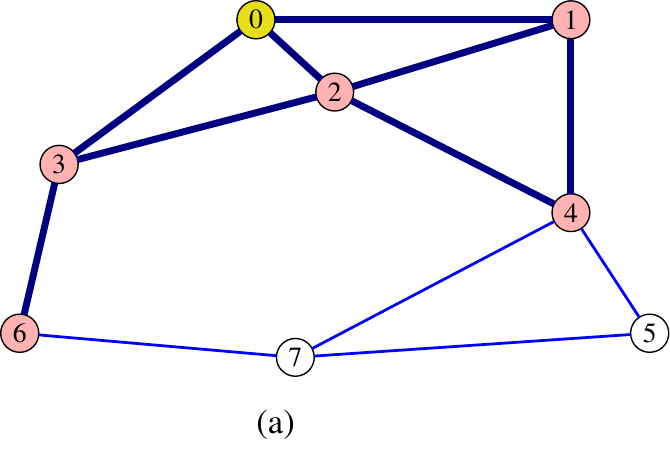}
\caption{The $K$-neighborhoods of vertex $0$ for the graph from Fig. \ref{fig:walks}, where: (a) $K=1$ and (b) $K=2$. The neighboring vertices are shaded.}
\label{fig:neighborhood}
\end{figure}

\bigskip \item \textit{A path}  is a special kind of walk whereby each vertex can be included only once, while the number of edges included in a path is referred to as  the \textit{path cardinality} or \textit{path length}, and the \textit{path weight} is defined as the sum of weights along these edges.

An \textit{Euler path} is a graph path that uses every edge of a graph exactly once. An Euler path for an unweighted  graph does exist if and only if at most two of its vertices are of an odd degree. An Euler path which starts and ends at the same vertex is referred to as an \textit{Euler circuit}, and it exists if and only if the degree of every vertex is even.

A \textit{Hamiltonian path} is a graph path between two vertices of a graph that visits each vertex in a graph exactly once, while a cycle that uses every vertex in a graph exactly once is called a \textit{Hamiltonian cycle}. 

\bigskip \item \textit{The distance}, $r_{mn}$, between two vertices $m$ and $n$ in an unweighed graph is equal to the minimum path length between these vertices. For example, for the graph in Fig. \ref{fig:walks},
the distance between  vertex 1 and vertex 5  is $r_{15}=2$.
 
\bigskip \item \textit{The diameter, $d$, of a graph}   is equal to the largest distance (number of edges) between  all pairs of its vertices, that is, $d=\max_{m,n \in \mathcal{V}}r_{mn}$. For example, the diameter of a complete graph is  $d=1$, while  the diameter of the graph in Fig. \ref{fig:walks} is $d=3$, with one of the longest paths being $6 \rightarrow 3 \rightarrow 2 \rightarrow 1$.

\bigskip \item \textit{Vertex closeness centrality}. The farness (remoteness) of a vertex is equal the sum of its distances to all other vertices, $f_n=\sum_{m\ne n}r_{nm}$. The vertex closeness is defined then as an inverse to the farness, $c_n=1/f_{n}$, and can be interpreted as a measure of how long it will take for data to sequentially shift from the considered vertex to all other vertices. For example, the vertex farness and closeness for the vertices $n=2$ and $n=5$ in Fig. \ref{GSPb_ex1a}(a) are respectively $f_2=10$, $f_5=14$, and $c_2=0.1$, $c_5=0.071$.

\bigskip \item \textit{Vertex or edge betweenness}. Vertex/edge betweenness of a vertex $n$ or edge $(m,n)$ is equal to the number of times that this vertex/edge acts as a bridge along the shortest paths between any other two vertices.  
	
\bigskip \item \textit{Spanning Tree and Minimum Spanning Tree.}
The spanning tree of a graph is a subgraph that is tree-shaped and connects all its vertices together. A tree does not have cycles and cannot be disconnected. \textit{The cost of the spanning tree} is the sum of the weights of all the edges in the tree.  \textit{The minimum spanning tree} is a spanning tree for which the cost is minimum among all possible spanning trees of a graph. Spanning trees are typically used in graph clustering analysis.

 In the literature on graph theory, it is commonly  assumed that the values of  edge weights in weighted graphs are proportional to the standard vertex distance, $r_{mn}$. However, this is not the case in data analytics on graphs, where the edge weights are typically defined as a function of vertex distance, for example, through a Gaussian kernel, $W_{mn} \sim \exp(-r_{mn}^2)$, or some other data similarity metric. The cost function to minimize for the Minimum Spanning Tree (MST) can be defined as a sum of  distances,  $r_{mn}=-2\ln W_{mn} $. A spanning tree for the graph from Fig. \ref{GSPb_ex2} is shown in Fig. \ref{fig:spanningtree}. The cost for  this spanning tree, calculated as a sum of all distances (log-weights), $r_{mn}$, is $15.67$.

\begin{figure}[tb]
	\centering
	\includegraphics[]{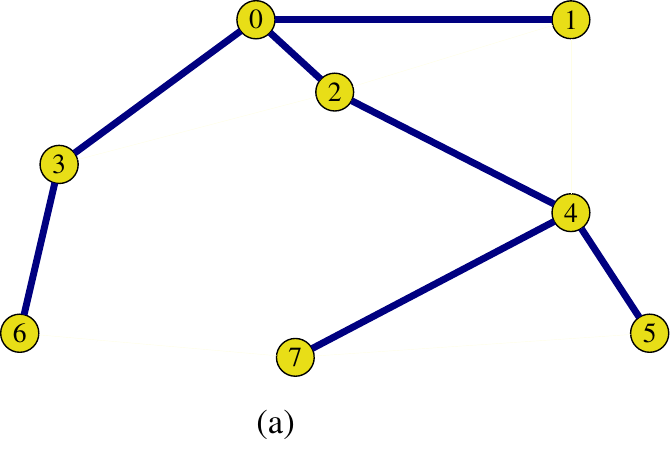} (a)
	
	\vfill
	
	\includegraphics[]{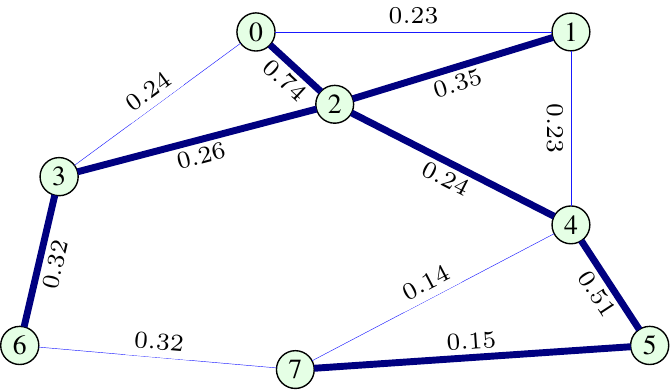}(b)
	\caption{Concept of the spanning tree for graphs. (a) A spanning tree for  the unweighted graph  from Fig. \ref{GSPb_ex1a}(a). (b) A spanning tree for the weighted graph from Fig. \ref{GSPb_ex2}, designated by thick blue edges. The graph edges in thin blue lines are not included in this spanning tree.}
	\label{fig:spanningtree}
\end{figure}

\bigskip \item  \textit{An undirected  graph is connected} if there exists a walk between each pair of its vertices.
\bigskip \item If \textit{the graph is not connected}, then it consists of two or more disjoint but locally connected subgraphs (\textit{graph components}). Back to mathematical formalism, such disjoint graphs produce a block-diagonal form of the adjacency matrix, $\mathbf{A}$, and the Laplacian, $\mathbf{L}$. For $M$ disjoint components (subgraphs) of a graph, these matrices take the form
\begin{gather}
\mathbf{A}
 = \begin{bmatrix}
\mathbf{A}_1 & \mathbf{0} &   \cdots & \mathbf{0}\\
\mathbf{0} & \mathbf{A}_2 &  \cdots & \mathbf{0}\\
\vdots &  \vdots  &\ddots & \vdots \\
\mathbf{0} & \mathbf{0} &   \cdots & \mathbf{A}_M \\
\end{bmatrix}  
\\
\quad \mathbf{L}
  = \begin{bmatrix}
\, \mathbf{L}_1   &   \mathbf{0} \,  &   \cdots & \mathbf{0}\\
\,  \mathbf{0} &   \mathbf{L}_2 \,  &  \cdots & \mathbf{0}\\
\,  \vdots  &  \vdots \,  &\ddots & \vdots \\
\,  \mathbf{0} &   \mathbf{0} \,  &   \cdots & \mathbf{L}_M \\
\end{bmatrix} \label{BlckDL}.
\end{gather}
Note that this block diagonal form is obtained only if the vertex numbering follows the subgraph structure.

\begin{figure}
	\centering
	\includegraphics[]{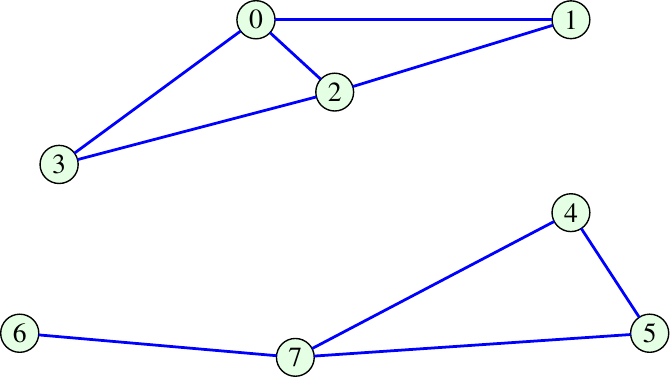}
	\caption{A disconnected graph which consists of two sub-graphs.}
	\label{GSPb_ex1b}
\end{figure}

\begin{Example} Consider a graph derived from Fig.~\ref{GSPb_ex1a}(a) by removing some
edges, as shown in Fig.~\ref{GSPb_ex1b}. The adjacency matrix for this graph is given by
\begin{equation}
\mathbf{A}=\begin{array}{cr}
	& \\
	{
		\color{blue}
		\begin{matrix}
			\text{\footnotesize 0}\\
			\text{\footnotesize 1}\\
			\text{\footnotesize 2}\\
			\text{\footnotesize 3}\\
			\text{\footnotesize 4}\\
			\text{\footnotesize 5}\\
			\text{\footnotesize 6}\\
			\text{\footnotesize 7}\\
		\end{matrix}
	} 
\left[
\begin{array}{cccc:cccc} 
 \cellcolor[gray]{0.9} 0 &   \cellcolor[gray]{0.9} \mathbf{1} &   \cellcolor[gray]{0.9} \mathbf{1} &   \cellcolor[gray]{0.9} \mathbf{1}&   0 &   0 &   0 &   0\\
 \cellcolor[gray]{0.9} \mathbf{1} &   \cellcolor[gray]{0.9} 0 &   \cellcolor[gray]{0.9} \mathbf{1} &   \cellcolor[gray]{0.9} 0 &   0 &   0 &   0 &   0\\
 \cellcolor[gray]{0.9} \mathbf{1} &  \cellcolor[gray]{0.9}  \mathbf{1} &  \cellcolor[gray]{0.9}  0 &  \cellcolor[gray]{0.9}  \mathbf{1} &   0 &   0 &   0 &   0\\
 \cellcolor[gray]{0.9} \mathbf{1} &   \cellcolor[gray]{0.9}  0 &   \cellcolor[gray]{0.9} \mathbf{1} &   \cellcolor[gray]{0.9} 0 &   0 &   0 &   0 &   0\\  \hdashline
0 &   0 &   0 &   0 &   \cellcolor[gray]{0.9} 0 &   \cellcolor[gray]{0.9} \mathbf{1} &   \cellcolor[gray]{0.9} 0 &  \cellcolor[gray]{0.9} \mathbf{1}\\
0 &   0 &   0 &   0 &  \cellcolor[gray]{0.9}  \mathbf{1} &   \cellcolor[gray]{0.9} 0 &   \cellcolor[gray]{0.9} 0 &  \cellcolor[gray]{0.9}  \mathbf{1}\\
0 &   0 &   0 &   0 &   \cellcolor[gray]{0.9} 0 &   \cellcolor[gray]{0.9} 0 &  \cellcolor[gray]{0.9}  0 &   \cellcolor[gray]{0.9} \mathbf{1}\\
0 &   0 &   0 &   0 &   \cellcolor[gray]{0.9} \mathbf{1} &   \cellcolor[gray]{0.9} \mathbf{1} &   \cellcolor[gray]{0.9} \mathbf{1}&   \cellcolor[gray]{0.9} 0
 \end{array}
\right]
\\
{
	\color{blue}
	\begin{matrix}
		& \	\text{\footnotesize \ \  0} &
		\text{\footnotesize 1 }  &
		\text{\footnotesize 2}  &
		\text{\footnotesize 3}  &
		\text{\footnotesize 4}  &
		\text{\footnotesize 5 }  &
		\text{\footnotesize 6}  &
		\text{\footnotesize 7 } &
	\end{matrix}
}
\end{array}\!\! \label{matAdisc}
\end{equation}
with the corresponding Laplacian
\begin{equation}
\mathbf{L}=
\left[
\begin{array}{rrrr:rrrr}
 \cellcolor[gray]{0.9} 3 &  \cellcolor[gray]{0.9} -1 &  \cellcolor[gray]{0.9}-1 &  \cellcolor[gray]{0.9} -1 & 0 & 0 & 0 & 0\\
 \cellcolor[gray]{0.9} -1 &  \cellcolor[gray]{0.9} 2 & \cellcolor[gray]{0.9} -1 &  \cellcolor[gray]{0.9} 0 & 0 & 0 & 0 & 0\\
 \cellcolor[gray]{0.9} -1 &  \cellcolor[gray]{0.9} -1 &  \cellcolor[gray]{0.9} 3 & \cellcolor[gray]{0.9} -1 & 0 & 0 & 0 & 0\\
 \cellcolor[gray]{0.9} -1 &  \cellcolor[gray]{0.9} 0 &  \cellcolor[gray]{0.9} -1 &  \cellcolor[gray]{0.9} 2 & 0 & 0 & 0 & 0\\ \hdashline
0 & 0 & 0 & 0 &  \cellcolor[gray]{0.9} 2 &  \cellcolor[gray]{0.9} -1 &  \cellcolor[gray]{0.9} 0 &  \cellcolor[gray]{0.9} -1\\
0 & 0 & 0 & 0 & \cellcolor[gray]{0.9} -1 &  \cellcolor[gray]{0.9} 2 &  \cellcolor[gray]{0.9} 0 & \cellcolor[gray]{0.9} -1\\
0 & 0 & 0 & 0 &  \cellcolor[gray]{0.9} 0 &  \cellcolor[gray]{0.9} 0 &  \cellcolor[gray]{0.9} 1 &  \cellcolor[gray]{0.9} -1\\
0 & 0 & 0 & 0 & \cellcolor[gray]{0.9} -1 & \cellcolor[gray]{0.9} -1 & \cellcolor[gray]{0.9} -1 &  \cellcolor[gray]{0.9} 3
 \end{array}
\right].
\end{equation}
Observe that, as elaborated above, these matrices are in a block-diagonal form with the two constituent blocks clearly separated. Therefore, for an isolated vertex in a graph, the corresponding row and column
of the matrices $\mathbf{A}$ and $\mathbf{L}$ will be zero-valued. 
\end{Example}

\bigskip \item For two graphs defined on the same  set of vertices, with the corresponding adjacency matrices $\mathbf{A}_1$ and $\mathbf{A}_2$,   \textit{the summation} operator produces a new graph, for which the adjacency matrix is given by
$$ \mathbf{A}= \mathbf{A}_1 +  \mathbf{A}_2. $$ 
To maintain the binary values $A_{mn} \in \{0,1\}$ in the resultant adjacency matrix, a logical (Boolean) summation rule, e.g., $1+1=1$, may be used for matrix addition. In this article, the arithmetic summation rule is assumed in data analytics algorithms, as for example, in equation (\ref{sumwalks}) in property $M_3$. 

\bigskip \item \textit{The Kronecker (tensor) product}  of two disjoint graphs $\mathcal{G}_1=(\mathcal{V}_1,\mathcal{B}_1)$ and $\mathcal{G}_2=(\mathcal{V}_2,\mathcal{B}_2)$ yields a new graph $\mathcal{G}=(\mathcal{V},\mathcal{B})$ where $\mathcal{V}=\mathcal{V}_1 \times \mathcal{V}_2$ is a direct product of the sets $\mathcal{V}_1$ and $\mathcal{V}_2$,   and  $\big((n_1,m_1),(n_2,m_2)\big) \in \mathcal{B}$ only if  $(n_1,n_2) \in \mathcal{B}_1$ and $(m_1,m_2) \in \mathcal{B}_2$.  

The adjacency matrix $\mathbf{A}$ of the resulting graph $\mathcal{G}$ is then equal to the Kronecker product of the individual adjacency matrices $\mathbf{A}_1$ and $\mathbf{A}_2$, that is
$$\mathbf{A}=\mathbf{A}_1 \otimes \mathbf{A}_2.$$

An illustration of the Kronecker product for two simple graphs is given in Fig. \ref{fig:cronecker}.

\bigskip \item \textit{The Cartesian product (graph product)}  of two disjoint graphs  $\mathcal{G}_1=(\mathcal{V}_1,\mathcal{B}_1)$ and $\mathcal{G}_2=(\mathcal{V}_2,\mathcal{B}_2)$ gives a new graph $\mathcal{G}=\mathcal{G}_1\square \mathcal{G}_2=(\mathcal{V},\mathcal{B})$,  where $\mathcal{V}=\mathcal{V}_1 \times \mathcal{V}_2$ is a direct product of the sets $\mathcal{V}_1$ and $\mathcal{V}_2$,  and  $\big((m_1,n_1),(m_2,n_2)\big) \in \mathcal{B}$, only if
\begin{align*}
m_1 & =m_2\text{ and }(n_1,n_2) \in \mathcal{B}_2\text{ or }\\
n_1 & =n_2\text{ and }(m_1,m_2) \in \mathcal{B}_1.
\end{align*}

The adjacency matrix of a Cartesian product of two graphs is then given by the Kronecker sum 
$$\mathbf{A}=\mathbf{A}_1 \otimes \mathbf{I}_{N_2}+\mathbf{I}_{N_1} \otimes \mathbf{A}_2 \, {\overset{def}{=}} \, \mathbf{A}_1 \oplus\mathbf{A}_2,$$
where $\mathbf{A}_1 $ and $\mathbf{A}_2$ are the respective adjacency matrices of graphs $\mathcal{G}_1$, $\mathcal{G}_2$, while $N_1$ and $N_2$ are the corresponding numbers of vertices in $\mathcal{G}_1$ and $\mathcal{G}_2$, with $\mathbf{I}_{N_1}$ and $\mathbf{I}_{N_2}$ being the identity matrices of orders $N_1$ and $N_2$. The Cartesian product of two simple graphs is illustrated in Fig. \ref{fig:cartesian}. Notice that a Cartesian product of two graphs that correspond to a two-dimensional space can be considered as a three-dimensional structure of vertices and edges (\textit{cf.}  tensors \cite{saito2018hypergraph}).

\begin{figure}
\centering
\includegraphics[]{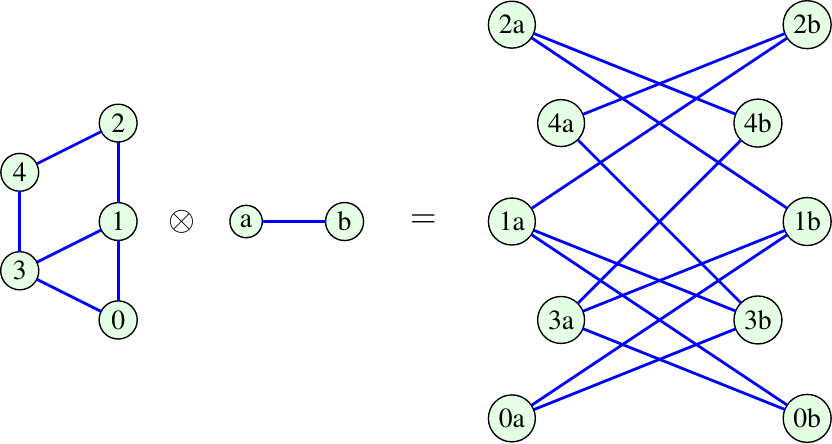}
\caption{Kronecker (tensor) product of two graphs.}
\label{fig:cronecker}
\end{figure}

\begin{figure}
\centering
\includegraphics[]{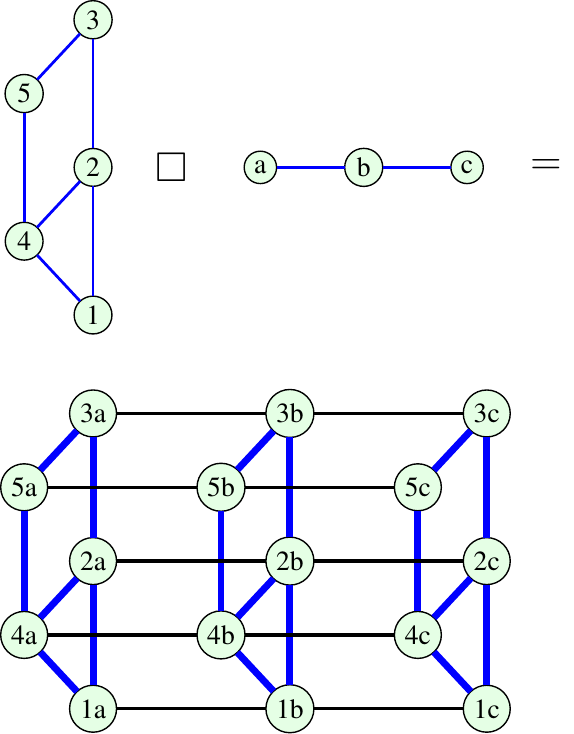}
\caption{Cartesian product of two graphs.}
\label{fig:cartesian}
\end{figure}

\end{enumerate}

   \section{Spectral Decomposition of Graph Matrices}
As a prerequisite for the optimization and data analytics on graphs, we next  introduce several connections between linear algebraic tools and graph topology \cite{bapat1996laplacian,o2016eigenvectors,fujiwara1995eigenvalues,maheswari2016some,cvetkovic1980spectra,brouwer2011spectra,Chung1997,jones2013spectra}.
 
   \subsection{Eigenvalue Decomposition of the Adjacency Matrix}

    Like any other general matrix, graph matrices can be analyzed  using eigenvalue decomposition. In this sense, a column vector $\mathbf{u}$ is an eigenvector of the adjacency matrix $\mathbf{A}$ if
   \begin{equation}
   \mathbf{A} \mathbf{u} = \lambda \mathbf{u}, \label{eiganal}
   \end{equation}
    where the constant $\lambda$, that  corresponds to the eigenvector  $\mathbf{u}$, is called the eigenvalue.  
    
    The above relation can be equally written as  $(\mathbf{A}  - \lambda \mathbf{I}) \mathbf{u}=\mathbf{0}$, and a nontrivial solution for $\mathbf{u}$ does exist if
$$\det|\mathbf{A}-\lambda\mathbf{I}|=0.$$
In other words, the problem turns into that of finding zeros of  $\det|\mathbf{A}-\lambda\mathbf{I}|$ as roots of a polynomial in $\lambda$, called \textit{the characteristic polynomial} of matrix $\mathbf{A}$, and given by
   \begin{equation}
P(\lambda)=\det|\mathbf{A}-\lambda\mathbf{I}|=\lambda^N+c_1\lambda^{N-1}+c_2\lambda^{N-2}+\cdots+c_N. \label{CharPol}
   \end{equation}

\begin{Remark} The order of the characteristic polynomial of graphs has the physical meaning of the number of vertices, $N$, within a graph while the eigenvalues represent the roots of the characteristic polynomial, that is, $
P(\lambda)=0
$. 
\end{Remark}

In general, for a graph with $N$ vertices, its adjacency matrix has $N$ eigenvalues, $\lambda_0$, $\lambda_1$, \ldots, $\lambda_{N-1}$. Some eigenvalues may also be repeated, which indicates that zeros of algebraic multiplicity higher than one exist in the characteristic polynomial.  The total number of roots of a characteristic polynomial, including their multiplicities, must be equal to its degree, $N$, whereby 
\begin{itemize}
	\item
The \textit{algebraic multiplicity of an eigenvalue}, $\lambda_k$, is equal to its multiplicity when considered as a root of the characteristic polynomial;
 \item 
\textit{The geometric multiplicity of an eigenvalue}, $\lambda_k$, is the number of linearly independent eigenvectors that can be associated with this eigenvalue. 
\end{itemize}

The geometric multiplicity of an eigenvalue (the number of independent eigenvectors for one eigenvalue) is always equal or lower than its algebraic multiplicity.

Denote the distinct eigenvalues in (\ref{CharPol}) by $\mu_1$, $\mu_2$, \ldots, $\mu_{N_m}$, and their corresponding algebraic multiplicities by $p_1$, $p_2$, \ldots, $p_{N_m}$, where $p_1+p_2+\cdots+p_{N_m}=N$ is equal to the order of the considered matrix/polynomial and $N_m\le N$ is the number of distinct eigenvalues.
The characteristic polynomial can now be rewritten in the form
$$
P(\lambda)=(\lambda-\mu_1)^{p_1}  (\lambda-\mu_2)^{p_2} \cdots (\lambda-\mu_{N_m})^{p_{N_m}}.
$$

\medskip\noindent\textit{Definition:} \textit{The minimal polynomial} of the considered  adjacency matrix, $\mathbf{A}$, is obtained from its characteristic polynomial by reducing the algebraic multiplicities of all eigenvalues to unity, and has the form
$$
P_{min}(\lambda)=(\lambda-\mu_1)  (\lambda-\mu_2) \cdots (\lambda-\mu_{N_m}).
$$

\subsubsection{Properties of the characteristic and minimal polynomial}
\begin{enumerate}[label=\subscript{P}{{\arabic*}}:] 
	
\item The order of the characteristic polynomial is equal to the number of vertices in the considered graph.

\item For $\lambda=0$, $P(0)=\det(\mathbf{A})=-\lambda_0 (-\lambda_1) \cdots (-\lambda_{N-1}).$
\item The sum of all the eigenvalues is equal to the sum of the diagonal elements   of the adjacency matrix, $\mathbf{A}$, that is, its trace, $\mathrm{tr}\{\mathbf{A}\}$. For the characteristic polynomial of the adjacency matrix, $P(\lambda)$, this means that  the value of $c_1$ in (\ref{CharPol}) is $c_1=\mathrm{tr}\{\mathbf{A}\}=0$. 
\item The coefficient $c_2$ in $P(\lambda)$ in (\ref{CharPol}) is equal to the number of edges multiplied by $-1$.

 This property, together with $P_3$, follows from the Faddeev–LeVerrier algorithm to calculate the coefficients of the characteristic polynomial of a square matrix, $\mathbf{A}$, as $c_1=-\mathrm{tr}\{\mathbf{A}\}$, $c_2=-\frac{1}{2}(\mathrm{tr} \{\mathbf{A}^2\}-\mathrm{tr}\{ \mathbf{A}^2\} )$, and so on. Since $\mathrm{tr}\{\mathbf{A}\}=0$ and the diagonal elements of $\mathbf{A}^2$ are equal to the number of edges connected to each vertex (vertex degree), the total number of edges is equal to $\mathrm{tr}\{ \mathbf{A}^2\}/2=-c_2$.  
\item The degree of the minimal polynomial, $N_m$, is strictly larger than the graph diameter, $d$.

\begin{Example} Consider a connected graph with $N$ vertices and only two distinct eigenvalues, $\lambda_0$ and $\lambda_1$. The order of minimal polynomial is then  $N_m=2$, while the diameter of this graph is $d=1$, which indicates a complete graph.
\end{Example}

\end{enumerate}

\begin{Example} For the graph from Fig.~\ref{GSPb_ex1a}(a),  the characteristic polynomial of its adjacency matrix, $\mathbf{A}$, defined in (\ref{matA1a}), is given by 
$$
P(\lambda)= \lambda^8        -12\lambda^6    -8\lambda^5    +36\lambda^4   + 36\lambda^3   -22 \lambda^2  -32 \lambda    -8,
$$
with the eigenvalues 
$$\lambda \in \{-2,   -1.741,   -1.285,\allowbreak   -0.677,\allowbreak   -0.411,\allowbreak    1.114,   1.809,    3.190\}. $$
 With all the eigenvalues different, the minimal polynomial is  equal to the characteristic polynomial, $P_{min}(\lambda)=P(\lambda)$.
\end{Example}

\begin{Example} The adjacency matrix for the disconnected graph from Fig.~\ref{GSPb_ex1b} is given in (\ref{matAdisc}), and its characteristic polynomial has the form
$$
P(\lambda)=  \lambda^8      -9\lambda^6    -6\lambda^5   + 21\lambda^4 +    26 \lambda^3  +  3\lambda^2    -4\lambda     
$$
with the eigenvalues 
$$\lambda \in \{
-1.5616,   -1.4812,   -1, \allowbreak  -1, \allowbreak       0,    0.3111,    2.1701,     2.5616
\}.$$
Observe that the eigenvalue $\lambda=-1$ is of multiplicity higher than $1$ (multiplicity of 2), so that the corresponding minimal polynomial becomes
$$
P_{min}(\lambda)=
\lambda^7 - \lambda^6 - 8\lambda^5 + 2\lambda^4 + 19\lambda^3 + 7\lambda^2 - 4\lambda.
$$
Although this graph is disconnected, the largest eigenvalue of its adjacency matrix, $\lambda_{\max}=2.5616$, is of multiplicity $1$. Relation between the graph connectivity and the multiplicity of eigenvalues will be discussed later.
\end{Example}

\subsection{Spectral Graph Theory}
If all the eigenvalues of $\mathbf{A}$ are distinct (of algebraic multiplicity 1), then the $N$ equations in the eigenvalue problem in (\ref{eiganal}), that is,
$\mathbf{A} \mathbf{u}_k = \lambda_k \mathbf{u}_k$, $k=0,1,\dots,N-1$,  
can be written in a compact form as one matrix equation with respect to the adjacency matrix, as
$$\mathbf{A}\mathbf{U}=\mathbf{U}\mathbf{\Lambda}$$
or
\begin{equation}
\mathbf{A}=\mathbf{U}\mathbf{\Lambda}\mathbf{U}^{-1}, \label{eigmatA}
\end{equation}
where $\boldsymbol{\Lambda}=\mathrm{diag}(\lambda_0,\lambda_1,\dots,\lambda_{N-1})$ is the diagonal matrix with the eigenvalues on its diagonal and $\mathbf{U}$ is a matrix composed of the eigenvectors, $\mathbf{u}_k$, as its columns. Since the eigenvectors, $\mathbf{u}$, are obtained by solving a homogeneous system of equations, defined by (\ref{eiganal}) and in the form $(\mathbf{A} - \lambda \mathbf{I}) \mathbf{u} =  \mathbf{0}$, one element of the eigenvector $\mathbf{u}$ can be arbitrarily chosen. The common choice is to enforce unit energy, $\left\Vert\mathbf{u}_k\right\Vert_2^2=1$, for every $k=0,1,\dots,N-1$. 

\begin{Remark} For an undirected graph, the adjacency matrix $\mathbf{A}$ is symmetric, $\mathbf{A}=\mathbf{A}^T$. Any symmetric matrix (i) has real-valued eigenvalues; (ii) is diagonalizable; and (iii) has orthogonal eigenvectors, and hence
 $$ \mathbf{U}^{-1}=\mathbf{U}^T.$$
\end{Remark} 

\begin{Remark} For directed graphs, in general, $\mathbf{A} \ne \mathbf{A}^T$. 
\end{Remark}

Recall that a square matrix is diagonalizable if all its eigenvalues are distinct (this condition is sufficient, but not necessary) or if the algebraic multiplicity of each eigenvalue is equal to its geometrical multiplicity.  

For some directed graphs, which exhibit the eigenvalues of their adjacency matrix with algebraic multiplicity higher than one, the matrix $\mathbf{A}$ may not be diagonalizable. In such cases, the algebraic multiplicity of the considered eigenvalue is higher than its geometric multiplicity  and the Jordan normal form may be used.

\medskip\noindent\textit{Definition:} The set of the eigenvalues of an adjacency matrix is called the \textit{graph adjacency spectrum}.

\begin{Remark} The spectral theory of graphs studies properties of graphs through the eigenvalues and eigenvectors of their associated adjacency and graph Laplacian matrices.
\end{Remark}

\begin{Example} For the graph presented in Fig. \ref{GSPb_ex1a}(a), the graph adjacency spectrum is given by $\lambda \in \{-2,   -1.741,   -1.285,\allowbreak   -0.677,\allowbreak   -0.411,\allowbreak    1.114,   1.809,    3.190\}$, and is shown in Fig. \ref{GSPb_spectrum2a}(top). 
	\begin{figure}
		\centering
		\includegraphics[]{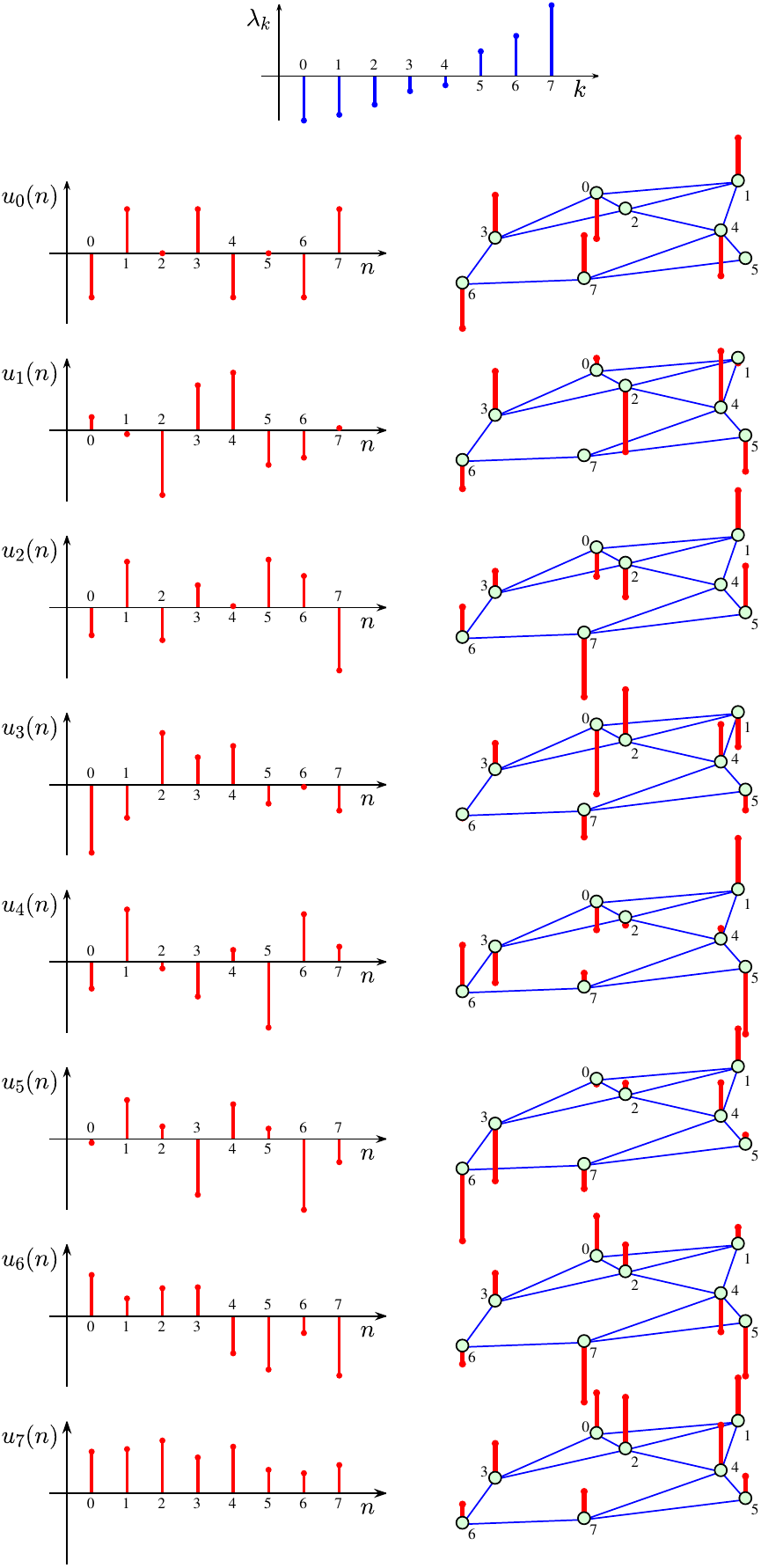}
		\caption{Eigenvalues, $\lambda_k$, for spectral indices (eigenvalue number) $k=0,1,\ldots,N-1$,  and elements of the corresponding eigenvectors, $u_k(n)$, as a function of the vertex index $n=0,1,\ldots,N-1$, for the adjacency matrix, $\mathbf{A}$, of the undirected graph presented in Fig.~\ref{GSPb_ex1a}(a). The distinct eigenvectors are shown both on the vertex index axis, $n$, (left) and on the graph itself (right).}
		\label{GSPb_spectrum2a}
	\end{figure}

\end{Example}

\begin{Example} The vertices of the graph presented in Fig. \ref{GSPb_ex1a}(a) are randomly reordered,  as shown in Fig. \ref{GSPb_spectrum2bb}. Observe that the graph adjacency spectrum, given in the same figure, retains the same values, with vertex indices of the eigenvectors reordered in the same way as the graph vertices, while the eigenvalues (spectra) retain the same order as in the original graph in Fig. \ref{GSPb_spectrum2a}. By a simple inspection we see that, for example, the eigenvector elements at the vertex index position $n=0$ in  Fig. \ref{GSPb_spectrum2a} are now at the vertex index position $n=3$ in all eigenvectors in  Fig. \ref{GSPb_spectrum2bb}. 
	\begin{figure}
		\centering
		\includegraphics[]{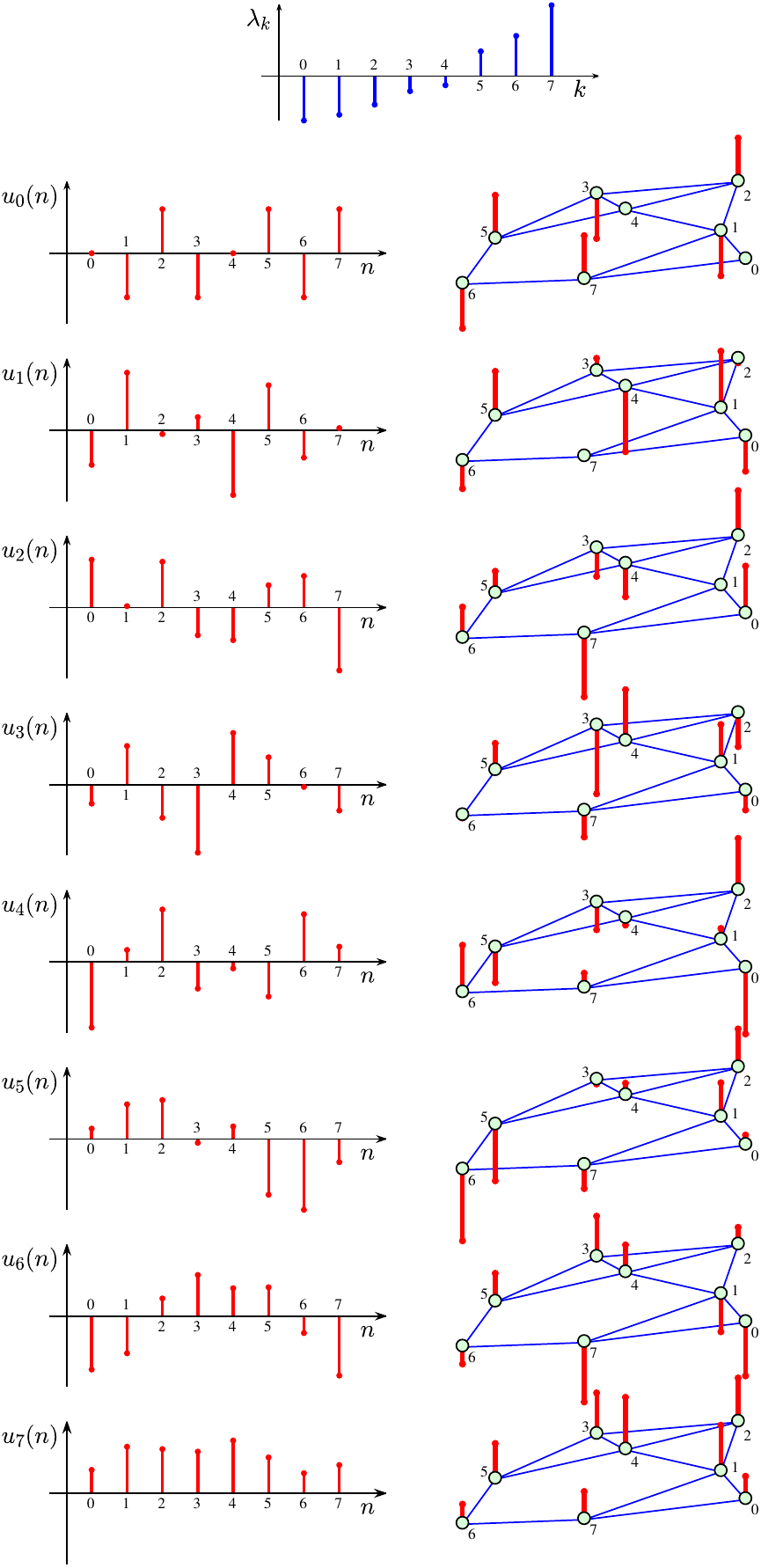}
		\caption{Eigenvalues, $\lambda_k$, for spectral indices (eigenvalue number) $k=0,1,\ldots,N-1$,  and elements of the corresponding eigenvectors, $u_k(n)$, as a function of the vertex index $n=0,1,\ldots,N-1$, for the adjacency matrix, $\mathbf{A}$, of the undirected graph presented in Fig.~\ref{GSPb_ex1a}(a) with index reordering according to the scheme $[0,1,2,3,4,5,6,7]\rightarrow[3, 2, 4, 5, 1, 0, 6, 7]$. The distinct eigenvectors are shown both on the vertex index axis, $n$, (left) and on the graph itself (right). Compare with the results for the original vertex ordering in Fig. \ref{GSPb_spectrum2a}.}
		\label{GSPb_spectrum2bb}
	\end{figure}

	\end{Example}

\begin{Remark}\label{RReordering} A unique feature of graphs is that vertex reindexing does not alter the eigenvalues of the adjacency matrix, while the corresponding eigenvectors of the reindexed adjacency matrix contain the same elements as the original eigenvectors, but reordered according to the vertex renumbering. This follows from the properties of the permutation matrix, as in relation (\ref{PermMat}).
\end{Remark}

\subsubsection{The DFT basis functions as a special case of eigenvectors of the adjacency matrix }
For  continuity with standard spectral analysis, we shall first consider directed circular graphs, as this graph topology encodes the standard time and space domains. 
 
Eigenvalue decomposition for the directed circular graph in Fig. \ref{fig:spec-graph}(g), assuming $N$ vertices, follows from the definition 
$\mathbf{A} \mathbf{u}_k = \lambda_k \mathbf{u}_k$, and the form of the adjacency matrix in (\ref{AdjMtxFirsDir}). Then, the  elements of vector $\mathbf{A} \mathbf{u}_k$ are  $u_k(n-1)$, while the elements of vector $\lambda_k \mathbf{u}_k$ are $\lambda_k u_k(n)$, to give  
\begin{equation} u_k(n-1) = \lambda_k u_k(n), \label{DFTdiffeq}
\end{equation}
where $u_k(n)$ are the elements of the eigenvector $\mathbf{u}_k$ for given vertex indices $n=0,1,\ldots,N-1$, and $k$ is the index of an eigenvector, $k=0,1,\ldots,N-1$.   This is a first-order linear difference equation, whose general form for  a discrete signal $x(n)$ is $x(n)=ax(n-1)$, and the solution of which is
\begin{equation}
u_k(n)=\frac{1}{\sqrt{N}}e^{j2\pi n k /N}  \textrm{ and }  \lambda_k=e^{-j2\pi  k /N},  \label{DFT_baisis}
\end{equation}
for $k=0,1,\dots,N-1.$ It is simple to verify that this solution satisfies the difference equation (\ref{DFTdiffeq}). Since the graph is circular, the eigenvectors  also exhibit circular behavior, that is, $u_k(n)=u_k(n+N)$.  For convenience, a unit energy condition is used to find the constants within the general solution of this first-order linear difference equation. Observe that the eigenvectors in (\ref{DFT_baisis}) correspond exactly to the standard DFT harmonic basis functions. 

\begin{Remark} Classic DFT analysis may be obtained as a special case of the graph spectral analysis in (\ref{DFT_baisis}), when  considering directed circular graphs. Observe that for circular graphs, the adjacency matrix plays the role of a shift operator, as seen in (\ref{DFTdiffeq}), with the elements of $\mathbf{A} \mathbf{u}_k$ equal to $u_k(n-1)$. This property will be used to define the shift operator on a graph in the following sections.   
\end{Remark}

 \subsubsection{Decomposition of graph product adjacency matrices}
 
 We have already seen in Fig. \ref{fig:cronecker}  and Fig. \ref{fig:cartesian} that complex graphs, for example those with a three-dimensional vertex space, may be obtained as a Kronecker (tensor) product or a Cartesian (graph) product of two disjoint graphs $\mathcal{G}_1$ and $\mathcal{G}_2$. Their respective adjacency matrices, $\mathbf{A}_1 $ and  $\mathbf{A}_2$, are correspondingly combined into the adjacency matrices of the Kronecker graph product, $\mathbf{A}_{\otimes}=\mathbf{A}_1 \otimes \mathbf{A}_2$
and the Cartesian graph product,  $\mathbf{A}_{\oplus}=\mathbf{A}_1 \oplus \mathbf{A}_2$, as described in properties $M_{14}$ and $M_{15}$. 

For the eigen-decomposition of the Kronecker product of matrices $\mathbf{A}_1 $ and  $\mathbf{A}_2$, the following holds
\begin{gather*}\mathbf{A}_{\otimes}=\mathbf{A}_1 \otimes \mathbf{A}_2=(\mathbf{U}_1\mathbf{\Lambda}_1\mathbf{U}_1^{H})\otimes(\mathbf{U}_2\mathbf{\Lambda}_2\mathbf{U}_2^{H})\\
=(\mathbf{U}_1\otimes\mathbf{U}_2)(\mathbf{\Lambda}_1\otimes\mathbf{\Lambda}_2)(\mathbf{U}_1\mathbf{U}_2)^{H},
\end{gather*}
or in other words, the eigenvectors of the adjacency matrix of the Kronecker product of graphs are obtained by a Kronecker product of the eigenvectors of the adjacency matrices of individual graphs, as
$\mathbf{u}_{k+lN_1}=\mathbf{u}^{(A_1)}_k\otimes \mathbf{u}^{(A_2)}_l$, $k=0,1,2,\dots,N_1-1$, $l=0,1,2,\dots,N_2-1$. 

\begin{Remark}The eigenvectors of the individual graph adjacency matrices, $\mathbf{u}^{(A_1)}_k$ and $\mathbf{u}^{(A_2)}_k$, are of much lower dimensionality than those of the adjacency matrix of the resulting graph Kronecker product. This property can be used to reduce computational complexity when analyzing data observed on this kind of graph. The eigenvalues of the resulting graph adjacency matrix are equal to the product of the eigenvalues of adjacency matrices of the constituent graphs, $\mathcal{G}_2$ and $\mathcal{G}_2$, that is,
	$$\lambda_{k+lN_1}=\lambda_k^{(A_1)}\lambda_l^{(A_2)}.$$   
\end{Remark}
 
 The eigen-decomposition of the adjacency matrix of the Cartesian product of graphs, whose respective adjacency matrices are $\mathbf{A}_1 $ and  $\mathbf{A}_2$, is of the form 
 \begin{gather*}\mathbf{A}_{\oplus}=\mathbf{A}_1 \oplus \mathbf{A}_2
 =(\mathbf{U}_1\otimes\mathbf{U}_2)(\mathbf{\Lambda}_1\oplus\mathbf{\Lambda}_2)(\mathbf{U}_1\mathbf{U}_2)^{H}.
 \end{gather*}
 with
$\mathbf{u}_k=\mathbf{u}^{(A_1)}_k\otimes \mathbf{u}^{(A_2)}_k$ and $\lambda_{k+lN_1}=\lambda_k^{(A_1)}+\lambda_l^{(A_2)}$, $k=0,1,2,\dots,N_1-1$, $l=0,1,2,\dots,N_2-1$. 

\begin{Remark}
The Kronecker product and the Cartesian product of graphs share the same eigenvectors of their adjacency matrices, while their spectra (eigenvalues) are different.  
\end{Remark}

\begin{Example} The basis functions of classic two-dimensional (image) 2D-DFT follow from the spectral analysis of a Cartesian graph product which is obtained as a product the circular directed graph from Fig. \ref{fig:spec-graph} with itself. Since from (\ref{DFT_baisis}), the eigenvector elements of each graph are $u_k(n)=e^{j2\pi n k /N}/\sqrt{N}$,  then the elements of the resulting basis functions are given by $$u_{k+lN}(m+nN)=\frac{1}{N}e^{j2\pi mk /N}e^{j2\pi n l /N},$$ for $k=0,1,\dots,N-1$, $l=0,1,\dots,N-1$, $m=0,1,\dots,N-1$, and $n=0,1,\dots,N-1$. Fig. \ref{torus} illustrates the Cartesian product of two circular undirected graphs with $N_1=N_2=8$.

\begin{figure}
	\centering
	\includegraphics[trim={2.5cm 3.0cm 2cm 2.5cm},clip]{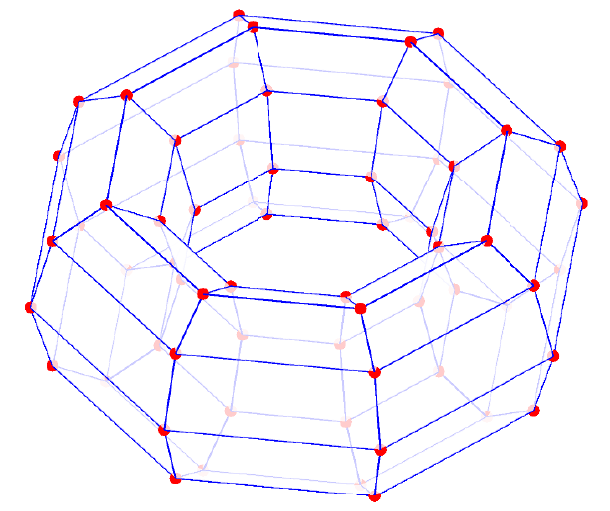}
	\caption{Cartesian graph product of two planar circular unweighted graphs, with $N=8$ vertices, produces a three-dimensional torus topology.} 
	\label{torus}
\end{figure}
\end{Example}

	\begin{Remark}
		Cartesian products of graphs may be used for a multidimensional extension of  vertex spaces and graph data domains, whereby the resulting eigenvectors (basis functions) can be efficiently calculated using the eigenvectors of the original graphs, which are of lower dimensionality.  
	\end{Remark}

\subsubsection{Decomposition of matrix powers and polynomials}
 
From the eigendecomposition of the adjacency matrix $\mathbf{A}$ in (\ref{eigmatA}), eigenvalue decomposition of the squared  adjacency matrix,
$\mathbf{A}\mathbf{A}=\mathbf{A}^2$, is given by  
$$\mathbf{A}^2=\mathbf{U}\mathbf{\Lambda}\mathbf{U}^{-1}\mathbf{U}\mathbf{\Lambda}\mathbf{U}^{-1}=\mathbf{U}\mathbf{\Lambda}^2\mathbf{U}^{-1},$$
under the assumption that $\mathbf{U}^{-1}$ exists. 
For an arbitrary natural number, $m$, the above result generalizes straightforwardly to \begin{equation}
\mathbf{A}^m=\mathbf{U}\mathbf{\Lambda}^m\mathbf{U}^{-1}. \label{matxpow}
\end{equation}
 Further, for any matrix function, $f(\mathbf{A})$, that can be written in a polynomial form, given by
 $$
f(\mathbf{A})=h_0\mathbf{A}^0+h_1\mathbf{A}^1+h_2\mathbf{A}^2+\cdots+h_{N-1}\mathbf{A}^{N-1},
 $$
its eigenvalue decomposition is, in general,  given by 
$$f(\mathbf{A})=\mathbf{U}f(\mathbf{\Lambda})\mathbf{U}^{-1}.$$
This is self-evident from the properties of eigendecomposition of matrix powers,  defined in (\ref{matxpow}), and the linearity of the matrix multiplication operator, $\mathbf{U}(h_0\mathbf{A}^0+h_1\mathbf{A}^1+h_2\mathbf{A}^2+\cdots+h_{N-1}\mathbf{A}^{N-1})\mathbf{U}^{-1}$.

\subsection{Eigenvalue Decomposition of the graph Laplacian}

Spectral analysis for graphs can also be performed based on the graph Laplacian, $\mathbf{L}$, defined in (\ref{LapDef}). For convenience, we here adopt the same notation for the eigenvalues and eigenvectors of the Laplacian, as we did for the adjacency matrix $\mathbf{A}$, although the respective eigenvalues and eigenvectors are not directly related. The Laplacian of an undirected graph can be therefore written  as      
$$\mathbf{L}=\mathbf{U}\mathbf{\Lambda}\mathbf{U}^{T} \,\,\, \text{ or }  \,\,\, \mathbf{L}\mathbf{U}=\mathbf{U}\mathbf{\Lambda}, $$
where $\boldsymbol{\Lambda}=\mathrm{diag}(\lambda_0,\lambda_1,\dots,\lambda_{N-1})$ is a diagonal matrix with the Laplacian eigenvalues and $\mathbf{U}$ the orthonormal matrix of its eigenvectors (in columns), with $\mathbf{U}^{-1}=\mathbf{U}^{T}$. Note that the Laplacian of an undirected graph is always diagonalizable,  since $\mathbf{L}$ is a real symmetric matrix. 

Then, every eigenvector,  $\mathbf{u}_k$, $k=0,1,\dots,N-1$, of a graph Laplacian, $\mathbf{L}$, satisfies
\begin{equation}
\mathbf{L} \mathbf{u}_k = \lambda_k \mathbf{u}_k.
\end{equation}

\medskip\noindent\textit{Definition:} The set of the eigenvalues, $\lambda_k$, $k=0,1,\dots,N-1$, of the graph Laplacian is referred to as \textit{the graph spectrum} or \textit{graph Laplacian spectrum} (\textit{cf.} graph adjacency spectrum based on $\mathbf{A}$).

 \begin{Example} 
 		The Laplacian spectrum of the undirected graph from Fig. \ref{GSPb_ex2}, is given by $$\lambda \in \{ 0,     0.29,    0.34,    0.79,    1.03,    1.31,\allowbreak    1.49,    2.21
 		\},$$  
 		 and shown in  Fig. \ref{GSPb_spectrum3a}, along with the corresponding eigenvectors. The Laplacian spectrum of the disconnected graph from Fig. \ref{GSPb_ex1bW}, is given by
 		$$\lambda \in \{0,   0,    0.22,    0.53,    0.86,    1.07,    1.16,    2.03 \},$$ 
 		and is illustrated in Fig. \ref{GSPb_spectrum3b}.  The disconnected graph is characterized by the zero eigenvalue of algebraic multiplicity $2$, that is, $\lambda_0=\lambda_1=0$.

\begin{figure}
	\centering
	\includegraphics[]{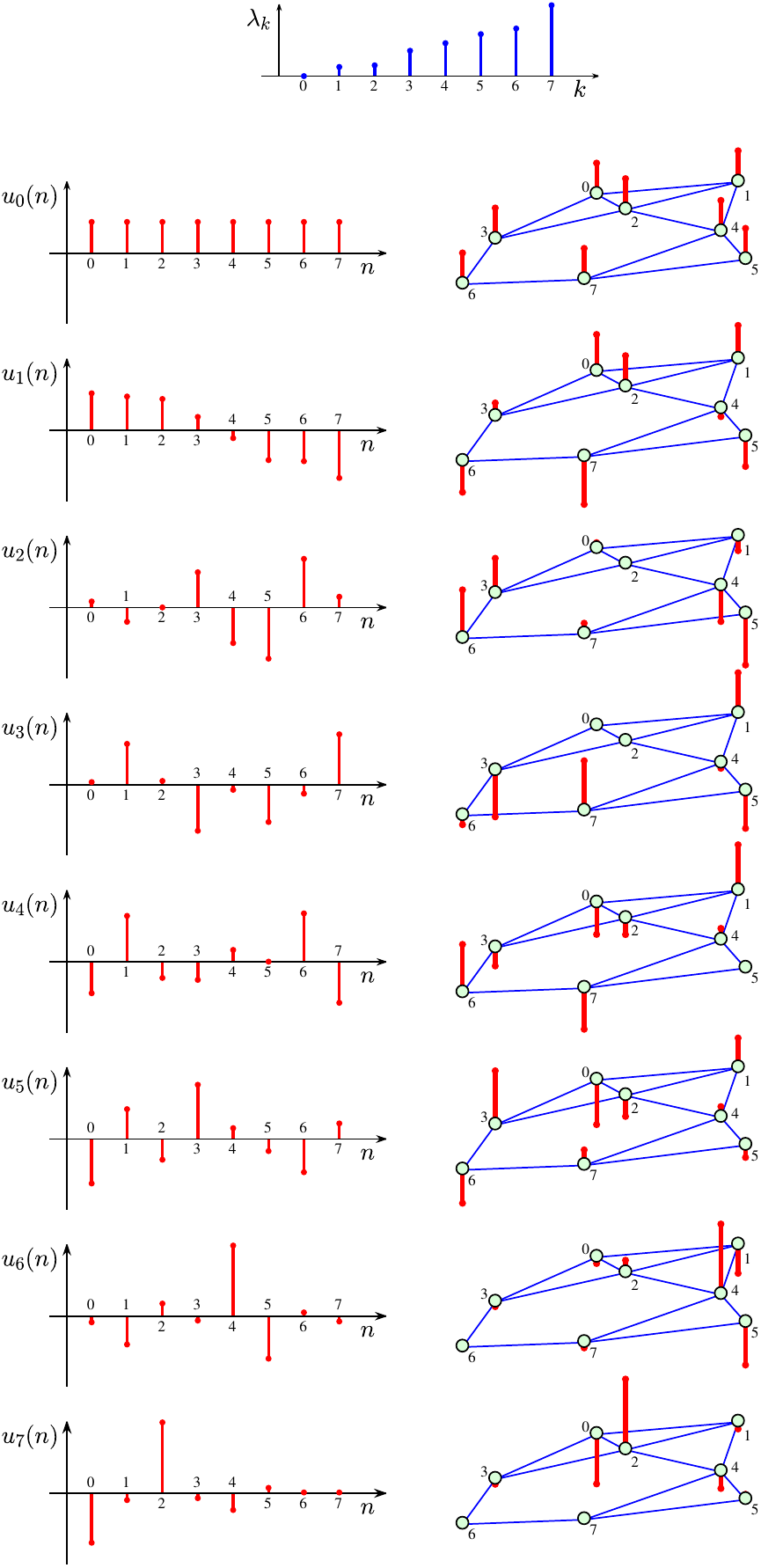}
	\caption{Eigenvalues, $\lambda_k$, for spectral indices (eigenvalue number) $k=0,1,\ldots,N-1$,  and elements of the corresponding eigenvectors, $u_k(n)$, as a function of the vertex index $n=0,1,\ldots,N-1$, for the Laplacian matrix, $\mathbf{L}$, of the undirected graph presented in Fig.~\ref{GSPb_ex2}. The distinct eigenvectors are shown both on the vertex index axis, $n$, (left) and on the graph itself (right). A comparison with the eigenvectors of the adjacency matrix in Fig. \ref{GSPb_spectrum2a}, shows that for the adjacency matrix the smoothest eigenvector corresponds to the largest eigenvalue, while for the graph Laplacian the smoothest eigenvector corresponds to the smallest eigenvalue, $\lambda_0$. } 
	\label{GSPb_spectrum3a}
\end{figure}

\begin{figure}
	\centering
	\includegraphics[]{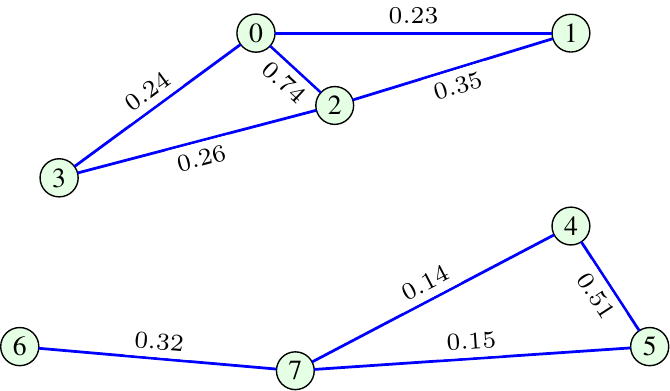}
	\caption{A disconnected weighted graph which consists of two sub-graphs.}
	\label{GSPb_ex1bW}
\end{figure}

\begin{figure}
	\centering
	\includegraphics[]{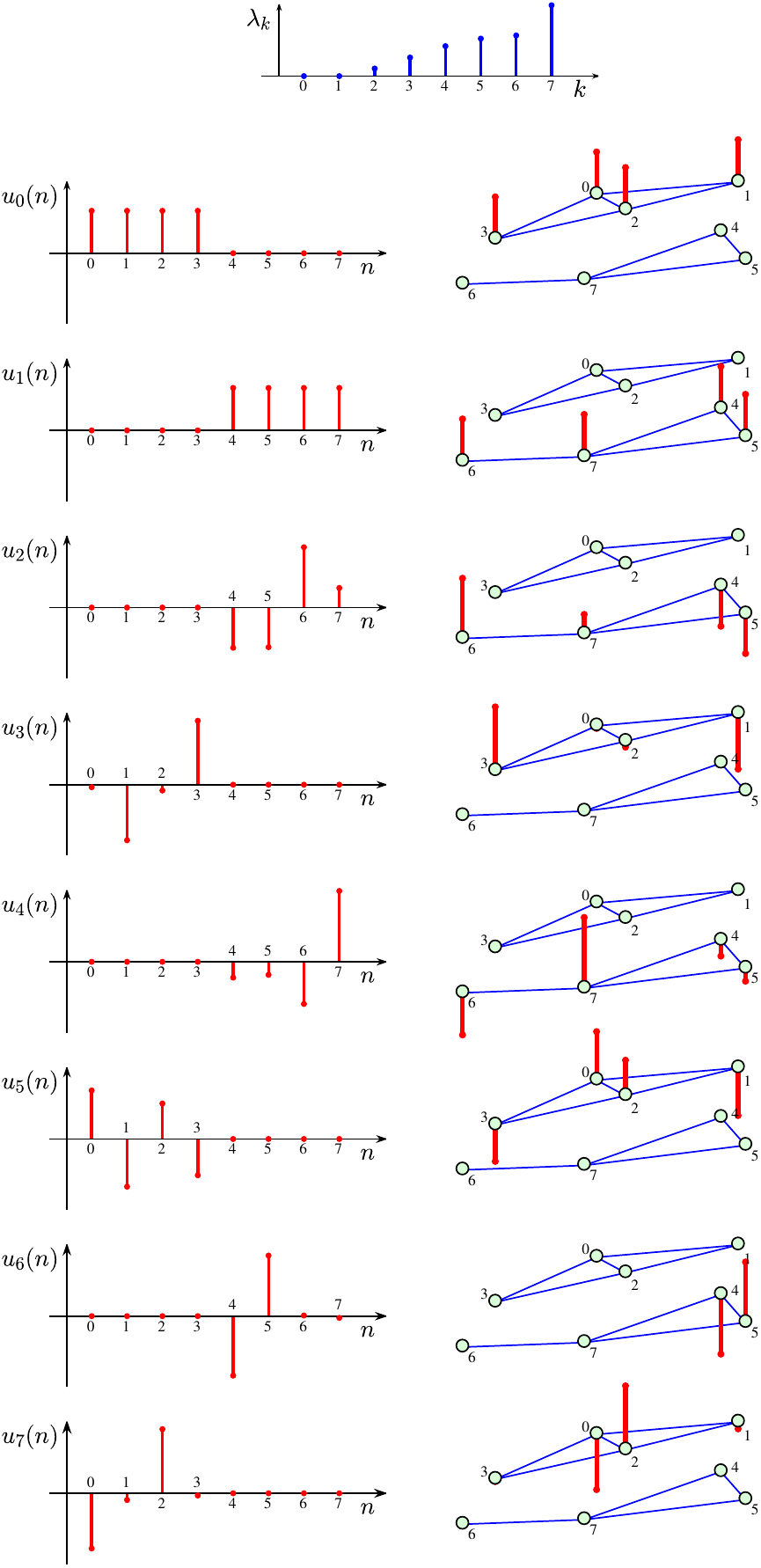}
	\caption{Eigenvalues, $\lambda_k$, for spectral indices (eigenvalue number) $k=0,1,\ldots,N-1$,  and elements of the corresponding  eigenvectors, $u_k(n)$, as a function of the vertex index $n=0,1,\ldots,N-1$, for the Laplacian matrix, $\mathbf{L}$, of the undirected graph presented in Fig. \ref{GSPb_ex1bW}. The distinct eigenvectors are shown both on the vertex index axis, $n$, (left) and on the graph itself (right).  This graph is characterized with the zero eigenvalue of algebraic multiplicity $2$, that is, $\lambda_0=\lambda_1=0$. Observe that for every spectral index, $k$, the corresponding eigenvectors take nonzero values on only one of the disconnected graph components.}
	\label{GSPb_spectrum3b}
\end{figure}
\end{Example}

\begin{Remark}
	Observe that when graph-component (sub-graph) based vertex indexing is employed, even though the respective graph spectra for the connected graph in  Fig. \ref{GSPb_spectrum3a} and the disconnected graph Fig. \ref{GSPb_spectrum3b} are similar, the eigenvectors for a given spectral index of a disconnected graph take nonzero values on only one of the individual disconnected graph components. 
\end{Remark}

\subsubsection{Properties of Laplacian eigenvalue decomposition}
\begin{enumerate}[label=\subscript{L}{{\arabic*}}:] 
\item 
The Laplacian matrix is defined in (\ref{LapDef}) in such a way that the sum of  elements in its each row (column) is zero. As a consequence, this enforces the inner products of every row of $\mathbf{L}$ with any constant vector, $\mathbf{u}$, to be zero-valued, that is, $\mathbf{Lu}=\mathbf{0}=0 \cdot \mathbf{u}$, for any constant vector $\mathbf{u}$.  This means that  at least one eigenvalue of the Laplacian is zero, $\lambda_0=0$,
and its corresponding constant unit energy eigenvector is given by $\mathbf{u}_0=[1,1,\ldots, 1]^T/\sqrt{N}=\mathbf{1}/\sqrt{N}$.  

\bigskip \item
The multiplicity of the eigenvalue $\lambda_0=0$ of the Laplacian is equal to the number of connected components (connected subgraphs) in the corresponding graph. 

This property follows from the fact that the Laplacian matrix of disconnected graphs can be written in a block diagonal form, as in (\ref{BlckDL}). \textit{The set of eigenvectors of a block-diagonal matrix is obtained by grouping together the sets of eigenvectors of individual block submatrices.} Since each subgraph of a disconnected graph behaves as an independent graph, then for each subgraph $\lambda_0=0$ is the eigenvalue of the corresponding block Laplacian submatrix, according to property $L_1$. Therefore, the multiplicity of the eigenvalue $\lambda_0=0$ corresponds to the number of disjoint components (subgraphs) within a graph. 

This property does not hold for the adjacency matrix, since \textit{there are no common eigenvalues}  in the adjacency matrices for the blocks (subgraphs) or arbitrary graphs, like in the case of $\lambda_0=0$  for the graph  Laplacian matrix and any graph. \textit{In this sense, the graph Laplacian matrix carries more physical meaning than the corresponding adjacency matrix.} 

\begin{Remark}\label{RL2} If $\lambda_0=\lambda_1=0$, then the graph is not connected. If $\lambda_2>0$, then there are exactly two individually connected but globally disconnected components in this graph. If $\lambda_1 \ne 0$ then this eigenvalue may be used to describe the so called \textit{algebraic connectivity of a graph}, whereby very small values of $\lambda_1$ indicate that the graph is weakly connected. This can be used as an indicator of the possibility of graph segmentation, as elaborated in Section \ref{ClustSEG}.  
\end{Remark}

\bigskip \item
As with  any other matrix, the sum of the eigenvalues of  the Laplacian matrix is equal to its trace . For the normalized Laplacian, the sum of its eigenvalues is equal to the number of vertices, $N$, if there are no isolated vertices.

\bigskip \item
The coefficient, $c_{N}$, in the characteristic polynomial of the graph Laplacian matrix
$$
P(\lambda)=\det|\mathbf{L}-\lambda\mathbf{I}|=\lambda^N+c_1\lambda^{N-1}+\cdots+c_{N-1}\lambda+c_N
$$
 is equal to $0$, since $\lambda=0$ is an eigenvalue for the Laplacian matrix. 
 
 For unweighted graphs, the coefficient $c_1$ is equal to the number of edges multiplied by $-2$. This is straightforward to show following the relations from property $P_4$ which state that $c_1=-\mathrm{tr}\{\mathbf{L}\}$. For unweighted graphs, the diagonal elements of the Laplacian are equal to the corresponding vertex degrees (number of edges). Therefore, the number of edges in an unweighted graph is equal to $-c_1/2$.

\begin{Example} The characteristic polynomial of the Laplacian for the graph from Fig.~\ref{GSPb_ex1a}(a) is given by
\begin{align*}
P(\lambda)&=
\lambda^8         -24\lambda^7     +    238\lambda^6    -1256\lambda^5   +     3777\lambda^4    \\   &   -6400\lambda^3  +5584 \lambda^2      -1920\lambda 
\end{align*}
with the eigenvalues $\lambda \in \{       0,
5.5616,
5,
4,
4,
3,
1,
1\}$.
Observe that the eigenvalues $\lambda=1$ and $\lambda=4$ are of multiplicity higher than one. The minimal polynomial therefore becomes
$P_{min}(\lambda)= \lambda^6   -19\lambda^5 +  139\lambda^4  -485\lambda^3 +  796\lambda^2  -480\lambda$.

For the disconnected graph in Fig.~\ref{GSPb_ex1b}, the characteristic polynomial of the Laplacian is given by
$$
P(\lambda)=
\lambda^8 - 18\lambda^7 + 131\lambda^6 - 490\lambda^5 + 984\lambda^4 - 992\lambda^3 + 384\lambda^2,
$$
with the eigenvalues
$
\lambda \in \{ 0, 0, 1, 2, 3, 4, 4, 4 \}
$.
The eigenvalue $\lambda=0$ is of algebraic multiplicity $2$ and the eigenvalue $\lambda=4$  of algebraic multiplicity $3$, so that the minimal polynomial takes the form
$$
P_{min}(\lambda)=
\lambda^5 - 10\lambda^4 + 35\lambda^3 - 50\lambda^2 + 24\lambda
$$
Since the eigenvalue $\lambda=0$ is of algebraic multiplicity $2$, property  $L_2$ indicates that this graph is disconnected, with two disjoint sub-graphs as its constituent components. 
\end{Example}

\bigskip \item Graphs with identical spectra are called \textit{isospectral} or \textit{cospectral} graphs. However, isospectral graphs are not necessary isomorphic, and construction of isospectral graphs that are not isomorphic is an important topic in graph theory. 

\begin{Remark}\label{RLC}
A complete graph is  \textit{uniquely determined by its Laplacian spectrum} \cite{van2003graphs}. The Laplacian spectrum of a complete unweighted graph, with $N$ vertices, is $\lambda_k \in \{0,
	N,
	N, \dots, N \}$.  Therefore, two complete isospectral graphs are also isomorphic. 
\end{Remark}

\bigskip \item
For a $\mathcal{J}$-regular graph, as in Fig. \ref{fig:spec-graph}(c), the eigenvectors of the Laplacian and the adjacency matrices  are identical, with the following relation for the eigenvalues,
$$\lambda^{(L)}_k=\mathcal{J}-\lambda^{(A)}_k,$$
where the superscript $L$ designates the Laplacian and superscript $A$ the corresponding adjacency matrix. This follows directly from $\mathbf{U}^T\mathbf{L}\mathbf{U}=\mathbf{U}^T(\mathcal{J}\mathbf{I}-\mathbf{A})\mathbf{U}$.

\bigskip \item 
Eigenvalues of the normalized Laplacian, $ \mathbf{L}_N=\mathbf{I}-\mathbf{D}^{-1/2}\mathbf{A}\mathbf{D}^{-1/2}$,  are nonnegative and upper-bounded by
$$0\le \lambda \le 2.$$
The equality for the upper bound holds if and only if the graph is a bipartite graph, as in Fig. \ref{fig:spec-graph}(b). This will be proven within the next property. 

\bigskip \item 
The eigenvalues and eigenvectors of the normalized Laplacian  of a bipartite graph, with the disjoint sets of vertices $\mathcal{E}$ and $\mathcal{H}$, satisfy the relation, referred to as \textit{the graph spectrum folding}, given by
\begin{gather}
\lambda_k=2-\lambda_{N-k} \label{BGeig} \\
\mathbf{u}_k=
\begin{bmatrix}
\mathbf{u}_{\mathcal{E}} \\
\mathbf{u}_{\mathcal{H}}
\end{bmatrix}
\,\,\,\,\,\, \textrm{ and } \,\,\,\,\,\,
\mathbf{u}_{N-k}=
\begin{bmatrix}
\mathbf{u}_{\mathcal{E}} \\
-\mathbf{u}_{\mathcal{H}}
\end{bmatrix},
\end{gather}
where $\mathbf{u}_k$ designates the $k$-th eigenvector of a bipartite graph,   $\mathbf{u}_{\mathcal{E}}$ is  its part indexed on the first set of vertices, $\mathcal{E}$, while $ \mathbf{u}_{\mathcal{H}}$ is the part of the eigenvector $\mathbf{u}_k$ indexed on the second set of vertices, $\mathcal{H}$.

In order to prove this property,  we shall write the adjacency and the normalized Laplacian matrices of an undirected bipartite graph in  their block forms 
$$
\mathbf{A}
=\begin{bmatrix}
\mathbf{0} &  \mathbf{A}_{\mathcal{E}\mathcal{H}}\\
\mathbf{A}_{\mathcal{E}\mathcal{H}}^T &  \mathbf{0}
\end{bmatrix}\quad
\text{ and }\quad
\mathbf{L}_N
=\begin{bmatrix}
\mathbf{I} &  \mathbf{L}_{\mathcal{E}\mathcal{H}}\\
\mathbf{L}_{\mathcal{E}\mathcal{H}}^T &  \mathbf{I}
\end{bmatrix}.
$$
The eigenvalue relation, $
\mathbf{L}_N \mathbf{u}_k=\lambda_k\mathbf{u}_k$, can now be evaluated as
$$
\mathbf{L}_N \mathbf{u}_k
=\begin{bmatrix}
\mathbf{u}_{\mathcal{E}} +  \mathbf{L}_{\mathcal{E}\mathcal{H}} \mathbf{u}_{\mathcal{H}}\\
 \mathbf{L}_{\mathcal{E}\mathcal{H}}^T \mathbf{u}_{\mathcal{E}} + \mathbf{u}_{\mathcal{H}} 
\end{bmatrix}
= \lambda_k 
\begin{bmatrix}
\mathbf{u}_{\mathcal{E}} \\
\mathbf{u}_{\mathcal{H}} 
\end{bmatrix}.
$$
From there,  we have $\mathbf{u}_{\mathcal{E}} +  \mathbf{L}_{\mathcal{E}\mathcal{H}} \mathbf{u}_{\mathcal{H}}=\lambda_k \mathbf{u}_{\mathcal{E}}$ and $ \mathbf{L}_{\mathcal{E}\mathcal{H}}^T \mathbf{u}_{\mathcal{E}} + \mathbf{u}_{\mathcal{H}}=\lambda_k \mathbf{u}_{\mathcal{H}}$, resulting in $ \mathbf{L}_{\mathcal{E}\mathcal{H}} \mathbf{u}_{\mathcal{H}} =(\lambda_k-1) \mathbf{u}_{\mathcal{E}}$ and $ \mathbf{L}_{\mathcal{E}\mathcal{H}}^T \mathbf{u}_{\mathcal{E}}=(\lambda_k-1) \mathbf{u}_{\mathcal{H}} $, to finally yield
$$\mathbf{L}_N
\begin{bmatrix}
\mathbf{u}_{\mathcal{E}} \\
-\mathbf{u}_{\mathcal{H}}
\end{bmatrix}
=(2-\lambda_k)
\begin{bmatrix}
\mathbf{u}_{\mathcal{E}} \\
-\mathbf{u}_{\mathcal{H}}
\end{bmatrix}.
$$
This completes the proof. 

Since for the graph Laplacian $\lambda_0=0$ always holds (see the property $L_1$), from $\lambda_k=2-\lambda_{N-k}$ in (\ref{BGeig}), it then follows that the largest eigenvalue is $\lambda_N=2$, which also proves the property $L_7$ for a bipartite graph. 
\end{enumerate}

\subsubsection{Fourier analysis as a special case of the Laplacian spectrum}

Consider the undirected circular graph from Fig.~\ref{fig:spec-graph}(e). Then, from the property $L_1$, the eigendecomposition relation for the Laplacian of this graph, $\mathbf{L}\mathbf{u}=\lambda \mathbf{u}$, admits a simple form
\begin{align}
-u(n-1)+2u(n)-u(n+1)=\lambda u(n). \label{sofde}
\end{align}
This is straightforward to show by inspecting the Laplacian for the  undirected circular graph from Fig.~\ref{fig:spec-graph}(e), with $N=8$ vertices for which the eigenvalue analysis is based on
\begin{equation}
\mathbf{Lu}=
\left[
\begin{array}{rrrrrrrr}
2 \! & \!-1 \! & \! 0 \! & \! 0 \! & \! 0 \! & \! 0 \! & \! 0 \! & \! -1\\
-1 \! & \! 2 \! & \!-1 \! & \! 0 \! & \! 0 \! & \! 0 \! & \! 0 \! & \! 0\\
0 \! & \!-1 \! & \! 2 \! & \!-1 \! & \! 0 \! & \! 0 \! & \! 0 \! & \! 0\\
0 \! & \! 0 \! & \!-1 \! & \! 2 \! & \! -1 \! & \! 0 \! & \! 0 \! & \! 0\\ 
0 \! & \! 0 \! & \! 0 \! & \! -1 \! & \! 2 \! & \!-1 \! & \! 0 \! & \! 0\\
0 \! & \! 0 \! & \! 0 \! & \! 0 \! & \!-1 \! & \! 2 \! & \! -1 \! & \! 0\\
0 \! & \! 0 \! & \! 0 \! & \! 0 \! & \! 0 \! & \! -1 \! & \! 2 \! & \! -1 \\
-1 \! & \! 0 \! & \! 0 \! & \! 0 \! & \! 0 \! & \! 0 & \! -1 \! & \! 2
\end{array}
\right]
\left[
\begin{array}{r}
u(0)\\
u(1)\\
u(2) \\
u(3) \\ 
u(4) \\
u(5) \\
u(6)  \\
u(7)
\end{array}
\right].
\end{equation}
This directly gives the term $-u(n-1)+2u(n)-u(n+1)$, while a simple inspection of the values $u(0)$ and $u(N)$ illustrates the circular nature of the eigenvectors; see also Remark \ref{ReCir}.  
 The solution to the second order difference equation in (\ref{sofde}) is 
 $u_k(n)=\cos( \frac{2 \pi k n}{N}+\phi_k)$, with $\lambda_{k}=2(1-\cos(\frac{2\pi k}{N})).$ Obviously, for every eigenvalue, $\lambda_k$ (except for $\lambda_0$ and for the last eigenvalue, $\lambda_{N-1}$, for an even $N$), we can choose to have two orthogonal eigenvectors with, for example, $\phi_k=0$ and $\phi_k=\pi /2$. This means that most of the eigenvalues are of algebraic multiplicity $2$, i.e., $\lambda_1=\lambda_2$, $\lambda_3=\lambda_4$, and so on. This eigenvalue multiplicity of two can be formally expressed as
$$\lambda_k=
\begin{cases}
2-2\cos(\pi (k+1)/N), & \text{ for odd } k=1,3,5,\dots,  \\
2-2\cos(\pi k/N), & \text{ for even } k=2,4,6,\dots.
\end{cases}
$$
 For an odd $N$, $\lambda_{N-2}=\lambda_{N-1}$, whereas for an even $N$ we have $\lambda_{N-1}=2$ which is of algebraic multiplicity $1$.

The corresponding eigenvectors $\mathbf{u}_0$, $\mathbf{u}_1$, \ldots, $\mathbf{u}_{N-1}$, then have the form
\begin{gather}
u_k(n)=
\begin{cases}
\sin(\pi (k+1) n /N,) & \text{ for odd }k,\ k<N-1 \\
\cos(\pi k n /N), & \text{ for even } k \\
\cos(\pi n), & \text{ for odd } k,\ k=N-1, \\
\end{cases} \label{GFT_lap}
\end{gather}
where $k=0,1,\ldots,N-1$ and $n=0,1,\ldots,N-1$.

Recall that an arbitrary linear combination of eigenvectors $\mathbf{u}_{2k-1}$ and $\mathbf{u}_{2k}$, $1\le k <N/2$, is also an eigenvector since the corresponding eigenvalues are equal (in this case their algebraic and the geometric multiplicities are both equal to $2$). With this in mind, we can rewrite the full set of the eigenvectors in an alternative compact form, given by
\begin{gather*}
u_k(n)=
\begin{cases}
1, & \text{ for } k=0 \\
\exp(j\pi (k+1) n /N), & \text{ for odd }k,\ k<N-1 \\
\exp(-j\pi k n /N),   & \text{ for even } k,\ k>0 \\
\cos(\pi n), & \text{ for odd } k,\ k=N-1, \\
\end{cases} 
\end{gather*}
where $j^2=-1$. It is now clearly evident that, as desired, this set of eigenvectors is orthonormal, and that the individual eigenvectors, $\mathbf{u}_k$, correspond to the standard harmonic basis functions within the standard temporal/spatial DFT.


\section{Vertex Clustering and Mapping}

\medskip\noindent\textit{Definition:}  Vertex clustering is a type of graph learning which aims to group together vertices from the set $\mathcal{V}$ into multiple disjoint subsets, $\mathcal{V}_i$, called clusters. Vertices which are clustered into a subset of vertices, $\mathcal{V}_i$, are expected to exhibit  a larger degree of within-cluster mutual similarity (in some sense) than with the vertices in other subsets, $\mathcal{V}_j$, $j\ne i$. 

While the clustering of graph vertices refers to the process of identifying and arranging the vertices of a graph into nonverlapping vertex subsets, with data in each subset expected to exhibit relative similarity in same sense, the \textit{segmentation} of a graph refers to its partitioning into graph segments (components). 
	
The notion of \textit{vertex similarity metrics} and their use to accordingly cluster the vertices into sets, $\mathcal{V}_i$, of “related” vertices in graphs, has been a focus of significant research effort in machine learning and pattern recognition; this has resulted in  a number of established vertex similarity measures and corresponding methods for  graph clustering \cite{schaeffer2007graph}. These can be considered within two main categories (i) \textit{clustering based on graph topology} and (ii) \textit{spectral (eigenvector-based) methods for graph clustering}.

Notice that in traditional clustering, a vertex is assigned to one cluster  only. The type of clustering where a vertex may belong to more than one cluster is referred to as \textit{fuzzy clustering} \cite{schaeffer2007graph,mordeson2012fuzzy},  an approach that is not yet widely accepted in the context of graphs.

\subsection{Clustering based on graph topology}\label{NormCutsSec} 

Among many such existing methods, the most popular ones are based on:
\begin{itemize}
	\item
Finding  the minimum set of edges whose removal would disconnect a graph in some \textquotedblleft optimal\textquotedblright \, way (\textit{minimum cut} based clustering). 
\item 
Designing clusters within a graph based on the disconnection  of vertices or edges  which belong to the highest numbers of shortest paths in the graph (\textit{vertex betweenness} and  \textit{edge betweenness} based clustering). 
\item The minimum spanning tree of a graph has been a basis for a number of widely used clustering methods  \cite{kleinberg2006algorithm,morris1986graph}. 
\item
Analysis of highly connected subgraphs (HCS) \cite{khuller1998approximation} has also been  used for graph clustering. 
\item 
Finally, \textit{graph data analysis} may be used for \textbf{machine learned graph clustering}, like for example, the $k$-means based clustering methods \cite{jain2010data,dhillon2004kernel}. 
\end{itemize}

\subsubsection{Minimum cut}  
We shall first briefly review the notion of graph cuts, as spectral methods for graph clustering may be introduced and interpreted  based on the analysis and approximation of the (graph topology-based) minimum cut clustering.
	 
\noindent\textit{Definition:} Consider an undirected graph which is defined by a set of vertices, $\mathcal{V}$, and the corresponding set of  edge weights, $\mathcal{W}$. Assume next that the vertices are grouped into $k=2$ disjoint subsets of vertices, $\mathcal{E} \subset \mathcal{V}$ and $\mathcal{H} \subset \mathcal{V}$, with  $\mathcal{E} \cup \mathcal{H}=\mathcal{V}$ and $\mathcal{E} \cap \mathcal{H}=\emptyset$. A cut of this graph, for the the given subsets of vertices, $\mathcal{E}$ and $\mathcal{H}$,  is equal to a sum of all weights that correspond to the edges which connect the vertices between the subsets, $\mathcal{E}$ and $\mathcal{H}$, that is 
\[
Cut(\mathcal{E},\mathcal{H})=\sum_{\substack{m \in \mathcal{E} \\ n \in \mathcal{H} }} W_{mn}.
\]   
\begin{Remark}
For clarity, we shall focus on the case with $k=2$ disjoint subsets of vertices. However, the analysis can be straightforwardly generalized to $k \ge 2$ disjoint subsets of vertices and the corresponding minimum $k$-cuts. 
\end{Remark}

\begin{Example}\label{ExampleCUT}
Consider the graph in Fig. \ref{GSPb_ex2}, and the sets of vertices $\mathcal{E}=\{0,1,2,3\}$ and $\mathcal{H}=\{4,5,6,7\}$, shown in Fig. \ref{GSPb_ex2CUT}. Its cut into the two components (sub-graphs), $\mathcal{E}$ and $\mathcal{H}$,  involves the weights of all edges which exist between these two sets, that is, $Cut(\mathcal{E},\mathcal{H})= 0.32+0.24+0.23=0.79.$ Such edges are shown  by thin red  lines in Fig. \ref{GSPb_ex2CUT}.
\begin{figure}[ptb]
	\centering
	\includegraphics[]{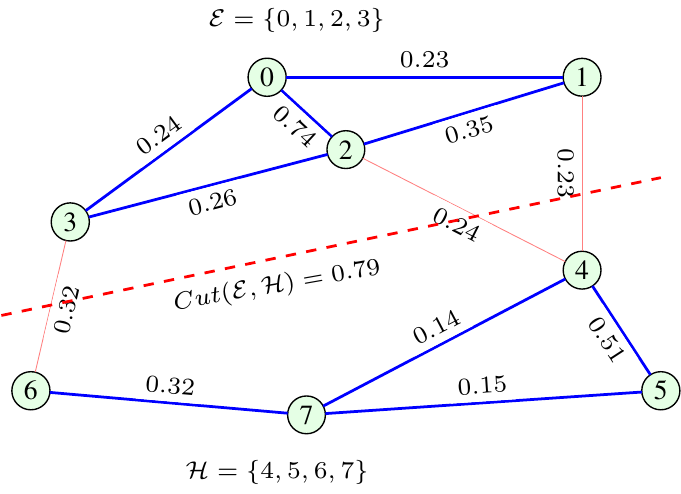}
	\caption{A cut for the weighted graph from Fig. \ref{GSPb_ex2}, with the disjoint subsets of vertices defined by $\mathcal{E}=\{0,1,2,3\}$ and $\mathcal{H}=\{4,5,6,7\}$. The edges between  the sets $\mathcal{E}$ and $\mathcal{H}$ are designated by thin red lines. The cut, $Cut(\mathcal{E},\mathcal{H})$, is equal to the sum of the weights that connect sets $\mathcal{E}$ and $\mathcal{H}$, and has the value $Cut(\mathcal{E},\mathcal{H})=0.32+0.24+0.23=0.79.$ }
	\label{GSPb_ex2CUT}
\end{figure}
\end{Example}

\noindent\textit{Definition:} A cut which exhibits the minimum value of the sum of weights between the disjoint subsets $\mathcal{E}$ and $\mathcal{H}$, considering all possible  divisions of the set of vertices, $\mathcal{V}$, is referred to as \textit{the minimum cut}. 
Finding the minimum cut of a graph in this way is a combinatorial  problem. 

\begin{Remark}\label{CutComb}
The number of all possible combinations to split an even number  of $N$ vertices into two disjoint subsets is given by 
$$C={N\choose 1}+{N\choose 2}+\dots+{N\choose N/2-1}+{N\choose N/2}/2.$$
 To depict the computational burden associated with this \textquotedblleft  brute force\textquotedblright\, graph cut approach, even for a relatively small graph with $N=50$ vertices, the number of combinations to split the vertices into two subsets is $C=5.6 \, \cdot 10^{14}$.
\end{Remark}

\begin{Example}
The minimum cut for the graph from Fig. \ref{GSPb_ex2CUT} is then $$Cut(\mathcal{E},\mathcal{H})\allowbreak=0.32+0.14+0.15=0.61$$
 for $\mathcal{E}=\{0,1,2,3,4,5\}$ and $\mathcal{H}=\{6,7\}$. This can be confirmed by considering all ${8\choose 1}+{8\choose 2}+{8\choose 3}+{8\choose 4}/2=127$ possible cuts, that is, all combinations of the subsets $\mathcal{E}$ and $\mathcal{H}$ for this small size graph or by using, for example, the Stoer-Wagner algorithm \cite{stoer1997simple}.  
\end{Example}

\subsubsection{Maximum-flow minimum-cut approach} 
This approach to the minimum cut problem employs the framework of \textit{flow networks}. 

\smallskip 

\noindent\textit{Definition:} A flow network is \textbf{a directed graph} with \textbf{two given  vertices }(nodes) called the source vertex, $s$, and the sink vertex, $t$, whereby the \textit{capacity of edges (arcs)} is defined by their weights. The flow (of information, water, traffic, ...) through an edge cannot exceed its capacity (the value of edge weight). For any vertex in the graph \textit{the sum of all input flows is equal to the sum of all its output flows} (except for the source and sink vertices). 

\bigskip

\noindent\textbf{Problem formulation.}  The maximum-flow minimum-cut solution to the graph partitioning aims to find the maximum value of \textit{flow} that can be passed through the graph (network flow) from the source vertex, $s$, to the sink vertex, $t$. The solution is based on the \textit{max-flow min-cut theorem} which states that the maximum flow through a graph from a given source vertex, $s$, to a given sink vertex, $t$, is equal to the minimum cut, that is, the minimum sum of those edge weights (capacities) which, if removed, would disconnect the source, $s$ from the sink, $t$ (minimum cut capacity). Physical interpretation of this theorem is obvious, since the maximum flow is naturally defined by the graph flow bottleneck between the source and sink vertices. The capacity of the bottleneck (maximum possible flow)   will then be equal to the minimum capacity (weight values) of the edges which, if removed, would disconnect the graph into two parts, one containing vertex $s$ and the other containing vertex $t$. Therefore, the problem of maximum flow is equivalent to the minimum cut (capacity) problem, under the assumption that the considered vertices, $s$ and $t$, must belong to  different disjoint subsets of vertices $\mathcal{E}$ and $\mathcal{H}$. \textit{This kind of cut, with predefined vertices $s$ and $t$, is called the $(s,t)$ cut. }

\begin{Remark}	In general, if the source and sink vertices are not given, the maximum flow algorithm should be repeated for all combinations of the source and sink vertices in order to find the minimum cut of a graph.
\end{Remark}

The most widely used  approach to solve the minimum-cut maximum-flow  problem is the \textit{Ford–Fulkerson method} \cite{kleinberg2006algorithm,kron1963diakoptics}.

\begin{Example}\label{FFAlg}
	\begin{figure}
		\centering
		\includegraphics[]{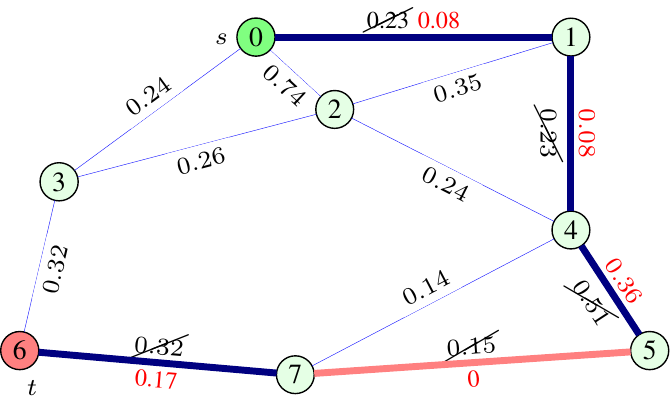} (a)
		\vspace{5mm}
		
		\includegraphics[]{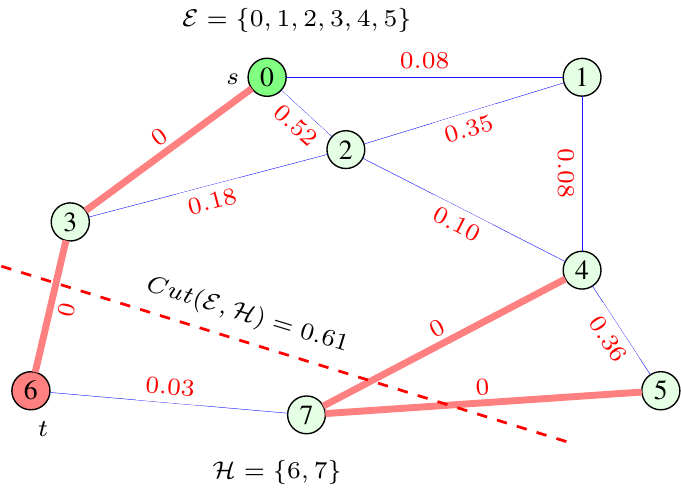} (b)
		
		\vspace{5mm}
		
		\includegraphics[]{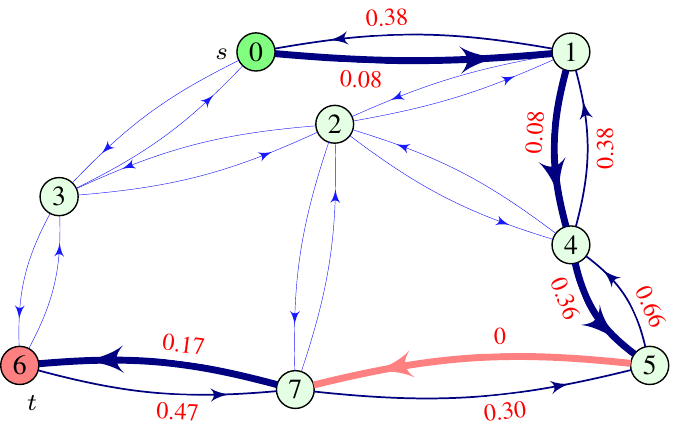} (c)
		\caption{Principle of the maximum flow minimum cut method. (a) The weighted graph from Fig. \ref{GSPb_ex2}, with the assumed source vertex $s=0$ and sink vertex $t=6$, and a path between these two vertices for which the maximum flow is equal to the minimum capacity (weight) along this path, $W_{57}=0.15$. This maximum flow value, $W_{57}=0.15$, is then subtracted from all the original edge capacities (weights) to yield the new \textit{residual edge capacities (weights)} which are shown in red. (b) The final edge capacities (weights) after the maximum flows are subtracted for all paths $0\rightarrow 3\rightarrow 6$, $0\rightarrow 2\rightarrow 4\rightarrow 7\rightarrow 6$, and $0\rightarrow 2\rightarrow 3\rightarrow6$, between vertices $s=0$ and $t=6$, with the resulting minimum cut now crossing only the  zero-capacity (zero-weight) edges with its value equal to the sum of their initial capacities (weights), shown in Panel (a) in black. (c) A directed form of the undirected graph from (a), with the same path and the residual capacities (weights) given for both directions.} 
		\label{fig-ff-1-2}
	\end{figure}
	Consider the weighted graph from Fig. \ref{GSPb_ex2}, with the assumed source and sink vertices, $s=0$ and $t=6$, as shown in Fig. \ref{fig-ff-1-2}(a). The Ford–Fulkerson method is based on the analysis of paths and the corresponding flows between the source and sink vertex. One such  possible path between $s$ and $t$, $0\rightarrow 1 \rightarrow  4\rightarrow  5 \rightarrow 7 \rightarrow 6$,  is designated by the thick line in Fig. \ref{fig-ff-1-2}(a).  Recall that the  \textit{maximum flow}, for a path connecting the vertices $s=0$ and $t=6$, is restricted by the minimum capacity (equal to the minimum weight) along the considered path. For the considered path $0\rightarrow 1 \rightarrow  4\rightarrow  5 \rightarrow 7 \rightarrow 6$ the maximum flow from $s=0$ to $t=6$ is therefore equal to  
	$$\maxflow_{0\rightarrow 1 \rightarrow  4\rightarrow  5 \rightarrow 7 \rightarrow 6}=\min\{0.23,0.23,0.51,0.15,0.32\}=0.15,$$ since  the minimum weight along this path is that connecting vertices $5$ and $7$, $W_{57}=0.15$. The value of this maximum flow is then subtracted from each capacity (weight) in the considered path, with the new \textit{residual edge capacities (weights)} designated in red in the \textit{residual graph} in Fig. \ref{fig-ff-1-2}(a). The same procedure is repeated for the remining possible paths $0\rightarrow 3\rightarrow 6$, $0\rightarrow 2\rightarrow 4\rightarrow 7\rightarrow 6$, and $0\rightarrow 2\rightarrow 3\rightarrow 6$, with appropriate corrections to the capacities (edge weights) after  consideration of each path. The final residual form of the graph, after zero-capacity edges are obtained in such a way that no new path with nonzero flow from $s$ to $t$  can be defined, is given in Fig. \ref{fig-ff-1-2}(b). For example, if we consider the path $0\rightarrow 1\rightarrow 2\rightarrow 3\rightarrow 6$ (or any other path), in the residual graph, then its maximum flow would be $0$, since the residual weight in the edge $3\rightarrow 6$ is equal to $0$. The minimum cut has now been obtained as that which separates the sink vertex, $t=6$, and its neighborhood from the the source vertex, $s=0$,   through the remaining zero-capacity (zero-weight) edges. This cut is shown in Fig. \ref{fig-ff-1-2}(b), and separates  the vertices $\mathcal{H}=\{6,7\}$  from the rest of vertices by cutting the edges connecting vertices $3\rightarrow 6$, $4\rightarrow 7$, and $5\rightarrow 7$. The original total weights of these edges are $Cut(\mathcal{E},\mathcal{H})\allowbreak=0.32+0.14+0.15=0.61$.

	We have so far considered  an undirected graph, but since the Ford–Fulkerson algorithm is typically applied to directed graphs, notice that an undirected graph can be considered as a directed graph with every edge being split into a pair of edges having the same weight (capacity), but with opposite directions. After an edge is used in one direction (for example, edge $5-7$ in Fig. \ref{fig-ff-1-2}(a)) with a flow equal to its maximum capacity of $0.15$ in the considered direction, the other flow direction (sister edge) becomes $0.30$, as shown in Fig. \ref{fig-ff-1-2}(c). The edge with opposite direction could be used (up the  algebraic sum of flows in both directions being equal to the total edge capacity) to form another path (if possible) from the source to the sink vertex. More specifically, the capacity of an edge  (from the pair) in the assumed direction is reduced by the same value of the considered flow, while the capacity of the opposite-direction edge (from the same pair) is increased by the same flow, and can be used to send the flow in reverse direction if needed. All residual capacities for the path from Fig. \ref{fig-ff-1-2}(a) are given in Fig. \ref{fig-ff-1-2}(c). For clarity, the edge weights which had not been changed by this flow are not shown in Fig. \ref{fig-ff-1-2}(c).       
\end{Example}

	%

\subsubsection{Normalized (ratio) minimum cut}  

A number of optimization approaches may be employed to enforce some desired properties on graph clusters. One such approach is  \textit{the normalized minimum cut}, which is commonly used in graph theory, and  is introduced by penalizing the value of $Cut(\mathcal{E},\mathcal{H})$ by an additional term (cost) to enforce the subsets $\mathcal{E}$ and $\mathcal{H}$ to be \textit{simultaneously as large as possible}. An obvious form of the normalized cut (\textit{ratio cut}) is given by \cite{hagen1992new}
\begin{gather}
CutN(\mathcal{E},\mathcal{H})=\Big(\frac{1}{N_{\mathcal{E}}}+\frac{1}{N_{\mathcal{H}}} \Big)\sum_{\substack{m \in \mathcal{E} \\ n \in \mathcal{H} }} W_{mn}, \label{CutN}
\end{gather}     
where $N_{\mathcal{E}}$ and $N_{\mathcal{H}}$ are the respective numbers of vertices in the sets  $\mathcal{E}$ and $\mathcal{H}$. Since $N_{\mathcal{E}}+N_{\mathcal{H}}=N$, the term $\frac{1}{N_{\mathcal{E}}}+\frac{1}{N_{\mathcal{H}}}$ reaches its minimum for  $N_{\mathcal{E}}=N_{\mathcal{H}}=N/2.$

\begin{figure}[ptb]
	\centering
	\includegraphics[]{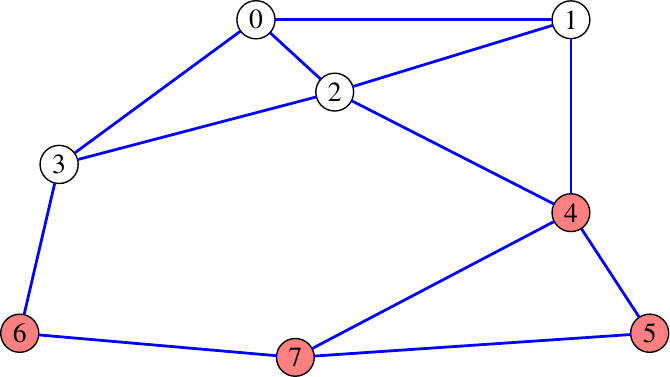}
	\caption{A clustering scheme  based on the minimum normalized cut of the vertices in the graph from Fig. \ref{GSPb_ex2} into two vertex clusters, $\mathcal{E}=\{0,1,2,3\}$ and $\mathcal{H}=\{4,5,6,7\}$. This cut corresponds to the arbitrarily chosen cut presented in Fig. \ref{GSPb_ex2CUT}.}
		\label{GSPb_ex2NCUTCL}
	\end{figure}

\begin{Example}\label{ExampNcut}
Consider again Example \ref{ExampleCUT},  and the graph from  Fig. \ref{GSPb_ex2CUT}. For the sets of vertices, $\mathcal{E}=\{0,1,2,3\}$ and $\mathcal{H}=\{4,5,6,7\}$, the normalized cut is calculated as $CutN(\mathcal{E},\mathcal{H})=(1/4+1/4)0.79=0.395$.  This cut  also represents the minimum normalized cut for this graph; this can be confirmed by checking all possible cut combinations of $\mathcal{E}$ and $\mathcal{H}$ in this (small) graph. Fig. \ref{GSPb_ex2NCUTCL} illustrates the clustering of vertices according to the minimum normalized cut.  Notice, however, that in general the minimum cut and the minimum normalized cut do not produce the same vertex clustering into  $\mathcal{E}$ and $\mathcal{H}$.
\end{Example}

\noindent\textbf{Graph separability.} Relevant to this section, the minimum cut value admits a physical interpretation as a \textit{measure of graph separability}. An ideal separability is possible if the minimum cut is equal to zero, meaning that there is no edges between subsets $\mathcal{E}$ and $\mathcal{H}$.  
 In Example \ref{ExampNcut}, the minimum cut value was $CutN(\mathcal{E},\mathcal{H})=0.395$, which is not close to $0$, and indicates that the segmentation of this graph into two subgraphs would not yield a close approximation of the original graph.

\subsubsection{Volume normalized minimum cut}  

A more general form of the normalized cut may also involve vertex weights when designing the size of subsets $\mathcal{E}$ and $\mathcal{H}$. By defining, respectively, \textit{the volumes} of these sets as $V_{\mathcal{E}}=\sum_{n \in \mathcal{E}}D_{nn}$ and $V_{\mathcal{H}}=\sum_{n \in \mathcal{H}}D_{nn}$, and using these volumes instead of the number of vertices $N_{\mathcal{E}}$ and $N_{\mathcal{H}}$  in the definition of the normalized cut  in  (\ref{CutN}), we arrive at \cite{shi2000normalized}
\begin{gather} 
CutV(\mathcal{E},\mathcal{H})=\Big(\frac{1}{V_{\mathcal{E}}}+\frac{1}{V_{\mathcal{H}}} \Big)\sum_{\substack{m \in \mathcal{E} \\ n \in \mathcal{H} }} W_{mn}, \label{CutV}
\end{gather} 
where $D_{nn}=\sum_{m \in \mathcal{V}}W_{mn}$ is the degree of a vertex $n$. \textit{The vertices with a higher degree, $D_{nn}$, are considered as structurally more important than the vertices with lower degrees. } 

The above discussion shows that finding the normalized minimum cut is also a combinatorial problem, for which an approximative spectral-based solution will be discussed later in this section.

\subsubsection{Other forms of the normalized cut}  
 In addition to the two presented forms of the normalized cut, based on the number of vertices and volume, other frequently used forms in open  literature include: 
		\begin{enumerate} 
			\item The \textit{sparsity of a cut} is defined by 
\begin{gather}
\rho(\mathcal{E})=\frac{1}{N_{\mathcal{E}}N_{\mathcal{V-E}}} \sum_{\substack{m \in \mathcal{E} \\ n \in \mathcal{V-E} }} W_{mn},
\end{gather}
where $\mathcal{V} -  \mathcal{E}$ is a \textit{set difference} of $\mathcal{V}$ and $\mathcal{E}$.
The sparsity of a cut, $\rho(\mathcal{E})$, is related to the normalized cut as $N\rho(\mathcal{E})= CutN(\mathcal{E},\mathcal{H})$, since $\mathcal{H}=\mathcal{V-E}$ and $N_{\mathcal{E}}+N_{\mathcal{V-E}}=N$. \textit{The sparsity of a graph is equal to the minimum sparsity of a cut.} It then follows that the  cut which exhibits minimum sparsity and the minimum normalized cut in (\ref{CutN}) produce the same set $\mathcal{E}$.
\item The \textit{edge expansion of a subset,} $\mathcal{E} \subset \mathcal{V}$,   is defined by
		\begin{gather}
		\alpha(\mathcal{E})=\frac{1}{N_{\mathcal{E}}}\sum_{\substack{m \in \mathcal{E} \\ n \in \mathcal{V} -  \mathcal{E} }} W_{mn},
		\end{gather}
		with $N_{\mathcal{E}} \le N/2$. Observe a close relation of edge expansion to the normalized cut in (\ref{CutN}). 
		\item The \textit{Cheeger ratio of a subset,} $\mathcal{E} \subset \mathcal{V}$, is defined as  
		\begin{gather} 
		\phi(\mathcal{E})=\frac{1}{\min\{V_{\mathcal{E}},V_{\mathcal{V-E}}\}} \sum_{\substack{m \in \mathcal{E} \\ n \in \mathcal{V-E} }} W_{mn}. \label{CheegerC}
		\end{gather}
		The minimum value of $\phi(\mathcal{E})$ is called \textit{the Cheeger constant} or  \textit{conductance} of a graph \cite{mohar1989isoperimetric}. This form is closely related to the  volume normalized cut in (\ref{CutV}). 
		\end{enumerate}

\subsection{Spectral methods for graph clustering} 

This class of methods is a modern alternative to the classical direct graph topology analysis, whereby vertex clustering is based on the eigenvectors of the graph Laplacian.  Practical spectral methods for graph clustering typically employ several smoothest eigenvectors of the graph Laplacian.

Simplified algorithms for vertex clustering  may even employ only one eigenvector, namely the second (Fiedler, \cite{fiedler1973algebraic}) eigenvector of the graph Laplacian,  $\mathbf{u}_1$,  to yield a \textit{quasi-optimal clustering} or partitioning scheme on a graph. These are proven to be efficient  in a range of applications, including data processing on graphs, machine learning, and computer vision \cite{malik2001contour}.  
Despite their simplicity, such algorithms are typically quite accurate, and a number of studies show that \textit{graph clustering and cuts based on the second eigenvector,  $\mathbf{u}_1$, give a good approximation to the optimal cut} \cite{ng2002spectral,spielman2007spectral}. Using more than one smooth eigenvector in  graph clustering and partitioning will increase the number of degrees of freedom to consequently yield more physically meaningful clustering, when required for practical applications in data analytics. 

For an enhanced insight we shall next review the smoothness index, before introducing the notions of graph spectral vectors and their distance, followed by the notions of similarity and clustering of vertices.    

\subsubsection{Smoothness of Eigenvectors on Graphs}
\medskip\noindent\textit{Definition:} \textit{The smoothness of an eigenvector}, $\mathbf{u}_k$, is introduced through its quadratic Laplacian  form, $\mathbf{u}_k^T\mathbf{L} \mathbf{u}_k$, with the smoothness index equal to the corresponding eigenvalue, $\lambda_k$, that is
\begin{equation}\mathbf{u}_k^T(\mathbf{L} \mathbf{u}_k)=\mathbf{u}_k^T (\lambda_k \mathbf{u}_k)=\lambda_k. \label{smmothLq}
\end{equation}
 To demonstrate physical intuition behind using the quadratic form, $\mathbf{u}_k^T\mathbf{L} \mathbf{u}_k$, as the smoothness metric of $\mathbf{u}_k$, consider
 \begin{gather*}
 \mathbf{u}_k^T \mathbf{L} \mathbf{u}_k = \mathbf{u}_k^T (\mathbf{D}-\mathbf{W}) \mathbf{u}_k. 
 \end{gather*}
 Then, an $n$-th element of the vector $ \mathbf{L} \mathbf{u}_k $ is given by 
 $$\sum_{m=0}^{N-1}  W_{nm}u_k(n) - \sum_{m=0}^{N-1}  W_{nm} u_k(m),$$ since $D_{nn}=\sum_{m=0}^{N-1}  W_{nm}$. Therefore, 
 \begin{gather}
 \mathbf{u}_k^T \mathbf{L} \mathbf{u}_k =\sum_{m=0}^{N-1}\ u_k(m) \sum_{n=0}^{N-1}  W_{mn}\Big(u_k(m) - u_k(n)\Big)\nonumber \\
 = \sum_{m=0}^{N-1}\ \sum_{n=0}^{N-1}  W_{mn}\Big(u_k^2(m) -u_k(m) u_k(n)\Big). \label{smoothQF} 
 \end{gather}
Owing to the symmetry of the weight matrix, $\mathbf{W}$ (as shown in (\ref{simW})), we can use  $W_{nm}=W_{mn}$ to replace the full summation of $u_k^2(n)$ over $m$ and $n$ with a half of the summations for both $u_k^2(m)$ and $u_k^2(n)$, over all $m$ and $n$. The same applies for the term $u(m) u(n)$. With that, we can write 
 \begin{gather}
 \mathbf{u}_k^T \mathbf{L} \mathbf{u}_k = \frac{1}{2} \sum_{m=0}^{N-1}\  \sum_{n=0}^{N-1}  W_{mn}\Big(u_k^2(m) -u_k(m) u_k(n)\Big) \nonumber \\
 +\frac{1}{2} \sum_{m=0}^{N-1}\  \sum_{n=0}^{N-1}W_{mn}\Big(u_k^2(n) -u_k(n) u_k(m)\Big) \nonumber  \\
 =\frac{1}{2} \sum_{m=0}^{N-1}\ \sum_{n=0}^{N-1} W_{mn}\Big(u_k(n) - u_k(m)\Big)^2 \ge 0.
 \label{eq:energijaLaplaciana}
 \end{gather}
 
 Obviously, a small $\mathbf{u}_k^T\mathbf{Lu}_k=\lambda_k$ implies that all terms $W_{nm}(u_k(n) - u_k(m))^2\le2\lambda_k$ are also small, thus indicating close values of $u_k(m)$ and $u_k(n)$ for vertices $m$ and $n$ with significant connections, $W_{mn}$. \textit{The eigenvectors corresponding to a small $\lambda_k$ are therefore slow-varying and smooth on a graph.}   

\medskip

\begin{Example} An exemplar of eigenvectors with a small, a moderate and a large smoothness index,  $\lambda_k$, is given on the three graphs in Fig. \ref{LS_VF_sig1}.
	
	\begin{figure*}[tbph]
		\centering
		\includegraphics[scale=1]{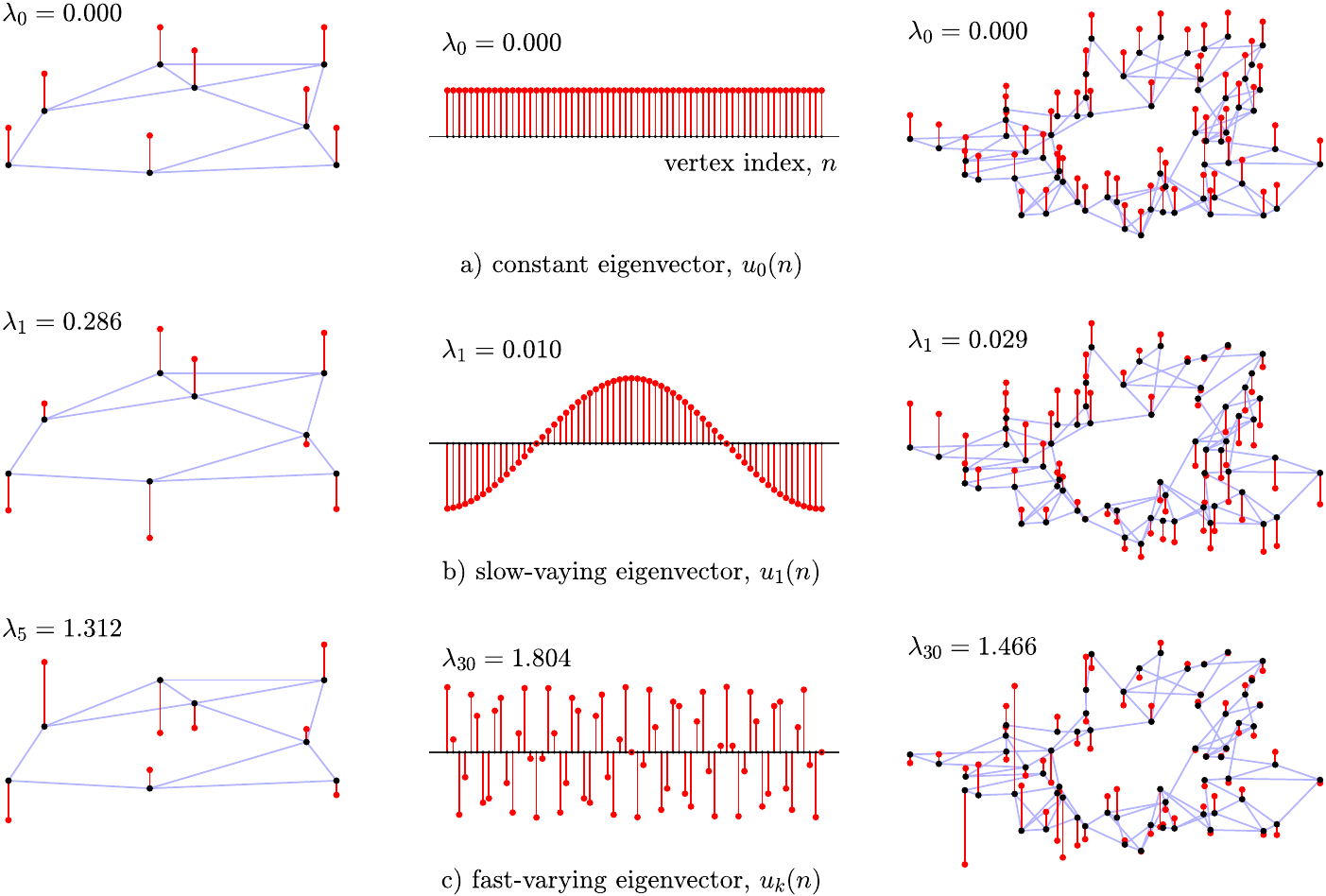}
		\caption{ Illustration of the concept of smoothness of the graph Laplacian eigenvectors for three different graphs: The graph from Fig. \ref{GSPb_ex2} (left), a path graph corresponding to classic temporal data analysis (middle), and an example of a more complex graph with $N=64$ vertices (right).   (a) Constant eigenvector, $u_0(n)$, shown on the three considered graphs. This is the smoothest possible eigenvector for which the smoothness index is $\lambda_0=0$. (b) Slow-varying Fiedler eigenvector (the smoothest eigenvector whose elements are not constant), $u_1(n)$, for the three graphs considered.  (c) Fast-varying eigenvectors, for $k=5$ (left), and $k=30$ (middle and right). Graph vertices are denoted by black circles, and the values of elements of the eigenvectors, $u_k(n)$, by red lines, for $n=0,1,\dots,N-1$. The  smoothness index, $\lambda_k$, is also given for each case.}
		\label{LS_VF_sig1}
	\end{figure*} 
In order to illustrate the interpretation of the smoothness index  in classical time-domain data processing, the time-domain form of the eigenvectors/basis functions in the real-valued Fourier analysis (\ref{GFT_lap}) is also shown in Fig. \ref{LS_VF_sig1} (middle).  In this case, the basis functions can be considered as the eigenvectors of a directed circular graph, where the vertices assume the role of time instants. 

Observe that in all three graphs the smooth eigenvectors, $\mathbf{u}_0$ and $\mathbf{u}_1$,  have similar elements on the neighboring vertices (in the case of a path graph -- time instants), and thus may be considered as smooth data on the corresponding graph domains. Such similarity does not hold for the fast-varying eigenvectors, $\mathbf{u}_5$ (left of Fig. \ref{LS_VF_sig1})  and $\mathbf{u}_{30}$ (middle and right of Fig. \ref{LS_VF_sig1}), which exhibit a much higher smoothness index.
\end{Example}

 \begin{Remark}  The eigenvector of the graph Laplacian which corresponds to $\lambda_0=0$ is constant (maximally smooth for any vertex ordering) and is therefore not appropriate as a template for vertex ordering. The next smoothest eigenvector is $\mathbf{u}_1$, which corresponds to the eigenvalue $\lambda_1$. 
 	\end{Remark}
 
 It is natural to order vertices within a graph in such a way so that the presentation of the sequence elements of the smoothest eigenvector, $\mathbf{u}_1$, as a function of the vertex index, $n$, is also maximally smooth. This can be achieved by sorting (rank ordering) the elements of the Fiedler vector, $\mathbf{u}_1$, in a nondecreasing order. Recall from Remark \ref{RReordering} that the isomorphic nature of graphs means that the reindexing of vertices does not change any graph property. The new order of graph vertices in the sorted $\mathbf{u}_1$  then corresponds to the smoothest sequence of elements of this vector along the vertex index line.  
 
 \textit{A unique feature of graphs, which renders them indispensable in modern data analytics on irregular domains, is that the ordering of vertices of a graph can be arbitrary,  an important difference from classical data analytics where the ordering is inherent sequential and fixed \cite{LNDM}. }
 Therefore, in general, any change in data ordering (indexing) would cause significant changes in the results of classical methods, while when it comes to graphs, owing to their topological invariance as shown in Fig.~\ref{GSPb_spectrum2a} and Fig.~\ref{GSPb_spectrum2bb} in the previous section, reordering of vertices would automatically imply the corresponding reordering of indices within each eigenvector, with no implication on the analysis results.
 However, the presentation  of data sensed at the graph vertices, along a line of vertex indices, as in Fig. \ref{GSPb_spectrum2a}(left), a common case for practical reasons, would benefit from an appropriate vertex ordering. Notice that vertex ordering in a graph is just a one-dimensional simplification of an important paradigm in graph analysis, known as graph clustering  \cite{mejia2017spectral,lu2014non,dong2012clustering,horaud2009short,hamon2016relabelling,masoumi2017spectral,masoumi2016spectral}.

\subsubsection{Spectral Space and Spectral Similarity of Vertices}\label{sepcvec}
For a graph with $N$ vertices, the orthogonal eigenvectors of its graph Laplacian form the basis of an $N$-dimensional space, called \textit{spectral space}, as shown in  Fig. \ref{Spectral_Vectors_Matrix}(a).  The elements $u_k(n)$ of the eigenvector $\mathbf{u}_k$, $k=0,1,2,\dots,N-1$, are assigned to vertices $n$, $n=0,1,2,\dots,\allowbreak N-1$. In other words, a set of elements, $u_0(n), u_1(n), u_2(n),\allowbreak\dots,\allowbreak u_{N-1}(n),$ is assigned to every vertex $n$. For every vertex, $n$, we can then group these elements into an $N$-dimensional \textit{spectral vector} $$\mathbf{q}_n\, {\overset{def}{=}} \,[u_0(n),u_1(n),\allowbreak\ldots,u_{N-1}(n)],$$
which is associated with the vertex $n$. Since the elements of the first eigenvector, $\mathbf{u}_0$, are constant, they do not convey any spectral difference to the graph vertices. Therefore, the elements of $\mathbf{u}_0$ are commonly omitted from the spectral vector for vertex $n$, to yield 
\begin{equation}
\mathbf{q}_n=[u_1(n),\ldots,u_{N-1}(n)], \label{spectvecqd}
\end{equation}
as illustrated in Fig. \ref{Spectral_Vectors_Matrix}(b).

\noindent\textbf{Vertex dimensionality in the spectral space.} Now that we have associated a unique spectral vector  $\mathbf{q}_n$ in (\ref{spectvecqd}), to every vertex $n=0,1,\dots,N-1$, it is important to note that this $(N-1)$-dimensional representation of every vertex in a graph  (whereby the orthogonal graph Laplacian eigenvectors, $\mathbf{u}_1$, $\mathbf{u}_2$, $\dots$, $\mathbf{u}_{N-1}$, serve as a basis of that representation) does not affect the graph itself; it just means that the additional degrees of freedom introduced through spectral vectors facilitate more sophisticated and efficient graph analysis. For example, we may now talk about vertex similarity in the spectral space, or about the spectral based graph cut, segmentation, and vertex clustering. 

An analogy with classical signal processing would be to assign a vector of harmonic basis function values at a time instant (vertex) $n$, to \textquotedblleft describe\textquotedblleft \, this instant, that is, to assign the $n$-th column of the Discrete Fourier transform matrix to the instant $n$. This intuition is illustrated in Fig. \ref{Spectral_Vectors_Matrix}(a) and \ref{Spectral_Vectors_Matrix}(b).

The spectral vectors shall next be used to define spectral similarity of vertices.  

\medskip\noindent\textit{Definition:}  Two vertices, $m$ and $n$, are called \textit{spectrally similar} if their distance in the spectral space  is within a small predefined threshold.  
 The spectral similarity between vertices $m$ and $n$ is typically measured through the Euclidean norm of their spectral space distance, given by
$$d_{mn}\, {\overset{def}{=}} \,\Vert \mathbf{q}_m-\mathbf{q}_n \Vert_2.$$

	\noindent \textbf{Spectral Manifold.} Once the graph is characterized by the original  $(N-1)$-dimensional spectral vectors, the so obtained vertex positions  in spectral vertex representation may reside near some well defined surface (commonly a hyperplane) of a reduced dimensionality $M<(N-1)$, such a hyperplane is called \textit{a spectral manifold}. The aim of spectral vertex mapping is then to map each spectral vertex representation from the original $N$-dimensional spectral vector space to a new spectral manifold which lies in a reduced $M$-dimensional spectral space, at a position closest to its original $(N-1)$-dimensional spectral position. This principle is related to the Principal Component Analysis (PCA) method, and  this relation will be discussed later in this section. \textit{An analogy with classical Discrete Fourier transform analysis, would mean to restrict the spectral analysis from the space of $N$ harmonics to the reduced space of the $M$ slowest-varying harmonics} (excluding the constant one).

These spectral dimensionality reduction considerations suggest to restrict the definition of spectral similarity to only a few lower-order (smooth) eigenvectors in the spectral space of reduced dimensionality. If the spectral similarity is restricted to the two smoothest eigenvectors, $\mathbf{u}_1$  and $\mathbf{u}_2$ (omitting $\mathbf{u}_0$), then the spectral vector for a vertex $n$ becomes
 $$\mathbf{q}_n=[u_1(n),u_2(n)],$$
 as illustrated in Fig. \ref{Spectral_Vectors_Matrix}(c) and Fig. \ref{Spectral_Vectors}(a).
If for two vertices, $m$ and $n$, the values of $u_1(m)$ are close to $u_1(n)$ and the values of $u_2(m)$ are close to $u_2(n)$, then these two vertices are said to be \textit{spectrally similar}, that is, they exhibit a small spectral distance, $d_{mn}=\Vert \mathbf{q}_m-\mathbf{q}_n \Vert_2$.   

Finally, the simplest spectral description uses only one (smoothest nonconstant) eigenvector to describe the spectral content of a vertex, so that the spectral vector reduces to a spectral scalar  
$$\mathbf{q}_n=[ q_n ]=[u_1(n)].$$
whereby the so reduced spectral space is a one-dimensional line.

\begin{Example}
The two-dimensional and three-dimensional spectral vectors, $\mathbf{q}_n=[u_1(n),u_2(n)]$ and $\mathbf{q}_n=[u_1(n),u_2(n),u_3(n)]$, of the graph from Fig. \ref{GSPb_ex2} are shown in Fig. \ref{Spectral_Vectors}, for $n=2$ and $n=6$.

\begin{figure*}
	\centering
	\includegraphics[scale=.85]{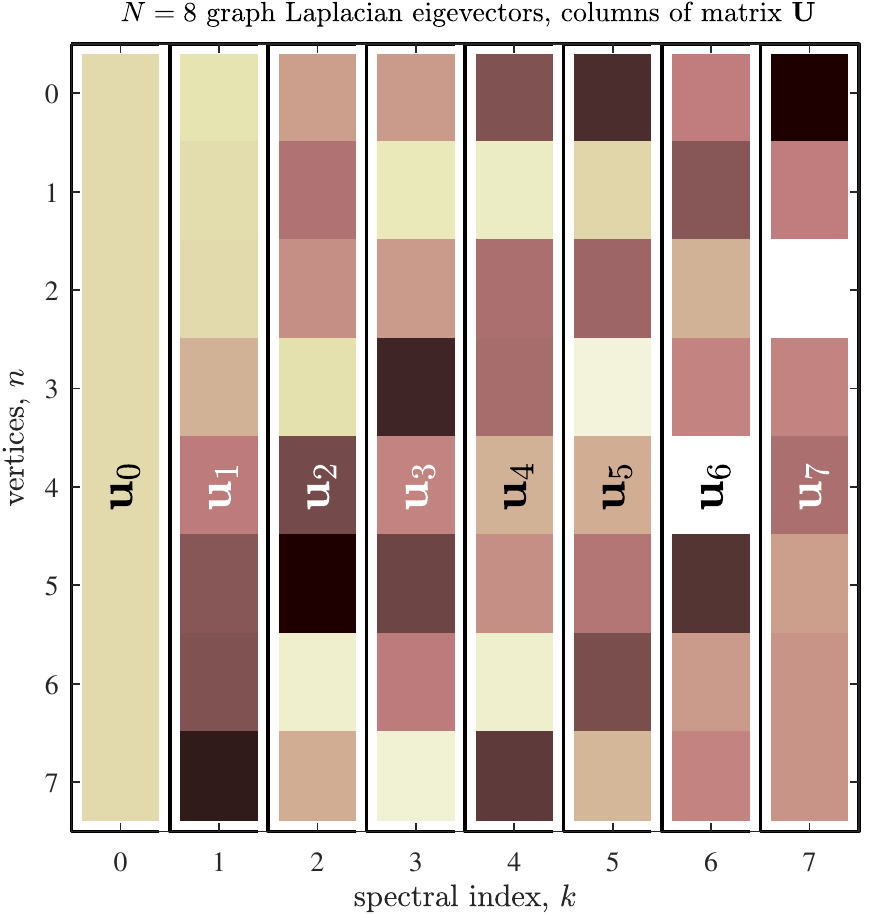}(a)
	\includegraphics[scale=.85]{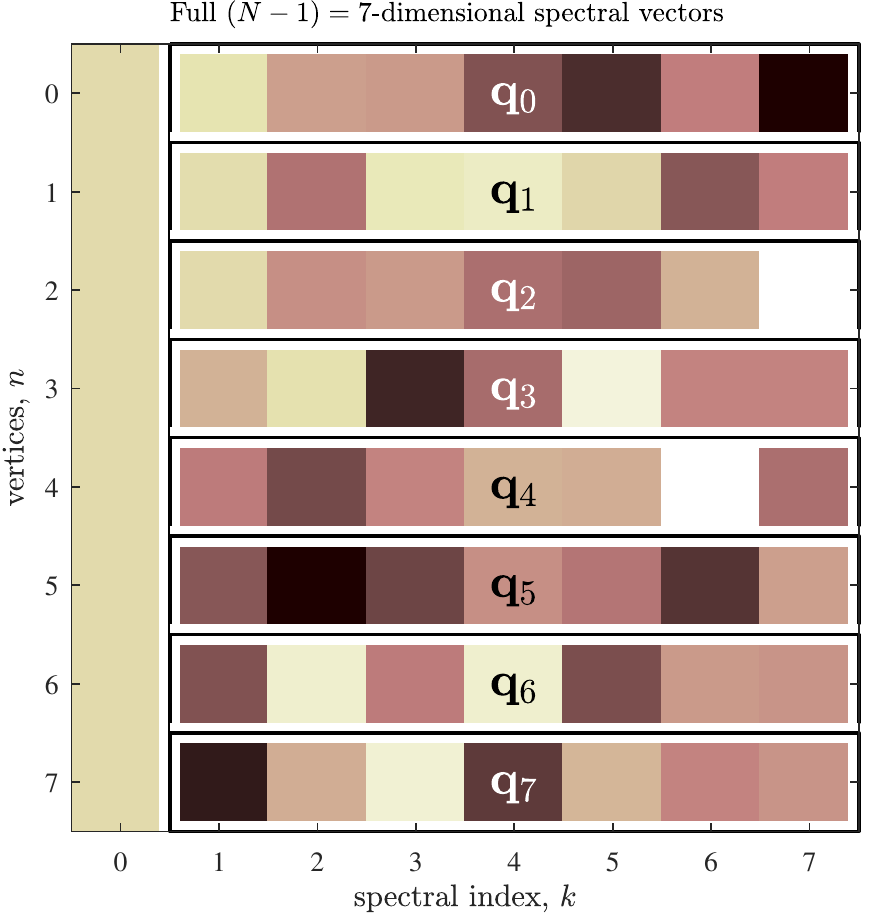}(b)
	
	\vfill
	\vspace{10mm}
	
	\includegraphics[scale=.85]{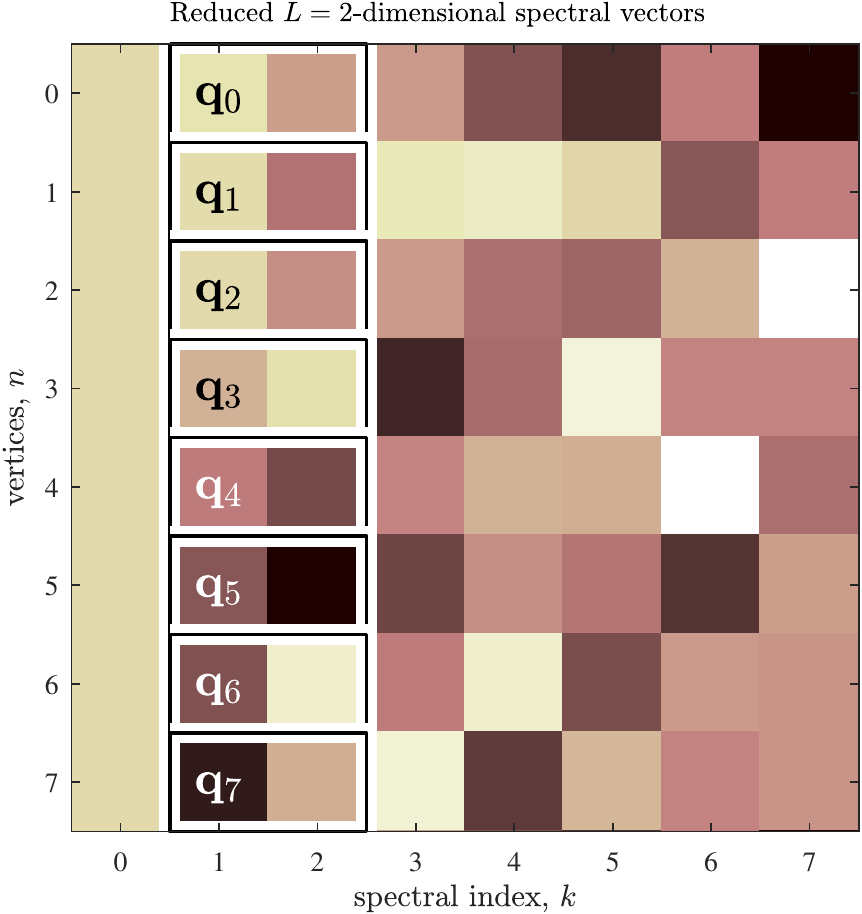}(c)
	\includegraphics[scale=.85]{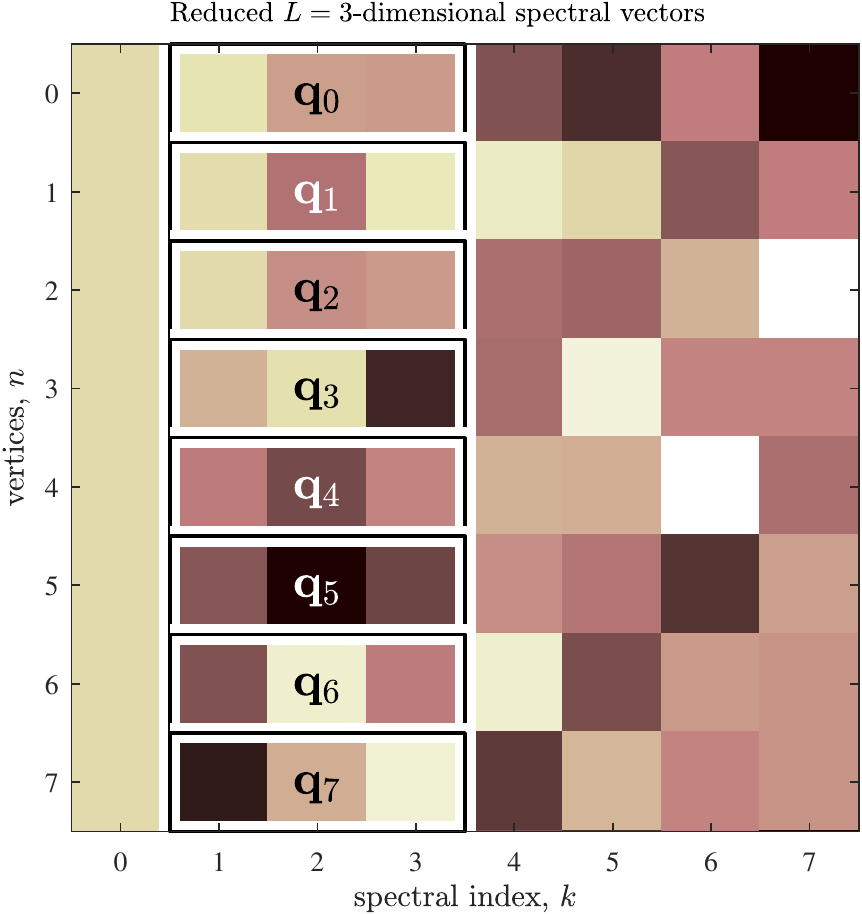}(d)
	\caption{Illustration of the spectral vectors for a graph from Fig. \ref {GSPb_ex2}, with $N=8$  vertices. For an intuitive analogy with the classical Discrete Fourier Transform, notice that the complex harmonic basis functions would play the role of eigenvectors $\mathbf{u}_k$, $k=0,1,\dots,8$. Then, the spectral vectors would be equal to the basis functions of the inverse Discrete Fourier transform  (excluding the first constant element).}
	\label{Spectral_Vectors_Matrix}
\end{figure*}

\begin{figure*}
	\centering
	\includegraphics[scale=.85]{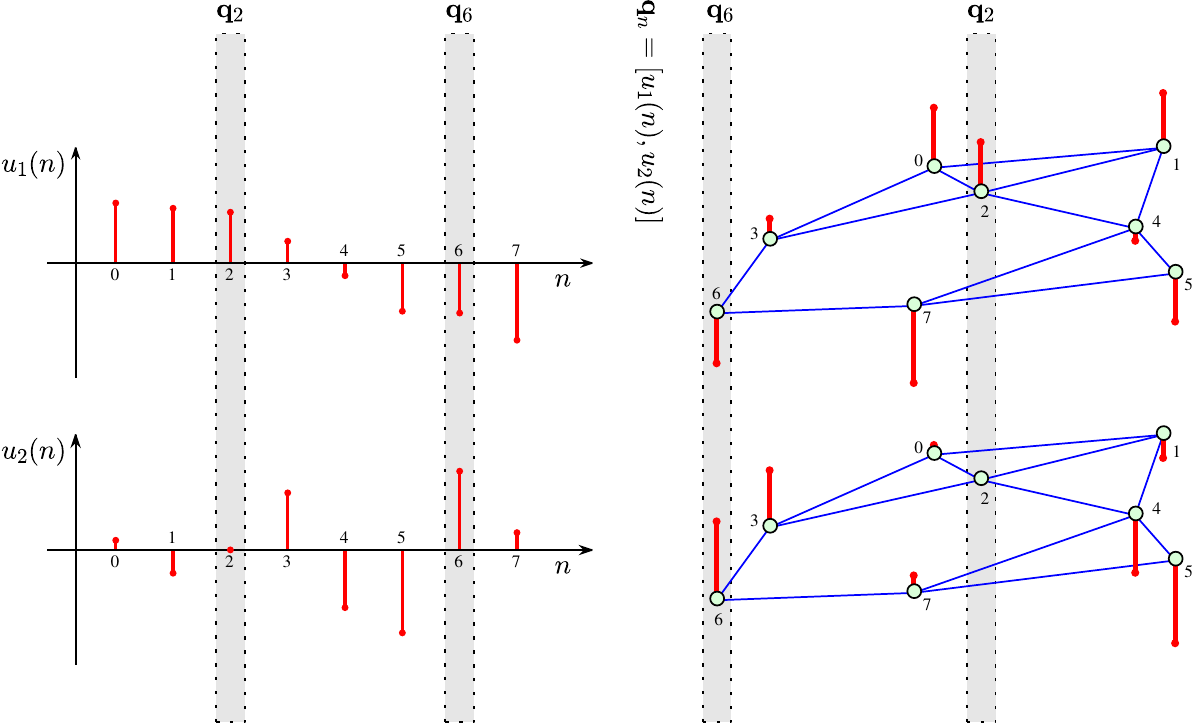}(a)
	\vfill
	\vspace{5mm}
	
	\includegraphics[scale=.85]{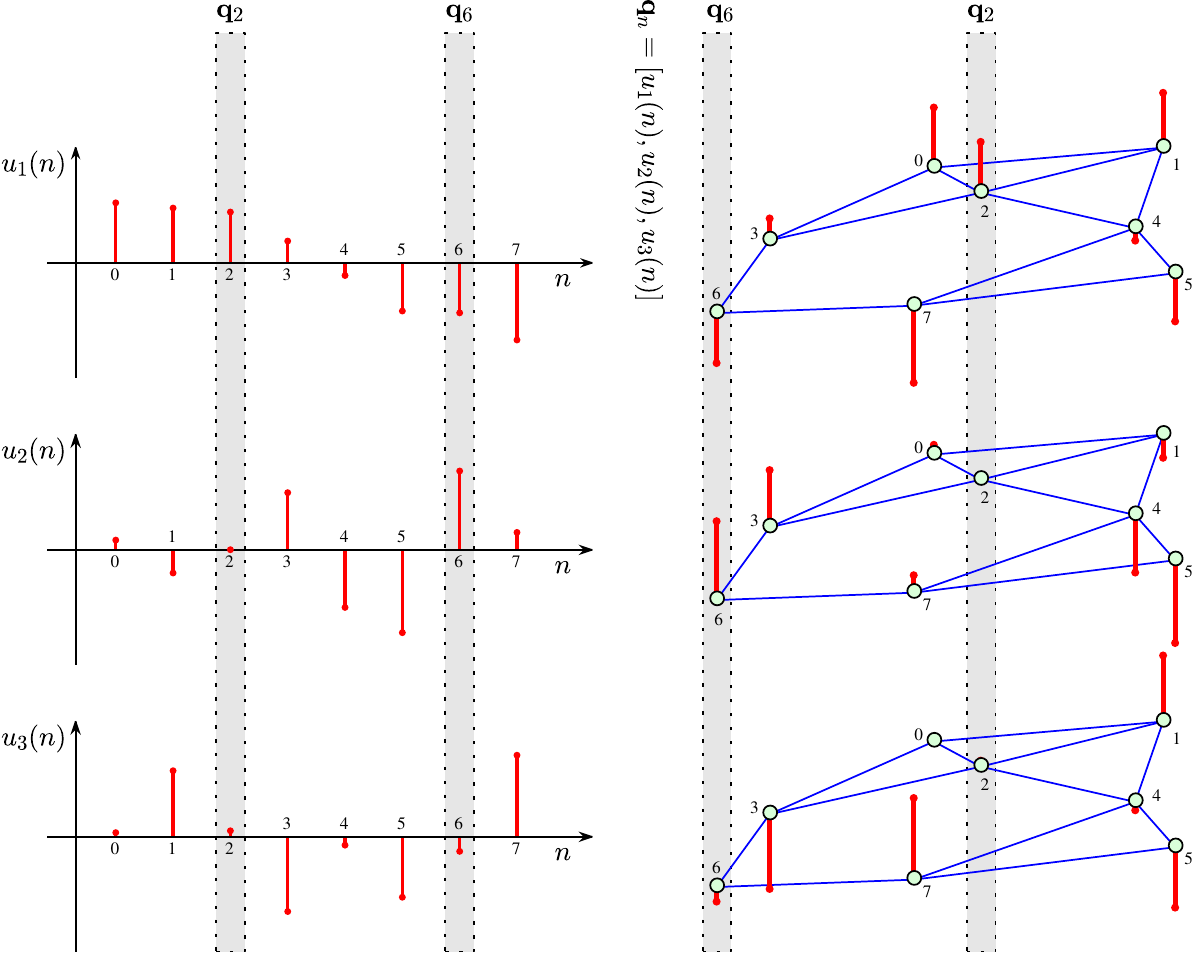}(b)
	\caption{Illustration of the spectral vectors, $\mathbf{q}_n=[u_1(n),u_2(n)]$ and $\mathbf{q}_n=[u_1(n),u_2(n),u_3(n)]$, for the Laplacian of the graph in Fig. \ref{GSPb_ex2}. (a) Two-dimensional spectral vectors, $\mathbf{q}_2=[u_1(2),u_2(2)]$ and $\mathbf{q}_6=[u_1(6),u_2(6)]$. (b)  Three-dimensional spectral vectors, $\mathbf{q}_2=[u_1(2),u_2(2),u_3(2)]$ and $\mathbf{q}_6=[u_1(6),u_2(6),u_3(6)]$.}
	\label{Spectral_Vectors}
\end{figure*}

\end{Example}

\noindent\textbf{Spectral embedding:} The mapping from the reduced  dimensionality spectral space  back onto the original vertices  is referred to as \textit{Embedding}. 

We can now proceed in two ways with  the reduced spectral vertex space  representation: (i) to assign the reduced dimension spectral vectors to the original vertex positions, for example, in the form of vertex coloring, as a basis for  subsequent vertex clustering (Section \ref{ClustSEG}),  or (ii) to achieve new vertex positioning in the reduced  dimensionality space of eigenvectors, using eigenmaps (Section \ref{EIGLMapigMaps}). Both yield similar information and can be considered as two sides of the same coin \cite{belkin2003laplacian}.   For visualization purposes, we will use  coloring to represent the spectral vector values in a reduced dimensionality spectral space. Vertices at the original graph positions will be colored according to the spectral vector values.

 \subsubsection{Indicator vector}\label{ClustSEG}

 Remark \ref{CutComb} shows that the combinatorial approach to minimum cut problem is computationally infeasible, as even for a graph with only $50$ vertices we have $5.6 \cdot 10^{14}$ such potential cuts.

 To break this \textit{Curse of Dimensionality} it would be very convenient to relate the  problem of the minimization of  normalized cut in (\ref{CutN}) and (\ref{CutV}) to that of eigenanalysis of graph Laplacian.
     To this end, we shall introduce the notion of an \textit{indicator vector} $\mathbf{x}$ on a graph, the elements of which are constant for vertices within each disjoint subset (cluster) of vertices, with these constants taking different values for different clusters of vertices  (\textit{subset-wise constant vector}). While this does not immediately reduce the computational burden (the same number of combinations remains  as in the brute force method), the elements of  $\mathbf{x}$  now uniquely reflect the assumed cut of the graph into disjoint subsets $\mathcal{E},\mathcal{H} \subset \mathcal{V}$.  
 
\textit{A further relation with only the smoothest eigenvector of the graph Laplacian would convert the original  combinatorial minimum cut problem into an algebraic eigenvalue problem, for which the computation complexity is of the $\mathcal{O}(N^3)$ order.}  Complexity of calculation can be additionally reduced through efficient  eigenanalysis methods, such as the \textit{Power Method} which sequentially  computes the desired number of largest eigenvalues and the corresponding eigenvectors, at an affordable $\mathcal{O}(N^2)$ computations per iteration, as shown in the Appendix.

 However, unlike the indicator vector, $\mathbf{x}$, the smoothest eigenvector (corresponding to the smallest nonzero eigenvalue) of graph Laplacian is  not subset-wise constant, and so such solution would be approximate, but computationally feasible.

\begin{Remark}
	The concept of indicator vector can be introduced through the analysis of a graph with an \textit{ideal minimum cut},  $$Cut(\mathcal{E},\mathcal{H})=\sum_{\substack{m \in \mathcal{E} \\ n \in \mathcal{H} }} W_{mn}=0,$$ 
	that is, a disjoint graph whereby $Cut(\mathcal{E},\mathcal{H})=0$ indicates  that there exist no edges between the subsets $\mathcal{E}$ and $\mathcal{H}$, when  $ W_{mn}=0$ for $m \in \mathcal{E}$, and $n \in \mathcal{H}$. This ideal case can be solved without resorting to the combinatorial approach, since this graph is already in the form of two disconnected subgraphs, defined by the sets of vertices $\mathcal{E}$ and $\mathcal{H}$. For such a disconnected graph, the second eigenvalue of the graph Laplacian is $\lambda_{1}=0$, as established in the graph Laplacian property $L_2$. When $\lambda_{1}=0$, then 
	$$2\mathbf{u}_1^T \mathbf{L} \mathbf{u}_1 
	=\sum_{m=0}^{N-1}\ \sum_{n=0}^{N-1} W_{mn}\Big(u_1(n) - u_1(m)\Big)^2=2\lambda_1=0,$$
	 which follows from (\ref{smmothLq}) and (\ref{eq:energijaLaplaciana}). Since all terms in the last sum are nonnegative, this implies that they must be zero-valued, that is, the eigenvector $\mathbf{u}_1$ is subset-wise constant, with $u_1(n) = u_1(m)=c_1$ for $m,n \in \mathcal{E}$ and $u_1(n) = u_1(m)=c_2$ for  $m,n, \in \mathcal{H}$. Since the eigenvector $\mathbf{u}_1$ is orthogonal to the constant eigenvector $\mathbf{u}_0$, then $\sum_{n=0}^{N-1}u_1(n)=0$. A possible solution for $u_1(n)$, that satisfies the subset-wise constant form and has zero mean, is $u_1(n)=c_1=1/N_{\mathcal{E}}$ for $n \in \mathcal{E}$ and $u_1(n)=c_2=-1/N_{\mathcal{H}}$ for $n \in \mathcal{H}$. We can conclude that the problem of finding an ideal minimum cut can indeed be solved by introducing an indicator vector $\mathbf{x}=\mathbf{u}_1$,  such that $x(n)=1/N_{\mathcal{E}}$ for $n \in \mathcal{E}$ and $x(n)=-1/N_{\mathcal{H}}$ for $n \in \mathcal{H}$. The membership of a vertex, $n$, to either the subset $\mathcal{E}$ or $\mathcal{H}$ of the ideal minimum cut is therefore uniquely defined by the sign of indicator vector $\mathbf{x}=\mathbf{u}_1$. This form of $\mathbf{x}$ is not normalized to unit energy, as its scaling by any constant would not influence  solution for vertex clustering into subsets $\mathcal{E}$ or $\mathcal{H}$.       
\end{Remark}

For a general graph,  and following the above reasoning, we here consider two specific subset-wise constant forms of the indicator vector, $\mathbf{x}$, based on
	 
	 \noindent (i) \textbf{The  number of vertices in disjoint subgraphs,}
	 \begin{gather}
	 x(n)=\begin{cases} \frac{1}{N_{\mathcal{E}}}, & \text{ for } n \in \mathcal{E}  \\ -\frac{1}{N_{\mathcal{H}}}, & \text{ for } n \in \mathcal{H},
	 \end{cases} 
	 \end{gather}
	  where $N_{\mathcal{E}}$ is the number of vertices in $\mathcal{E}$, and $N_{\mathcal{H}}$ is the number of vertices in $\mathcal{H}$, and 
	 
	 \noindent (ii) \textbf{The volumes of the disjoint subgraphs,}
	 \begin{gather}
	 x(n)=\begin{cases} \frac{1}{V_{\mathcal{E}}}, & \text{ for } n \in \mathcal{E}  \\ -\frac{1}{V_{\mathcal{H}}}, & \text{ for } n \in \mathcal{H},
	 \end{cases} 
	 \end{gather}
	  where \textit{the volumes of the sets},  $V_{\mathcal{E}}$ and $V_{\mathcal{H}}$, are defined as the sums of all vertex degrees, $D_{nn}$, in the corresponding subsets, $V_{\mathcal{E}}=\sum_{n \in \mathcal{E}}D_{nn}$ and $V_{\mathcal{H}}=\sum_{n \in \mathcal{H}}D_{nn}$.

	 Before proceeding further with the analysis of these two forms of indicator vector (in the next two remarks), it is important to note that if we can find the vector  $\mathbf{x}$ which minimizes the normalized cut, $CutN(\mathcal{E}, \mathcal{H})$ in (\ref{CutN}), then the elements of vector  $\mathbf{x}$ (their signs, $\mathrm{sign}(x(n))=1$ for $n \in \mathcal{E}$ and $\mathrm{sign}(x(n))=-1$ for $n \in \mathcal{H}$) may be used to decide whether to associate a vertex, $n$, to either the set $\mathcal{E}$ or  $\mathcal{H}$ of the minimum normalized cut.     
	
	\begin{Remark}\label{ProofCutN}
		The normalized cut, $CutN(\mathcal{E}, \mathcal{H})$, defined in (\ref{CutN}), for the indicator vector  $\mathbf{x}$ with the elements $x(n)=1/N_{\mathcal{E}}$ for $n \in \mathcal{E}$ and $x(n)=-1/N_{\mathcal{H}}$ for $n \in \mathcal{H}$, is equal to the  \textit{Rayleigh quotient} of matrix $\mathbf{L}$ and vector $\mathbf{x}$, that is
		\begin{equation}
		CutN(\mathcal{E}, \mathcal{H})=\frac{\mathbf{x}^T\mathbf{L}\mathbf{x}}{\mathbf{x}^T\mathbf{x}}. \label{LqfCut}
		\end{equation} 
		
		To prove this relation we shall rewrite (\ref{eq:energijaLaplaciana}) as 
		\begin{gather}
		\mathbf{x}^T \mathbf{L} \mathbf{x}
		=\frac{1}{2} \sum_{m=0}^{N-1}\ \sum_{n=0}^{N-1} W_{mn}\Big(x(n) - x(m)\Big)^2.
		\label{eq:CutNL}
		\end{gather}
		For all vertices $m$ and $n$, such that $m\in \mathcal{E}$ and $n\in \mathcal{E}$, the elements of vector  $\mathbf{x}$ are therefore the same and equal to $x(m)=x(n)=1/N_{\mathcal{E}}$, so that the terms $(x(n) - x(m))^2$ in (\ref{eq:CutNL}) are zero-valued. The same holds for any two vertices belonging to the set $\mathcal{H}$, that is, for  $m\in \mathcal{H}$ and $n\in \mathcal{H}$. Therefore, only the terms corresponding to the edges which define the cut, when $m\in \mathcal{E}$ and $n\in \mathcal{H}$, and vice versa, remain in the sum, and they are constant and equal to $(x(n) - x(m))^2=(1/N_{\mathcal{E}}-(-1/N_{\mathcal{H}}))^2$, to yield 
		\begin{gather}
		\mathbf{x}^T \mathbf{L} \mathbf{x}
		=\Big(\frac{1}{N_{\mathcal{E}}}+\frac{1}{N_{\mathcal{H}}} \Big)^2\sum_{\substack{m \in \mathcal{E} \\ n \in \mathcal{H} }} W_{mn} \nonumber \\
		=\Big(\frac{1}{N_{\mathcal{E}}}+\frac{1}{N_{\mathcal{H}}} \Big)	CutN(\mathcal{E}, \mathcal{H}),
		\label{eq:CutNLIz}
		\end{gather}
		where the normalized cut, $CutN(\mathcal{E}, \mathcal{H})$, is defined in (\ref{CutN}).  Finally, from the energy of the indicator vector, $\mathbf{x}^T\mathbf{x}=e^2_x$,   \begin{equation}\mathbf{x}^T\mathbf{x}=||\mathbf{x}||_2^2=e^2_x=\frac{N_{\mathcal{E}}}{N^2_{\mathcal{E}}}+\frac{N_{\mathcal{H}}}{N^2_{\mathcal{H}}}=\frac{1}{N_{\mathcal{E}}}+\frac{1}{N_{\mathcal{H}}},\label{cutEne} \end{equation}
		which proves (\ref{LqfCut}). 
	\end{Remark}
	
	The same analysis holds if \textit{the indicator vector is normalized to unit energy}, whereby
	$x(n)=1/(N_{\mathcal{E}}e_x)$ for $n \in \mathcal{E}$ and $x(n)=-1/(N_{\mathcal{H}}e_x)$ for $n \in \mathcal{H}$, with $e_x$ defined in (\ref{cutEne}) as $e_x=||\mathbf{x}||_2.$
	
	We can therefore conclude that the indicator vector, $\mathbf{x}$, which solves the problem of minimization of the normalized cut, is also the solution to (\ref{LqfCut}). This minimization problem, for the unit energy form of the indicator vector,  can also be written as 
	\begin{equation}
		\min\{\mathbf{x}^T\mathbf{L}\mathbf{x} \} \text{ subject to  } \mathbf{x}^T\mathbf{x}=1. \label{idicativeMIN}
	\end{equation}
	 In general, this is again a combinatorial problem, since all possible combinations of subsets of vertices, $\mathcal{E}$ and $\mathcal{H}$, together with the corresponding indicator vectors, $\mathbf{x}$, should be considered. 
	
 For a moment we shall put aside the very specific (subset-wise constant) form of the indicator vector and consider the general minimization problem in (\ref{idicativeMIN}). This problem can be solved using the method of Lagrange multipliers, with the corresponding cost function
	$$\mathcal{L}(\mathbf{x})=\mathbf{x}^T\mathbf{L}\mathbf{x}-\lambda(\mathbf{x}^T\mathbf{x}-1).$$
	From $\partial \mathcal{L}(\mathbf{x})/\partial \mathbf{x}^T=\mathbf{0}$, it follows that $\mathbf{L}\mathbf{x}= \lambda \mathbf{x}$, which is precisely the eigenvalue/eigenvector relation for the graph Laplacian $\mathbf{L}$, the solution of which is $\lambda=\lambda_k$ and $\mathbf{x}=\mathbf{u}_k$, for $k=0,1,\dots,N-1$. In other words, upon replacing vector $\mathbf{x}$ by $\mathbf{u}_k$ into the term $\min\{\mathbf{x}^T\mathbf{L}\mathbf{x} \}$ above, we obtain $\min_k\{\mathbf{u}_k^T\mathbf{L}\mathbf{u}_k \}=\min_k\{\lambda_k \}.$ After neglecting the trivial solution $\lambda_0=0$, which  produces a constant eigenvector $\mathbf{u}_0$, we next arrive at $\min_k\{\lambda_k \}=\lambda_1$ and $\mathbf{x}=\mathbf{u}_1$. Note that this solution yields a general form of vector $\mathbf{x}$ that minimizes (\ref{LqfCut}). However, such a form does not necessarily correspond to a subset-wise constant indicator vector, $\mathbf{x}$.

\subsubsection{Bounds on the minimum cut} 	
	
 In general, the subset-wise constant indicator vector, $\mathbf{x}$, can be written as a linear combination of the eigenvectors, $\mathbf{u}_k$, $k=1,2,\dots,N-1$, given by 
	\begin{equation}
	\mathbf{x}=\alpha_1\mathbf{u}_1+\alpha_2\mathbf{u}_2+\dots+\alpha_{N-1}\mathbf{u}_{N-1}. \label{indveceigv}
	\end{equation}
	This kind of vector expansion onto the set of eigenvectors  shall be considered in Part 2 of this monograph. Note that the constant vector $\mathbf{u}_0$ is omitted since the indicator vector is zero-mean (orthogonal to a constant vector) by definition. The calculation of coefficients $\alpha_i$  would require the indicator vector (that is, the sets $\mathcal{E}$ and $\mathcal{H}$) to be known, leading again to the combinatorial problem of vertex set partitioning. It is interesting to note that the quadratic form of indicator vector, $\mathbf{x}$, given by (\ref{indveceigv}) is also equal to $\mathbf{x}^T\mathbf{L}\mathbf{x}=\alpha^2_1\lambda_1+\alpha^2_2\lambda_2+\dots+\alpha^2_{N-1}\lambda_{N-1}$ and that it assumes the minimum value for $\alpha_1=1$, $\alpha_2=\dots=\alpha_{N-1}=0$, when  $\mathbf{x}=\mathbf{u}_1$, which corresponds to the normalized energy condition,  $\mathbf{x}^T\mathbf{x}=\alpha^2_1+\alpha^2_2+\dots+\alpha^2_{N-1}=1$, being  imposed. In other words, this means that 
	$$\lambda_1 \le \mathbf{x}^T\mathbf{L}\mathbf{x}=CutN(\mathcal{E}, \mathcal{H}).$$
	Observe that this inequality corresponds to the lower Cheeger bound  for the minimum normalized cut in (\ref{CutN}).

	\begin{Remark}
		If  the space of approximative solutions for the indicator vector, $\mathbf{x}$, is relaxed to allow for vectors that are not subset-wise constant (while avoiding the constant eigenvector of the graph Laplacian, $\mathbf{u}_0$), the approximative solution is $\mathbf{x}=\mathbf{u}_1$ (as previously shown and illustrated in Example \ref{Exampl15}).  The  analysis so far indicates that  this solution is  \textit{quasi-optimal}, however, despite its simplicity, the graph cut based on only the second graph Laplacian eigenvector,  $\mathbf{u}_1$, typically produces a good approximation to the optimal (minimum normalized) cut.
		\end{Remark}

The value of the true normalized minimum cut in (\ref{CutN}), when the form of indicator vector $\mathbf{x}$ is subset-wise constant, is bounded on both sides (upper and lower) with the constants which are proportional to the smallest nonzero eigenvalue, $\mathbf{u}_1^T\mathbf{L}\mathbf{u}_1=\lambda_1$, of the graph Laplacian. 
The simplest form of these (Cheeger's) bounds for the cut defined by (\ref{CheegerC}), has the form \cite{chung2005laplacians,trevisan2013lecture}
	\begin{equation}\frac{\lambda_1}{2} \le \phi(\mathcal{V})  {\overset{def}{=}} \min_{\mathcal{E} \subset \mathcal{V}} \{\phi(\mathcal{E})\} \le \sqrt{2 \lambda_1}.\label{CheegerB}
	\end{equation}

	 Therefore, the eigenvalue $\lambda_1$ is also a good \textit{measure of a graph separability and consequantly the quality of spectral clustering} in the sense of a minimum normalized cut.  The value of the minimum normalized cut of a graph (also referred to as \textit{Cheeger's constant, conductivity, or isoperimetric number} of a graph)  may also be considered as a numerical measure of whether or not a graph has a  \textit{\textquotedblleft bottleneck\textquotedblright.}  
	
\subsubsection{Indicator vector for normalized graph Laplacian}

	\begin{Remark}\label{VolNindicv}
	The volume normalized cut, $CutV(\mathcal{E}, \mathcal{H})$, defined in (\ref{CutV}), is equal to 
	\begin{equation}
	CutV(\mathcal{E}, \mathcal{H})=\frac{\mathbf{x}^T\mathbf{L}\mathbf{x}}{\mathbf{x}^T\mathbf{D}\mathbf{x}}, \label{LqfCutV}
	\end{equation} 
	where the corresponding, subset-wise constant, indicator vector has the values $x(n)=1/V_{\mathcal{E}}$ for $n \in \mathcal{E}$ and $x(n)=-1/V_{\mathcal{H}}$ for $n \in \mathcal{H}$, while the volumes of the sets,  $V_{\mathcal{E}}$ and $V_{\mathcal{H}}$, are defined in (\ref{CutV}). 
	 	
	The proof is identical that presented in Remark \ref{ProofCutN}. For the normalized indicator vector, we have $\mathbf{x}^T\mathbf{D}\mathbf{x}=1$, so that the minimization problem in (\ref{LqfCutV}) reduces to
	\begin{equation}
	\min\{\mathbf{x}^T\mathbf{L}\mathbf{x} \} \text{ subject to }  \mathbf{x}^T\mathbf{D}\mathbf{x}=1. \label{cutD}
	\end{equation}
	If the solution space is restricted to the space of \textit{ generalized eigenvectors of the graph Laplacian}, defined by $$\mathbf{L}\mathbf{u}_k=\lambda_k\mathbf{D}\mathbf{u}_k,$$
	 then the solution to (\ref{cutD}) is given by
	$$ \mathbf{x}=\mathbf{u}_1,$$
	where $\mathbf{u}_1$ is the generalized eigenvector of the graph Laplacian that corresponds to the lowest nonzero eigenvalue. 
	
	The \textit{eigenvectors of the normalized Laplacian } $\mathbf{L}_N=\mathbf{D}^{-1/2}\mathbf{L}\mathbf{D}^{-1/2}$,  may also be used in optimal cut approximations since the minimization problem in (\ref{LqfCutV}) can be rewritten using the normalized Laplacian, through a change of the variable in the minimization problem, to yield $$\mathbf{x}=\mathbf{D}^{-1/2}\mathbf{y},$$ 
	which allows us to arrive at the following form  \cite{ng2002spectral} 
	\begin{gather}
\min\{\mathbf{y}^T\mathbf{D}^{-1/2}\mathbf{L}\mathbf{D}^{-1/2}\mathbf{y} \}=\min\{\mathbf{y}^T\mathbf{L}_N\mathbf{y} \}, \nonumber \\  \text{
	 subject to } \mathbf{y}^T\mathbf{y}=1. \label{NorLCut}
	\end{gather}
	   If the space of  solutions to this minimization problem is  relaxed to the  eigenvectors, $\mathbf{v}_k$, of the normalized graph Laplacian, $\mathbf{L}_N$, then  $\mathbf{y}=\mathbf{v}_1$. For more detail on the various forms of the eigenvalues and eigenvectors of graph Laplacian, we refer to Table \ref{tab:1}.
	   
	   It is obvious now from (\ref{cutD}) and (\ref{NorLCut}) that the relation of the form $\mathbf{x}=\mathbf{D}^{-1/2}\mathbf{y}$ also holds for the corresponding eigenvectors of the normalized graph Laplacian, $\mathbf{v}_k$, and  the generalized eigenvectors of the Laplacian, $\mathbf{v}_k$, that is,
	    $$\mathbf{u}_k=\mathbf{D}^{-1/2}\mathbf{v}_k.$$

	\end{Remark}

It is important to note that, in general, results of clustering based on the three forms of eigenvectors, 
\begin{itemize}[label={}]
	\item
(i) the smoothest \textit{graph Laplacian eigenvector}, 
\item 
(ii) the smoothest \textit{generalized eigenvector of the Laplacian}, and 
\item 
(iii) the smoothest \textit{eigenvector of the normalized Laplacian}, 
\end{itemize}
are different.
 While the  method (i) favors the clustering into  subsets with (almost) equal number of vertices, the  methods (ii) and (iii) favor subsets with (almost) equal volumes (defined as sums of the vertex degrees in the subsets). Also note that the methods (i) and (ii) approximate the indicator vector in different eigenvector subspaces. All three methods will produce the same clustering result for unweighted regular graphs, for which the volumes of subsets are proportional to the number of their corresponding vertices, while the eigenvectors for all the three Laplacian forms are the same  in regular graphs, as shown in (\ref{regulGGGG}). 

\smallskip
	 \noindent \textbf{Generalized eigenvectors of the graph Laplacian and eigenvectors of the normalized Laplacian.} Recall that the matrix $\mathbf{D}^{-1/2}$ is of a diagonal form, and with positive elements. Then, the solution to (\ref{cutD}), which is equal to the  generalized eigenvector of the graph Laplacian, and the solution to (\ref{NorLCut}), equal to the eigenvector of the normalized Laplacian, are related as $\mathrm{sign}(\mathbf{y})=\mathrm{sign}(\mathbf{x})$ or $\mathrm{sign}(\mathbf{v}_1)=\mathrm{sign}(\mathbf{u}_1)$. This means that if the sign of the corresponding eigenvector is used for the minimum cut approximation (clustering), both results are the same.

\subsection{Spectral clustering implementation}

Spectral clustering is implemented using only a low-dimensional spectral vector. The simplest case is when a one-dimensional spectral vector is used as indicator vector for the clustering. More degrees of freedom are achieved when clustering schemes use two or three Laplacian eigenvectors. These spectral clustering schemes will be discussed next.

\subsubsection{Clustering based on only one (Fiedler) eigenvector}
From the analysis in the previous section, we can conclude that only the smoothest eigenvector, $\mathbf{u}_1$, can produce a good (quasi-optimal) approximation to the problem of minimum normalized cut graph clustering into two subsets of vertices, $\mathcal{E}$ and $\mathcal{H}$. Within the concept of spectral vectors, presented in Section \ref{sepcvec}, this means that the simplest  form of spectral vector,  $\mathbf{q}_n=u_1(n)$, based on just one (the smoothest) Fiedler eigenvector, $\mathbf{u}_1$, can be used for efficient  \textit{spectral vertex clustering}. Since the spectral vector $\mathbf{q}_n=u_1(n)$ is used as an approximative solution to the indicator vector for the minimum normalized cut definition, its values may be normalized. The normalization  
\begin{equation}\mathbf{y}_n=\mathbf{q}_n/||\mathbf{q}_n||_2 \label{spectvectnorm}
\end{equation} 
yields a two-level form of the spectral vector $$\mathbf{y}_n=[u_1(n)/||u_1(n)||_2]=[\textrm{sign}(u_1(n))],$$ 
and represents a step before clustering, as proposed in \cite{ng2002spectral}. This is justified based on the original form of the indicator vector, whose sign indicates the vertex association to the subsets, $\mathcal{E}$ or $\mathcal{H}$.  For illustrative representation of the normalized spectral vector, we may use a simple two-level colormap and assign one of two colors to each vertex. Such a simple algorithm for clustering is given in Algorithm \ref{AlgNN1} (for an algorithm with more options for clustering and representation see the Appendix (Algorithm \ref{Norm0Alg}) and Remarks \ref{clusteringRem} and \ref{RemRecalc}).

	\begin{algorithm}[!tbh]
	\caption{\!\!\textbf{.} \, Clustering using the graph Laplacian.}
	\label{Norm0Alg}
	\begin{algorithmic}[1]
		\smallskip
		\Input
		\Statex
		\begin{itemize}	
			\item Graph vertices $\mathcal{V}=\{0,1,\dots,N-1\}$
			\item Graph Laplacian $\mathbf{L}$
				\end{itemize}
		\Statex
		\State $[\mathbf{U},\mathbf{\Lambda}] \gets \mathrm{eig}(\mathbf{L})$
		\State $y_n \gets U(2,n)$
		\State $\mathcal{E} \, \gets \{n \, | \,y_n>0 \}, \,\,\,\, \mathcal{H}  \gets \{n \, | \,y_n \le 0 \}$	
		\begin{itemize}
			\Output
			\item Vertex clusters $\mathcal{E}$ and $\mathcal{H}$
		\end{itemize} 
	\end{algorithmic}
\label{AlgNN1}
\end{algorithm}

\begin{Example}\label{Exampl15}
Consider the graph from Fig. \ref{GSPb_ex2} and its Laplacian eigenvector, $\mathbf{u}_1$, from Fig. \ref{GSPb_spectrum3a}.   The elements of this single eigenvector,  $\mathbf{u}_1$, are  used to encode the vertex colormap, as shown in Fig. \ref{coloring1}(a). Here, the minimum element of  $\mathbf{u}_1$ was used to select the red color (vertex 7), while the white color at vertex 0 was designated by the maximum value of this eigenvector.  Despite its simplicity, this scheme immediately allows us to threshold $\mathbf{u}_1$ and identify two possible \textit{graph clusters}, $\{0,1,2,3\}$,  and $\{4,5,6,7\}$, as illustrated in Fig. \ref{coloring1}(b). The same result would be obtained if the sign of $\mathbf{u}_1$ was used to color the vertices, and this would correspond to the minimum normalized cut clustering in Fig. \ref{GSPb_ex2NCUTCL}.

The true indicator vector, $\mathbf{x}$, for the minimum normalized cut of this graph is presented in Fig. \ref{idicative_vector_eig}(a). This vector is obtained by checking all the 127 possible cut combinations of $\mathcal{E}$ and $\mathcal{H}$ in this small graph, together with the corresponding $x(n)$. The signs of this vector indicate the way for optimal clustering into the subsets $\mathcal{E}=\{0,1,2,3\}$  and $\mathcal{H}=\{4,5,6,7\}$, while the minimum cut value is $CutN(\mathcal{E}, \mathcal{H})=\mathbf{x}^T\mathbf{L}\mathbf{x}=0.395.$ Fig. \ref{idicative_vector_eig}(b) shows an approximation of the indicator vector within the space of the graph Laplacian eigenvectors, $\mathbf{u}_1$. The quadratic form of the eigenvector, $\mathbf{u}_1$, is equal to $\mathbf{u}_1^T\mathbf{L}\mathbf{u}_1=\lambda_1=0.286$. Note that the true indicator vector, $\mathbf{x}$, can be decomposed into the set of all graph Laplacian eigenvectors, $\mathbf{u}_k$, and written as their linear combination.

The generalized Laplacian eigenvector, $\mathbf{u}_1=[0.37, \,    0.24,\allowbreak \,    0.32, \,    0.13, \,   -0.31, \,   -0.56,\allowbreak \,   -0.34, \,   -0.58]$, which is an approximation of the indicator vector for the minimum volume normalized cut in (\ref{CutV}), is presented in Fig. \ref{idicative_vector_eig}(c). In this case, the generalized eigenvector indicates the same clustering subsets, $\mathcal{E}=\{0,1,2,3\}$  and $\mathcal{H}=\{4,5,6,7\}$. The eigenvector of the normalized Laplacian, $\mathbf{v}_1$, is shown in Fig. \ref{idicative_vector_eig}(d).

\begin{figure}
	\centering
	\includegraphics[scale=.99]{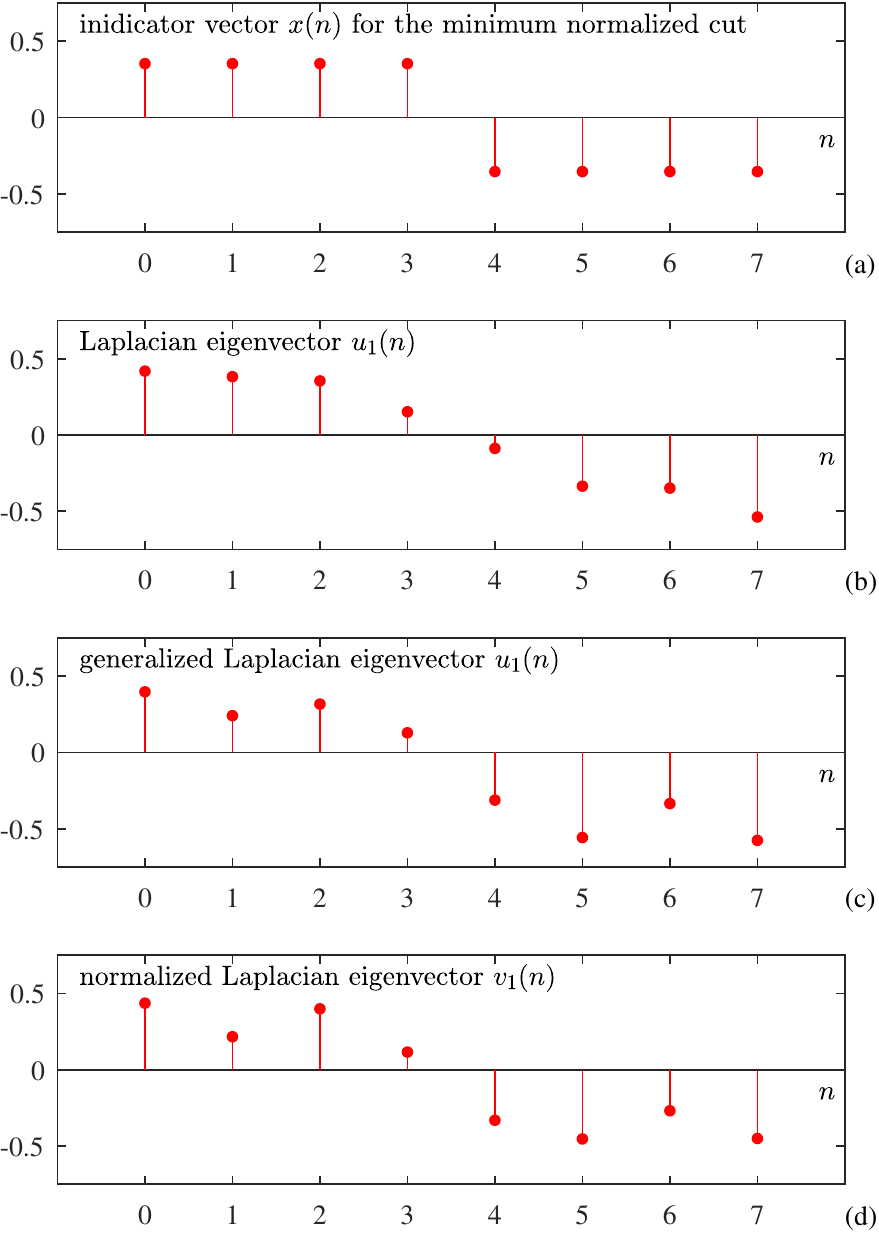}
	\caption{ Principle of the minimum normalized cut based clustering and its spectral (graph Laplacian eigenvector) based approximation; all vectors are plotted against the vertex index $n$. (a) The ideal indicator vector for a minimum normalized cut, $CutN(\mathcal{E}, \mathcal{H})$, normalized to unite energy. (b) The graph Laplacian eigenvector, $\mathbf{u}_1$. (c) The generalized eigenvector of the Laplacian, $\mathbf{u}_1$. (d) The eigenvector of the normalized Laplacian, $\mathbf{v}_1$. The eigenvectors in (c) and (d) are related as $\mathbf{u}_1=\mathbf{D}^{-1/2}\mathbf{v}_1$. In this case, the sign values of the indicator vector and the eigenvectors, $\mathrm{sign}(\mathbf{x})$, $\mathrm{sign}(\mathbf{u}_1)$, and $\mathrm{sign}(\mathbf{v}_1)$ are the same in all the four vectors. The signs of these vectors then all define the minimum normalized cut based clustering into $\mathcal{E}$  and $\mathcal{H}$, that is, the association of a vertex, $n$, to either the subset $\mathcal{E}$ or subset $\mathcal{H}$ .
	}  
	\label{idicative_vector_eig}
\end{figure}

\end{Example}

\begin{Example}\label{EX17} Consider the graph from Fig. \ref{GSPb_ex2}, with the weight matrix, $\mathbf{W}$, in (\ref{WeightMatr}),  and the graph Laplacian eigenvector $\mathbf{u}_1$ (shown in Fig. \ref{GSPb_spectrum3a},  Fig. \ref{LS_VF_sig1}(b)(left), and Fig. \ref{idicative_vector_eig}(b)). When this eigenvector is thresholded to only two intensity levels, $\mathrm{sign}(\mathbf{u}_1)$, two graph clusters are obtained, as shown in Fig. \ref{coloring1} (right). In an ideal case, these clusters may even be considered as independent graphs (graph segmentation being the strongest form of clustering); this can be achieved by redefining the weights as $W_{nm} = 0$, if $m$ and $n$ are in different clusters, and $W_{nm} = W_{nm}$ otherwise \cite{ng2002spectral}, for the corresponding disconnected (segmented) graph, whose weight matrix, $\mathbf{\hat{W}}$, is given by  
	\begin{equation}
	\small
	\setlength{\arraycolsep}{4pt}
	\mathbf{\hat{W}}=\begin{array}{cr}
	& \\
	{
		\color{blue}
		\begin{matrix}
		\text{\footnotesize 0}\\
		\text{\footnotesize 1}\\
		\text{\footnotesize 2}\\
		\text{\footnotesize 3}\\
		\text{\footnotesize 4}\\
		\text{\footnotesize 5}\\
		\text{\footnotesize 6}\\
		\text{\footnotesize 7}\\
		\end{matrix}
	} 
	\left[
	\begin{array}{rrrrrrrr}  
	\cellcolor[gray]{0.9}  0  &  \cellcolor[gray]{0.9} 0.23 & \cellcolor[gray]{0.9}  0.74  &  \cellcolor[gray]{0.9}  0.24&   0   &   0   &   0   & 0 \\
	\cellcolor[gray]{0.9}  0.23 &  \cellcolor[gray]{0.9}  0   & \cellcolor[gray]{0.9}  0.35  & \cellcolor[gray]{0.9}  0 	& 0  &   0   &   0   & 0  \\
	\cellcolor[gray]{0.9}  0.74 & \cellcolor[gray]{0.9}  0.35  &  \cellcolor[gray]{0.9}  0   & \cellcolor[gray]{0.9}  0.26 &   0&   0   &   0   &   0   \\
	\cellcolor[gray]{0.9}  0.24 & \cellcolor[gray]{0.9}  0 	 & \cellcolor[gray]{0.9}  0.26  &   \cellcolor[gray]{0.9} 0  & 0 	&   0   &  0 & 0  \\
	0  &  0 & 0  & 0 	&  \cellcolor[gray]{0.9}  0    & \cellcolor[gray]{0.9}  0.51  & \cellcolor[gray]{0.9}  0     & \cellcolor[gray]{0.9}  0.14 \\
	0  &    0  &   0   & 0 	& \cellcolor[gray]{0.9}  0.51  &  \cellcolor[gray]{0.9}  0   &  \cellcolor[gray]{0.9}  0   &  \cellcolor[gray]{0.9}  0.15   \\
	0  &    0  &   0   & 0 & \cellcolor[gray]{0.9}  0     & \cellcolor[gray]{0.9} 0     &   \cellcolor[gray]{0.9}  0   &   \cellcolor[gray]{0.9}  0.32   \\
	0    & 0 	 &  0    & 0    & \cellcolor[gray]{0.9}  0.14	& \cellcolor[gray]{0.9}  0.15 	&  \cellcolor[gray]{0.9}  0.32 &   \cellcolor[gray]{0.9}  0   \\
	\end{array}
	\right].
	\\
	{
		\color{blue}
		\begin{matrix*}
		\text{\footnotesize 0 \hspace{3.0mm}} &
		\text{\footnotesize 1 \hspace{3.0mm}}  &
		\text{\footnotesize 2 \hspace{3.0mm}} &
		\text{\footnotesize 3 \hspace{3.0mm}} &
		\text{\footnotesize 4 \hspace{3.0mm}} &
		\text{\footnotesize 5 \hspace{3.0mm}} &
		\text{\footnotesize 6 \hspace{3.0mm}} &
		\text{\footnotesize 7 \hspace{-8mm}}
		\end{matrix*}
	}
	\end{array}\label{WappSeg}
	\end{equation}

	\begin{figure}
		\centering
		(a)\includegraphics[scale=.88]{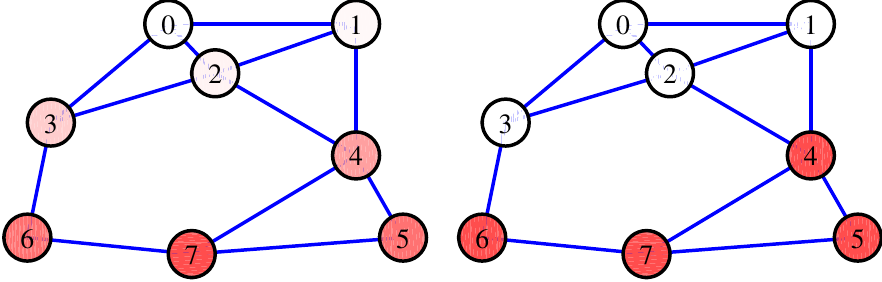}(b)
		\caption{Vertex coloring for the graph from Fig. \ref{GSPb_ex2}, with its spectrum shown in Fig. \ref{GSPb_spectrum3a}. (a) The   eigenvector, $\mathbf{u}_1$, of the Laplacian matrix of this graph, given in (\ref{LaplacianSCe}), is normalized and is used to define the red color intensity levels within the colormap for every vertex. For this example, $\mathbf{u}_1=[0.42, \,   0.38, \,   0.35, \,  0.15, \,  -0.088, \,  -0.34, \,  -0.35, \,   -0.54]^T$.  The largest element of this eigenvector is $u_1(0)=0.42$ at vertex $0$, which indicates that this vertex should be colored by the lowest red intensity (white), while the smallest  element is $u_1(7)=-0.54$, so that vertex $7$ is colored with the strongest red color intensity. (b) Simplified two-level coloring based on the sign of the elements   of eigenvector  $\mathbf{u}_1$.}  
		\label{coloring1}
	\end{figure}
	
\end{Example}

\subsubsection{\textquotedblleft Closeness\textquotedblright  of the segmented and original graphs}\label{SEC:closseg} 

The issue of how \textquotedblleft close\textquotedblright \, the behavior of the weight matrix of the segmented graph, $\mathbf{\hat{W}}$, in (\ref{WappSeg}) (and the corresponding $\mathbf{\hat{L}}$) is to the original  $\mathbf{{W}}$ and $\mathbf{{L}}$, in (\ref{WeightMatr}) and (\ref{LaplacianSCe}), is usually considered within matrix perturbation theory.

It can be shown that a good measure of the \textquotedblleft closeness\textquotedblright\ is the so-called \textit{eigenvalue gap}, $\delta=\lambda_2-\lambda_1$, \cite{ng2002spectral},  between the eigenvalue $\lambda_1$ associated with the eigenvector $\mathbf{u}_1$, that is used for segmentation, and the next eigenvalue, $\lambda_2$, in the graph spectrum of the normalized graph Laplacian (for an illustrative explanation see Example \ref{PHExamp}). For the obvious reason of analyzing the eigenvalue gap at an appropriate scale, we suggest to consider \textit{the relative eigenvalue gap }
	\begin{equation}
	\delta_r=\frac{\lambda_2-\lambda_1}{\lambda_2}=1-\frac{\lambda_1}{\lambda_2} \label{relativeGap}.
	\end{equation}
The relative eigenvalue gap value is within the interval $0\le \delta_r \le 1$, since the eigenvalues are nonnegative real-valued numbers sorted into a nondecreasing order. A value of this gap may be considered as large if it is close to the maximum eigengap value, $\delta_r=1$.
		
		\begin{Example} The Laplacian eigenvalues for the graph in Fig. \ref{coloring1} are $\lambda \in \{0, \,    0.29, \,    0.34, \,    0.79, \,    1.03, \,    1.31, \,    1.49, \,    2.21 \}$,
			with the relative eigenvalue gap, $\delta_r=(\lambda_2-\lambda_1)/\lambda_2=0.15$, which is not large and indicates that the segmentation in Example \ref{EX17} is not \textquotedblleft close\textquotedblright.    
		
		As an illustration, consider three hypothetical but practically relevant scenarios: (i) $\lambda_2=0$ and $\lambda_3=1$, (ii) $\lambda_2=0$ and $\lambda_3=\varepsilon$, (iii) $\lambda_2=1$ and $\lambda_3=1+\varepsilon$, where $\varepsilon$ is small positive number and close to $0$.  According to Remark \ref{RL2}, the graph in case (i) consists of exactly two disconnected components, and the subsequent clustering and segmentation is appropriate, with $\delta_r=1$. For case (ii), the graph consists of more than two almost disconnected components and the clustering in two sets can be performed in various ways, with $\delta_r=1/\varepsilon$. Finally, in the last scenario the relative gap is very small, $\delta_r=\varepsilon$, thus indicating that the behavior of the segmented graph is not \textquotedblleft close\textquotedblright to the original graph, that is, $\mathbf{\hat{L}}$ is not \textquotedblleft close\textquotedblright to $\mathbf{L}$, and thus any segmentation into two disconnected subgraphs would produce inadequate results.   
	\end{Example}
 
 \begin{Remark}
 	The thresholding of elements of the Fiedler vector, $\mathbf{u}_1$, of the normalized Laplacian, $\mathbf{L}_N=\mathbf{D}^{-1/2}\mathbf{L}\mathbf{D}^{-1/2}$, performed in order to cluster the graph is referred to as the \textit{Shi -- Malik algorithm} \cite{shi2000normalized, weiss1999segmentation}. Note that similar results would have been obtained if clustering was based on the thresholding of  elements of the smoothest eigenvector (corresponding to the second largest  eigenvalue) of the normalized weight matrix, $\mathbf{W}_N=\mathbf{D}^{-1/2}\mathbf{W}\mathbf{D}^{-1/2}$  (\textit{Perona --
 		Freeman algorithm} \cite{perona1998factorization,weiss1999segmentation}).  This becomes  clear after recalling that the relation between the normalized weight and graph Laplacian matrices is given by 
 	\begin{gather}
 	\mathbf{L}_N=\mathbf{D}^{-1/2}\mathbf{L}\mathbf{D}^{-1/2}=\mathbf{I}-\mathbf{D}^{-1/2}\mathbf{W}\mathbf{D}^{-1/2}, \nonumber \\
 	\mathbf{L}_N=\mathbf{I}-\mathbf{W}_N.  \label{relaWNLN}
 	\end{gather}
 The eigenvalues of these two matrices are therefore related as $\lambda_k^{(L_N)}=1-\lambda_k^{(W_N)}$, while they share the same corresponding eigenvectors.
 \end{Remark}  

\subsubsection{Clustering based on more than one eigenvector}

More complex clustering schemes can be achieved when using more than one Laplacian eigenvector. In turn, vertices with similar values of several  slow-varying eigenvectors, $\mathbf{u}_k$, would exhibit high spectral similarity. 

 The principle of \textbf{using more than one eigenvector in vertex clustering} and possible subsequent graph segmentation was first introduced by Scott and Longuet-Higgins \cite{scott1990feature}. They used $k$ eigenvectors of the weight matrix $\mathbf{W}$ to form a new $N\times k$ matrix $\mathbf{V}$, for which an additional row normalization was performed. The vertex clustering is then performed based on the elements of matrix $\mathbf{V}\mathbf{V}^T$.

For the normalized weight matrix, $\mathbf{W}_N$, the Scott and Longuet-Higgins algorithm reduces to the corresponding analysis with $k$ eigenvectors of the normalized graph Laplacian, $\mathbf{L}_N$. Since  $\mathbf{W}_N$ and  $\mathbf{L}_N$ are related by (\ref{relaWNLN}),   they thus have the same eigenvectors.

\begin{Example}
Consider two independent normalized cuts of a graph, where the first cut splits the graph into the sets of vertices $\mathcal{E}_1$ and $\mathcal{H}_1$,  and the second cut further splits all vertices into the sets $\mathcal{E}_2$ and $\mathcal{H}_2$, and define this two-level cut as
\begin{gather}
CutN2(\mathcal{E}_1,\mathcal{H}_1,\mathcal{E}_2,\mathcal{H}_2) = CutN(\mathcal{E}_1,\mathcal{H}_1)+CutN(\mathcal{E}_2,\mathcal{H}_2) \label{CutN2}
\end{gather}
 where both $CutN(\mathcal{E}_i,\mathcal{H}_i)$, $i=1,2$, are defined by (\ref{CutN}).

If we now introduce two indicator vectors, $\mathbf{x}_1$ and  $\mathbf{x}_2$, for the two corresponding cuts, then, from (\ref{LqfCut}) we may write
\begin{gather}
CutN2(\mathcal{E}_1,\mathcal{H}_1,\mathcal{E}_2,\mathcal{H}_2) = \frac{\mathbf{x}_1^T\mathbf{L}\mathbf{x}_1}{\mathbf{x}_1^T\mathbf{x}_1}+\frac{\mathbf{x}_2^T\mathbf{L}\mathbf{x}_2}{\mathbf{x}_2^T\mathbf{x}_2}.\label{CutN2xxx}
\end{gather}
As mentioned earlier, finding the indicator vectors, $\mathbf{x}_1$ and  $\mathbf{x}_2$, which minimize (\ref{CutN2xxx}) is a combinatorial problem. 
However, if the space of solutions for the indicator vectors is now relaxed from the subset-wise constant form to the space spanned by the eigenvectors of the graph Laplacian, then the approximative minimum value of two cuts,  $CutN2(\mathcal{E}_1,\mathcal{H}_1,\mathcal{E}_2,\mathcal{H}_2)$, is obtained for  $\mathbf{x}_1=\mathbf{u}_1$ and $\mathbf{x}_2=\mathbf{u}_2$, since $\mathbf{u}_1$ and $\mathbf{u}_2$ are maximally smooth but not constant (for the proof see (\ref{min2dl})-(\ref{min2dl2}) and for  the illustration see Example \ref{HighDimCut}). 

For the case of two independent cuts, for convenience, we may form the indicator $N \times 2$ matrix 
$\mathbf{Y}=[\mathbf{x}_1, \mathbf{x}_2]$, so that the corresponding  matrix of the solution (within the graph Laplacian eigenvectors space)  to the two normalized cuts minimization problem, has the form $$\mathbf{Q}=[\mathbf{u}_1, \mathbf{u}_2].$$
The rows of this matrix, $\mathbf{q}_n=[u_1(n), u_2(n)]$, are the spectral vectors assigned to each vertex, $n$ . 

The same reasoning can be followed for the cases of three or more independent cuts, to  obtain $N \times M$ indicator matrix $\mathbf{Y}=[\mathbf{x}_1, \mathbf{x}_2, \dots,  \mathbf{x}_M]$ with corresponding eigenvector approximation, $\mathbf{Q}$, whose rows are the spectral vectors $\mathbf{q}_n=[u_1(n),\allowbreak u_2(n), \dots,\allowbreak u_M(n)]$. 
\end{Example}

\begin{Remark}
	\textit{Graph clustering} in the spectral domain may be performed by assigning the spectral vector, $$\mathbf{q}_n = [u_1(n), \ldots, u_{M}(n)]$$ in (\ref{spectvecqd}), to each vertex, $n$, and subsequently grouping the vertices with similar spectral vectors into the corresponding  clusters \cite{ng2002spectral,belkin2003laplacian}. 
\end{Remark} 

Low dimensional spectral vectors (up to $M=3$) can be represented by color coordinates of, for example, standard RGB coloring system. To this end, it is common to use different vertex colors, which represent different spectral vectors, for the visualization of spectral domain clustering.

\begin{Example}\label{HighDimCut} Fig. \ref{LS_VF_coloring1} illustrates the clustering  for the graph in Fig. \ref {LS_VF_sig1} (right), based on the three smoothest eigenvectors $\mathbf{u}_1$, $\mathbf{u}_2$, and $\mathbf{u}_3$, shown in Figs.  \ref{LS_VF_coloring1}(a), (c), and (e), respectively.  Clustering based on the eigenvector $\mathbf{u}_1$, with $\mathbf{q}_n = [u_1(n)]$, is given in Fig. \ref{LS_VF_coloring1}(b), the clustering using the eigenvector $\mathbf{u}_2$ only, with $\mathbf{q}_n = [u_2(n)]$, is shown Fig. \ref{LS_VF_coloring1}(d), while Fig. \ref{LS_VF_coloring1}(e) gives the clustering based on the eigenvectors $\mathbf{u}_3$, when $\mathbf{q}_n = [u_3(n)]$.  Clustering based on the combination of the two smoothest eigenvectors  $\mathbf{u}_1$, and $\mathbf{u}_2$, with spectral vectors $\mathbf{q}_n = [u_1(n), u_{2}(n)]$, is shown in  Fig. \ref{LS_VF_coloring1}(g), while  Fig. \ref{LS_VF_coloring1}(h) illustrates clustering based on the  three smoothest vectors,  $\mathbf{u}_1$, $\mathbf{u}_2$, and $\mathbf{u}_3$ with the spectral vector $\mathbf{q}_n = [u_1(n), u_{2}(n),u_{3}(n)]$. In all cases, two-level colormaps were used for each eigenvector.  The smallest eigenvalues were $\lambda_0=0$,  $\lambda_1=0.0286$, $\lambda_2=0.0358$, $\lambda_3=0.0899$, $\lambda_4=0.104$, and $\lambda_5=0.167$, so that the largest relative gap is obtained if  $\mathbf{u}_1$, and $\mathbf{u}_2$ are used for clustering, with the corresponding eigenvalue gap  of $\delta_r=1-\lambda_2/\lambda_3=0.6$.

	\begin{figure*}
		\centering
		\includegraphics[]{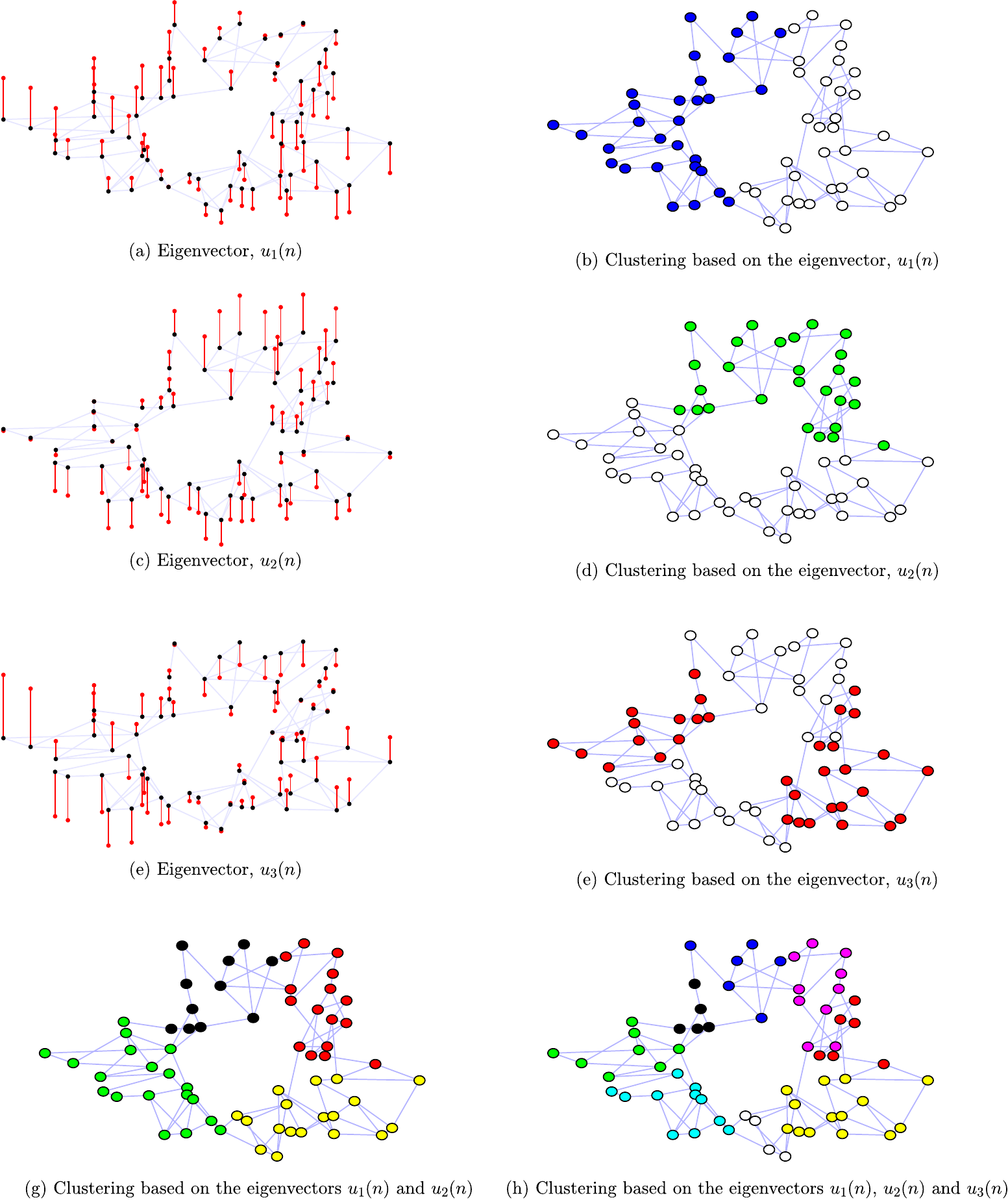}
		\caption{Spectral vertex clustering schemes for the graph from Fig. \ref{LS_VF_sig1}. (a) The   eigenvector, $\mathbf{u}_1$, of the Laplacian matrix (plotted in red lines on vertices designated by black dots) is first normalized and is then used  to designate (b) a two-level blue colormap intensity  for every vertex (blue-white circles). (c) The   eigenvector, $\mathbf{u}_2$, of the Laplacian matrix is normalized and is then used to provide (d)  a two-level green colormap intensity  for every vertex. (e) The   eigenvector, $\mathbf{u}_3$, of the Laplacian matrix is normalized and  used as (f)  a two-level red colormap intensity  for every vertex. (g) Clustering based on the combination of the eigenvectors $\mathbf{u}_1$ and $\mathbf{u}_2$. (h) Clustering based on the combination of the eigenvectors $\mathbf{u}_1$, $\mathbf{u}_2$, and $\mathbf{u}_3$. Observe an increase in degrees of freedom with the number of eigenvectors used; this is reflected in the number of detected clusters, starting from two clusters in (b) and (d), via four clusters in (g), to 8 clusters in (h). }  
		\label{LS_VF_coloring1}
	\end{figure*}
	
\end{Example}

	\begin{Remark}\label{clusteringRem}
	\textbf{k-means algorithm.} The above clustering is based on the quantized levels of spectral vectors which can be refined using \textit{the $k$-means algorithm}, that is, through postprocessing in the form of unsupervised learning and in the following way,
	
	\noindent (i) After the initial vertex clustering is performed by grouping the vertices into $\mathcal{V}_i$, $i=1,2,\dots,k$ nonoverlapping vertex subsets, a new spectral vector centroid, $\mathbf{c}_i$, is calculated as 
	$$\mathbf{c}_i=\mathrm{mean}_{n \in \mathcal{V}_i}\{\mathbf{q}_n\},$$ for each cluster of vertices $\mathcal{V}_i$; 
	
	\noindent (ii) Every vertex, $n$, is then reassigned to its nearest (most similar) spectral domain centroid, $i$, where the spectral distance (spectral similarity) is calculated as $\Vert \mathbf{q}_n-\mathbf{c}_i \Vert_2$. 
	
	\noindent This two-step algorithm is iterated until no vertex changes clusters. Finally, all vertices in one cluster are colored based on the corresponding common spectral vector $\mathbf{c}_i$ (or visually, a color representing $\mathbf{c}_i$). 

	Clustering refinement using the $k$-means algorithm is illustrated later in Example \ref{VariousEigM}.    
	\end{Remark}	

\begin{Example}\label{PHExamp} 
	Graphs represent quite a general mathematical formalism, and we will here provide only one possible physical interpretation of graph clustering. Assume that each vertex represents one out of the set of $N$ images, which exhibit both common elements and individual differences. If the edge weights are calculated so as to represent mutual similarities between these images, then spectral vertex analysis can be interpreted as follows. 
	 If the set is complete and with very high similarity among all vertices, then $W_{mn}=1$, and $\lambda_0=0, \lambda_1=N,  \lambda_2=N, \dots, \lambda_{N-1}=N$, as shown in Remark \ref{RLC}. The relative eigenvalue gap is then $\delta_r=(\lambda_2-\lambda_1)/\lambda_2=0$ and the segmentation is not possible. 
	 
	 Assume now that the considered set of images consists of two connected subsets with the respective numbers of $N_1$ and $N_2\ge N_1$ of very similar photos within each subset. In this case, the graph consists of two complete components (sub-graphs). According to Remarks \ref{RL2} and \ref{RLC}, the graph Laplacian eigenvalues are now $\lambda_0=0, \lambda_1=0, \lambda_2=N_1, \dots, \lambda_{N_1}=N_1, \lambda_{N_1+1}=N_2,\dots, \lambda_{N-1}=N_2$. Then, the graph can be well segmented into two components (sub-graphs) since the relative eigenvalue gap is  now large, $\delta_r=(\lambda_2-\lambda_1)/\lambda_2=1$.  Therefore, this case can be used for \textit{collaborative data processing} within each of these subsets. The analysis can be continued and refined for cases with more than one eigenvector and more than two subsets of vertices. Note that the segmentation represents a \textquotedblleft hard-thresholding\textquotedblright \,   operation of cutting the connections between vertices in different subsets, while the clustering represents  just a grouping of vertices, which exhibit some similarity, into subsets, while keeping their mutual connections. 
\end{Example}

\begin{Example}	
\label{Ex:images}
For enhanced intuition, we next consider a real-world dataset with  8 images, shown in Fig. \ref{image-grpaph}. The connectivity weights were  calculated using the structural similarity index (SSIM), \cite{wang2003multiscale},  with an appropriate threshold.  The so obtained weight matrix, $\mathbf{W}$, is given by   
\begin{equation}
\small
\setlength{\arraycolsep}{5pt}
\mathbf{W}= \!\!
\begin{array}{cr}
& \\
{
	\color{blue}
	\begin{matrix}
	\text{\footnotesize 0}\\
	\text{\footnotesize 1}\\
	\text{\footnotesize 2}\\
	\text{\footnotesize 3}\\
	\text{\footnotesize 4}\\
	\text{\footnotesize 5}\\
	\text{\footnotesize 6}\\
	\text{\footnotesize 7}\\
	\end{matrix}
} &  \!\! \!\!
\begin{bmatrix*}[r]
  0  &  0.49 & 0.33  & 0.29 & 0.31  &   0   &   0  & 0     \\
0.49 &   0   & 0.32  & 0    & 0.30  &  0    &   0  & 0.29  \\
0.33 & 0.33  &   0   & 0.37 & 0.30  &  0    &   0  & 0     \\
0.29 & 0     & 0.37  &   0  & 0,31 	&   0   &   0  & 0     \\
0.31 &  0.30 & 0.30  & 0.31	&  0    & 0.31  & 0.30 & 0.29  \\
  0  & 0     & 0     & 0 	& 0.31  &   0   & 0.40 & 0.48  \\
 0   & 0     &   0   & 0    & 0.30  & 0.40  &   0  & 0.64  \\
0    & 0.29  & 0     & 0    & 0.29	& 0.48 	& 0.64 &  0   \\
\end{bmatrix*},
 \\
& 
{
	\color{blue}
	\begin{matrix*}
	\text{\footnotesize 0 \hspace{3.3mm}} &
	\text{\footnotesize 1 \hspace{3.3mm}}  &
	\text{\footnotesize 2 \hspace{3.3mm}} &
	\text{\footnotesize 3 \hspace{3.3mm}} &
	\text{\footnotesize 4 \hspace{3.3mm}} &
	\text{\footnotesize 5 \hspace{3.3mm}} &
	\text{\footnotesize 6 \hspace{3.3mm}} &
	\text{\footnotesize 7 \hspace{2.5mm}}
	\end{matrix*}
}
\end{array}
\end{equation}  
while the standard graph form for this real-world scenario is shown in Fig. \ref{imageGrpaph}, with the corresponding image/vertex indexing. Notice the almost constant background in all $8$ images (the photos are taken in the wild \textquotedblleft by hand-held device\textquotedblright), and that the only differences between the images are in that the model gradually moved her head position from the left profile (bottom left) to the right profile (top right). Therefore, the two frontal face positions, at vertices $n=4$ and $n=0$, exhibit higher vertex degrees than the other head orientations, which exemplifies physical meaningfulness of graph representations. The normalized spectral vectors $\mathbf{q}_n=[u_1(n)]/||[u_1(n)]||_2$ and $\mathbf{q}_n=[u_1(n), u_2(n)]/||[u_2(n)]|_2$ for this graph were obtained as the generalized eigenvectors of the graph Laplacian, and are used to define the coloring scheme for the graph clustering in Fig. \ref{imageGraphColoring}. Similar vertex colors indicate spectral similarity of the images assigned to the corresponding vertices. 

\begin{figure}
	\centering
	\includegraphics[scale=0.7]{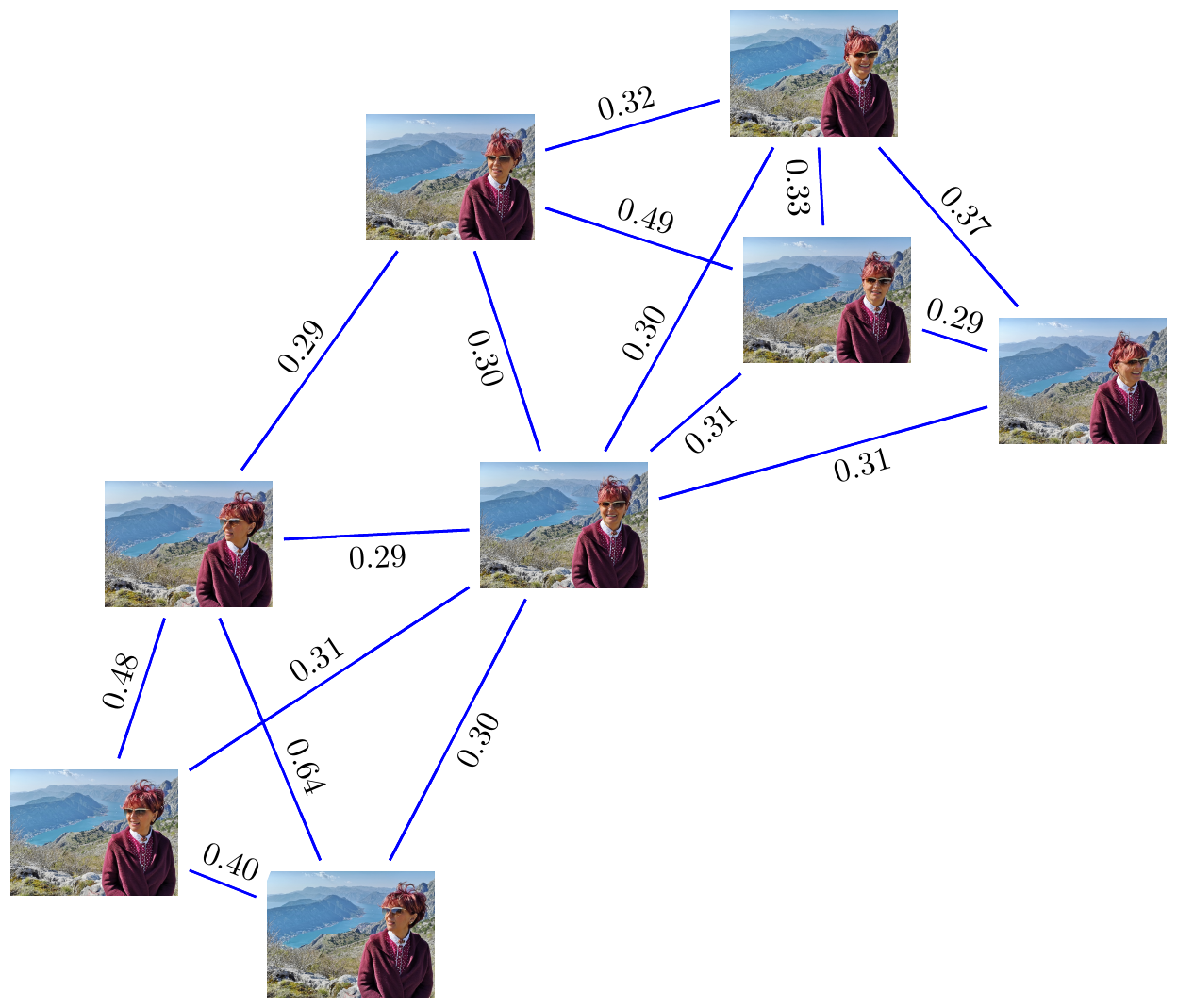}
	\caption{A graph representation of a set of the real-world images which exhibit an almost constant background but different head orientation, which moves gradually from the left profile (bottom left) to the right profile (top right). The images serve as vertices, while the edges and the corresponding weight matrix are defined through the squared structural similarity index (SSIM) between images, with $W_{mn}=\mathrm{SSIM}_T^2(m,n)$, and hard thresholded at $0.28$ to account for the contribution of the background to the similarity index, that is,  $\mathrm{SSIM}_T(m,n)=\mathrm{hard}(\mathrm{SSIM}(m,n),0.53)$.}  
	\label{image-grpaph}
\end{figure}

\begin{figure}
	\centering
	\includegraphics[scale=0.9]{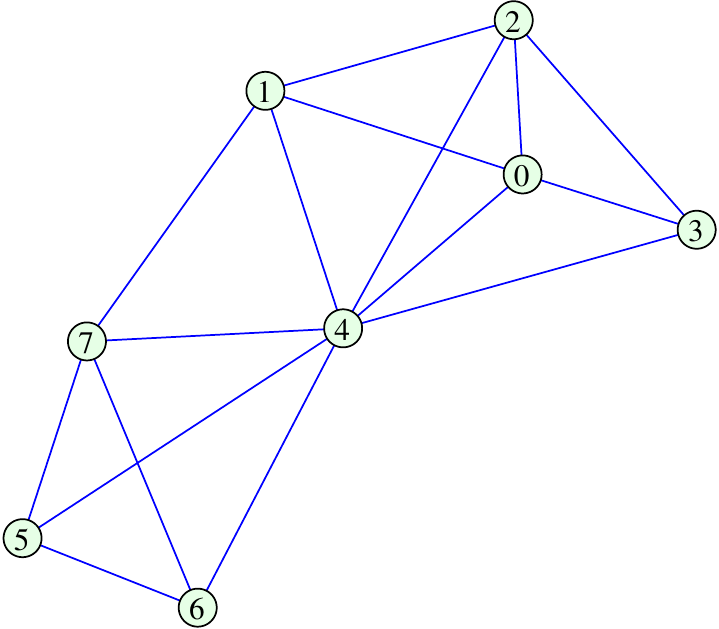}
	\caption{Graph topology for the real-world images from Fig. \ref{image-grpaph}. } 
	\label{imageGrpaph}
\end{figure}

\begin{figure}
	\centering
	(a)\includegraphics[scale=0.9]{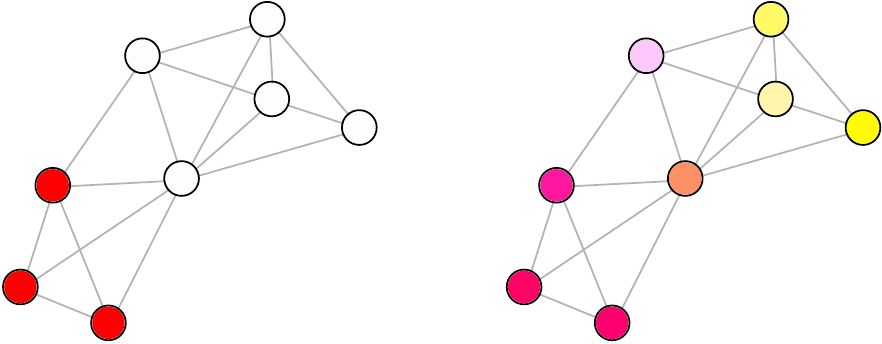}(b)
	\caption{Graph  clustering  structure for the images from Fig. \ref{image-grpaph}. (a) Vertices are clustered (colored) using the row-normalized spectral Fiedler eigenvector $\mathbf{u}_1$, $\mathbf{q}_n=[u_1(n)]/||[u_1(n)]||_2$. (b) Clustering when spectral values of vertices are calculated using two smoothest eigenvectors, $\mathbf{q}_n=[u_1(n), u_2(n)]$, which are then used to designate colormap for the vertices. Recall that similar vertex colors indicate spectral similarity of the images from Fig. \ref{image-grpaph}, which are assigned to the corresponding vertices. } 
	\label{imageGraphColoring}
\end{figure}

 The eigenvalues of the generalized eigenvectors of the graph Laplacian for this example are $\lambda_k \in \{0, \ 0.32,\allowbreak \        0.94, \         1.22, \          1.27,\allowbreak \      1.31, \   1.39, \  1.55\}.$ The largest relative eigenvalue gap is therefore between $\lambda_1=0.42$ and $\lambda_2=1.12$, and indicates that the best clustering is obtained in a one-dimensional spectral space (with clusters shown in Fig. \ref{imageGraphColoring}(a)). However, the value of such cut would be large, $Cut(\{0,1,\allowbreak 2,3,4\},\{5,6,7\})=1.19$, while the value of normalized cut, 
 $$CutN(\{0,1,2,3,4\},\{5,6,7\})  \sim \lambda_1=0.42,$$
  indicates that the connections between these two clusters are too significant for a segmented graph to produce a \textquotedblleft close\textquotedblright \, approximation of the original graph with only two components (disconnected subgraphs).  Given the gradual change in head orientation, this again conforms with physical intuition, and the subsequent clustering based on two smoothest eigenvectors, $\mathbf{u}_1$ and $\mathbf{u}_2$, yields three meaningful clusters of vertices corresponding to the \textquotedblleft  left head orientation\textquotedblright  \, (red), \textquotedblleft frontal head orientation \textquotedblright \, (two shades of pink), and \textquotedblleft right head orientation\textquotedblright  \,  (yellow).

%
	
\end{Example}

\begin{Example} \textbf{ Minnesota roadmap graph.} Three eigenvectors of the graph Laplacian matrix, $\mathbf{u}_2$, $\mathbf{u}_3$,  and $\mathbf{u}_4$,  were used as the coloring templates to represent the spectral similarity and clustering in the commonly used \textit{Minnesota roadmap graph}, shown in Fig. \ref{coloring2}. The eigenvectors $\mathbf{u}_0$ and $\mathbf{u}_1$ were omitted, since their corresponding eigenvalues are $\lambda_0=\lambda_1=0$ (due to an isolated vertex in the graph data which behaves as a graph component, see Remark \ref{RL2}). The full (nonquantized) colormap scale was used to color the vertices (represent there-dimensional spectral vectors). Regions where the vertices visually assume similar colors are also spectrally similar, with similar behavior of the corresponding slow-varying eigenvectors. 

\begin{figure}
	\centering
	\includegraphics[scale=0.75]{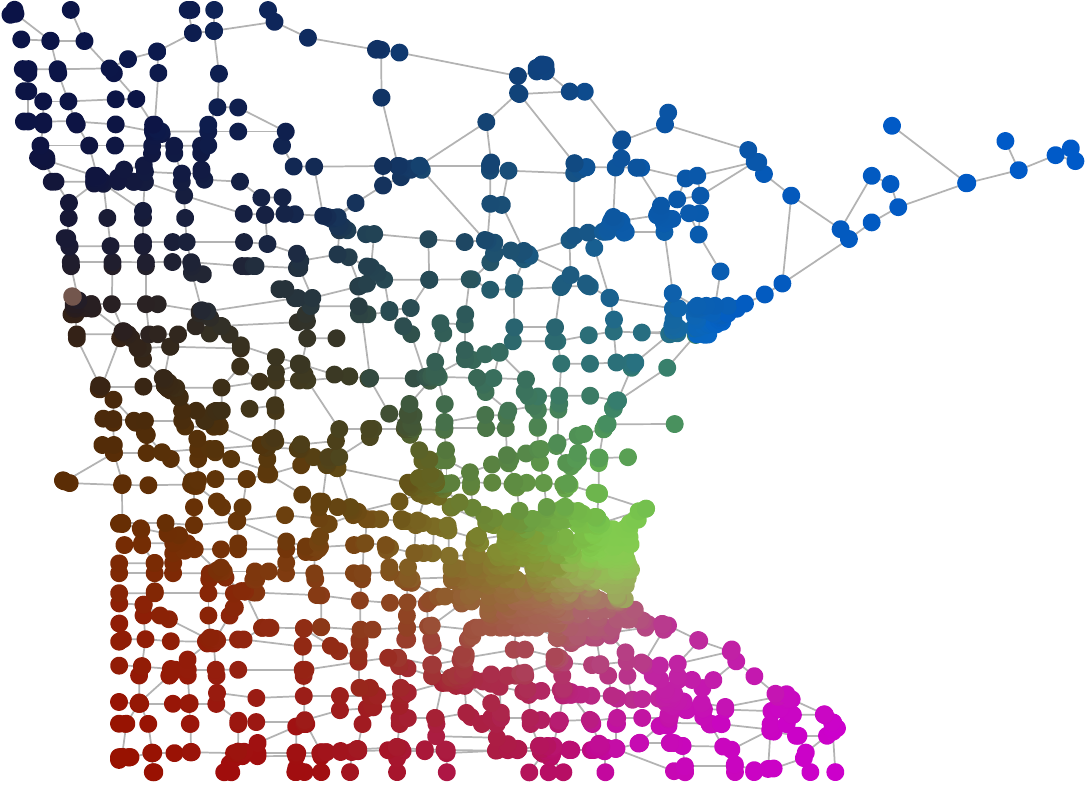}
	\caption{Vertex coloring in the benchmark Minnesota road-map graph using the three smoothest  Laplacian eigenvectors \{$\mathbf{u}_2$,$\mathbf{u}_3$,$\mathbf{u}_4$\},  as coordinates in the standard RGB coloring system (a three-dimensional spectral space with the spectral vector $\mathbf{q}_n=[u_2(n),u_3(n),u_4(n)]$ for every vertex, $n$). The vertices with similar colors are therefore also considered spectrally similar.  Observe three different clusters, characterized by the shades of predominantly red, green, and blue color that correspond to intensities defined by the eigenvectors $u_2(n),u_3(n)$, and $u_4(n)$.  } 
	\label{coloring2}
\end{figure}
\end{Example}

\begin{Example}\textbf{ Brain connectivity graph.}  Fig. \ref{brain} shows 
	the benchmark Brain Atlas connectivity graph \cite{BGRAPH,RUBINOV20101059}, where the
data is given in two matrices: \textquotedblleft Coactivation matrix\textquotedblright, $\mathbf{\hat{W}}$, and \textquotedblleft Coordinate matrix\textquotedblright. The \textquotedblleft Coordinate matrix\textquotedblright contains the vertex coordinates in a three-dimensional Euclidean space, whereby the coordinate of a vertex $n$ is defined by the $n$-th row of the \textquotedblleft Coordinate matrix\textquotedblright, that is, $[x_n,y_n,z_n]$. 

In our analysis, the graph weight matrix, $\mathbf{W}$, was empirically formed by: 

(i) Thresholding the \textquotedblleft Coactivation matrix\textquotedblright, $\mathbf{\hat{W}}$, to preserve only the strongest connections within this brain atlas, for example, those greater than $ 0.1\max\{\hat{W}_{mn}\}$, as recommended in \cite{RUBINOV20101059};

(ii) Only the edges between the vertices $m$ and $n$, whose Euclidean distance satisfies $d_{mn}\le 20$ are kept in the graph representation.  

The elements, $W_{mn}$, of the brain graph weight matrix, $\mathbf{W}$, are therefore obtained from the corresponding elements, $\hat{W}_{mn}$, of the \textquotedblleft Coactivation matrix\textquotedblright \, as   
\begin{equation}W_{mn}=\begin{cases} \hat{W}_{mn}, & \text{ if }   \hat{W}_{mn} > 0.1\max\{\hat{W}_{mn}\}  \text{ and }  d_{mn}\le 20 \\
0,  & \text{  elsewhere}.\end{cases}\end{equation} 	

Finally, the brain connectivity graph with the so defined weight matrix, $\mathbf{W}$, is presented in Fig. \ref{brain} (bottom),

	 The three smoothest generalized eigenvectors, $\mathbf{u}_1$, $\mathbf{u}_2$ and $\mathbf{u}_3$, of the corresponding graph Laplacian matrix, $\mathbf{L}=\mathbf{W}-\mathbf{D}$,   were then used to define the spectral vectors $$\mathbf{q}_n=[u_1(n),u_2(n),u_3(n)]$$ for each vertex $n=0,1,\dots,N-1$. The elements of this spectral vector, $\mathbf{q}_n$, were used to designate the corresponding RGB coordinates to color the vertices of the brain graph, as shown in Fig. \ref{brain}.

\begin{figure}
	\centering
	\fbox{
		\includegraphics[trim={3.5cm 4.5cm 3cm 4cm},clip,width=8.4cm]{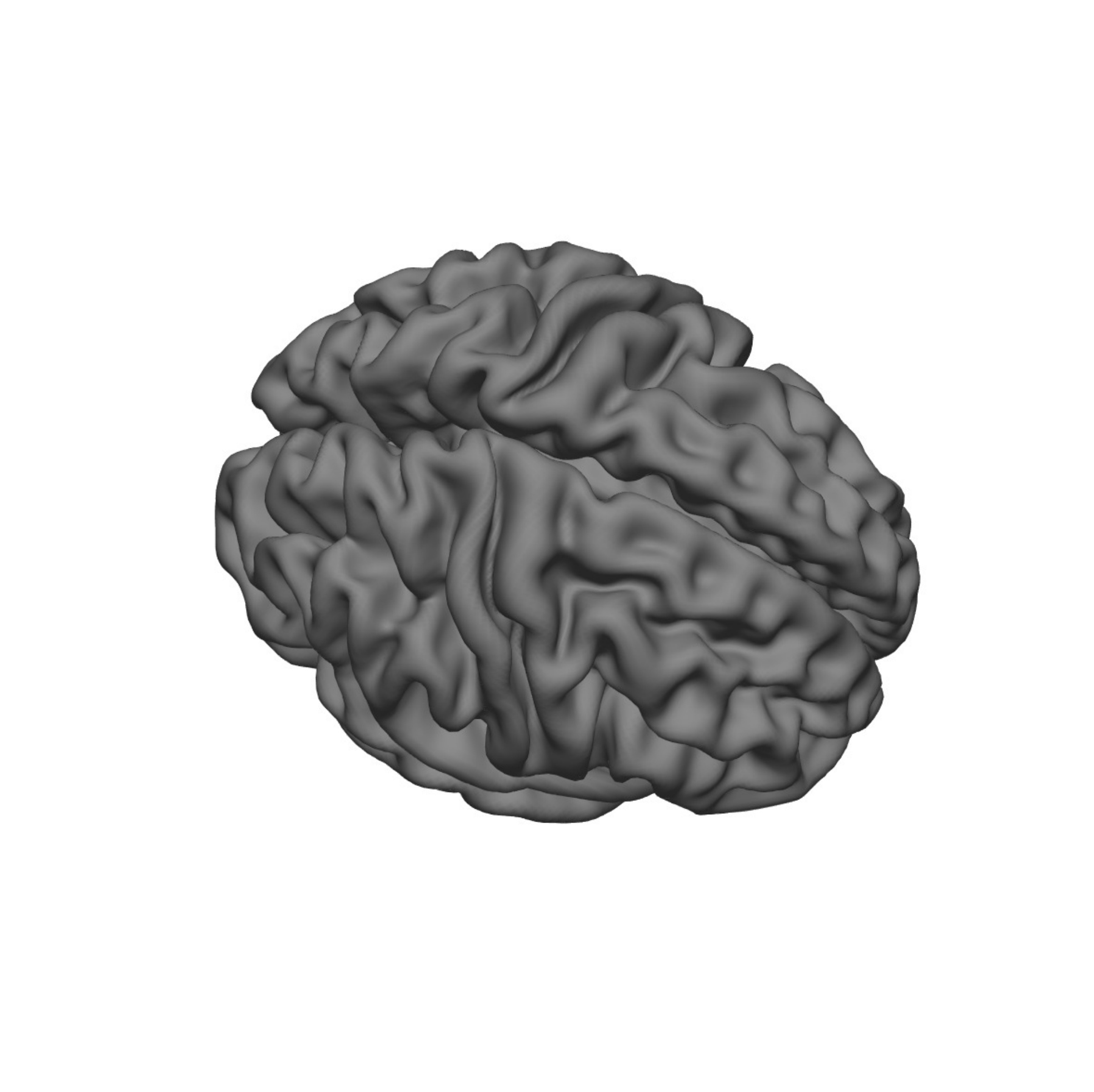}
	}
	
	\fbox{\includegraphics[trim={3.5cm 4.5cm 3cm 4cm},clip,width=8.5cm]{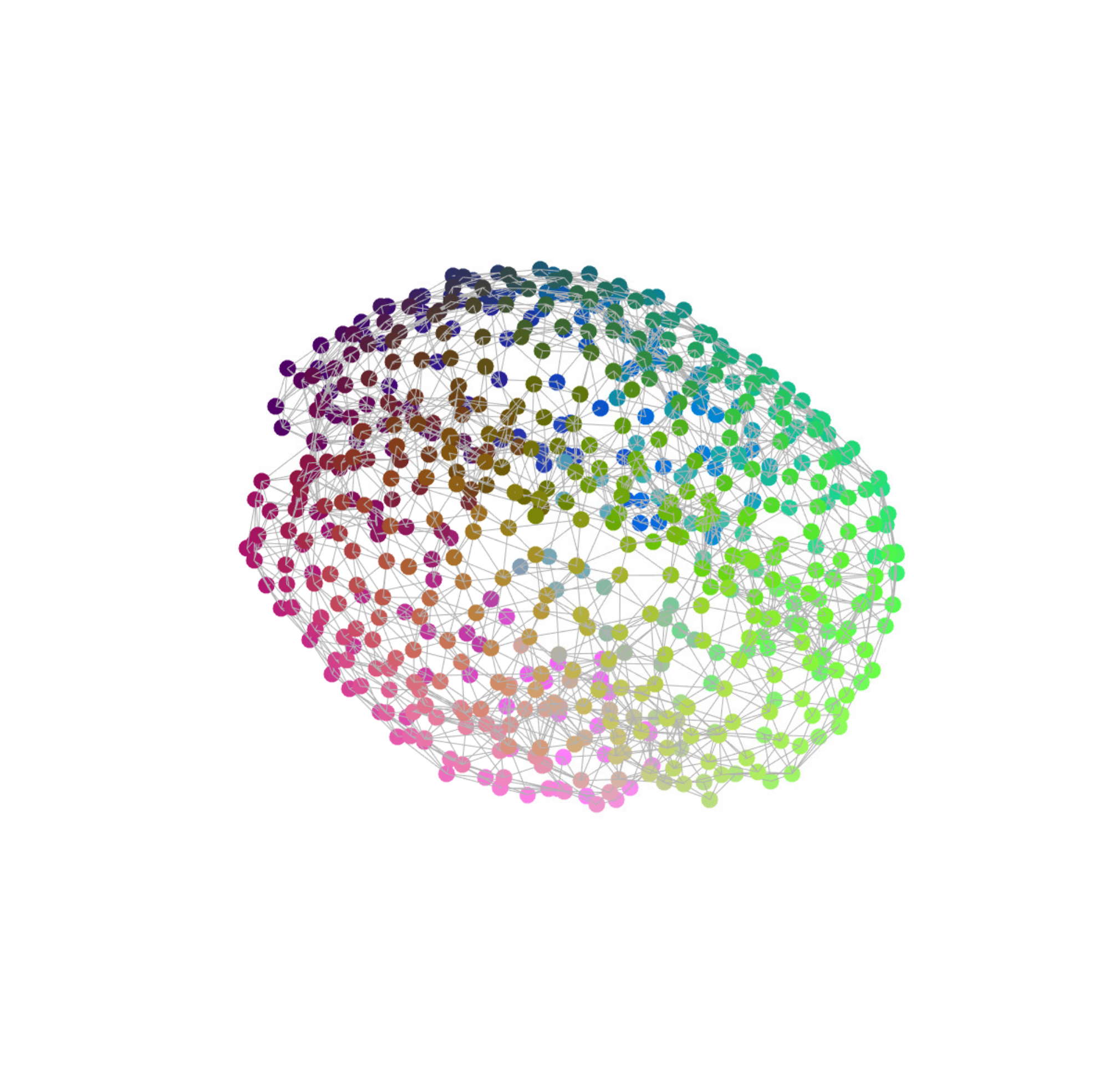}
	}
	\caption{Brain atlas (top) and its graph, with vertex coloring based on three smoothest generalized eigenvectors, $\mathbf{u}_1$, $\mathbf{u}_2$, and $\mathbf{u}_3$,  of graph Laplacian, with the spectral vector, $\mathbf{q}_n=[u_1(n),u_2(n),u_3(n)]$, employed as  coordinates for the RGB coloring scheme (bottom), \cite{BGRAPH,RUBINOV20101059}.}
	\label{brain}
\end{figure}
\end{Example}

\subsection{Vertex Dimensionality Reduction Using the Laplacian Eigenmaps}\label{EIGLMapigMaps}
	We have seen that graph clustering can be used for collaborative processing on the set of data which is represented by the vertices within a cluster. In general, \textit{any form of the presentation of a graph and its corresponding vertices that employs the eigenvectors of the graph Laplacian may be considered as the Laplacian eigenmap}.  The idea which underpins eigenmap-based approaches presented here is to employ spectral vectors, $\mathbf{q}_n$, to define the new positions of the original vertices in such a \textquotedblleft transform-domain\textquotedblright \, space so that spectrally similar vertices appear spatially closer than in the original vertex space.  

\begin{Remark}  The Laplacian eigenmaps may also be employed  for  \textit{vertex dimensionality reduction}, while at the same time preserving the local properties and natural connections within the original graph  \cite{belkin2003laplacian}. 
\end{Remark}

Consider a vertex $n$, $n=0,1,\dots,N-1$, which resides in an $L$-dimensional space $\mathbb{R}^L$, at the position defined by $L$-dimensional vector $\mathbf{r}_n$. A spectral vector for vertex $n$ is then defined in a new low-dimensional ($M$-dimensional) space by keeping the $M$ smoothest eigenvectors of graph Laplacian, $\mathbf{u}_0$, $\mathbf{u}_1$, $\dots$, $\mathbf{u}_{M-1}$, and omitting the constant eigenvector, $\mathbf{u}_0$, to give the new basis
\begin{equation}
\mathbf{q}_n=[u_1(n),\ldots,u_{M-1}(n)], \label{EIGLMap}
\end{equation}
with $M<L$, thus providing the desired dimensionality reduction of the vertex space.

\begin{Example}\textbf{Vertex dimensionality reduction.} 
Consider a set of $N=70$ students and their marks in 40 lecture courses. Every student can be considered as a vertex located in $L=40$ dimensional space at position $\mathbf{r}_n$ where $r_n(k)$ is a mark for $n$-th student at $k$-th course.  Assume that the marks are within the set $\{2,3,4,5\}$ and that some students have affinity on certain subset of courses (for example, social sciences, natural sciences and skills). This set-up can be represented in a tabular ($70\times40$) compact form as in Fig. \ref{fig:ocjene-graf}(a) where the columns contain the marks for every student (the marks are color coded).

 \begin{figure*}[ptb]
\centering
\includegraphics[]{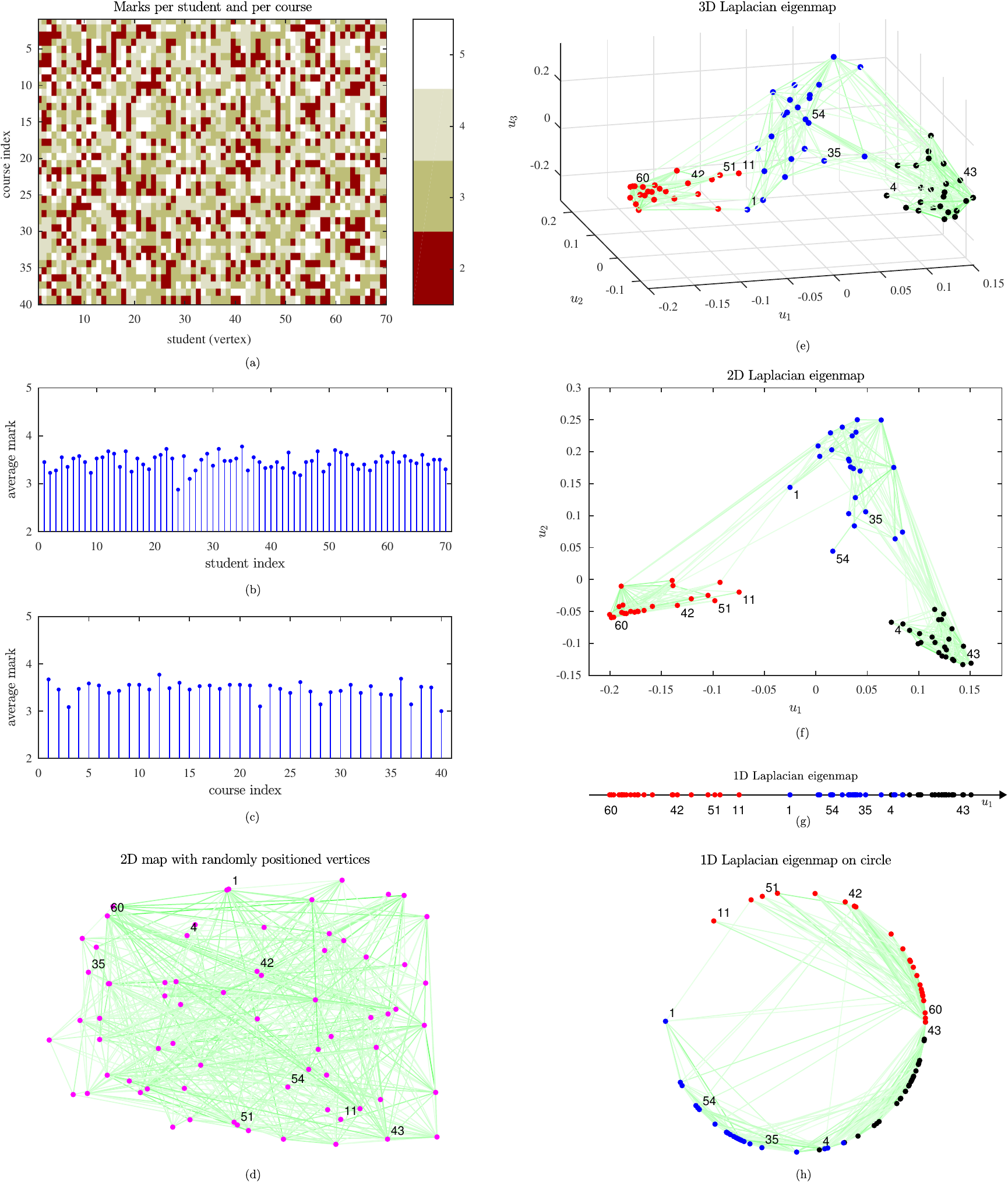}
\caption{
Dimensionality reduction example of exam marks for a cohort of students. (a) Each of the 70 columns is a 40 dimensional vector with student marks. (b) Average mark per student. (c) Average mark per course. (d) Two-dimensional graph representation of the matrix in a), where the individual students are designated with randomly positioned vertices in a plane. To perform vertex (student) dimensionality reduction we can use spectral vectors to reduce their $L=40$ dimensional representation space to (e) $M=3$, (f) $M=2$, and (g) $M=1$ dimensional representation spaces.  (h) Vertices from path graph g) positioned on a circle (by connecting the ends of the line) which allows us to also show the edges.
}
 	\label{fig:ocjene-graf}
 \end{figure*}  

The average marks per student and per course are shown in Fig.  \ref{fig:ocjene-graf}(b) and \ref{fig:ocjene-graf}(b)(c). Observe that average marks cannot be used to determine a student affinities. 

We can now create graph by connecting with edges students with similar marks. In our example, the edge weights were determined through a distance in the $40$-dimensional feature (marks) space, as
$$
W_{mn}=\begin{cases}
e^{-\Vert \mathbf{r}_m-\mathbf{r}_n \Vert_2^2/70} & \text{for } \Vert \mathbf{r}_m-\mathbf{r}_n \Vert_2  \ge 7 \\
0 & \text{otherwise.}
\end{cases}
$$
With the so obtained connectivity, the graph  presented in Fig. \ref{fig:ocjene-graf}(d), whereby the vertices (students) are randomly positioned in a plane and connected with edges.
We shall now calculate normalized Laplacian eigenvectors and remap vertices according to three-dimensional, two-dimensional and one-dimensional spectral vectors $\mathbf{q}_n$ defined by (\ref{EIGLMap}) that is, $M=3$, $M=2$, and $M=1$. In this way, the original vertex dimensionality is reduced from $L=40$ to a much lower $M\ll L$. The corresponding graph representations are respectively given in Figs. \ref{fig:ocjene-graf}(e), (f), and (g). 
For $M=2$ and $M=3$ we can divide students into three affinity groups (marked with vertex colors).
Although the obtained groups are logically ordered even in the one-dimensional case in \ref{fig:ocjene-graf}(g), we cannot use $M=1$ for precise grouping since there is no enough gap between groups. However, even in this case, if we put vertices on circle instead on a line (by connecting two ends of a line), and draw connecting edges (the same edges as in Figs. \ref{fig:ocjene-graf}(d), (e) and (f)) we can see the benefit of a graph representation even after such a radical dimensionality reduction.

The dimensionality reduction principle can also be demonstrated based on Example \ref{Ex:images}, whereby
each vertex is a $640\times480$ color image which can be represented as a vector in $L=640\times 480\times 3 = 921600$ dimensional space. 
Indeed, using spectral vectors with $M=2$, this graph can be presented in a two-dimensional space as in Fig.  \ref{image-grpaph}.

\end{Example}

Within the Laplacian eigenmaps method, we may use any of the presented  three forms of graph Laplacian eigenvectors introduced in Section \ref{ClustSEG}. Relation among these three presentations is explained in Section \ref{ClustSEG} and Table \ref{tab:1}. A unified algorithm for all three variants of the Laplacian eigenmaps, and corresponding clustering methods, is given in Algorithm \ref{Norm0Alg} in the Appendix. 

\begin{Remark}
The  Laplacian eigenmaps are optimal in the sense that they minimize an objective function which penalizes for the distance between the neighboring vertices in the spectral space. This ensures that if the vertices at the positions $\mathbf{r}_m$ and $\mathbf{r}_n$ in the original high-dimensional $L$-dimensional spectral space are \textit{“close” in some data association metric},  then they will also be  \textit{close in the Euclidean sense} in the reduced $M$-dimensional space, where these positions are defined by the corresponding spectral vectors, $\mathbf{q}_m$ and $\mathbf{q}_n$.
\end{Remark}

\subsubsection{Euclidean distances in the space of spectral vectors.}  We shall prove the \textquotedblleft distance preserving\textquotedblright \, property of this mapping in an inductive way. Assume that a graph is connected, i.e., $\lambda_1\ne 0$. The derivation is based on the quadratic form in (\ref{eq:energijaLaplaciana}) $$\mathbf{u}_k^T\mathbf{L}\mathbf{u}_k=\frac{1}{2}\sum_{m=0}^{N-1}\sum_{n=0}^{N-1}\Big(u_k(m)-u_k(n)\Big)^2W_{mn}$$
which states that  $\mathbf{u}_k^T\mathbf{L}\mathbf{u}_k$ is equal to the weighted sum of squared Euclidean distances between elements of the $m$-th and $n$-th eigenvector at vertices $m$ and $n$, for all $m$ and $n$. Recall that $\mathbf{u}_k^T\mathbf{L}\mathbf{u}_k$  is also equal to  $\lambda_k$, by definition (see the elaboration after (\ref{smmothLq})).

\noindent \textbf{Single-dimensional case.}  To reduce the original $L$-dimensional vertex space to a single-dimensional  path graph with vertex coordinates $\mathbf{q}_n=u_k(n)$, the minimum sum of the weighted squared distances between the vertices $m$ and $n$, that is
 \begin{gather*}
 \frac{1}{2}\sum_{m=0}^{N-1}\sum_{n=0}^{N-1}||\mathbf{q}(m)-\mathbf{q}(n)||_2^2W_{mn}\\
 =\frac{1}{2}\sum_{m=0}^{N-1}\sum_{n=0}^{N-1}\Big(u_k(m)-u_k(n)\Big)^2W_{mn}=\lambda_k
 \end{gather*}
  will be obtained  with the new positions of vertices, designated by $\mathbf{q}_n=[u_1(n)]$, with $k=1$, since $\min_{k, \lambda_k\ne 0}\{\lambda_k\}=\lambda_1$ is the smallest nonzero eigenvalue. 
 
  \noindent \textbf{Two-dimensional case.} If we desire to reduce the $L$-dimensional vertex space to a two-dimensional space, designated by $\mathbf{q}_n=[u_k(n),u_l(n)]$ and defined through any two eigenvectors of the graph Laplacian, $\mathbf{u}_k$ and $\mathbf{u}_l$,  then the minimum sum of the weighted squared distances between all vertices, $m$ and $n$, given by 
\begin{gather}
\frac{1}{2}\sum_{m=0}^{N-1}\sum_{n=0}^{N-1}||\mathbf{q}_m-\mathbf{q}_n||_2^2W_{mn} \nonumber\\
=\frac{1}{2}\sum_{m=0}^{N-1}\sum_{n=0}^{N-1}\Big(u_k(m)-u_k(n)\Big)^2W_{mn}+ \nonumber \\ \frac{1}{2}\sum_{m=0}^{N-1}\sum_{n=0}^{N-1}\Big(u_l(m)-u_l(n)\Big)^2W_{mn} \nonumber\\
=\mathbf{u}_k^T\mathbf{L}\mathbf{u}_k+\mathbf{u}_l^T\mathbf{L}\mathbf{u}_l=\lambda_k+\lambda_l \label{min2dl}
\end{gather}
 will be obtained with the new positions, $\mathbf{q}_n=[u_k(n), u_l(n)]$, such that $\mathbf{q}_n=[u_1(n), u_2(n)]$,  since 
 \begin{equation}\min_{k,l,k\ne l, kl \ne 0}\{\lambda_k+\lambda_l\}=\lambda_1+\lambda_2 \label{min2dl2}
 \end{equation}
  for nonzero $k$ and $l$, and having in mind that $\lambda_1 \le \lambda_2 \le \lambda_3 \le \dots \le \lambda_{N-1}$.  
 The same reasoning holds for three- and higher-dimensional new representation spaces for the vertices, which yields (\ref{EIGLMap}) as optimal vertex positions in the reduced $M$-dimensional vertex space.

 The same relations hold for both the generalized eigenvectors of the Laplacian, defined by $\mathbf{L}\mathbf{u}_k=\lambda_k\mathbf{D}\mathbf{u}_k$, and the eigenvectors of the normalized Laplacian, defined by $\mathbf{D}^{-1/2}\mathbf{L}\mathbf{D}^{-1/2}\mathbf{v}_k=\lambda_k\mathbf{v}_k$. The only difference is in their normalization conditions, $\mathbf{u}^T_k\mathbf{D}\mathbf{u}_k$ and  $\mathbf{v}^T_k\mathbf{v}_k$. 
 
 The relation between the eigenvectors of the normalized graph Laplacian, $\mathbf{v}_k$, and the   generalized eigenvectors of the graph Laplacian, $\mathbf{u}_k$, in the form $\mathbf{u}_k=\mathbf{D}^{-1/2}\mathbf{v}_k$, follows from their definitions  (see Remark \ref {VolNindicv}). Since the elements $u_1(n)$ and $u_2(n)$ are obtained by multiplying the elements $v_1(n)$ and $v_2(n)$ by the same value, $1/D_{nn}$, that is, $[u_1(n),u_2(n)]=[v_1(n),v_2(n)]/D_{nn}$, their normalized forms are identical,  $$\frac{\mathbf{q}_n}{||\mathbf{q}_n||_2}=\frac{[u_1(n),u_2(n)]}{||[u_1(n),u_2(n)]||_2}=\frac{[v_1(n),v_2(n)]}{||[v_1(n),v_2(n)]||_2}.$$
 
 \subsubsection{Examples of graph analysis in spectral space}
 \begin{Example}\label{ExForder} The graph from Fig. \ref{GSPb_ex2}, where the vertices reside in a two-dimensional plane, is shown in Fig. \ref{GSPb_ex1a-bLine}(a), while Fig. \ref{GSPb_ex1a-bLine}(b) illustrates its reduced vertex dimensionality along a line. The positions on the line are defined by the spectral vector, $\mathbf{q}_n=[u_1(n)]$, with  $\mathbf{u}_1=[0.42, \,   0.38, \,   0.35, \,  0.15, \,  -0.088, \,  -0.34,\allowbreak \,  -0.35, \,   -0.54]^T$.
 	
 \begin{figure}[ptb]
 	\centering
 \hspace{8mm}  \includegraphics[]{GSPb_ex2} \hfill 	(a)  
 	
 	\vfill
 	\vspace{8mm}
 	
 	\includegraphics[scale=.80]{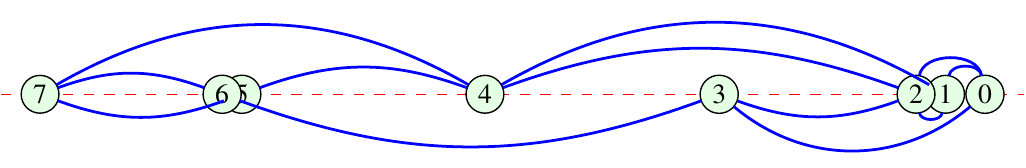}(b)
 	
 	\caption{Principle of vertex dimensionality reduction  based on the spectral vectors.  (a) The weighted graph from  Fig. \ref{GSPb_ex2} with its vertices in a two-dimensional space. (b) The graph from (a) with its vertices located along a line (one-dimensional vertex space), whereby the positions on the line are defined by the one-dimensional spectral vector, $\mathbf{q}_n=[u_1(n)]$, with $\mathbf{u}_1=[0.42, \,   0.38, \,   0.35, \,  0.15, \,  -0.088, \,  -0.34, \,  -0.35, \,   -0.54]^T$. Observe that  this dimensionality reduction method may be used for clustering, based on vertex position on the line.}
 	\label{GSPb_ex1a-bLine}
 \end{figure}  
\end{Example}

 	\begin{Remark}\label{RemRecalc}
	After the vertices are reordered according to the Fiedler eigenvector, $\mathbf{u}_1$, Example \ref{ExForder} inspires clustering refinement through the recalculation of normalized cuts. For the set of vertices $\mathcal{V}=\{0,1,2,\dots,N-1\}$, Fig. \ref{GSPb_ex1a-bLine}(b) illustrates their ordering along a line, with the new order $\{v_1,v_2,\dots,v_{N}\}=\{7,6,5,4,3,2,1,0\}$. Instead of using the sign of  $\mathbf{u}_1$ to cluster the vertices, we can recalculate the normalized cuts, $CutN(\mathcal{E}_p,\mathcal{H}_p)$, with this sequential vertex order, where $\mathcal{E}_p=\{v_1,v_2,\dots,v_p\}$ and $\mathcal{H}_p=\{v_{p+1},v_{p+2},\dots,v_N\}$, for $p=1,2,\dots,N-1$. The estimation of the minimum normalized cut  then becomes 
	$$(\mathcal{E}_p,\mathcal{H}_p)=\arg \min_p \{CutN(\mathcal{E}_p,\mathcal{H}_p)\}.$$ 
	This method is computationally efficient since only $N-1$ cuts, $CutN(\mathcal{E}_p,\mathcal{H}_p)$, need to be calculated. In addition, the cuts  $CutN(\mathcal{E}_p,\mathcal{H}_p)$ can be calculated recursively, using the previous $CutN(\mathcal{E}_{p-1},\mathcal{H}_{p-1})$ and the connectivity parameters (degree, $D_{pp}$, and weights, $W_{pm}$) of vertex $p$.  Any normalized cut form presented in  Section \ref{NormCutsSec} can also be used instead of $CutN(\mathcal{E}_p,\mathcal{H}_p)$. When the Cheeger ratio, defined in (\ref{CheegerC}), is used in this minimization,  then a simple upper bound on the normalized cut can be obtained as  \cite{trevisan2013lecture} \begin{equation}\min_p\{{\phi(\mathcal{E}_p)}\} \le \sqrt{2\lambda_1} \le 2\sqrt{\phi(\mathcal{V})},
	\end{equation}
	where $\phi(\mathcal{V})$ denotes the combinatorial (true) minimum cut, with bounds given in (\ref{CheegerB}).
\end{Remark}

\begin{Example}\label{VariousEigM} We shall now revisit the graph in Fig. \ref{LS_VF_coloring1} and examine the clustering scheme based on  (i) standard Laplacian eigenvectors in (Fig.  \ref{LS_VF_coloring1a}), (ii) generalized eigenvectors of graph Laplacian (Fig. \ref{LS_VF_coloring1aGen}), and (iii) eigenvectors of the normalized Laplacian (Fig. \ref{LS_VF_coloring1aNor}). Fig. \ref{LS_VF_coloring1a}(b) illustrates  Laplacian eigenmaps based dimensionality reduction for the graph from Fig. \ref{LS_VF_coloring1}(g), with the two eigenvectors, $\mathbf{u}_1$ and $\mathbf{u}_2$, serving  as new vertex coordinates, and using the same vertex coloring scheme as in  Fig. \ref{LS_VF_coloring1}(g). While both the original and the new vertex space are two-dimensional, we can clearly see that in the new vertex space the vertices belonging to the same clusters are also spatially closer, which is both physically meaningful and exemplifies the practical value of the eigenmaps. Fig. \ref{LS_VF_coloring1a}(c) is similar to  Fig. \ref{LS_VF_coloring1a}(b) but is presented using the normalized spectral space coordinates, $\mathbf{q}_n=[u_1(n),u_2(n)]/||[u_1(n),u_2(n)]||_2$. In Fig. \ref{LS_VF_coloring1a}(d) the clusters are refined using the $k$-means algorithm, as per Remark \ref{clusteringRem}. The  same representations are repeated and shown in Fig. \ref{LS_VF_coloring1aGen}(a)-(d) for the representation based on the generalized eigenvectors of the graph Laplacian, obtained as a solution to  $\mathbf{L}\mathbf{u}_k=\lambda_k\mathbf{D}\mathbf{u}_k$. Finally, in Fig. \ref{LS_VF_coloring1aNor}(a)-(d), the Laplacian eigenmaps and clustering are produced based on the eigenvectors of the normalized graph Laplacian, $\mathbf{L}_N=\mathbf{D}^{-1/2}\mathbf{L}\mathbf{D}^{-1/2}$. 
 As expected, the eigenmaps obtained using the generalized Laplacian eigenvectors, in Fig. \ref{LS_VF_coloring1aNor}(b), and the eigenvectors of the normalized Laplacian, in \ref{LS_VF_coloring1aGen}(b), are different; however, they reduce to the same eigenmaps after spectral vector normalization, as shown Fig. \ref{LS_VF_coloring1aNor}(c) and Fig. \ref{LS_VF_coloring1aGen}(c).  After the $k$-means based clustering refinement was applied, in all three cases two vertices switched their initial color (cluster), as shown in Fig. \ref{LS_VF_coloring1a}(d), Fig. \ref{LS_VF_coloring1aGen}(d), and Fig. \ref{LS_VF_coloring1aNor}(d).
 	
Observe that the eigenmaps obtained  with the normalized forms of the generalized eigenvectors of the Laplacian and the eigenvectors of the normalized Laplacian are the same, and in this case their clustering performances are similar to those based on the eigenmaps produced with eigenvectors of the  original Laplacian.

	\begin{figure}
		\centering
		\includegraphics[scale=0.75]{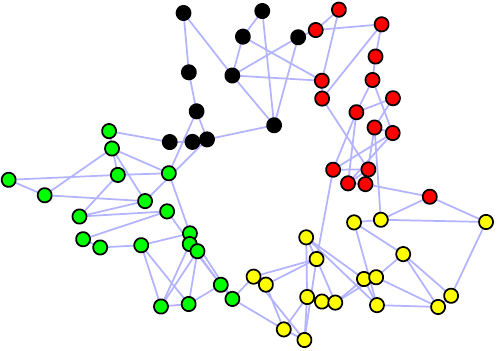}(a)
		\hfill
		\includegraphics[scale=0.75]{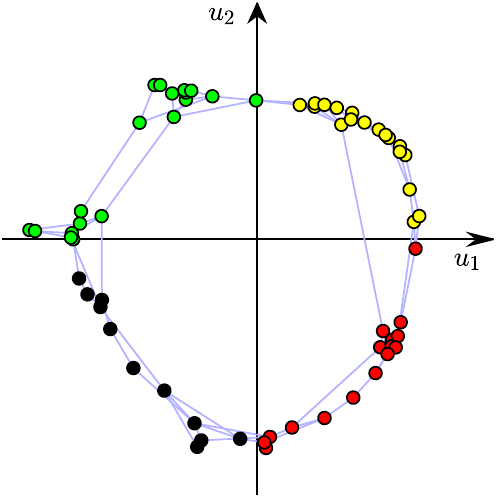}(b)
		\vfill \vspace{5mm}
		\includegraphics[scale=0.75]{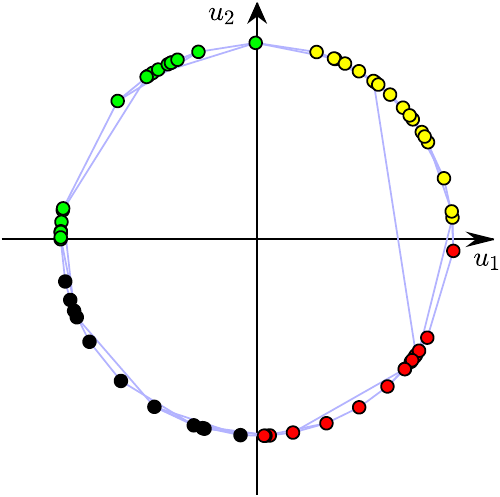}(c)
		\hfill
		\includegraphics[scale=0.75]{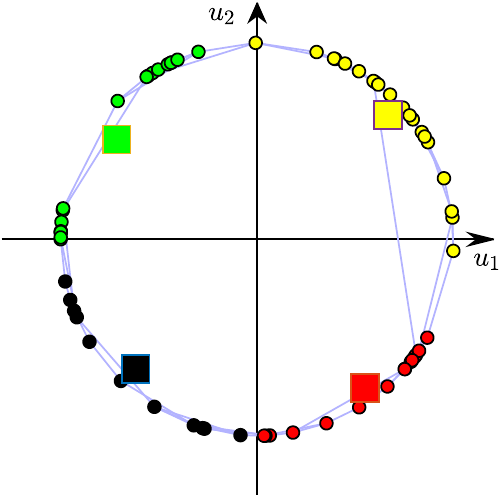}(d)
		
		\caption{Principle of Laplacian eigenmaps and clustering based on the eigenvectors of the graph Laplacian, $\mathbf{L}$. (a) The original graph from Fig. \ref{LS_VF_coloring1}, with  the spectral vector $\mathbf{q}_n=[u_1(n),u_2(n)]$, defined by the graph Laplacian eigenvectors \{$\mathbf{u}_1$,$\mathbf{u}_2$\}, used to cluster (color) vertices. (b) Two-dimensional vertex positions obtained through Laplacian eigenmaps, with the spectral vector $\mathbf{q}_n=[u_1(n),u_2(n)]$ serving as the vertex coordinates (the 2D Laplacian eigenmap). While both the original and this new vertex space are  two-dimensional, the new eigenmaps-based space is advantageous in that it emphasizes vertex spectral similarity in a spatial way (physical closeness of spectrally similar vertices). (c) The graph from (b) but produced using normalized spectral space coordinates $\mathbf{q}_n=[u_1(n),u_2(n)]/||[u_1(n),u_2(n)]||_2$, as in (\ref{spectvectnorm}). (d) The graph from (c) with clusters refined using the $k$-means algorithm, as in Remark \ref{clusteringRem}. The centroids of clusters are designated by squares of the same color. The complexity of graph presentation is also significantly reduced, with most of the edges between strongly connected vertices being very short and located along a circle. } 
		\label{LS_VF_coloring1a}
	\end{figure} 

\begin{figure}
	\centering
	\includegraphics[scale=0.75]{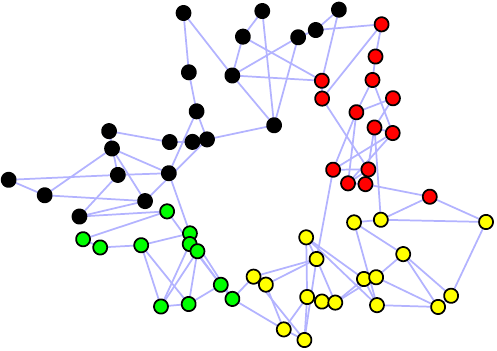}(a)
	\hfill
	\includegraphics[scale=0.75]{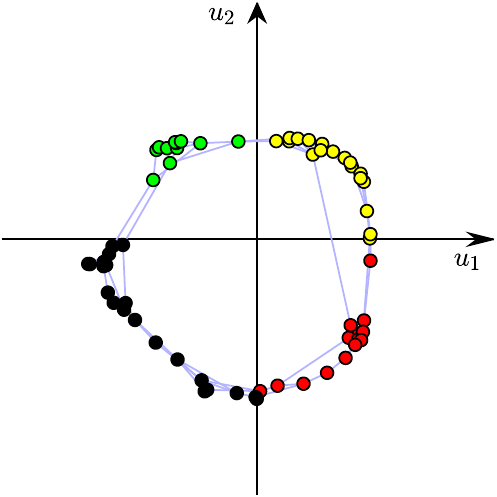}(b)
	\vfill \vspace{5mm}
	\includegraphics[scale=0.75]{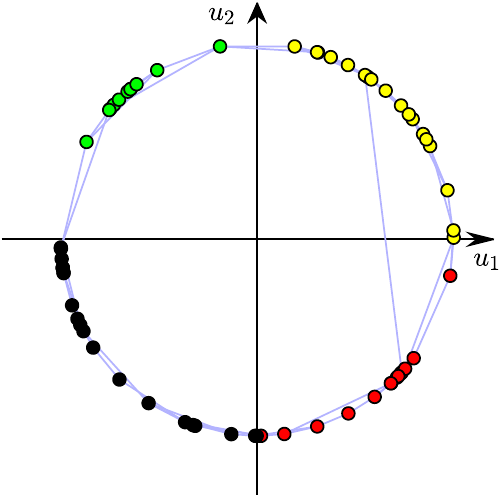}(c)
	\hfill
	\includegraphics[scale=0.75]{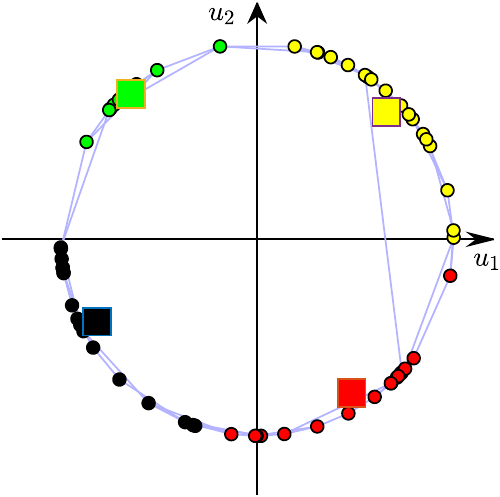}(d)
	\caption{Principle of Laplacian eigenmaps and clustering based on the \textit{generalized eigenvectors of the graph Laplacian}, obtained as a solution to  $\mathbf{L}\mathbf{u}_k=\lambda_k\mathbf{D}\mathbf{u}_k$. Vertex coloring was produced using the same procedure as in Fig. \ref{LS_VF_coloring1a}.}
	\label{LS_VF_coloring1aGen}
\end{figure} 
	
		\begin{figure}
				\centering
			\includegraphics[scale=0.75]{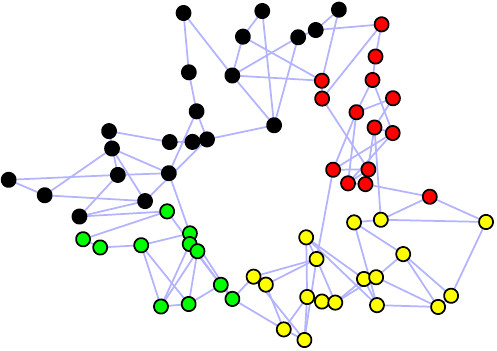}(a)
			\hfill
			\includegraphics[scale=0.75]{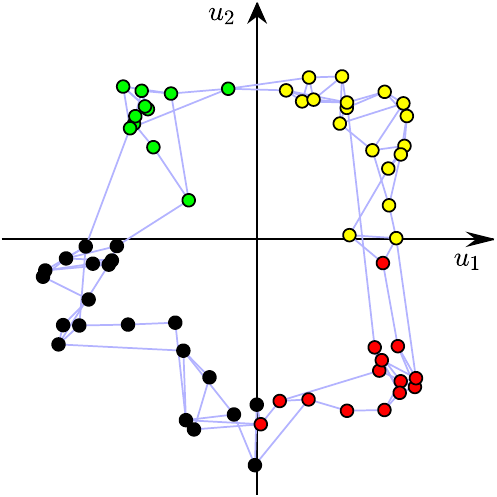}(b)
			\vfill \vspace{5mm}
			\includegraphics[scale=0.75]{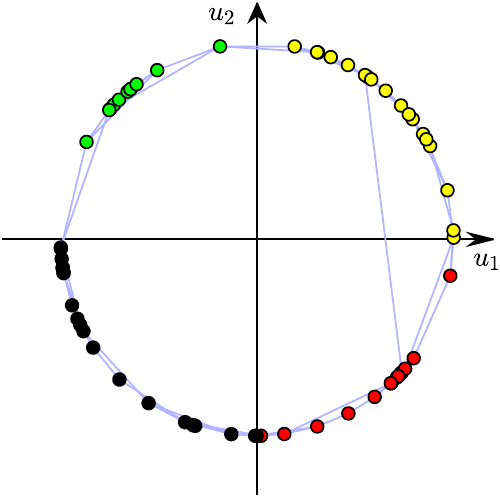}(c)
			\hfill
			\includegraphics[scale=0.75]{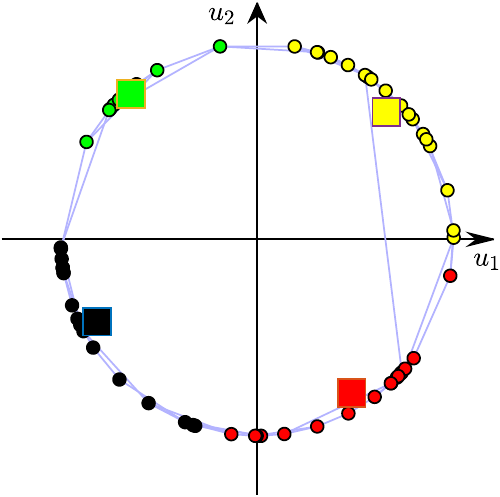}(d)
		\caption{Principle of Laplacian eigenmaps and clustering based on the \textit{eigenvectors of the normalized graph Laplacian}, $\mathbf{L}_N=\mathbf{D}^{-1/2}\mathbf{L}\mathbf{D}^{-1/2}$. Vertex coloring was performed using the same procedure as in Fig. \ref{LS_VF_coloring1a}. The eigenvectors of the normalized graph Laplacian, $\mathbf{v}_k$, are related to the   generalized eigenvectors of the graph Laplacian, $\mathbf{u}_k$, through $\mathbf{u}_k=\mathbf{D}^{-1/2}\mathbf{v}_k$, as stated in Remark \ref {VolNindicv}. This means that the signs of these two eigenvectors are the same, $\mathrm{sign}(\mathbf{u}_k)=\mathrm{sign}(\mathbf{v}_k)$. Since in order to obtain  $u_1(n)$ and $u_2(n)$, the elements $v_1(n)$ and $v_2(n)$ are multiplied by the same value, $1/D_{nn}$,  then $[u_1(n),u_2(n)]/||[u_1(n),u_2(n)]||_2=[v_1(n),v_2(n)]/||[v_1(n),v_2(n)]||_2$, thus yielding the same graph forms in (c) and (d) in both this figure and in Fig. \ref{LS_VF_coloring1aGen}. } 
		\label{LS_VF_coloring1aNor}
	\end{figure}

\end{Example}

\begin{Remark}
In general,  an independent quantization of two smoothest  eigenvectors of the graph Laplacian, $\mathbf{u}_1$ and $\mathbf{u}_2$, will produce four clusters. However, that will not be the case if we analyze the graph with an almost ideal eigenvalue gap (unit value) between $\lambda_2$ and $\lambda_3$.  In other words, when the gap $\delta_r=1-\lambda_2/\lambda_3$ tends to $1$, that is, $\lambda_2 \rightarrow 0$ and $\lambda_1<\lambda_2 \rightarrow 0$, then this case corresponds to a graph with exactly two disjoint subgraph components, with vertices belonging to the disjoint sets $\mathcal{E}$, $\mathcal{H}$, and $\mathcal{K}$. Without loss of generality, assume $N_{\mathcal{E}}>N_{\mathcal{H}}>N_{\mathcal{K}}$.  The minimum normalized cut, $CutN(\mathcal{E},\mathcal{H}\cup\mathcal{K})$ is then obtained with the first indicator vector  $x_1(n)=c_{11}$ for $n \in \mathcal{E}$ and  $x_1(n)=c_{12}$ for $n \in \mathcal{H}\cup\mathcal{K}$. The second indicator vector will produce the next minimum normalized cut, $CutN(\mathcal{E}\cup \mathcal{K},\mathcal{H})$ with $x_2(n)=c_{21}$ for $n \in \mathcal{E}\cup \mathcal{K}$ and  $x_2(n)=c_{22}$ for $n \in \mathcal{H}$. Following the same analysis as in the case of one indicator vector and the cut of graph into two disjoint subsets of vertices, we can immediately conclude that the two smoothest eigenvectors, $\mathbf{u}_1$ and $\mathbf{u}_2$, which correspond to $\lambda_2 \rightarrow 0$ and $\lambda_1 \rightarrow 0$, can be used to form an indicator matrix
$\mathbf{Y}=[\mathbf{x}_1, \mathbf{x}_2]$, so that the corresponding  matrix of the solution (within the graph Laplacian eigenvectors space) to the minimization problem of two normalized cuts, has the form $[\mathrm{sign}(\mathbf{u}_1), \mathrm{sign}(\mathbf{u}_2)].$ The elements of indicator vectors,  $[\mathrm{sign}(u_1(n)), \mathrm{sign}({u}_2(n))]$, have therefore a subset-wise constant  vector form, assuming exactly three different vector values that correspond to individual disjoint sets $\mathcal{E}$, $\mathcal{H}$, and $\mathcal{K}$. 

This procedure can be generalized up to every individual vertex becoming a cluster (no clustering). To characterize $N$ independent disjoint sets we will need $N-1$ spectral vectors, if the constant eigenvector,  $\mathbf{u}_0$, is omitted.   
\end{Remark}

\begin{Example} The Laplacian eigenmap for the Minnesota roadmap graph in the two-dimensional case, with $M=2$, is given in Fig. \ref{coloring2a}. In this new space, the spectral vectors  $\mathbf{q}_n=[u_2(n),u_3(n)]$, are used as the coordinates of new vertex positions,   $\mathbf{u}_2$, $\mathbf{u}_3$. Here, two vertices with similar slow-varying eigenvectors  are located close to one another in the new coordinate system defined by $\mathbf{u}_2$, $\mathbf{u}_3$.  This illustrates that the eigenmaps can be considered as a basis for \textquotedblleft scale-wise \textquotedblright  graph representation.

\begin{figure}
	\centering
	\includegraphics[scale=0.75]{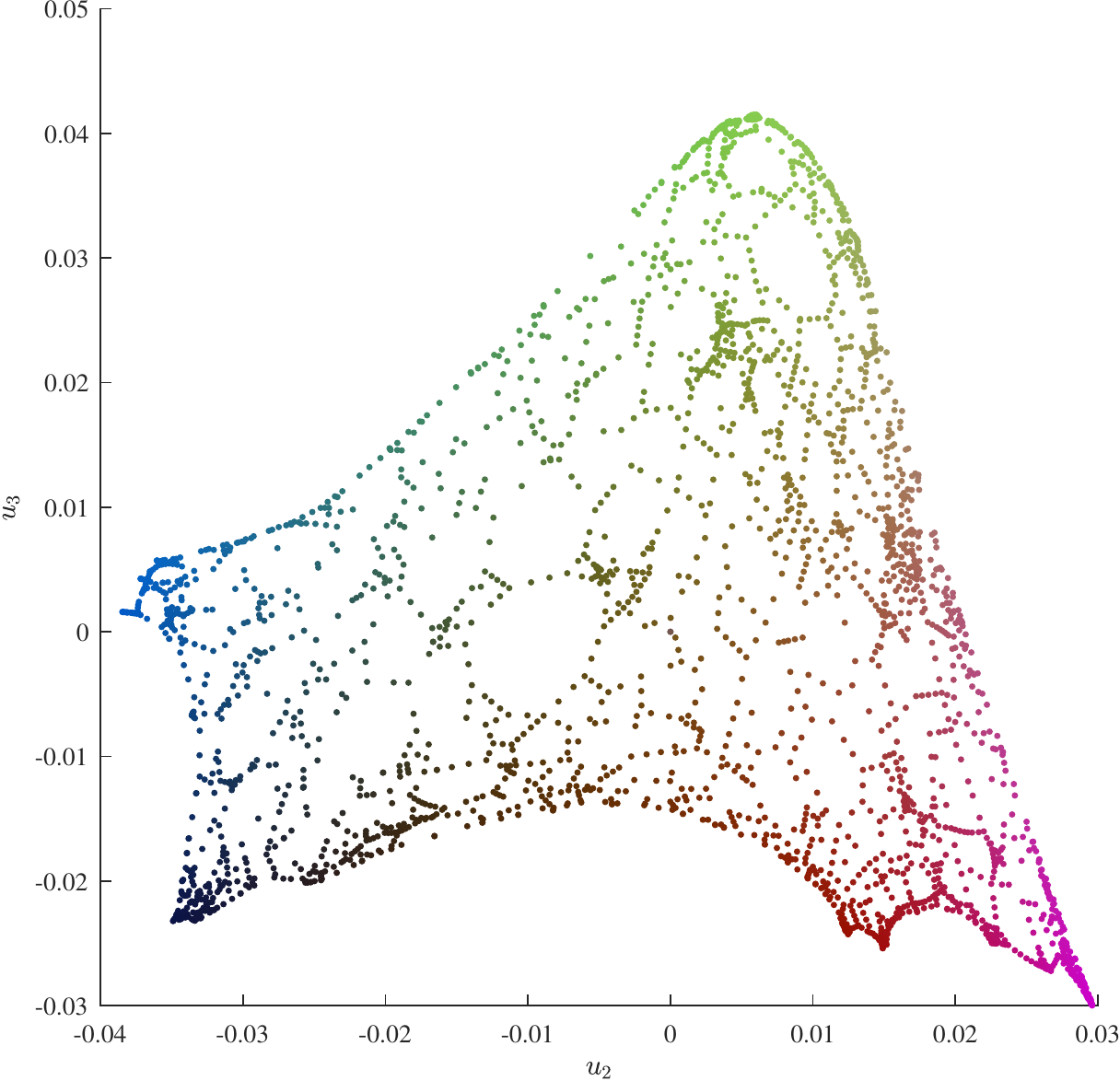}
	\caption{Laplacian eigenmaps for the Minnesota road-map graph, produced based on new two-dimensional vertex positions defined by the Laplacian eigenvectors \{$\mathbf{u}_2$,$\mathbf{u}_3$\} as the vertex coordinates (the 2D Laplacian eigenmap).}
	\label{coloring2a}
\end{figure}

\end{Example}

\begin{Example} The Laplacian eigenmaps of the Brain Atlas graph from Fig. \ref{brain}, whose original vertex locations reside in an $L=3$ dimensional space, is presented in a new reduced $M=2$ dimensional space  defined based on the two smoothest eigenvectors, $\mathbf{u}_1$ and $\mathbf{u}_2$. This is an example of vertex dimensionality reduction. This new graph, with new vertex locations but with  the original edges kept, is shown in Fig. \ref{brainLE}.
	
	\begin{figure}[tbh]
		\centering
		\includegraphics{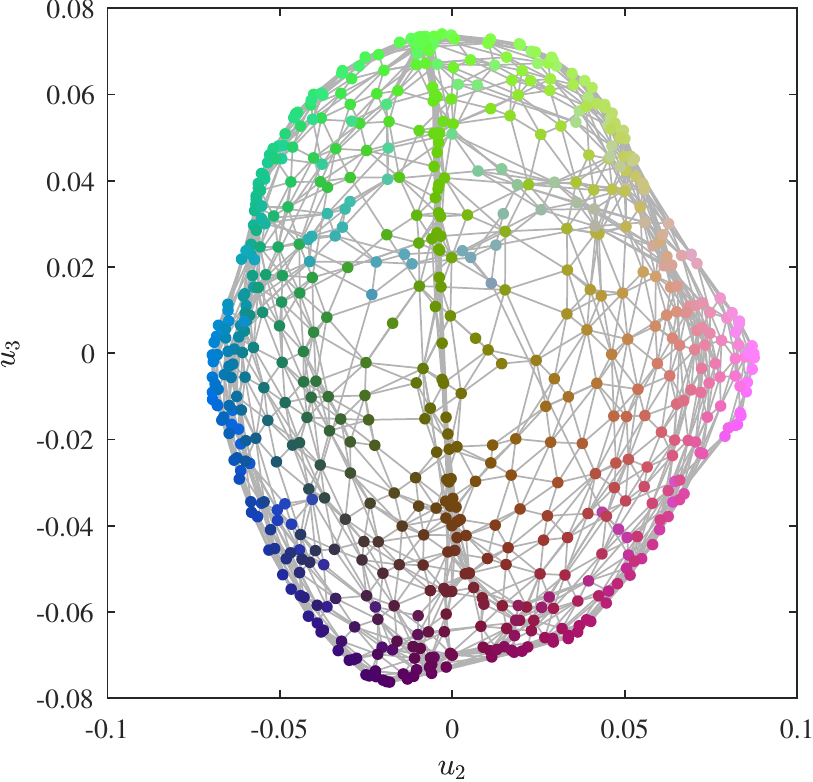}(a)
		
		\vfill
		
		\includegraphics{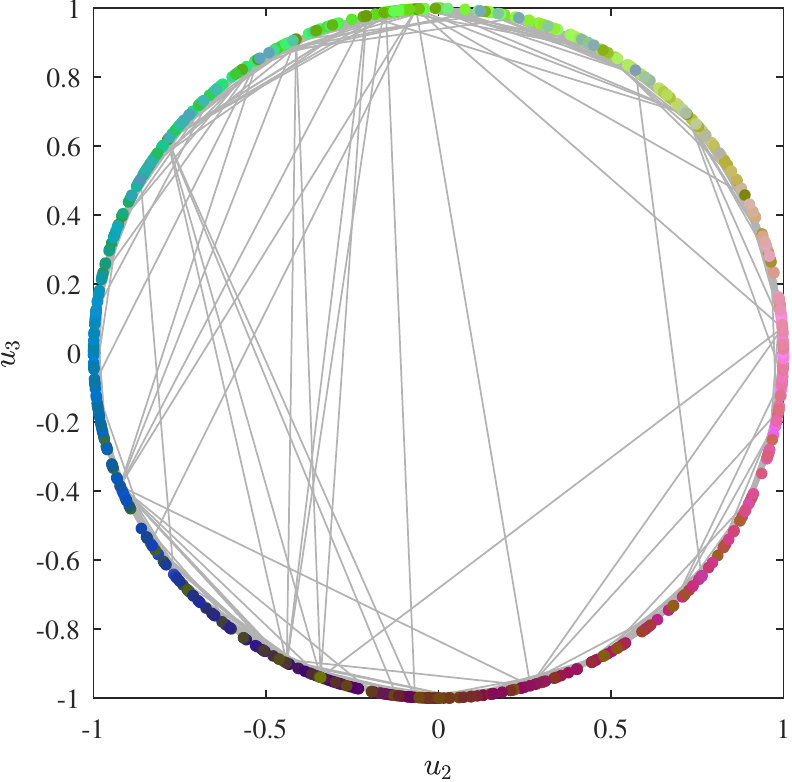}(b) 
		
		\caption{Brain atlas representation based on  normalized spectral vectors. (a) A two-dimensional Laplacian eigenmap based on the generalized  Laplacian eigenvectors. The  original $L=3$ dimensional graph from Fig. \ref{brain} is reduced to a two-dimensional representation based on the  two smoothest eigenvectors, $u_1(n)$ and $u_2(n)$, which both serve as spectral coordinates and define color templates in the colormap, as in Fig. \ref{brain}. (b) Eigenmaps from (a) but in the space of normalized spectral space coordinates, $\mathbf{q}_n=[u_2(n),u_3(n)]/||[u_2(n),u_3(n)]||_2$, with the complexity of graph representation now significantly reduced. Observe that most edges only exists between strongly connected vertices located along the circle. }
		\label{brainLE}
	\end{figure}
	
	The generalized eigenvectors of the graph Laplacian, $\mathbf{u}_k$, for $k=1,2,3,4,5,6$, are shown in Fig. \ref{ppp_Eig1_6}(a) using the standard colormap in both the original three-dimensional and the reduced two-dimensional space, as shown in Fig. \ref{ppp_Eig1_6}(b). 
	
		\begin{figure*}
		\centering
\fbox{\includegraphics[trim={2.2cm 2.2cm 1.8cm 4.5cm},clip,width=5.5cm]{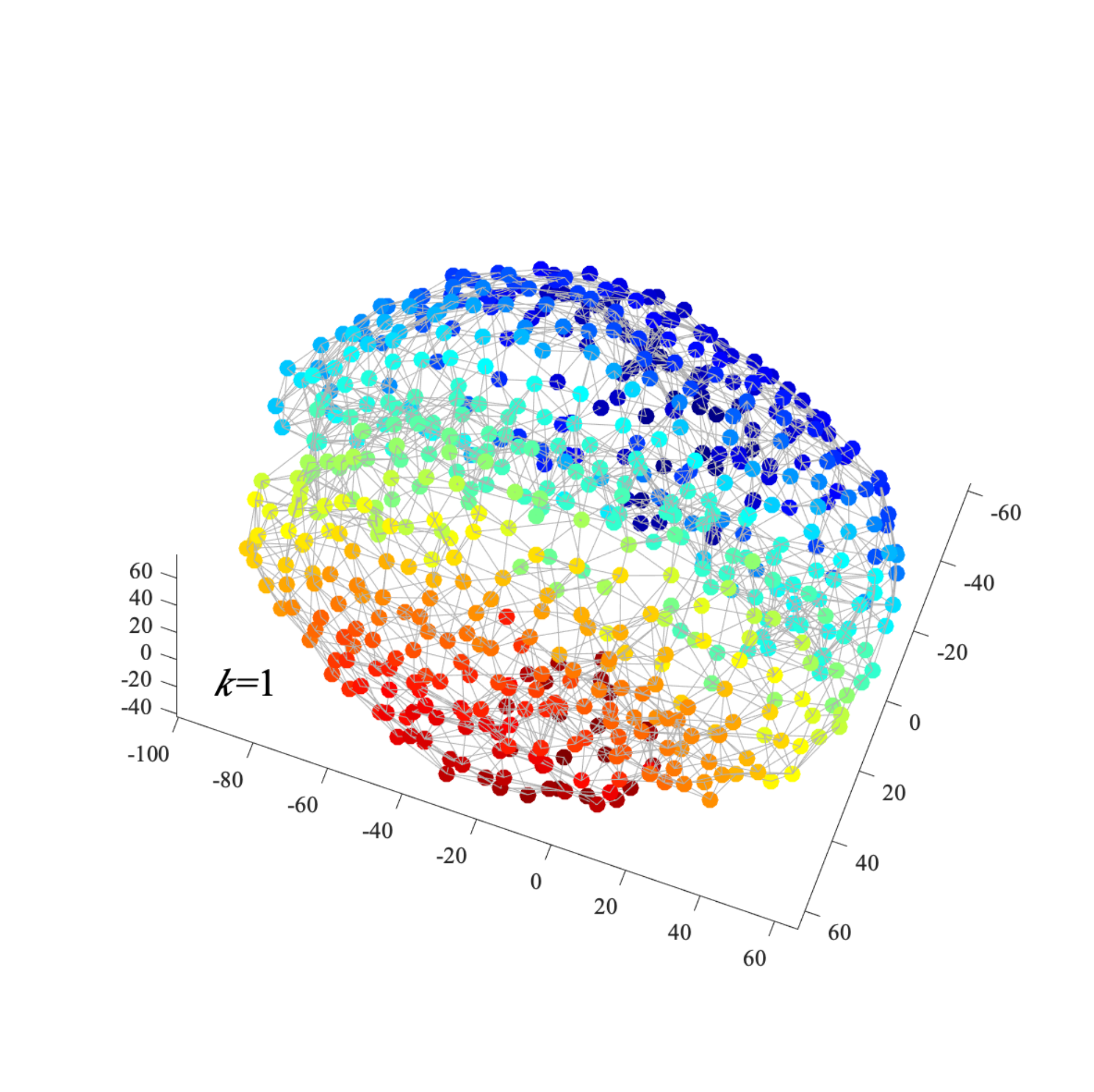}}
\fbox{\includegraphics[trim={2.2cm 2.2cm 1.8cm 4.5cm},clip,width=5.5cm]{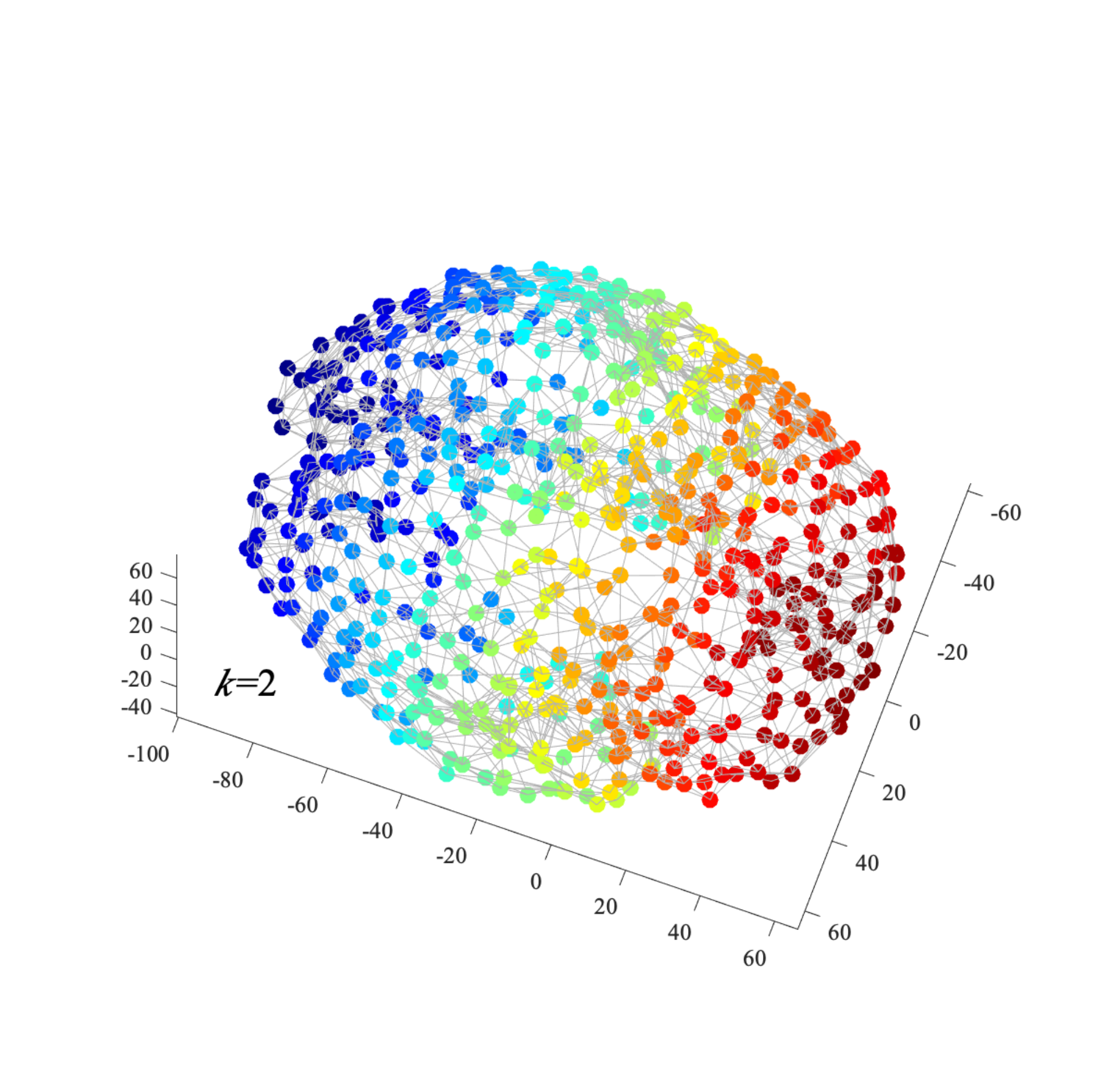}}
\fbox{\includegraphics[trim={2.2cm 2.2cm 1.8cm 4.5cm},clip,width=5.5cm]{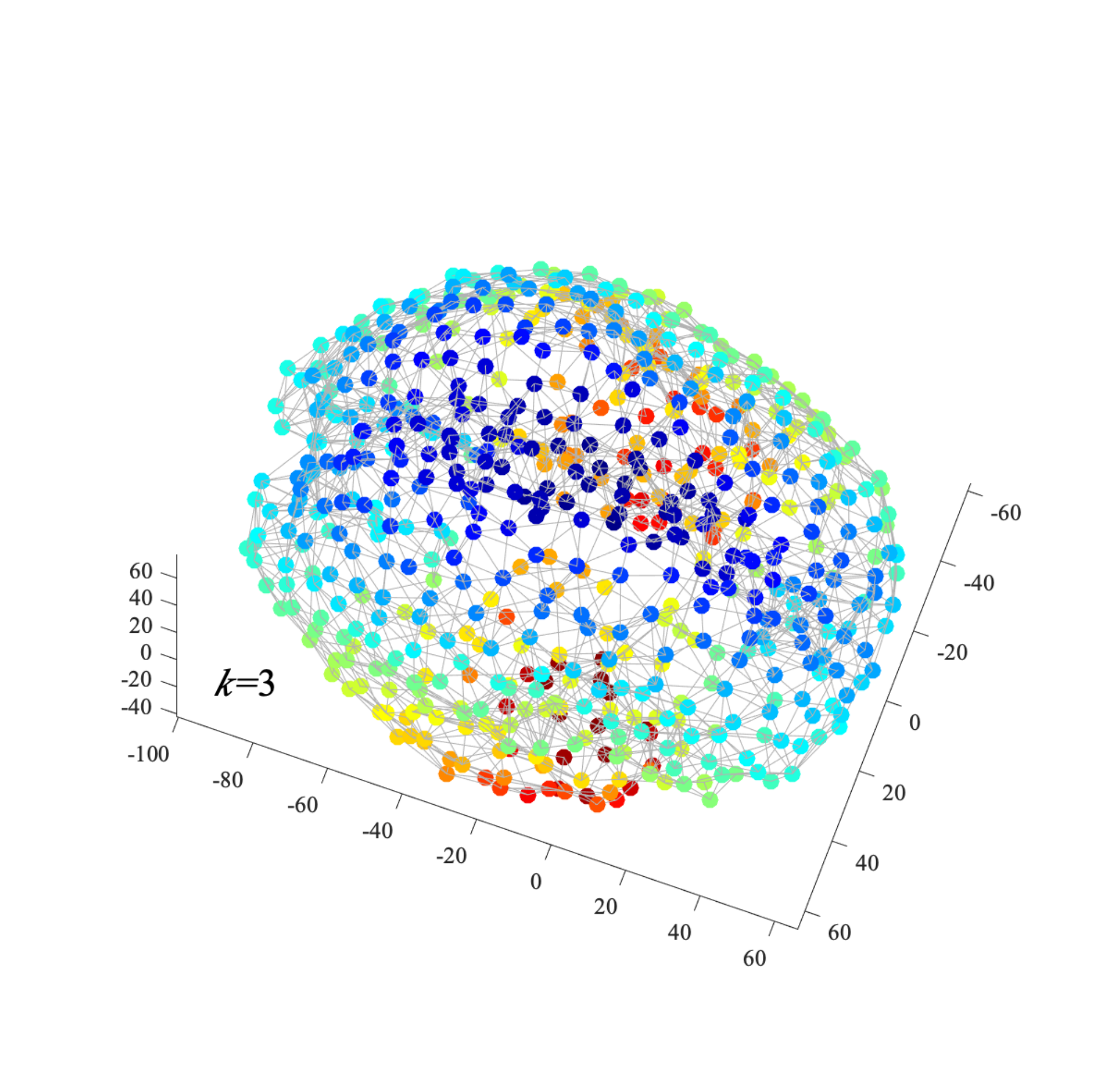}}

		\vfill
		
\fbox{\includegraphics[trim={2.2cm 2.2cm 1.8cm 4.5cm},clip,width=5.5cm]{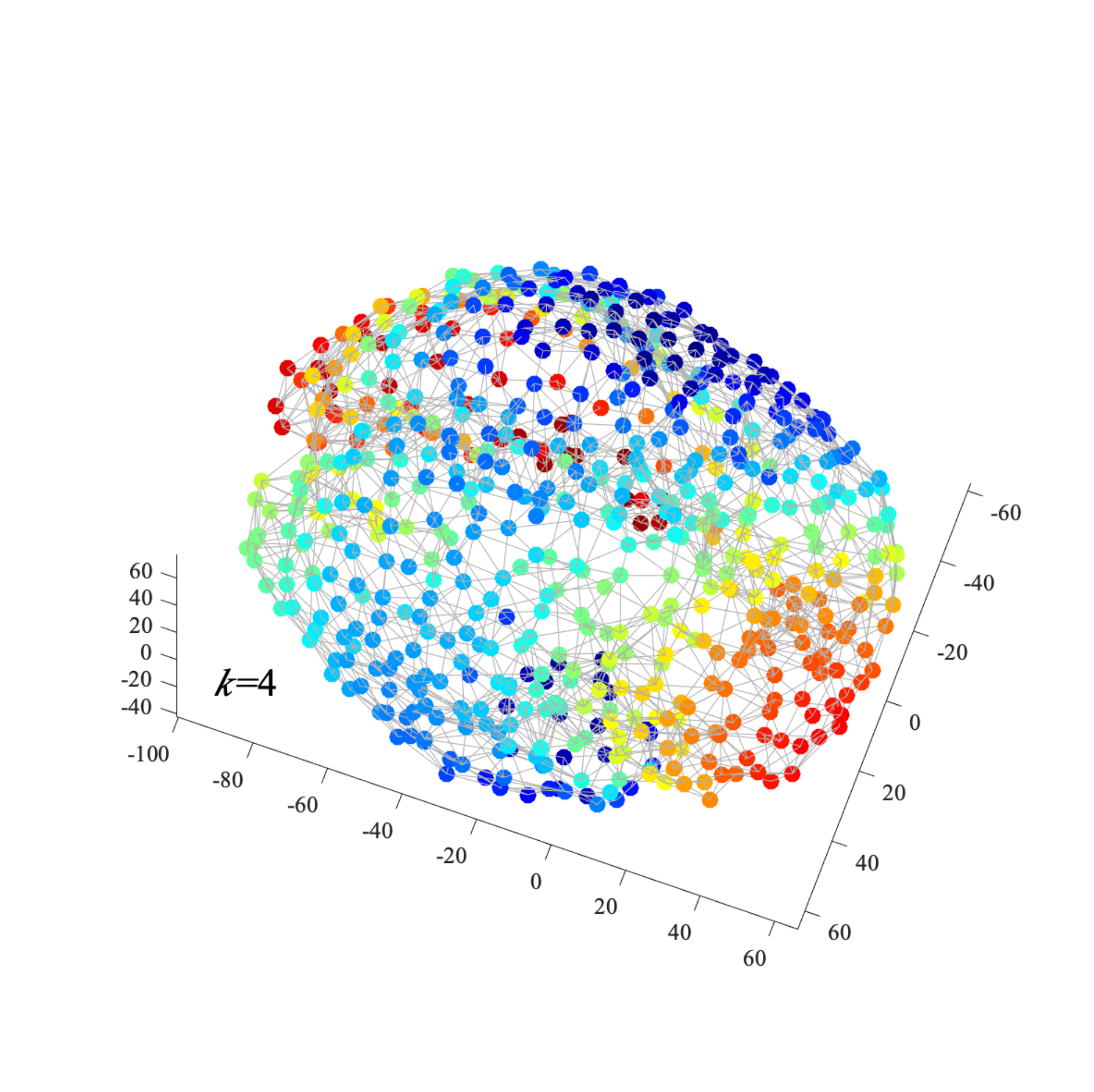}}
\fbox{\includegraphics[trim={2.2cm 2.2cm 1.8cm 4.5cm},clip,width=5.5cm]{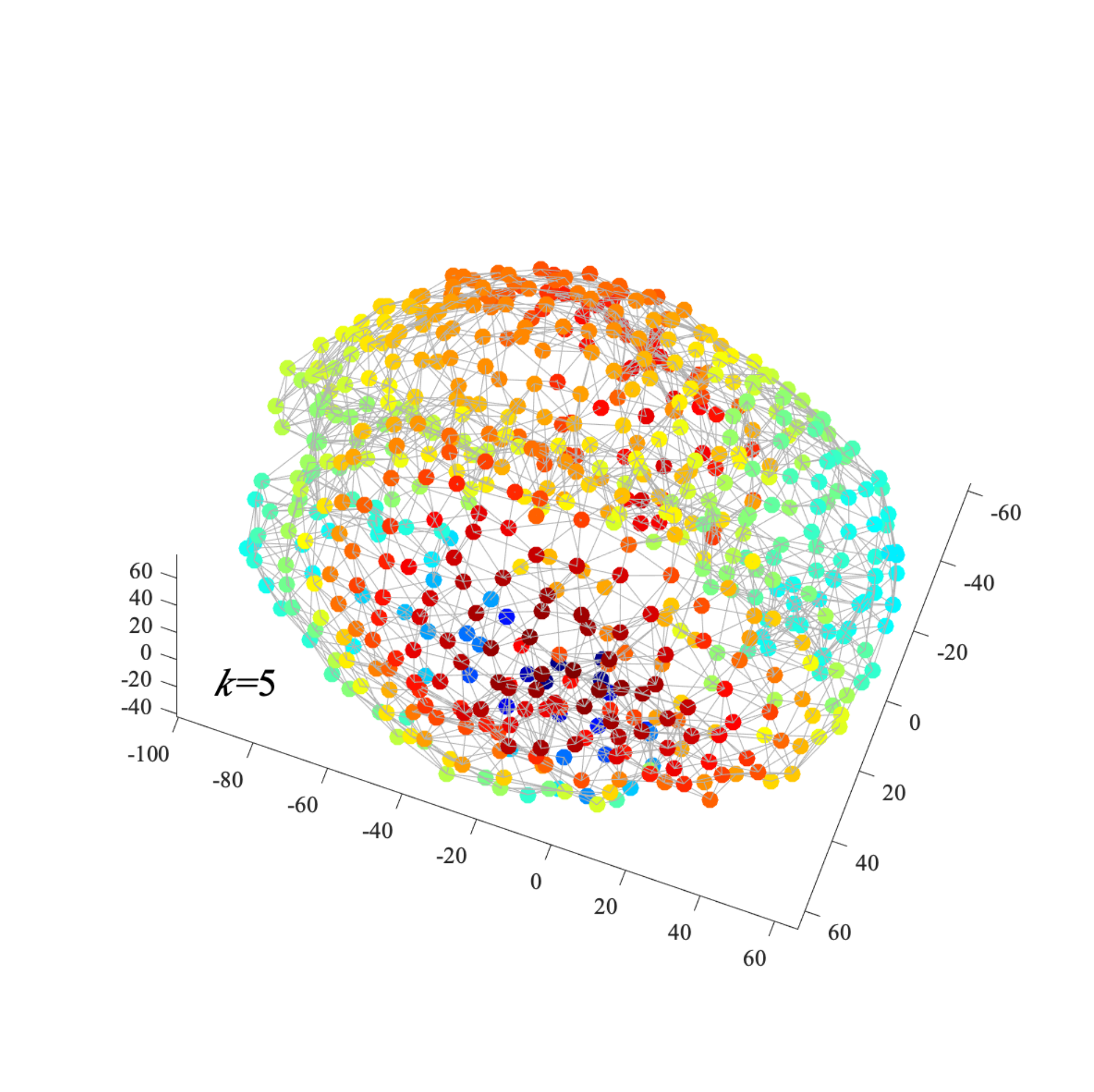}}
\fbox{\includegraphics[trim={2.2cm 2.2cm 1.8cm 4.5cm},clip,width=5.5cm]{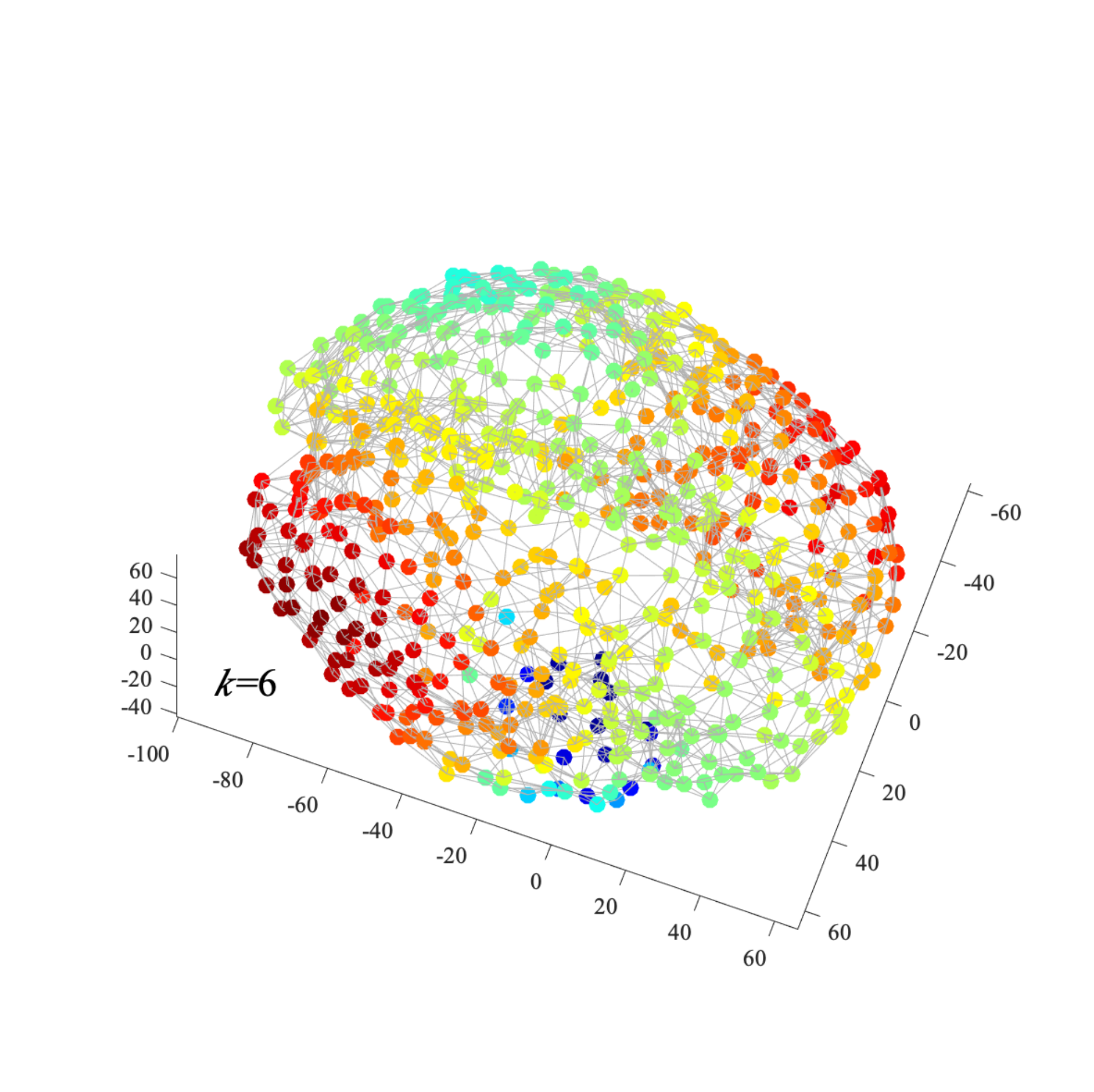}}
	
\caption{Generalized eigenvectors, $\mathbf{u}_k$, $k=1,2,3,4,5,6$, of the graph Laplacian of the Brain Atlas graph, shown using  vertex coloring in the original three-dimensional vertex space. Each panel visualizes a different $\mathbf{u}_k$, $k=1,2,3,4,5,6$.}
\label{ppp_Eig1_6}
\end{figure*}

	\begin{figure*}
	\centering		
\fbox{\includegraphics[width=5.5cm]{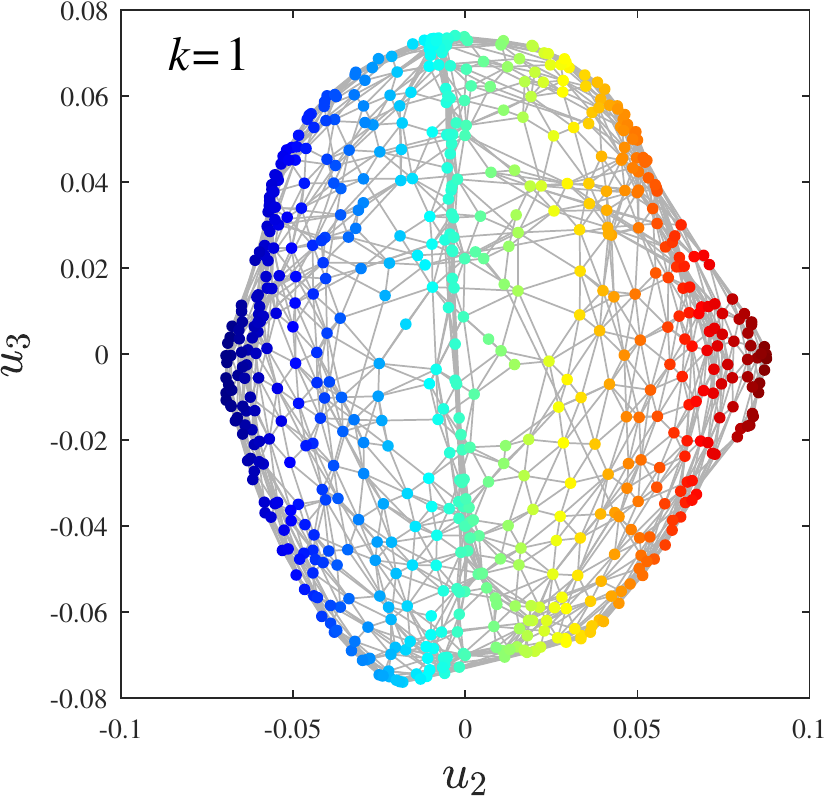}}
\fbox{\includegraphics[width=5.5cm]{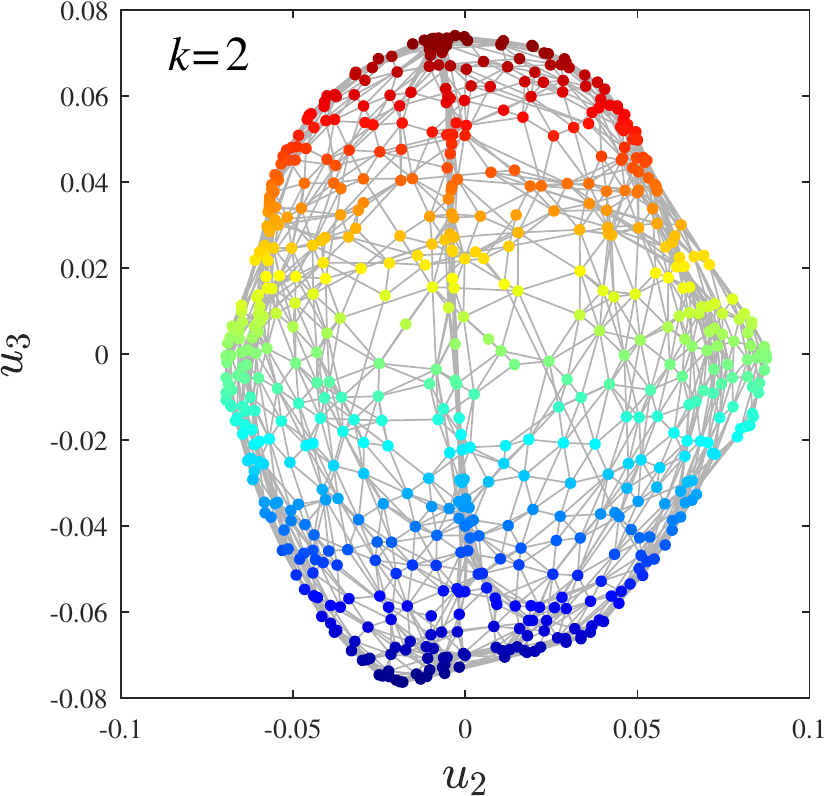}}
\fbox{\includegraphics[width=5.5cm]{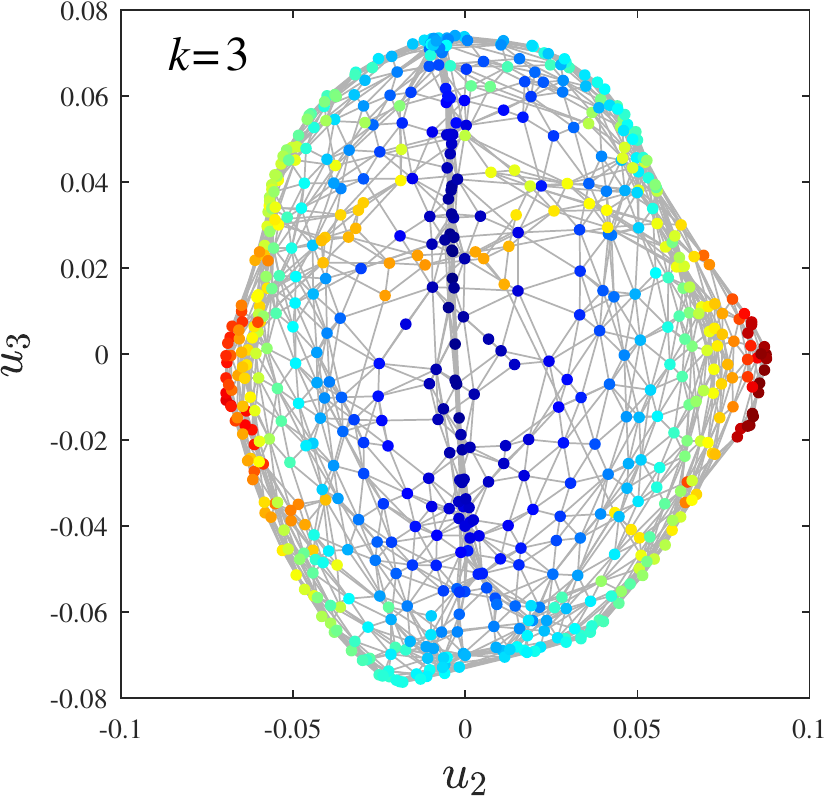}}

\vfill

\fbox{\includegraphics[width=5.5cm]{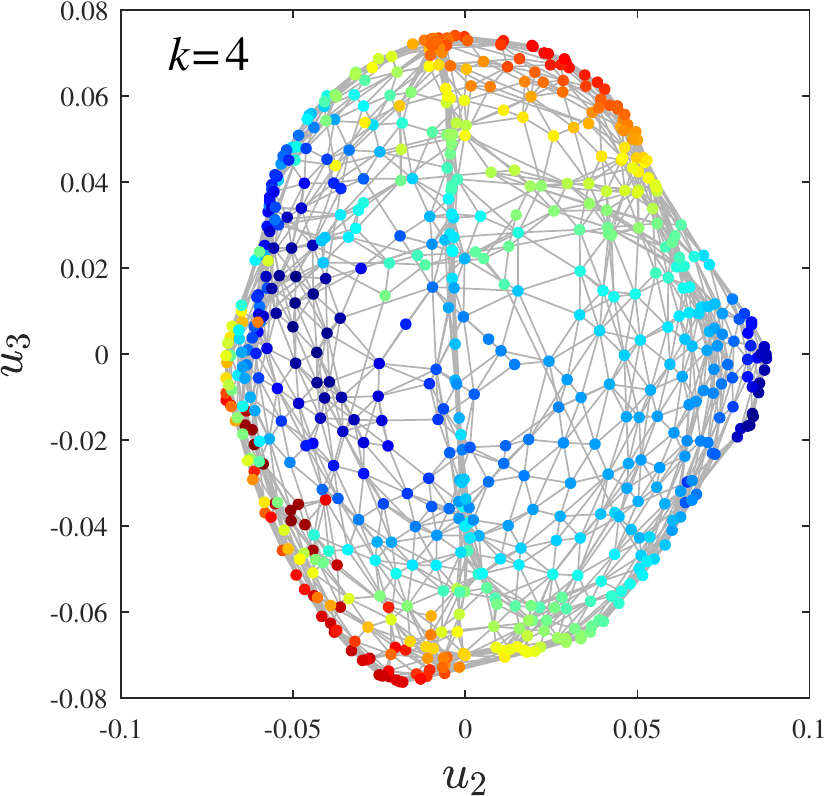}}
\fbox{\includegraphics[width=5.5cm]{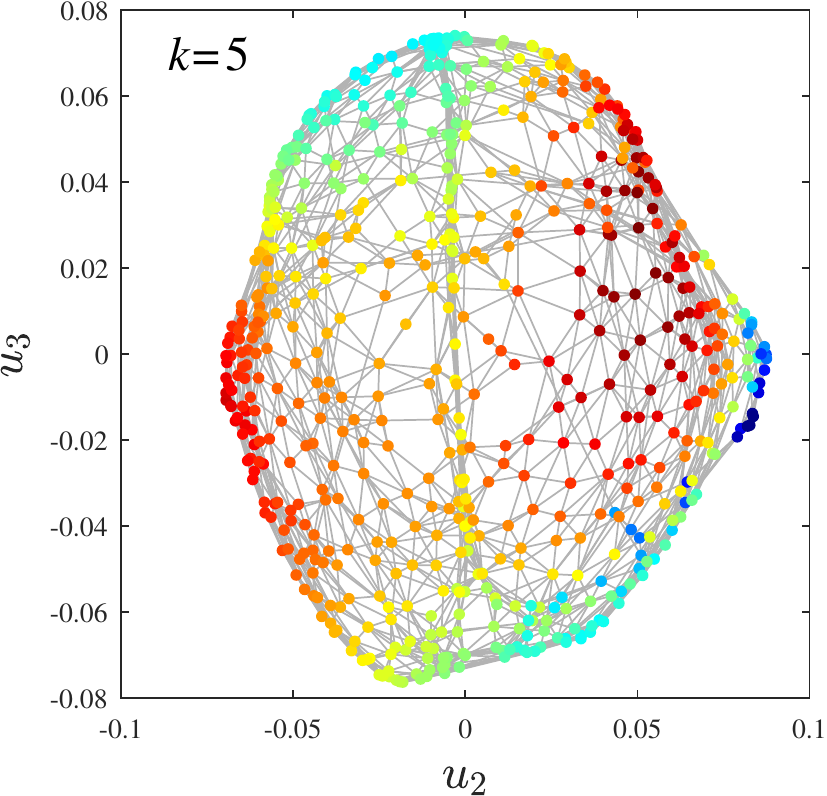}}
\fbox{\includegraphics[width=5.5cm]{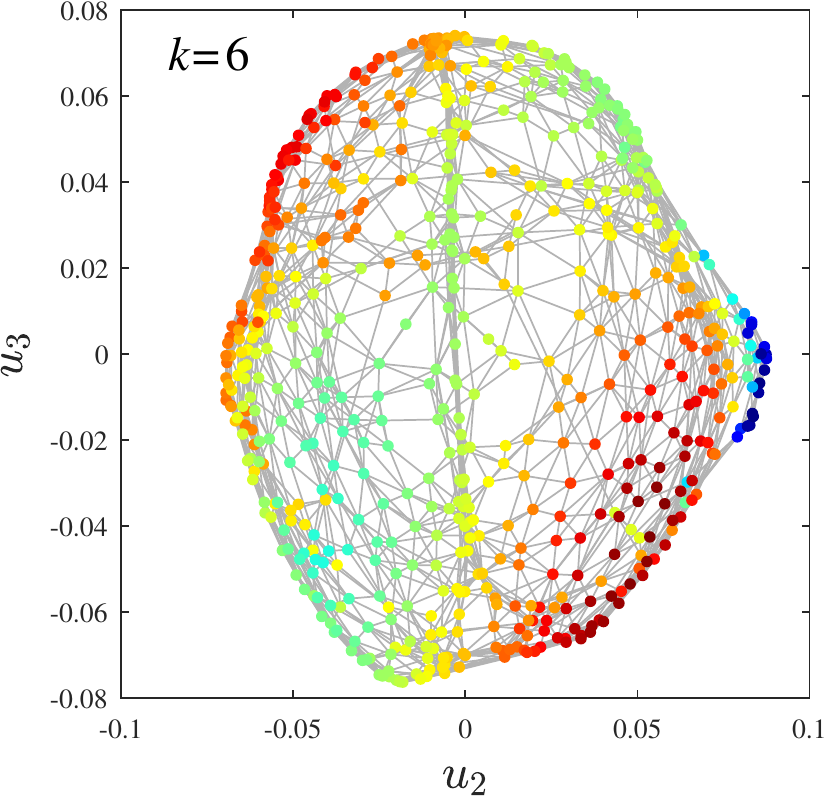}}
	
		\caption{Laplacian eigenmaps of the Brain Atlas graph in the reduced two-dimensional space defined by two smoothest generalized eigenvectors of the graph Laplacian, $\mathbf{u}_1$ and $\mathbf{u}_2$. The panels each visualize a different generalized eigenvector, $\mathbf{u}_k$, $k=1,2,3,4,5,6$. }
		\label{brain_LE_1_6}
	\end{figure*}

\end{Example}

\begin{Example} Vertices of a three-dimensional Swiss roll graph are shown in Fig. \ref{SwissRollGraph}(a). The vertex locations in this $L=3$ dimensional space are calculated as $x_n=\alpha_n\cos(\alpha_n)/(4 \pi)$
$y_n=\beta_n$, and $z_n=\alpha_n\sin(\alpha_n)/(4 \pi)$, $n=0,1,2,\dots,N-1$, with $\alpha_n$ randomly taking values between $\pi$ and $4\pi$, and $\beta_n$ from $-1.5$ to $1.5$. The edge weights are calculated using $W_{mn}=\exp(-d_{mn}^2/(2\kappa^2))$, where $d_{mn}$ is the square  Euclidean distance between the vertices $m$ and $n$, and $W_{mn}=0$ if $d_{mn}\ge 0.15$ with $\kappa=0.1$. The resulting three-dimensional Swiss roll graph is shown in Fig. \ref{SwissRollGraph}(b), while Fig. \ref{SwissRollGraph}(c) shows the same graph but with vertices colored (clustered) using the normalized graph Laplacian eigenvectors, $u_1(n)$ and $u_2(n)$, as a colormap. The same vectors are then used as the new coordinates in the reduced two-dimensional Laplacian eigenmap  vertex space for the Swiss roll graph, given in  Fig. \ref{SwissRollGraph}(d). 

\begin{figure*}
	\centering
	\includegraphics[]{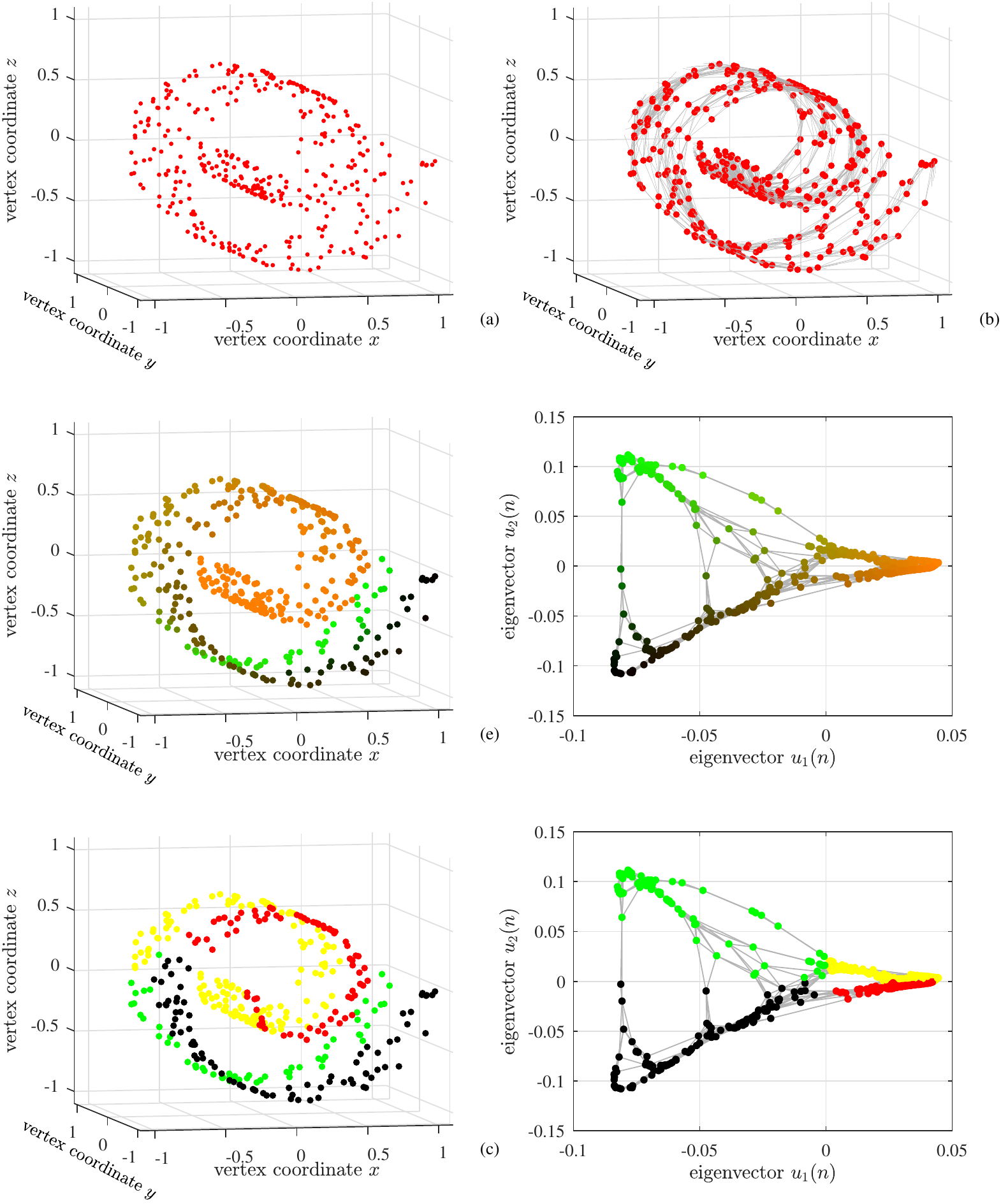}
	\caption{Laplacian eigenmaps based dimensionality reduction for the Swiss roll graph. (a) Vertex locations for the Swiss roll graph in the original $L=3$ dimensional space with $N=500$ points (vertices). (b) The the Swiss roll graph with edges whose weights are calculated based on the Euclidean distances between vertices. (c) The Swiss roll graph with vertices colored using the normalized graph Laplacian eigenvectors, $u_1(n)$ and $u_2(n)$, as a colormap. (d) The same vectors are used as the new coordinates (spectral vectors) in a reduced two-dimensional Laplacian eigenmap vertex space ($M=2$). The vertices with high similarity (similar values of the smoothest eigenvectors) are located close to one another, thus visually indicating the expected similarity of data observed at these vertices. (e) Clustering of the  Swiss roll graph, in the original $L=3$ dimensional space, using two the smoothest eigenvectors, $u_1(n)$ and $u_2(n)$.  (f) Clustering of the  Swiss roll graph using the two smoothest eigenvectors, $u_1(n)$ and $u_2(n)$, presented in the $M=2$ Eigenmap space, where for every vertex its spatial position (quadrant of the coordinate system) indicates the cluster  where it belongs.}
	\label{SwissRollGraph}
\end{figure*}

\end{Example}

\subsection{Pseudo-inverse of Graph Laplacian-Based Mappings}
The graph Laplacian is a singular matrix (since $\lambda_0=0$) for which an inverse does not exist. To deal with this issue, the pseudo-inverse of the graph Laplacian, $\mathbf{L}^{+}$, is defined as a  matrix that satisfies the property
\begin{equation}
\mathbf{L}\mathbf{L}^{+}=
\begin{bmatrix*}[l]
0 & \mathbf{0}_{1\times(N-1)} \\ \mathbf{0}_{(N-1)\times 1} & \mathbf{I}_{(N-1)\times(N-1)} 
\end{bmatrix*}.
\end{equation}
The eigenvalues of the   graph Laplacian pseudo-inverse are the inverses of the original eigenvalues, $\{0,1/\lambda_1, \dots, 1/\lambda_{N-1} \}$, while it  shares the same eigenvectors with the original  graph Laplacian, $\mathbf{u}_0$, $\mathbf{u}_1$, $\dots$, $\mathbf{u}_{N-1}$. The eigenmaps for which the corresponding spectral  coordinates are scaled based on the eigenvalues of the pseudo-inverse of graph Laplacian can be interpreted within the Principal Component Analysis (PCA) framework in the following way.

The eigenmaps based on the pseudo-inverse of the Laplacian are the same as for the original graph Laplacian since they share the same eigenvectors. If the spectral vectors $\mathbf{q}_n=[u_1(n),u_2(n),\allowbreak \dots,u_M(n)]$ are scaled  with the square roots  of the eigenvalues of Laplacian pseudo-inverse we obtain 
$$\mathbf{q}_n=[\frac{u_1(n)}{\sqrt{\lambda_1}},\frac{u_2(n)}{\sqrt{\lambda_2}},\dots,\frac{u_M(n)}{\sqrt{\lambda_{M}}}]$$    
The elements of the spectral vector are now equal to the first $M$ elements (omitting $0 \cdot u_0(n)$) of the full-scale spectral vector
\begin{equation}\mathbf{q}_n=[u_1(n),u_2(n),\dots,u_{N-1}(n)]\mathbf{\bar\Lambda}^{-1/2},  \label{CTmapp}
\end{equation}
where $\mathbf{\bar\Lambda}$ is a diagonal matrix with elements $\lambda_1,\lambda_2,\dots, \lambda_{N-1}.$

\medskip

\subsubsection{Commute time mapping} 
The physical meaning of these new positions in the spectral space is related to the notion of \textit{commute time},  which is a property of a diffusion process on a graph \cite{qiu2007clustering,horaud2012short}. The commute time, $CT(m,n)$ between vertices $m$ and $n$ is defined as the expected time for the random walk to reach vertex $n$ starting from vertex $m$, and then to return. Its is proportional to the Euclidean distance of vertices, with the positions in the new space defined by $\mathbf{q}_n$ in (\ref{CTmapp}), that is
$$CT^2(m,n)=V_{\mathcal{V}}||\mathbf{q}_m-\mathbf{q}_n||^2_2=V_{\mathcal{V}}\sum_{i=1}^{N-1}(q_i(m)-q_i(n))^2,$$
where $V_{\mathcal{V}}$ is the volume of the whole graph, $V_{\mathcal{V}}=\sum_{n=0}^{N-1}D_{nn}$. 

To put this into perspective, in a graph representation of a \textit{resistive electric circuit/network}, for which the edge weights are equal to the conductances (inverse resistances, see Part 3), the commute time, $CT(m,n)$, is defined by the equivalent resistance between the electric circuit nodes (vertices) $m$ and $n$ \cite{chandra1996electrical}. 

The covariance matrix of the scaled spectral vectors is given by 
$$\mathbf{S}=\frac{1}{N}\sum_{n=0}^{N-1}\mathbf{q}^T_n\mathbf{q}_n=\frac{1}{N}\mathbf{\bar{\Lambda}}^{-1}.$$
In other words, the reduced dimensionality  space of $M$ eigenvectors, $\mathbf{u}_1$, $\mathbf{u}_2$, $\ldots$, $\mathbf{u}_M$, is the space of which the principal directions correspond to the maximum
variance of the graph embedding, since $1/\lambda_1>1/\lambda_2>,\cdots,>1/\lambda_M$, which in turn directly corresponds to \textit{ principal component analysis (PCA)}. 

\medskip

\begin{Remark}\label{QuanCT} \textbf{Two-dimensional case comparison.}
The two-dimensional spectral space of the standard graph Laplacian eigenvectors is defined by $\mathbf{u}_1$ and $\mathbf{u}_2$, while the spectral vector in this space is given by
$$\mathbf{q}_n=[u_1(n),u_2(n)].$$   
In the case of commute time mapping, the two-dimensional spectral domain of the vertices is defined as 
\begin{equation}
\mathbf{q}_n=[\frac{u_1(n)}{\sqrt{\lambda_1}},\frac{u_2(n)}{\sqrt{\lambda_2}}].\label{2DCtMap}
\end{equation}

\textit{The commute time mapping is related to the graph Laplacian mapping through an appropriate axis scaling. }

We can conclude that  when $\lambda_1 \approx \lambda_2$ these two mappings are almost the same, when normalized.

However, when $\lambda_1 \ll \lambda_2$, the relative eigenvalue gap between the one dimensional and two-dimensional spectral space is large, since $\delta_r=1-\lambda_1/\lambda_2$ is close to $1$. This means that the segmentation into two disjoint subgraphs will be close to the original graph, while at the same time this also indicates that the eigenvector $\mathbf{u}_2$ does not contribute to a new close segmentation (in the sense of Section \ref{SEC:closseg}), since its gap $\delta_r=1-\lambda_2/\lambda_3$ is not small. Therefore, the influence of $\mathbf{u}_2$ should be reduced, as compared to the standard graph Laplacian spectral vector where both  $\mathbf{u}_1$ and $\mathbf{u}_2$  are used with equal unity weight, 
$\mathbf{q}_n=[u_1(n),u_2(n)]$. This reduction of the influence of the irrelevant vector $\mathbf{u}_2$, when $\lambda_1 \ll \lambda_2$, is exactly what is achieved in the commute time mapping, since $\mathbf{q}_n=[\frac{u_1(n)}{\sqrt{\lambda_1}},\frac{u_2(n)}{\sqrt{\lambda_2}}] = \frac{1}{\sqrt{\lambda_1}} [u_1(n),u_2(n)\sqrt{\frac{\lambda_1}{\lambda_2}}] \sim [u_1(n),0]$.

For the graph from Example \ref{VariousEigM}, shown in Fig. \ref{LS_VF_coloring1a}(a), the commute time mapping will produce the same presentation as in Fig. \ref{LS_VF_coloring1a}(b), which is obtained with the eigenvectors of the graph Laplacian, when the vertical, $\mathbf{u}_2$, axis is scaled by $$\sqrt{\frac{\lambda_1}{\lambda_2}}=\sqrt{\frac{0.0286}{
0.0358}}=0.8932.$$
This eigenmap will be very close to the eigenmap in Fig. \ref{LS_VF_coloring1a}(b), produced based on the graph Laplacian eigenvectors and spectral vector $\mathbf{q}_n=[u_1(n),u_2(n)]$.
\end{Remark}

\subsubsection{Diffusion (Random Walk) Mapping}

Finally, we shall now relate the commute time mapping  to the diffusion mapping. 

\noindent\textit{Definition:} Diffusion on a graph deals with the problem of propagation along the edges of a graph, whereby at the initial step, $t = 0$, the random walk starts at a vertex $n$. At  the next  step $t=1$, the walker moves from its current vertex $n$ to one of its neighbors $l$, chosen uniformly at random from the neighbors of $n$.  The probability of going from vertex $n$ to vertex $l$ is  equal to the ratio of the weight $W_{nl}$ and the sum of all possible edge weights from the vertex $n$, that is
\begin{equation}P_{nl}=\frac{W_{nl}}{\sum_{l}W_{nl}}=\frac{1}{D_{nn}}W_{nl}.\label{RWP}\end{equation}
When considering all vertices together, such probabilities can be written in a matrix form, within the weight of a random walk matrix, defined as in (\ref{LapRW}), by
\begin{equation}
\mathbf{P}=\mathbf{D}^{-1}\mathbf{W}.\label{rWdefinit}
\end{equation}


\noindent\textbf{Diffusion distance. }\textit{The diffusion distance} between the vertices $m$ and $n$, denoted by $D_f(m,l)$, is equal to the distance between the vector ($N$-dimensional ordered set) of probabilities for a random walk to move from a vertex $m$ to all other vertices (as in (\ref{RWP})), given by
$$\mathbf{p}_m=[P_{m0},P_{m1},\dots,P_{m(N-1)}]$$ 
and the corresponding vector of probabilities for a random walk to move from a vertex $n$ to all other vertices, given by
$$\mathbf{p}_n=[P_{n0},P_{n1},\dots,P_{n(N-1)}],$$ 
that is
\begin{gather*}
D_f^2(m,n)=|| (\mathbf{p}_m-\mathbf{p}_n)\mathbf{D}^{-1/2}||^2_2V_{\mathcal{V}}\\
={\sum_{i=0}^{N-1}\Big(P_{mi}-P_{ni}\Big)^2\frac{1}{D_{ii}}}V_{\mathcal{V}}
\end{gather*}
where $V_{\mathcal{V}}=\sum_{n=0}^{N-1}D_{nn}$ is constant for a given graph, and is equal to the sum of degrees (volume) of all graph vertices in ${\mathcal{V}}$.

\begin{Example}\label{ExDEMap}
	For the graph from Fig. \ref{GSPb_ex2}, with its weight matrix, $\mathbf{W}$, and the degree matrix, $\mathbf{D}$, given respectively in (\ref{WeightMatr}) and (\ref{DFTegMatrix}), the random walk weight matrix in (\ref{rWdefinit}) is of the form
	\begin{equation}
	\small
	\setlength{\arraycolsep}{5pt}
	\mathbf{P}= \!\!
	\begin{array}{cr}
	& \\
	{
		\color{blue}
		\begin{matrix}
		\text{\footnotesize $\mathbf{p}_0$}\\
		\text{\footnotesize $\mathbf{p}_1$}\\
		\text{\footnotesize $\mathbf{p}_2$}\\
		\text{\footnotesize $\mathbf{p}_3$}\\
		\text{\footnotesize $\mathbf{p}_4$}\\
		\text{\footnotesize $\mathbf{p}_5$}\\
		\text{\footnotesize $\mathbf{p}_6$}\\
		\text{\footnotesize $\mathbf{p}_7$}\\
		\end{matrix}
	} &  \!\! \!\! \!\! 
	\left[ \!
	\begin{array}{rrrrrrrr}  
	0 &  0.19  &  0.61  &  0.20  & 0 & 0 & 0 & 0 \\
	\cellcolor[gray]{0.9} 0.28  & \cellcolor[gray]{0.9} 0 &  \cellcolor[gray]{0.9} 0.43  & \cellcolor[gray]{0.9} 0 &  \cellcolor[gray]{0.9} 0.28  & \cellcolor[gray]{0.9} 0 & \cellcolor[gray]{0.9} 0 & \cellcolor[gray]{0.9} 0 \\
	0.47  &  0.22  & 0 &  0.16  &  0.15  & 0 & 0 & 0 \\
	\cellcolor[gray]{0.9} 0.29  & \cellcolor[gray]{0.9} 0 &  \cellcolor[gray]{0.9} 0.32  & \cellcolor[gray]{0.9} 0 & \cellcolor[gray]{0.9} 0 & \cellcolor[gray]{0.9} 0 &  \cellcolor[gray]{0.9} 0.39  & \cellcolor[gray]{0.9} 0 \\
	0  &  0.21  &  0.21  & 0 & 0 &  0.46  & 0 &  0.12  \\
	0  & 0 & 0 & 0 &  0.77  & 0 & 0 &  0.23  \\
	0  & 0 & 0 &  0.50  & 0 & 0 & 0 &  0.50  \\
	0  & 0 & 0 & 0 &  0.23  &  0.25  &  0.52  & 0
	\end{array} \!  \right]
	\\
	& 
	{
		\color{blue}
		\begin{matrix*}
		\text{\footnotesize 0 \hspace{3.3mm}} &
		\text{\footnotesize 1 \hspace{3.3mm}}  &
		\text{\footnotesize 2 \hspace{3.3mm}} &
		\text{\footnotesize 3 \hspace{3.3mm}} &
		\text{\footnotesize 4 \hspace{3.3mm}} &
		\text{\footnotesize 5 \hspace{3.3mm}} &
		\text{\footnotesize 6 \hspace{3.3mm}} &
		\text{\footnotesize 7 \hspace{2.5mm}}
		\end{matrix*}
	}
	\end{array}\label{RWWeightMatr}
	\end{equation}  
	with $V_{\mathcal{V}}=7.46$.

	Therefore, the diffusion distance between, for example, the vertices $m=1$ and $n=3$, for the $t=1$ step, is 
	$$D_f(1,3)=||(\mathbf{p}_1-\mathbf{p}_3)\mathbf{D}^{-1/2}||_2\sqrt{V_{\mathcal{V}}}=1.54,$$
	while the diffusion distance between the vertices $m=6$ and $n=3$ is $D_f(6,3)= 2.85$. From this simple example, we can see that the diffusion distance is larger for vertices $m=6$ and $n=3$ than for the neighboring vertices $m=1$ and $n=3$. This result is in a perfect accordance with the clustering scheme (expected similarity) in Fig. \ref{coloring1}(b), where the vertices $m=1$ and $n=3$ are grouped into the same cluster, while the vertices $m=6$ and $n=3$ belong to different clusters.
\end{Example}

The probability vectors $\mathbf{p}_n$ are called the \textit{diffusion clouds} (in this case for step $t=1$), since they resemble a cloud around a vertex $n$. The diffusion distance can then be considered as a distance between the diffusion clouds (sets of data) around a vertex $m$ and a vertex $n$. If the vertices are well connected (approaching a complete graph structure) then this distance is small, while for vertices with long paths between them, this distance is large.  

The diffusion analysis can be easily generalized to any value of the diffusion step, $t$,  whereby after  $t$  steps, the  matrix of probabilities in (\ref{rWdefinit}) becomes  $$\mathbf{P}^t=(\mathbf{D}^{-1}\mathbf{W})^t.$$ 
The elements of this matrix, denoted by ${P}_{mn}^{(t)}$, are equal to the probabilities that a random walker moves from a vertex $m$ to a vertex $n$, in $t$ steps. The $t$-step diffusion distance between the vertices $m$ and $n$, is accordingly defined as
$$D^{(t)}_f(m,n)=||(\mathbf{p}^{(t)}_m-\mathbf{p}^{(t)}_n)\mathbf{D}^{-1/2}||_2\sqrt{V_{\mathcal{V}}},$$
where $$\mathbf{p}^{(t)}_m=[P^{(t)}_{m0},P^{(t)}_{m1},\dots,P^{(t)}_{m(N-1)}]$$ 
and
$$\mathbf{p}^{(t)}_n=[P^{(t)}_{n0},P^{(t)}_{n1},\dots,P^{(t)}_{n(N-1)}].$$ 

It can be shown that \textit{the diffusion distance is equal to the Euclidean distance between the considered vertices when they are presented in a new space of their generalized Laplacian eigenvectors, which are then scaled by their corresponding eigenvalues;} \textbf{this new space is referred to as the diffusion maps} (\textit{cf.} eigenmaps).

The eigenanalysis relation for the random walk weight matrix for the state $t=1$ now becomes 
$$(\mathbf{D}^{-1}\mathbf{W})\, \mathbf{u}_k=\lambda^{(P)}_k \, \mathbf{u}_k.$$
Since the weight matrix can be written as  $\mathbf{W}=\mathbf{D}-\mathbf{L}$, this yields
$\mathbf{D}^{-1}(\mathbf{D-L})\mathbf{u}_k=\lambda^{(P)}_k \mathbf{u}_k$, or $$(\mathbf{I}-\mathbf{D}^{-1}\mathbf{L})\mathbf{u}_k=\lambda^{(P)}_k \mathbf{u}_k,$$ 
to finally produce the generalized graph Laplacian equation,  $$\mathbf{L}\mathbf{u}_k=\lambda_k \mathbf{D}\mathbf{u}_k,$$ with $\lambda_k=(1-\lambda^{(P)}_k)$.
\textit{This relation indicates that a one-step diffusion mapping is directly obtained from the corresponding generalized graph Laplacian mapping. }

After  $t$  steps, the  random walk matrix (of probabilities) becomes  $$\mathbf{P}^t=(\mathbf{D}^{-1}\mathbf{W})^t,$$ 
for which the eigenvalues are $\lambda^{(P)t}_k=(1-\lambda_k)^t$, while the (right) eigenvectors remain the same as for the graph Laplacian, see  (\ref{matxpow}). 

The spectral space for vertices, for a $t$-step diffusion process (\textit{diffusion mapping}),  is then defined based on  the spectral vector
$$\mathbf{q}_n=[u_1(n),u_2(n),\dots,u_{N-1}(n)](\mathbf{I}-\mathbf{\bar{\Lambda}})^t,$$
and is equal to the generalized Laplacian spectral space mapping, whereby the axis vectors $\mathbf{q}_n=[u_1(n),u_2(n),\dots,\allowbreak u_{N-1}(n)]$ are multiplied by the corresponding eigenvalues, $(1-\lambda_k)^t$.

It can be shown that the diffusion distance between vertices in the new diffusion map  space is equal to their Euclidean distance \cite{coifman2006diffusion}, that is 
\begin{equation}D^{(t)}_f(m,n)=\sqrt{V_{\mathcal{V}}}||\mathbf{q}_m-\mathbf{q}_n||_2.\label{DiffDistEigV}
\end{equation}
\begin{Example}For the graph from Fig. \ref{GSPb_ex2}, whose weight matrix, $\mathbf{W}$, and the degree matrix, $\mathbf{D}$, are defined in (\ref{WeightMatr}) and (\ref{DFTegMatrix}), the diffusion distance between the vertices  $m=1$ and $n=3$ can be calculated using (\ref{DiffDistEigV}) as
	$$D_f^{(1)}(1,3)=\sqrt{V_{\mathcal{V}}}||(\mathbf{q}_1-\mathbf{q}_3)||_2=1.54,$$
	where the spectral vectors, $\mathbf{q}_1=[u_1(1)(1-\lambda_1)^1,\dots,u_N(1)(1-\lambda_N)^1]$ and $\mathbf{q}_3=[u_1(3)(1-\lambda_1)^1,\dots,u_N(3)(1-\lambda_N)^1]$  are obtained using the generalized graph Laplacian eigenvectors, $\mathbf{u}_k$, and the corresponding eigenvalues, $\lambda_k$, from $\mathbf{L}\mathbf{u}_k=\lambda_k \mathbf{D}\mathbf{u}_k$. This is the same diffusion distance value, $D_f(1,3)$, as in Example \ref{ExDEMap}. 
\end{Example}

\noindent\textbf{Dimensionality reduced diffusion maps.} Dimensionality of the vertex representation space can be reduced in diffusion maps by keeping only the eigenvectors that correspond to the $M$ most significant eigenvalues, $(1-\lambda_k)^t$, $k=1,2,\dots,M$, in the same way as for the Laplacian eigenmaps, 

For example, the two-dimensional spectral domain of the vertices in the diffusion mapping is defined as 
$$\mathbf{q}_n=[u_1(n)(1-\lambda_1)^t,u_2(n)(1-\lambda_2)^t],$$
while the \textit{analysis and intuition for the diffusion mapping is similar to that for the commute time mapping}, presented in Remark \ref{QuanCT}, diffusion maps have an additional degree of freedom, the step $t$.

\begin{Example} For the graph in Fig \ref{image-grpaph}, which corresponds to  a set of real-world images,  the commute time two-dimensional spectral vectors in (\ref{2DCtMap}), normalized by the first eigenvector value through a multiplication of its coordinates by $\sqrt{\lambda_1}$, assume the form 
	$$\mathbf{q}_n=\Big[u_1(n),\frac{\sqrt{\lambda_1}}{\sqrt{\lambda_2}}u_2(n)\Big]=[u_1(n),0.62u_2(n)].$$
	The corresponding vertex colors indicate diffusion-based clustering, as shown in Fig. \ref{imageGraphColoring_Commute_Diffusion}(a). 
	Fig. \ref{imageGraphColoring_Commute_Diffusion}(b) shows the vertices of this graph, colored with the two-dimensional diffusion map spectral vectors, normalized by  $(1-\lambda_1)$, to yield
	$$\mathbf{q}_n=\Big[u_1(n),\frac{{1-\lambda_2}}{{1-\lambda_1}}u_2(n)\Big]=[u_1(n),0.09u_2(n)].$$

	\begin{figure}
		\centering
		(a)\includegraphics[scale=0.9]{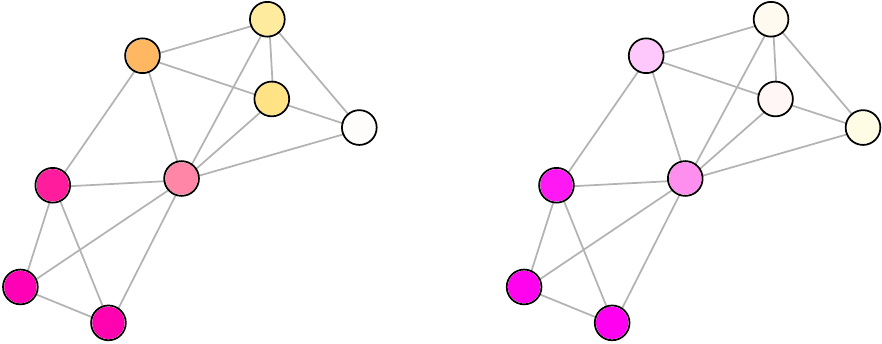}(b)
		\caption{Graph  structure for the images from Fig. \ref{image-grpaph}, with vertex color embedding which corresponds to the two-dimensional normalized spectral vectors in (a) the commute time representation, $\mathbf{q}_n=[u_1(n),0.62u_2(n)]$, and (b) the spectral eigenvectors of the diffusion process, $\mathbf{q}_n=[u_1(n),0.09u_2(n)]$,  with $t=1$. For the commute time presentation in (a), the graph Laplacian eigenvectors, $\mathbf{u}_1$ and $\mathbf{u}_2$,  are used, while for the diffusion process presentation in (b) the generalized Laplacian eigenvectors, $\mathbf{u}_1$ and $\mathbf{u}_2$, are used.} 
		\label{imageGraphColoring_Commute_Diffusion}
	\end{figure}
	
\end{Example}

Finally, the sum over all steps, $t=0,1,2,\dots$, of the diffusion space yields  
$$\mathbf{q}_n=[u_1(n),u_2(n),\dots,u_{N-1}(n)]\mathbf{\bar{\Lambda}}^{-1},$$
since the sum of a geometric progression is  equal to $$\sum_{t=0}^{\infty}(\mathbf{I}-\mathbf{\bar{\Lambda}})^t=\mathbf{\bar{\Lambda}}^{-1}.$$
This mapping also corresponds to the cumulative diffusion distance, given by 
$$D_c(n,l)=\sum_{t=0}^{\infty}D^{(t)}_f(n,l).$$

\textit{The diffusion eigenmaps can be therefore obtained by appropriate axis scaling of the standard eigenmaps, produced by the generalized eigenvectors of the graph Laplacian. }

\begin{Remark}
	The commute time and the diffusion process mappings are related in the same way as the mappings based on the graph Laplacian eigenvectors and the generalized eigenvectors of the graph Laplacian.   
\end{Remark}

\subsection{Summary of Embedding Mappings}

\begin{table*}[htb]

	\centering
	\renewcommand{\arraystretch}{1.25}
	\begin{tabular}{|l|l|l|}
		\hline\hline
		\textbf{Mapping} & \textbf{Eigen-analysis relation} & \textbf{Reduced dimensionality spectral vector} \\[5pt]
		\hline\hline 
		Graph Laplacian mapping & $\mathbf{L}\mathbf{u}_k=\lambda_k\mathbf{u}_k$ &  $\mathbf{q}_n=[u_1(n),u(2),\dots,u_M(n)]$ \\[5pt]
		\hline\hline
		Generalized eigenvectors  &   &   \\[-4pt] of Laplacian mapping  &  $\mathbf{L}\mathbf{u}_k=\lambda_k\mathbf{D}\mathbf{u}_k$ &  $\mathbf{q}_n=[u_1(n),u(2),\dots,u_M(n)]$ \\[5pt]
		\hline\hline
		Normalized Laplacian mapping & $\big(\mathbf{D}^{-1/2}\mathbf{L}\mathbf{D}^{-1/2}\big)\mathbf{u}_k=\lambda_k\mathbf{u}_k$ &  $\mathbf{q}_n=[u_1(n),u(2),\dots,u_M(n)]$ \\[5pt]
		\hline\hline
		Commute time mapping &  $\mathbf{L}\mathbf{u}_k=\lambda_k\mathbf{u}_k$ &  $\mathbf{q}_n=[\frac{u_1(n)}{\sqrt{\lambda_1}},\frac{u_2(n)}{\sqrt{\lambda_2}},\dots,\frac{u_M(n)}{\sqrt{\lambda_M}}]$ \\[5pt]
		\hline\hline
		Diffusion (random walk) mapping &  $\mathbf{L}\mathbf{u}_k=\lambda_k\mathbf{D}\mathbf{u}_k$ &  $\mathbf{q}_n=[u_1(n)(1-\lambda_1)^t,\dots,u_M(n)(1-\lambda_M)^t]$ \\[5pt]
		\hline\hline
		Cumulative diffusion mapping &  $\mathbf{L}\mathbf{u}_k=\lambda_k\mathbf{D}\mathbf{u}_k$ &  $\mathbf{q}_n=[\frac{u_1(n)}{{\lambda_1}},\frac{u_2(n)}{{\lambda_2}},\dots,\frac{u_M(n)}{{\lambda_M}}]$ \\[5pt]
		\hline\hline
	\end{tabular}
	\renewcommand{\arraystretch}{1}
	\caption{Summary for graph embedding mappings. The \textit{Graph Laplacian mapping}, the \textit{Generalized eigenvectors of the Laplacian mapping}, the \textit{Normalized Laplacian mapping}, the \textit{Commute time mapping}, the \textit{Diffusion mapping}, and the \textit{Cummulative diffusion mapping}.  }
	\label{tab:1}

\end{table*}

A summary of the embedding mappings considered is given in Table \ref{tab:1}. Various normalization schemes may be used to obtain the axis vectors, $\mathbf{y}_n$, from the spectral vectors, $\mathbf{q}_n$, for the presentation (see Algorithm \ref{Norm0Alg}). 
  
These examples of dimensionality reduction reveal close connections with spectral clustering algorithms developed in standard machine learning and computer vision; in this sense, the notions of	\textit{ dimensionality reduction and clustering can be considered as two sides of the same coin}   \cite{belkin2003laplacian}. In addition to the reduction of dimensionality for visualization purposes, the resulting vertex space of lower dimensionality may be used to mitigate the complexity and accuracy issues experienced with classification algorithms, or in other words to bypass the course of dimensionality.

\section{Graph Sampling Strategies}

In the case of extremely large graphs, subsampling and down-scaling of graphs is a prerequisite for their analysis \cite{leskovec2006sampling}.  For a given large (in general directed) graph, $\mathcal{G}$, with $N$ vertices,
the resampling aims to find a much simpler graph which retains most of the properties of the original graph, and is both less complex and more physically and computationally meaningful. The  similarity between the original large graph $\mathcal{G}$, and the down-scaled graph, $\mathcal{S}$, with $M$ vertices, where $M \ll N$, is defined with respect to the set of parameters of interest, like for example, the connectivity or distribution on a graph. Such criteria may also be related to the spectral behavior of graphs. 

Several methods exist for graph down-scaling, some of which are listed below. 

\begin{itemize}
	\item 
	The simplest method for graph down-sampling is the \textit{random vertex or random node (RN) selection method}, whereby a random subset of vertices is used for the analysis and representation of large graphs and data observed on such large graphs.  Even though the vertices are selected with equal probabilities, this method produces good results in practical applications.     
	\item
	Different from the RN method, where the vertices are selected with a uniform probability, \textit{the random degree vertex/node (RDN) selection method} is based on the probability of vertex selection that is proportional to the vertex degree. In other words, vertices with more connections, thus having larger $D_n=\sum_m W_{nm}$, are selected with higher probability. This makes the RDN approach  biased with respect to highly connected vertices.   
	\item
	\textit{The PageRank method} is similar to the RDN, and is based on the vertex rank. The PageRank is defined by the importance of the vertices connected to the considered vertex $n$. Then, the probability that a vertex $n$ will be used in a down-scaled graph is proportional to the PageRank of this vertex. This method is also known as the \textit{random PageRank vertex (RPN) selection}, and is biased with respect to the highly connected vertices (with a high PageRank).
	\item
	A method based on a random selection of edges that will remain in the simplified graph is called \textit{the random edge (RE) method}. This method may lead to graphs that are not well connected, and which exhibit large diameters.
	\item
	The RE method may be combined with random vertex selection to yield \textit{a combined RNE method}, whereby a random vertex selection is followed by a random selection of one of the edges corresponding to the selected vertex. 
	\item
	In addition to these methods, more sophisticated methods based on  random vertex selection and \textit{random walk (RW) analysis} may be defined. For example, we can randomly select a small subset of vertices and form several random walks starting from each selected vertex. The  \textit{Random Walk (RW)}, \textit{Random Jump (RJ)} and \textit{Forest Fire graph} down-scaling strategies are all defined in this way.
\end{itemize}

\section{Conclusion}
Although within the graph data analytics paradigm,  graphs have been present in various forms for centuries, the advantages of a graph framework for data analytics on graphs, as opposed to the optimization of the graphs themselves, have been recognized rather recently. In order to provide a comprehensive and Data Analytics friendly introduction to graph signal processing, an overview of graphs from this specific practitioner-friendly  signal processing point of view is a prerequisite. 

In this part of our  article, we have introduced  graphs as irregular signal domains, together with their properties relevant for data analytics applications which rest upon the estimation of signals on graphs. This has been achieved in a systematic and example rich way and by highlighting links with classic matrix analysis and linear algebra.  Spectral analysis of graphs has been elaborated upon in detail, as this is a main underpinning methodology for  efficient data analysis, the ultimate goal in Data Science. Both the adjacency matrix and the Laplacian matrix have been used in this context, along with their spectral decompositions. Finally, we have highlighted important aspects of graph segmentation and Laplacian eigenmaps, and have emphasized their role as the foundation for advances in Data Analytics on graphs.  

Part 2 will address theory and methods of processing data on graphs, while Part 3 is devoted to unsupervised graph topology learning, from the observed data.   

 \section{Appendix: Power Method for Eigenanalysis}\label{AppPM}
 Computational complexity of the eigenvalue and eigenvector calculation for a symmetric matrix is of the order of $\mathcal{O}(N^3)$, which is computationally prohibitive for very large graphs, especially when only a few the smoothest eigenvectors are needed, like in spectral graph clustering. To mitigate this computational bottleneck, an efficient iterative approach, called the Power Method, may be employed.

 Consider the normalized weight matrix, $$\mathbf{W}_N=\mathbf{D}^{-1/2}\mathbf{W}\mathbf{D}^{-1/2},$$
  and assume that the eigenvalues of $\mathbf{W}_N$ are $|\lambda_0|>|\lambda_1|>\dots>|\lambda_{M-1}|$, with the corresponding eigenvectors, $\mathbf{u}_1, \mathbf{u}_2,\allowbreak  \dots, \mathbf{u}_{M-1}$. 
 Consider an arbitrary linear combination of the eigenvectors,  $\mathbf{u}_n$, through the coefficients $\alpha_n$,
 
 $$\mathbf{x}=\alpha_1 \mathbf{u}_1+ \alpha_2 \mathbf{u}_2+ \dots + \alpha_{M-1} \mathbf{u}_{M-1}.$$ 
 Further multiplication of the vector $\mathbf{x}$ by the normalized weight matrix, $\mathbf{W}_N$, results in 
 \begin{gather*}\mathbf{W}_N\mathbf{x}=\alpha_1 \mathbf{W}_N\mathbf{u}_1+ \alpha_2 \mathbf{W}_N\mathbf{u}_2+ \dots + \alpha_{M-1} \mathbf{W}_N\mathbf{u}_{M-1} \\
 =\alpha_1 \lambda_1 \mathbf{u}_1+ \alpha_2 \lambda_2  \mathbf{u}_2+ \dots + \alpha_{M-1} \lambda_{M-1}  \mathbf{u}_{M-1}.
 \end{gather*}
 A repetition of this multiplication $k$ times yields
 \begin{gather*}\mathbf{W}^k_N\mathbf{x}
 =\alpha_1 \lambda^k_1 \mathbf{u}_1+ \alpha_2 \lambda^k_2  \mathbf{u}_2+ \dots + \alpha_{M-1} \lambda^k_{M-1}  \mathbf{u}_{M-1} \\
 =\alpha_1 \lambda^k_1 \Big(\mathbf{u}_1+ \alpha_2 \frac{\lambda^k_2}{\lambda^k_1}  \mathbf{u}_2+ \dots + \alpha_{M-1} \frac{\lambda^k_{M-1}}{\lambda^k_1}  \mathbf{u}_{M-1} \Big) \\ \approxeq \alpha_1 \lambda^k_1 \mathbf{u}_1.
 \end{gather*}
  In other words, we have just calculated the first eigenvector of $\mathbf{W}_N$, given by   $$\mathbf{u}_1=\mathbf{W}^k_N\mathbf{x}/||\mathbf{W}^k_N\mathbf{x}||_2$$
  through only matrix products of $\mathbf{W}_N$ and $\mathbf{x}$ \cite{trevisan2013lecture,tammen2018complexity}.  Convergence of this procedure depends on $\frac{\lambda_2}{\lambda_1}$ and requires that $\alpha_1$ is not close to $0$. Note that $\mathbf{W}_N$ is a highly sparse matrix, which significantly reduces the calculation complexity. 
  
    After the eigenvector $\mathbf{u}_1$ is obtained, the corresponding eigenvalue can be calculated as its smoothing index, $\lambda_1=\mathbf{u}^T_1 \mathbf{W}_N \mathbf{u}_1$. 
  
  After $\mathbf{u}_1$ and $\lambda_1$ are calculated, we can remove their contribution from the normalized weight matrix,$\mathbf{W}_N$ through deflation, as $\mathbf{W}_N- \lambda_1 \mathbf{u}_1 \mathbf{u}^T_1$, and then continue to calculate the next largest eigenvalue and its eigenvector, $\lambda_2$ and $\mathbf{u}_2$. The procedure can be repeated iteratively  until the desired number of eigenvectors is found.
  
  The relation of the normalized weight matrix, $\mathbf{W}_N$, with the normalized graph Laplacian, $\mathbf{L}_N$, is given by
   $$\mathbf{L}_N=\mathbf{I}-\mathbf{W}_N,$$
   while the relation between the eigenvalues and eigenvectors of $\mathbf{L}$ and $\mathbf{W}_N$ follows from $\mathbf{W}_N=\mathbf{U}^T\mathbf{{\Lambda}}\mathbf{U}$, to yield
   $$\mathbf{L}_N=\mathbf{I}-\mathbf{U}^T\mathbf{{\Lambda}}\mathbf{U}=\mathbf{U}^T(\mathbf{I}-\mathbf{{\Lambda}})\mathbf{U}.$$
   The eigenvalues of $\mathbf{L}_N$ and $\mathbf{W}_N$ are therefore related as $\lambda^{(L)}_n=1-\lambda_n$, and share the same corresponding eigenvectors, $\mathbf{u}_n$, of the normalized graph Laplacian  and the normalized weight matrix. 
   This means that $\lambda_1=1$ corresponds to $\lambda^{(L)}_0=0$ and that the second largest eigenvalue of $\mathbf{W}_N$ produces the Fiedler vector of the normalized Laplacian. 
   
   \textit{Note that the second largest eigenvector of $\mathbf{W}_N$ is not necessarily $\lambda_2$ since the eigenvalues of $\mathbf{W}_N$ can be negative.}    
  
  \begin{Example}
  	The weight matrix $\mathbf{W}$ from (\ref{WeightMatr}) is normalized by the degree matrix from (\ref{DFTegMatrix}) to arrive at $\mathbf{W}_N=\mathbf{D}^{-1/2}\mathbf{W}\mathbf{D}^{-1/2}$. The power algorithm is then used to calculate  the four largest eigenvalues and the corresponding eigenvectors of $\mathbf{W}_N$ in 200 iterations, to give  $\lambda_n \in \{1.0000   -0.7241   -0.6795,    0.6679\}$. These are very close to the four exact largest eigenvalues of $\mathbf{W}_N$, $\lambda_n \in \{1.0000   -0.7241   -0.6796,    0.6677\}$. Note that the Fiedler vector of the normalized graph Laplacian is associated with $\lambda_4 = 0.6679$ as it corresponds to the second largest eigenvalue of $\mathbf{W}_N$, when the eigenvalue signs are accounted for. Even when calculated using the approximative power method, the Fiedler vector is close to its exact value, as shown in Fig. \ref{idicative_vector_eig}(d), with the maximum relative error of its elements being $0.016$.   
  \end{Example}
  
  Notice that it is possible to calculate the Fiedler vector of a graph Laplacian even without using the weight matrix. Consider a graph whose eigenvalues of the Laplacian are $\lambda_0=0>\lambda_1>\lambda_2>\cdots>\lambda_{N-1}$, where $\lambda_1$ corresponds to the largest value of the sequence $\lambda_0=0,1/\lambda_1,1/\lambda_2,,\dots,1/\lambda_{N-1}$. These are also the eigenvalues of the pseudo-inverse of the graph Laplacian, $\mathbf{L}^+=\mathrm{pinv}(\mathbf{L})$. Now, since the pseudo-inverse of the graph Laplacian, $\mathbf{L}^+$, and the graph Laplacian, $\mathbf{L}$, have the same eigenvectors, we may also apply the power method to the pseudo-inverse of the graph Laplacian, $\mathbf{L}^+$, and the eigenvector corresponding to the largest eigenvalue is the Fiedler vector.    
  
 	\begin{algorithm}[!tbh]
 	\caption{\!\!\textbf{.} \, Power Method for eigenanalysis.}
 	\label{POWERM}
 	\begin{algorithmic}[1]
 		\smallskip
 		\Input
 		\Statex
 		\begin{itemize}
 			\item Normalized weight matrix $\mathbf{W}_N$
 			\item Number of iterations, $It$
 			\item Number of the desired largest eigenvectors, $M$	
 		\end{itemize}
 		\Statex
 		\State for $m=1$ to $M$ do
 		\smallskip
 		\State $\mathbf{u}_m \in \{-1,1\}^{M}$,  drawn randomly (uniformly)
 		\smallskip
 		\State for $i=1$ to $It$
 		\smallskip
 		\State $\mathbf{u}_m \gets \mathbf{W}_N\mathbf{u}_m/||\mathbf{W}_N\mathbf{u}_m||_2$ 
 		\smallskip
 		\State $\lambda_m \gets \mathbf{u}^H_m \mathbf{W}_N \mathbf{u}_m$
 		\smallskip
 		\State end do 
 		\smallskip
 		\State $\mathbf{W}_N \gets \mathbf{W}_N-\lambda_m \mathbf{u}_m \mathbf{u}^H_m$
 		\smallskip
 		\State end do
 		\Output
 		\Statex
 		\begin{itemize}
 			\item Largest $M$ eigenvalues $|\lambda_0|>|\lambda_1|>\dots>|\lambda_{M-1}| $ and the corresponding eigenvectors $\mathbf{u}_1, \dots, \mathbf{u}_{M-1}$   
 			\item Fiedler vector of the normalized graph Laplacian is the eigenvector $\mathbf{u}_{n_1}$ of the second largest eigenvalue, $\lambda_{n_1}$,  $\lambda_0=1>\lambda_{n_1}>\dots>\lambda_{n_{M-1}}$.    
 		\end{itemize} 
 		\end{algorithmic}
 \end{algorithm}

 \section{Appendix: Algorithm for Graph Laplacian Eigenmaps}
The algorithm for the Laplacian eigenmap and spectral clustering based on the eigenvectors of the graph Laplacian, the generalized eigenvectors of the graph Laplacian, and the eigenvectors of the normalized Laplacian, is given in the pseudo-code form in Algorithm \ref{AlgAp2}.

		\begin{algorithm}[!tbh]
	\caption{\!\!\textbf{.} \, Graph Laplacian Based Eigenmaps.}
	\label{Norm0Alg}
	\begin{algorithmic}[1]
		\smallskip
		\Input
		\Statex
		\begin{itemize}
			\item Vertex $\mathcal{V}=\{0,1,\dots,N-1\}$ positions, rows of $\mathbf{X}$
			\item Weight matrix $\mathbf{W}$, with elements $W_{mn}$
			\item Laplacian eigenmap dimensionality, $M$
			\item Position, mapping, normalization, and coloring indicators $P, Map, S, C$	
		\end{itemize}
		\Statex
		\State $\mathbf{D} \gets \mathrm{diag}(D_{nn}=\sum_{m=0}^{N-1}W_{mn}, n=0,1,\dots,N-1)$
		\State $\mathbf{L} \gets \mathbf{D} - \mathbf{W}$
		\State $[\mathbf{U},\mathbf{{\Lambda}}] \gets \mathrm{eig}(\mathbf{L})$
		\State $u_k(n) \gets U(n,k), \text{ for } k=1,\dots,M$, $n=0,1,\dots,N\!-\!1$. 
		\State $\mathbf{M} \gets \max_n(U(n,1:L))$, $\mathbf{m} \gets \min_n(U(n,1:L))$
		\State $\mathbf{q}_n \gets [u_1(n),u_2(n),\dots,u_L(n)], \text{ for all } n$
		\State \textbf{If} Map=1, $\mathbf{q}_n \gets \mathbf{q}_n\mathbf{\bar\Lambda}^{-1/2}$, \textbf{end}
		\State \textbf{If} Map=2, $\mathbf{q}_n \gets \mathbf{q}_n(\mathbf{I-\bar\Lambda})^{t}$, \textbf{end}
		\State $\mathbf{y}_n \gets 
		\begin{cases}
		\mathbf{q}_n, & \text{ for } S =0, \\
		\mathbf{q}_n/||\mathbf{q}_n||_2, & \text{ for } S=1, \\
		\mathrm{sign}(\mathbf{q}_n), & \text{ for } S=2, \\
		\mathrm{sign}(\mathbf{q}_n-(\mathbf{M}+\mathbf{m})/2),& \text{ for } S=3, \\ 
		(\mathbf{q}_n-\mathbf{m})./(\mathbf{M}-\mathbf{m}), & \text{ for } S=4 
		\end{cases}$  
		\State $\mathbf{Y} \gets \mathbf{y}_n$, as the rows of $\mathbf{Y}$
		\State $\mathbf{Z} \gets 
		\begin{cases}
		\mathbf{X}, \text{ for } P=0, \\
		\mathbf{Y}, \text{ for } P=1 
		\end{cases}$ 
		\State $\mathbf{ColorMap} \gets 
		\begin{cases}
		\mathbf{Constant}, \text{ for } C=0, \\
		(\mathbf{Y}+1)/2, \text{ for } C=1 
		\end{cases}$,   					
		\State $\mathrm{GraphPlot}(\mathbf{W},\mathbf{Z},\mathbf{ColorMap})$
		\State Cluster the vertices according to $\mathbf{Y}$ and refine using the $k$-means algorithm (Remark \ref{clusteringRem}) or the normalized cut recalculation algorithm (Remark \ref{RemRecalc}).
		\Statex
		\Output
		\Statex
		\begin{itemize}
			\item New graph 
			\item Subsets of vertex clusters 
		\end{itemize}

\noindent_____________________________________________________________________________________

\footnotesize
	\noindent\textbf{Comments on the Algorithm:} For the \textit{normalized Laplacian}, Line 2 should be replaced by $\mathbf{L} \gets \mathbf{I} - \mathbf{D}^{-1/2}\mathbf{W}\mathbf{D}^{-1/2}$, while for the \textit{generalized eigenvectors} Line 3 should be replaced by $[\mathbf{U},\mathbf{{\Lambda}}] \gets \mathrm{eig}(\mathbf{L},\mathbf{D})$, see also Table \ref{tab:1}. The indicator values of vertex positions in the output graph are: $P=0$, for the original vertex space, and $P=1$, for the spectral vertex space. The indicator of mapping is: $Map=1$, for the \textit{commute time} mapping (matrix $\mathbf{\bar\Lambda}$ is obtained from $\mathbf{\Lambda}$, by omitting the trivial element $\lambda_0=0$), and $Map=2$, for the \textit{diffusion mapping} (in this case the generalized eigenvectors must be used in Line 3, $[\mathbf{U},\mathbf{{\Lambda}}] \gets \mathrm{eig}(\mathbf{L},\mathbf{D})$ and the diffusion step $t$ should be given as an additional input parameter), otherwise $Map=0$. The indicator of the eigenvectors normalization is:  $S=0$, for the case without normalization, $S=1$, for two-norm normalization, $S=2$, for the case of binary normalization, $S=3$, for binary normalization with the mean as a reference,  and $S=4$, for marginal normalization. The indicator of vertex coloring is: $C=0$, for the same color for all vertices is used, and  $C=1$, when the spectral vector defines the vertex colors.
	\end{algorithmic}
\label{AlgAp2}
\end{algorithm}

\section*{References}

\bibliographystyle{elsarticle-num}
\bibliography{graph-signal-processing}

%
%



\end{document}